\documentclass [12pt,oneside]{mnthesis_2}
\newcommand{\CONT}{\noindent}
\newcommand{\FIG}{Fig.\ }
\newcommand{\FIGS}{Figs.\ }

\newcommand{\TAB}{Table }
\newcommand{\TABS}{Tables }
\newcommand{\EQ}{Eq.\ }
\newcommand{\EQS}{Eqs.\ }

\newcommand{\BR}{{\cal{B}}}

\newcommand{\etal}{{\it et al. }}
	\usepackage{lscape}
	\usepackage{graphicx}
	\usepackage{epsfig}
        \usepackage{amsmath}

\begin{document}
	\phd
	\title{\bf Measurement of the Total Charm Cross Section by
	Electron-Positron Annihilation at Energies Between 3.97-4.26 GeV  }
	\author{Brian William Lang}
	\campus{University of Minnesota} 
	\program{Physics} 
	\director{Professor Yuichi Kubota, Ph.D. and Professor Ronald Poling, Ph.D.} 
        \words{64}    % number of words in the abstract
 	\copyrightpage % Do you want copyright protection?

	\abstract{
Using the CLEO-c detector, we have measured the charm hadronic cross
sections for $e^{+}e^{-}$ annihilations at a total of thirteen
center-of-mass energies between 3.97 and 4.26 GeV.  Observed cross
sections for the production of $D\bar{D}$, $D^{*}\bar{D}$,
$D^{*}\bar{D}^{*}$, $D_{s}\bar{D}_{s}$, $D_{s}^{*}\bar{D}_{s}$,
and $D_{s}^{*}\bar{D}_{s}^{*}$, in addition to the
total charm cross section are presented.  Observed cross sections were
radiatively corrected to obtain tree-level cross sections and R.
}

        \dedication{
\vspace*{2.0 in}
\begin{center}
To my wife, my family, and my friends for all of their love,
support and most importantly their patience.
\end{center}
}
	\acknowledgements{
Acknowledgement - [ak-nol-ij-muh nt] An expression of appreciation or gratitude.
I have been so lucky during my time at the University of Minnesota:
lucky to have received supportive advisors,
lucky to have been part of an innovative, collaborative group,
and damn lucky to have such a loving network of friends and family.
I would like to express my appreciation and gratitude on these first pages of my dissertation.
\begin{itemize}
\item{I would like to thank my advisors Yuichi Kubota and Ron Poling for their support over the years. Their
continual patience, insight, and suggestions have no doubt helped me get to this point in my career. Their enthusiasm, curiosity and passion for physics and of tackling problems is truly inspirational. It was certainly a pleasure to have worked with and learn from such fine physicists.}

\item {I would also like to thank the other University of Minnesota group members both past and present: Valery
Frolov, Kaiyan Gao, Datao Gong, Selina Li, Alexander Scott, and Chris Stepaniak.  Tim Klein, who was always willing to lend a hand, in addition to making the best
mojito in town.   Prof. Dan Cronin-Hennessy's willingness and ability to add insight to perplexing and complicated questions and problems.  Alex Smith, for alway having his door, or cubicle, open for questions and discussions.  In particular, I would like to send
thanks to Justin Hietala and Pete Zweber for their many discussions with me about my analysis and of physics in general.  Additional thanks to Pete for, among these discussions, he was willing to be my chauffeur, tour guide,
and social director while in Ithaca, New York.}

\item{I would like to thank the many people that have supported me during this long process.  I can only work
so much, right?!?  I would like to thank Jim and Kara Melichar, Tom O'Connor, Justin Holzman, Mark McCarthy, Brian
Gulden, Maggie Edmiston, Steve Udycz, Joe Skinner, Jake Kern, Nathan Moore, Sarah and Paul Way, Andy Cady, Erik Beall, Paul Barsic, Pete Vassy, Christine Crane, Adrienne Zweber, Norm
Lowrey, Curtis Jastremsky, Paras Naik, Mike Watkins, Laura Adams, and Lynde Klein.  Special kudos to my softball team, Quit Yer Pitchin', for a painful and slightly embarrassing, yet enjoyable few seasons.  These people, and countless others,
have made my graduate school days in Minnesota fulfilling and adventurous.  As a result, I have many stories and
memories to keep with me always.}

\item{I would also like to send thanks for the overwhelming love and support of my family.  My parents, John
and Kathy Lang, have always taught me the value of hard work and persistence. I would also like to thank my grandmothers, Betty Walchli and Colletta Lang, my brother, Mike, sister-in-law, Brandy, and nephew, Michael, and my Aunt and Uncle Tricia and Tom Sweeney, as well as my adoring mother-in-law, Cheryl Clever (C.L.).

I would also like to take a moment to thank those family members who had supported me throughout these many years,
and who are now celebrating with me in spirit.  Thank you to my grandfathers, William F. Walchli and Russell Lang
and to my favorite little brother (in-law), Adam Clever.}

\item{Finally, but most certainly not least, I would like to thank my wife, Sarah, for whom I am extremely thankful.  Her love, friendship, patience, encouragement, and sense of humor, have given my life immense purpose, joy, and meaning.  The finished work that is to follow would have no meaning or satisfaction without having her with me to share it. I love you Angle!}

\end{itemize}
}

	\beforepreface 
	\figurespage
	\tablespage
	\afterpreface            
\chapter{Introduction}

This dissertation is devoted to the study of charm-meson production in
$e^+e^-$ annihilations at thirteen center-of-mass energies between 3.97 GeV
and 4.26 GeV. Specifically, we have used the CLEO-c detector to measure
exclusive cross sections for several final states and the inclusive cross 
sections for the production of the charmed mesons $D^{0}$, $D^{+}$ and 
$D_{s}$.  To provide the background for this research I first review some 
of the foundations of elementary particle physics, including the Standard 
Model of quarks and leptons.

During the past century, our understanding of nature
and structure of matter has been dramatically transformed, beginning with the
discovery that atoms are divisible structures composed of
subatomic particles.  With this
discovery and others the modern picture began to form of
atoms consisting of a positively charged nucleus surrounded by
negatively charged electrons.

Elementary particle physics is the study of the fundamental constituents of matter and their interactions.  It began with the discoveries of the electron (Thomson), atomic nucleus (Rutherford) and the neutron (Chadwick), developing slowly at the same time that the theoretical foundations of quantum mechanics were being laid.  Meanwhile, investigation of
cosmic rays and the development of accelerator-based experiments in the 1940's, 1950's, and the 1960's led to the discovery of many
more particles, particles that do not exist in ordinary matter. With the seemingly endless additions of new particles, the sentiment grew that there must be another level of substructure to explain this particle ``zoo''.  This sentiment ultimately led to the invention, by Gell-Mann and Zweig \cite{GellMannZweig}, of a model that describes 
most observed particles as being composed of the more fundamental particle known as the quark.

\section{The Standard Model}
The modern framework which describes the
fundamental particles and interactions is called the Standard Model.  
The Standard Model incorporates the quarks and the leptons
(the electron and its relatives) into a successful framework that has
proven to be very rugged and reliable over the years.
The quarks and leptons are said to be fundamental because they are
structureless: they are point-like and indivisible.  Even
though they are structureless, however, they still possess intrinsic
properties such as spin, charge, color, etc.  All particles but one
can be arranged into three groups: quarks, leptons, and force
carriers or mediators.  The remaining particle, the Higgs boson, has
a special job in the model.  The Higgs is the last remaining particle
of the Standard Model yet to be discovered.  It plays the key role
in explaining the mass of the other particles, specifically
the large mass difference between the photon, the vector bosons and the quarks.

The leptons come in six types which are listed in \TAB \ref{tab:leptons}
along with their corresponding masses and charges.  The lightest
charged lepton is the familiar electron.  The muon ($\mu$) and tau
($\tau$) have the same general properties as the electron, but
they have larger masses.  These heavier versions are unstable and
therefore not found in ordinary matter. Each charged lepton is
accompanied by a weakly interacting electrically neutral neutrino: $\nu_{e}$,
$\nu_{\mu}$, and $\nu_{\tau}$. Together each lepton and its accompanying
neutrino form a family, sometimes referred to as a generation.  All
leptons have intrinsic angular momentum, or spin, of
$\frac{1}{2}\hbar$ and are therefore fermions.
\begin{table}[!htbp]
\begin{center}
\caption{The six types of leptons in the Standard Model. The masses are
taken from Ref. \cite{pdg}. The electric charges are given in units
of $|e|$, where $e$ is the charge of an electron ($|e|=1.602\times10^{-19}$
Coulombs).  The neutrino's masses are given at $95\%$ confidence level.}
\vspace{0.2cm}
\label{tab:leptons}
\begin{tabular}{c|c|c|c}
\hline
{Family}& {Name} &{Electric} & {Mass} \\
{}& {} &{Change $\frac{Q}{|e|}$} & {} \\
\hline
{I}&{$e$}&$-1$&511 keV \\
   &{$\nu_{e}$}&$0$&$<$3 eV \\ \hline
{II}&{$\mu$}&$-1$&106 MeV \\
   &{$\nu_{\mu}$}&$0$&$<$0.19 MeV \\ \hline
{III}&{$\tau$}&$-1$&1.78 GeV \\
   &{$\nu_{\tau}$}&$0$&$<$18.2 MeV \\ \hline
\end{tabular}
\end{center}
\end{table}

Similarly, the quarks also come in six types, which can also be split
into three generations. The first generation of quarks consists of the up ($u$) and down
($d$). Together with the electron the first generation of quarks make up all the ordinary matter around us. The second
generation (strange ($s$) and charm ($c$) quarks) and the third generation
(bottom ($b$) and top ($t$) quarks) make-up the rest of the known quarks. Since
they have intrinsic angular momentum of $\frac{1}{2}\hbar$, all quarks
are fermions.
Some properties of the quarks are shown in \TAB \ref{tab:quarks}. 
Quarks are never individually observed but rather are seen only in combinations called hadrons. There are two types of hadrons: mesons,
which are bound states
of a quark and an anti-quark,\footnote{For example, a $\pi^+$ meson is a bound
state of an up-quark and an anti-down-quark ($u\bar{d}$)} and
baryons, which are bound states of three quarks or anti-quarks.\footnote{For example, a proton
($p$) is a bound state of two up-quarks and a single down-quark
($uud$).}

\begin{table}[!htbp]
\begin{center}
\caption{The six types of quarks in the Standard Model. The masses are
taken from Ref. \cite{pdg}. The electric charges are given in units
of $|e|$, where $e$ is the charge of an electron ($|e|=1.602\times10^{-19}$
Coulombs).  The quark masses are approximate since the complicated
strong interactions that take place inside the hadrons make it
difficult to define and measure the individual quark masses. However,
because of its short lifetime, the top quark does not
have time to bind into hadrons, and therefore it has the smallest fractional
uncertainty on its mass.}
\vspace{0.2cm}
\label{tab:quarks}
\begin{tabular}{c|c|c|c}
\hline
{Family}& {Name} &{Electric} & {Mass} \\
{}& {} &{Change $\frac{Q}{|e|}$} & {} \\
\hline
{I}&{$u$}&$+\frac{2}{3}$&1-4 MeV \\
   &{$d$}&$ -\frac{1}{3}$&4-8  MeV \\ \hline
{II}&{$c$}&$ +\frac{2}{3}$&1.15-1.35 GeV \\
   &{$s$}&$ -\frac{1}{3}$&80-130 MeV \\ \hline
{III}&{$t$}&$+\frac{2}{3} $&174 GeV \\
   &{$b$}&$-\frac{1}{3} $&4.1-4.4 GeV \\ \hline
\end{tabular}
\end{center}
\end{table}

The interactions among these fundamental particles are the four forces: strong, weak, electromagnetic, and
Gravitational.  Of these, Gravity is by far the weakest,
plays no role in the interactions described in this dissertation, and
is not part of the Standard Model.  In the Standard Model
the three remaining forces are represented as an exchange of gauge bosons between interacting particles.  All gauge bosons have integer spin, and some of their properties are shown in \TAB \ref{tab:gauge}. Of the
fundamental fermions, only the quarks interact through the action of all three forces. The charged leptons experience only
the weak and electromagnetic forces, and the neutrinos interact only
via the weak force.   
\begin{table}[!htbp]
\begin{center}
\caption{The gauge bosons in the Standard Model. The masses are
taken from Ref. \cite{pdg}. The electric charges are giving in units
of $|e|$, where $e$ is the charge of an electron ($|e|=1.602\times10^{-19}$
Coulombs).}
\vspace{0.2cm}
\label{tab:gauge}
\begin{tabular}{c|c|c|c}
\hline
{Force}& {Name} &{Electric} & {Mass} \\
{}& {} &{Change $\frac{Q}{|e|}$} & {} \\
\hline
{strong}&{gluons~$g$}&$0$&0 \\ \hline
 
{weak}&{W-Bosons $W^{\pm}$}&$\pm1$&80.4 GeV \\
{}&{Z-Boson $Z^{0}$}&$0$&91.2 GeV \\ \hline
{electromagnetic}&{Photon $\gamma$}&$0$&0 \\ \hline
\end{tabular}
\end{center}
\end{table}

The electromagnetic force is responsible for the binding of
electrons to the nucleus inside the atom. The mediator of this force,
the photon, couples to electrically charged particles. The quantum
theory describing the electromagnetic interaction is called Quantum
Electrodynamics (QED) and has been extensively tested. QED is the most
accurate physical theory constructed.

The weak interaction was first discovered in nuclear decays. All
fundamental fermions, including the neutrinos, participate in the weak
interaction.  The mediators of the weak force are the vector
bosons, the $W^{\pm}$ and the $Z$, which are massive, explaining its very short range, since the range is proportional to $1/M$.
Interactions involving the charged $W$-bosons involve changes the
flavor of the particles involved.  These changes can lead to spontaneous
decays of the type $\mu^-\rightarrow{e^-}\bar\nu_e\nu_{\mu}$.  Since
the decay involves the charged $W$-boson, these
types of interactions are known as ``charged current'' interactions.

The strong force involves the exchange of gluons between particles
possessing ``color charge'' \cite{Greenberg}, which is analogous to the electric charge.
The only fundamental particles possessing color charge are the quarks
and gluons.  Therefore, the leptons do not participate in strong interactions.  
Each quark carries one of the three colors, usually denoted red
($r$), green ($g$), and blue ($b$),  while the anti-quarks carry
anti-color: $\bar{r}$, $\bar{g}$, and $\bar{b}$.  The quantum theory
which describes the strong force is known as quantum chromodynamics
(QCD).

One interesting aspect of QCD is color confinement, the feature that
no free quarks exist in nature.  Rather, quarks must bind into the composite
particles known as hadrons, are color-neutral.  Therefore the
anti-quark of the meson will have the anti-color of its quark partner,
while for a baryon each quark will possess a different color.\footnote{While the color of the quarks has nothing to do with color in visual perception, the similarity of the rules of combinations to those of color theory make the label an apt one.}  The
origin of the confinement is related to the fact that gluons themselves
carry color and therefore interact with each other.  As two quarks are pulled apart,
the gluon-gluon interactions form a narrow tube resulting in a constant
force.  As the separation of the two quarks increases, so does the
energy, until at some distance it
becomes energetically favorable to ``pop'' out quarks from the vacuum
to ``dress'' the quarks into hadrons. Confinement is
not a feature of QED since the photon is electrically neutral.  That is,
as two charged particles get pulled apart, the force between them decreases
allowing atoms to ionize.

Another interesting aspect of QCD is asymptotic freedom \cite{asymptotic}.  As the
momentum transferred in an interaction increases, the strength, or
coupling, in the interaction decreases.  Therefore, in these types of
interactions, perturbative methods can be applied to the calculations, as they are in QED.
However, because the strength of the interaction increases at lower
momenta, QCD calculations are very difficult in most cases.

Since this thesis is devoted to the production of charm mesons, a
dedicated review of charm follows in the next section.

\section{Charm}

The quark model, when it was introduced by Gell-Mann and independently by Zweig in 1964, explained all hadrons matter as being combinations of three kinds of quarks: up ($u$), down ($d$), and
strange ($s$).  Initially, there was no experimental reason for any
additional quarks, but Bjorken and Glashow \cite{BjorkenandGlashow}, as a way of
making nature more symmetrical at a time when there were three known quarks
and four known leptons, predicted the existence of the yet-to-be
discovered charm ($c$) quark. The Glashow-Iliopoulos-Maiani (GIM) mechanism \cite{GIM,Griffiths} made the case for
charm more compelling, explaining the experimentally unobserved strangeness-changing neutral currents (SCNC) in the theory by adding charm in a particular way to the weak hadronic current which would otherwise be expected.

\begin{figure}[!t]
\begin{center}
\hspace{2.5pt}
\includegraphics[width=14.5cm]{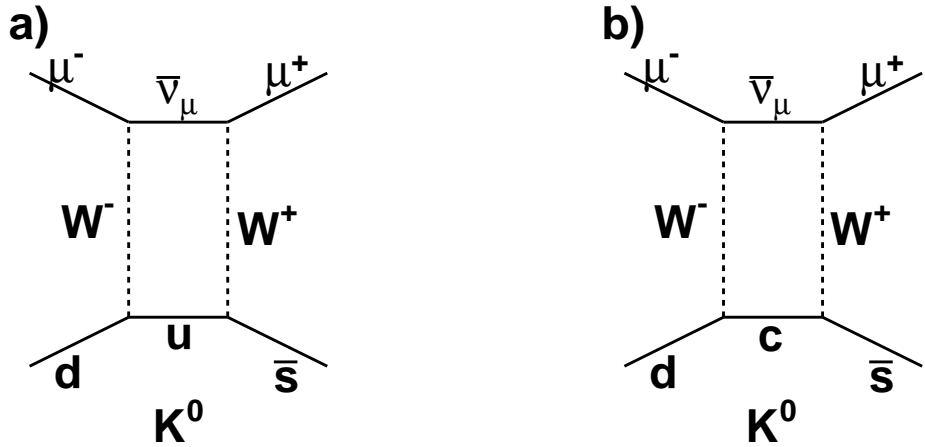}
\caption{The decay of $K^{0}\rightarrow\mu^+\mu^-$ and the GIM
mechanism.  The only difference between a) and b) is the virtual
quark, that is the $u$ quark is replaced by the $c$ when going from a)
to b).  The decay vanishes with the addition of the charm quark,
thus eliminating strangeness-changing neutral currents as demanded by
the data.}
\vspace{0.2cm}
\label{fig:kmumu}
\end{center}
\end{figure}

In 1963, Cabibbo had noticed that the decay
rates of strange particles differed from those of non-strange
particles \cite{Cabibbo}.  For example, the decay rate for
$K^{-}\rightarrow\mu^{-}\bar\nu_{\mu}$ was different from
$\pi^{-}\rightarrow\mu^{-}\bar\nu_{\mu}$, even after correcting for phase space.  Since the quark content of the particles
was different, $\bar{u}d$ for the $\pi^{-}$ as compared to $\bar{u}s$ for the $K^{-}$,
Cabbibo hypothesized that, in the weak interaction, the $u$ couples to
a $d^{'}$ quark, which is a superposition of the physical $d$ and $s$
quarks.  To be explicit, $d^{'} = d\cos\theta_{c} + s\sin\theta_{c}$,
where $\theta_{c} \approx 13^o$ is known as the Cabibbo angle. The transition
probability of a $d$ quark changing to a $u$ quark is proportional to
$\cos^2\theta_{c}$ whereas it is $\sin^2\theta_{c}$, for a $s$ quark
changing to a $u$ quark. This leads to the difference in the decay rates.

At the time, and assuming no charm quark, the decay
$K^{0}\rightarrow\mu^+\mu^-$ was calculated to have a rate much larger
($\propto \sin\theta_{c}\cos\theta_{c}$) than the experimental limits
would allow ($K_{s}\rightarrow\mu^{+}\mu^{-}<3.2\times10^{-7}$
\cite{pdg}).  However, with the addition of the charm quark, another
diagram, or term, enters into the calculation, canceling the decay in the limit of SU(4) symmetry.\footnote{The cancellation is not exact because the masses
of the $u$ and $c$ quarks are not the same.}
Therefore, not only do the $s$ and $d$ quark couple to the $u$ quark,
but they now also couple to the $c$ quark, where $s^{'} =
s\cos\theta_{c} - d\sin\theta_{c}$.  Now, the transition
probability of a $s$ quark changing to a $c$ quark is proportional to
$\sin^2\theta_{c}$, whereas it is $\cos^2\theta_{c}$ for a $s$ quark
changing to a $c$ quark.\footnote{This idea has
been expanded to incorporate all three generation of quarks by
Kobayashi and Maskawa \cite{KM}.}  Notice the $c$ quark diagram contribution ($\propto -\sin\theta_{c}\cos\theta_{c}$) to the amplitude of $K_{s}\rightarrow\mu^{+}\mu^{-}$, meaning that SCNC are
canceled with the addition of the $c$ quark.  Figure
\ref{fig:kmumu} shows the two diagrams for the decay
$K^{0}\rightarrow\mu^+\mu^-$.\footnote{A nice review can be found in
D. Griffiths ``Introduction to Elementary Particles'' \cite{Griffiths}.}
In addition to the charm quark
explaining the cancellation of SCNC, it
also explained the smallness of the $K_S$-$K_{L}$ mass difference,
which arises because of mixing. This difference,
gives one an idea, in broken SU(4), of the mass of the charm quark,
about 1.5 GeV \cite{LeeandGaillard}. Now, assuming the binding energy is
small relative to the mass of the charm quark, a meson composed of a
charm quark should also have a mass close to the charm quark mass.

The so-called ``November
Revolution'' followed the 1974 discovery of the $J/\psi$ resonance, and the subsequent exploration of the charm sector resoundingly validated the quark model. The computational power of the quark model was further verified in that the closed-charm ($c\bar{c}$)
states $J/\psi$ and $\psi{'}$ had widths in agreement with the
Okubo-Zweig-Iizuka (OZI) suppression hypothesis \cite{OZI}.

\subsection{$D$-mesons}

\begin{figure}[!h]
\begin{center}
\hspace{2.5pt}
\includegraphics[width=14.5cm]{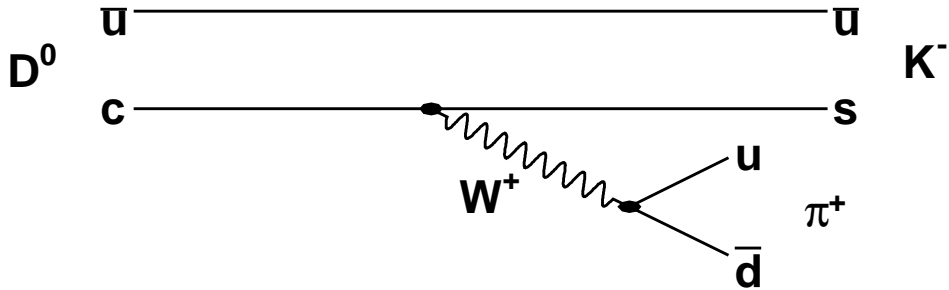}
\caption{Quark diagram for $D^0\rightarrow{K^-\pi^+}$. The light $\bar{u}$ quark does not participate in the decay and is referred to as a ``spectator''.}
\vspace{0.2cm}
\label{fig:D0kpi_diagram}
\end{center}
\end{figure}

\begin{figure}[!h]
\begin{center}
\hspace{2.5pt}
\includegraphics[width=14.5cm]{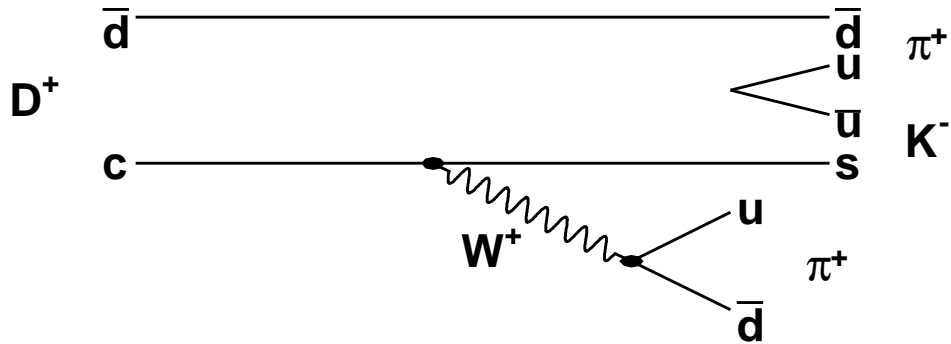}
\caption{Quark diagram for $D^+\rightarrow{K^-\pi^+\pi^+}$. The light $\bar{d}$ quark does not participate in the decay and is referred to as a ``spectator''.}
\vspace{0.2cm}
\label{fig:Dpkpipi_diagram}
\end{center}
\end{figure}

The theory of weak decays of quarks, as formulated by Cabibbo and extended by GIM, predicted that most of the charmed particles should have a $K$ meson in the final state.  This can
be viewed as the charm quark transforming into a strange quark followed by the dressing of the quarks into mesons. Gaillard, Lee, and Rosner \cite{LeeGaillardandRosner} proposed that charmed mesons,
either $D^{0}$($c\bar{u}$) or $D^{+}$($c\bar{d}$), could be observed
by looking for peaks in the invariant-mass spectrum of $K^{-}\pi^{+}$
or $K^{-}\pi^{+}\pi^{+}$.  Diagrams of these decays are shown in \FIGS
\ref{fig:D0kpi_diagram} and \ref{fig:Dpkpipi_diagram}.

The first detection of a $D$-meson decay was made by the Mark I
collaboration \cite{DMeson} in the modes $D^0\rightarrow{K}^{-}\pi^{+}$,
$D^0\rightarrow{K}^{-}\pi^{+}\pi^-\pi^+$, and
${D}^+\rightarrow{K}^{-}\pi^{+}\pi^+$. In addition to these charm
mesons, the Mark I collaboration also observed the lowest charm-meson excited states of $D^{*0}$ and $D^{*+}$ \cite{DStarMeson}.

In addition to a charm quark combining with an up or a down quark to
form a meson, it can also combine with a strange quark to form what is
referred to as a $D_{s}$ meson.  The decay of this type of meson can be
characterized by the conversion of the charm quark into a strange quark
with an emission of a virtual $W$-boson.  The simplest possible
final products result when the newly-formed strange quark pairs with
a strange quark to form a $\phi$ or an $\eta$ meson and the
virtual $W$-boson forms a charged pion. Therefore, a search for an excess of
$\eta\pi$ or $\phi\pi$ production would be evidence of $D_{s}$ mesons.
The discovery of $D_{s}$ (initially called $F$) was presented by the CLEO collaboration in 1983 by observing a peak in the invariant mass spectrum for the decay mode $\phi\pi$ \cite{DsMeson}, following a period of considerable experimental uncertainty \cite{Stone}.

\section{Motivation for this Measurement}

The years following the charm discovery provided few opportunities for its detailed study, especially the $D_{s}$.  The CLEO-c experiment was proposed in 2001 as to provide the first high-statistics study of charmed particles with a state-of-the--art detector.  Initial CLEO-c running beginning in late 2003, focused on studies of $D^0$ and $D^+$, with the expectation that $D_{s}$ studies would follow.

From August until October 2005, the CLEO-c collaboration carried out a scan 
of the center-of-mass energy range from 3970 to 4260~MeV.  The main purpose 
of this scan was to determine the optimal running point for CLEO-c 
studies of $D_{s}$-meson decays.  Secondary objectives included detailed 
measurements of the properties of charm production in this region, tests of 
previous theoretical predictions, and confirmation and additional studies
of the $Y(4260)$ state reported by the BaBar collaboration 
\cite{babar_4260}.  In this dissertation, measurements with the scan data of 
the cross sections for inclusive hadron production, charmed-hadron production,
and the production of specific final states including charmed mesons
are presented.  We present results both without and with radiative
corrections.  We also report initial studies with detailed features of
these events and present comparisons with past measurements and
theoretical expectations. All of which will add to our present understanding for systems containing heavy and light quarks. 

\subsection{Previous Experimental Results}
The total hadronic cross section in $e^+e^-$ annihilations is generally
presented in terms of $R$, which is defined as follows:
\begin{equation}
\label{eq:R}
	R=\frac{\sigma(e^+e^-\rightarrow{hadrons})}{\sigma(e^+e^-\rightarrow\mu^+\mu^-)}.
\end{equation}
\CONT Incorporating the well-known $\mu$-pair cross section with its $1/s$ 
energy dependence gives  
\begin{equation}
\label{R_eq}
	\sigma(e^+e^-\rightarrow{hadrons}) =
	R\sigma(e^+e^-\rightarrow\mu^+\mu^-) = 	R\frac{86.8 {\rm{~nb~GeV}}^{2}}{E_{\rm{cm}}^2}.
\end{equation}
\CONT If strong interactions are ignored:
\begin{equation}
R=3\sum_{i}q^{2}_{i}
\end{equation}
\CONT where the summation is over all kinematically allowed
quark flavors at the energy of interest, $q_{i}$ is the quark's
corresponding charge (either $-1/3$ or $2/3$), and the factor are 3
is due to the fact that there is 3 different colors available.  This
approximation only holds in energy regions well above the $q\bar{q}$
bound states.

Equation \ref{eq:R}, counts the number of possible quarks
that are available at a particular energy.  In the energy region below,
the $J/\psi$ resonance, $M_{J/\psi} = 3.097$ \cite{pdg}, only the
$u,~d,$ and $s$ quarks contribute to $R$:
\begin{equation}
R=3[(\frac{2}{3})^2+(\frac{1}{3})^2+(\frac{1}{3})^2] = 2
\end{equation}
\CONT Above this energy the $c$ quark can also contribute to the ratio.

\begin{figure}[!htbp]
\begin{center}
\hspace{2.5pt}
\includegraphics[width=14.5cm]{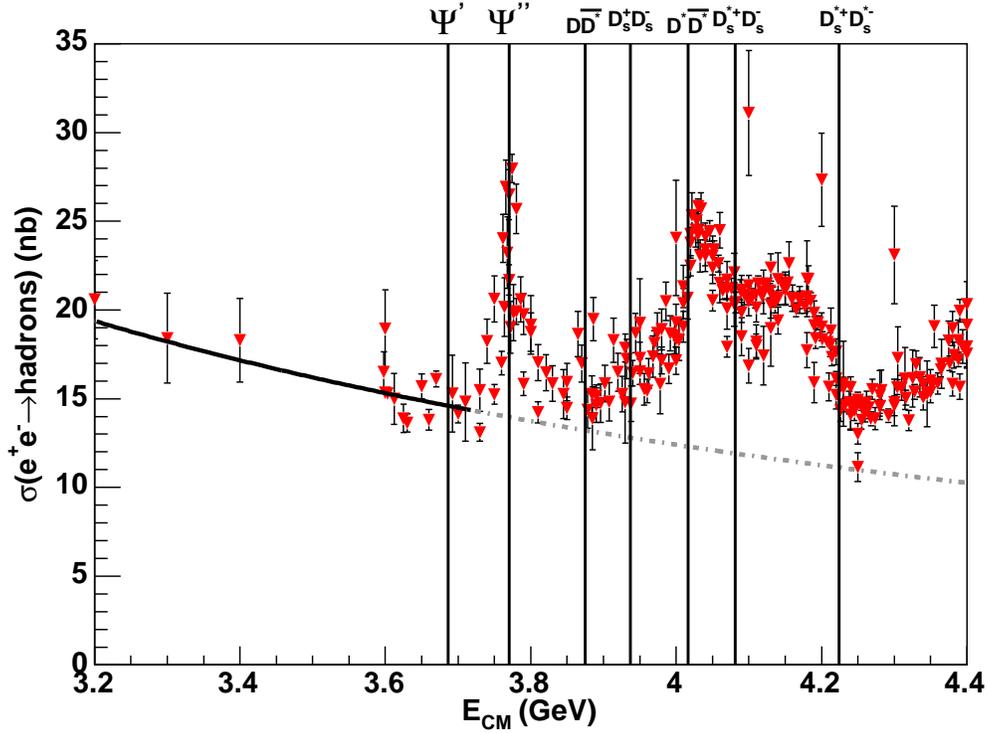}
\caption{$\sigma(e^+e^-\rightarrow{hadrons})$ from the 2005 PDG
\protect \cite{pdg}. The solid
line is a $\frac{1}{s}$ fit to the data and the dotted line is the
extension of the fit to higher energies. The vertical lines indicate
the various thresholds and resonances present in the region.}
\vspace{0.2cm}
\label{fig:PDG}
\end{center}
\end{figure}

$R$ has been measured over a very wide energy range by many experiments 
\cite{pdg}, including recent measurements with the Beijing Spectrometer (BES)
\cite{BES_R} in the energy range of interest for 
CLEO-c.  \FIG \ref{fig:PDG} shows the current state of knowledge of $R$ at 
center-of-mass energies between 3.2 and 4.4~GeV. 
\CONT The data in \FIG \ref{fig:PDG} has been radiatively corrected. The 
correction depends on the center-of-mass energy and the behavior of the 
cross section at lower energies.  There is a rich structure in this energy
region, reflecting the production of $c {\bar c}$ resonances and the crossing
of thresholds for specific charmed-meson final states.  Some of these 
``landmarks'' are highlighted with vertical lines in \FIG \ref{fig:PDG}. 
The specific energies corresponding to these thresholds are tabulated in
\TAB \ref{tab:thres_table}.
\begin{table}[!htbp]
\begin{center}
\caption{Thresholds of interest in the $e^+e^-$ center-of-mass energy
range of the CLEO-c scan. Threshold values are computed from the
PDG \cite{pdg}, and where needed charged and neutral states of the $D$ 
were averaged.}
\vspace{0.2cm}
\label{tab:thres_table}
\begin{tabular}{|c|c|c|c|c|c|}
\hline
{Center-of-Mass}& {3875~MeV} &{3937~MeV} & {4017~MeV}& {4081~MeV}&
{4224~MeV} \\
{Energy}&&&&&
\\ \hline
{State} & \(D^{*}\bar{D}\) &  \(D_{s}\bar{D}_{s}\) & \(D^{*}\bar{D}^{*}\) &
\(D_{s}^{*}\bar{D}_{s}\) &  \(D_{s}^{*}\bar{D}_{s}^{*}\)
\\ \hline
\end{tabular}
\end{center}
\end{table}

There are two interesting features in the hadronic cross section
between 3.9 and 4.2~GeV. There is a large enhancement at $\sim4$~GeV
corresponding to the $D^*\bar{D}^*$ threshold.\footnote{Throughout
this paper charge-conjugate modes are implied}  Next, there is a fairly
large plateau that begins at the $D_{s}^{*+}D_{s}^{-}$
threshold. There is considerable theoretical interest and little
experimental information about the specific composition of these enhancements.
\begin{figure}[!htbp]
\begin{center}
\hspace{2.5pt}
\includegraphics[width=14.5cm]{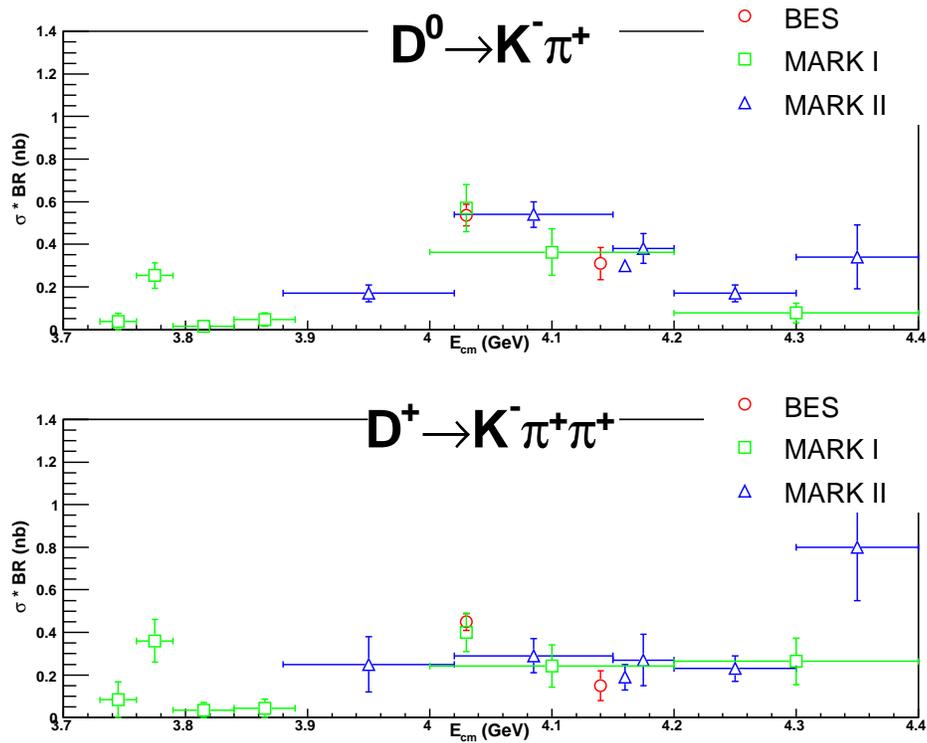}
\caption{The production cross section times branching ratio for
$D^{0}\rightarrow{K^-\pi^+}$ and $D^{+}\rightarrow{K^-\pi^+\pi^+}$ as
a function of energy.}
\vspace{0.2cm}
\label{fig:MARK_BES}
\end{center}
\end{figure}

Prior to the CLEO-c scan run there were insufficient data on $D_s$ production 
for an informed decision about the best energy at which to undertake 
CLEO-c studies 
of $D_s$ decays.  BES measured the inclusive $D_s$ production cross section 
times the $D_{s}^{+}\rightarrow{\phi}\pi^{+}$ branching ratio at the 
center-of-mass energy $4030$~MeV to be ($11.2\pm2.0\pm2.5$)~pb \cite{BES95}.  
Since there is only one accessible final state with $D_s$ at this energy, 
this measurement suggested a cross section of about 0.3~nb for 
$D_{s}^{+}{D}_{s}^{-}$.  The Mark~III collaboration previously measured the 
same quantity at a center-of-mass energy of $4140$~MeV to be 
($26\pm6\pm5$)~pb \cite{MARK89}.  $D_s$ production had previously been 
demonstrated to be dominated at this energy by $D^{*+}_{s}D_{s}^{-}$ 
\cite{MARK87}.  The ability to do $D_s$ physics with CLEO-c depends both 
on the quantity of $D_s$ production and on the complexity of the events.  It 
was therefore essential to measure all accessible final states and carefully
assess the physics reach for future $D_s$ studies under the conditions
prevailing at each energy.

Studies of $D_s$ production, when combined with measurements of $D^0$ and 
$D^+$ production, would constitute a comprehensive analysis of all
charm production in the region just above threshold.  There are
more previous measurements of $D^{0}$ and $D^{+}$ production than of
$D_s$, but here too the information is limited.  BES and MARK~II made 
measurements of cross section time branching ratio for 
$D^{0}\rightarrow{K^-\pi^+}$ and $D^{+}\rightarrow{K^-\pi^+\pi^+}$ \cite{MARK77,MARK79,MARK82,BES00}, 
which are shown in \FIG \ref{fig:MARK_BES}.  
Interpretation of these data points is 
complicated by the presence of several possible final states: $D\bar{D}$, 
$D^*\bar{D}$, and $D^*\bar{D}^*$ in both charged states.  Strong interaction
theory provides predictions of the overall cross sections and proportions of
the various final states, some of which are described in the next section.
Detailed measurements at several points would allow more rigorous 
testing of models of charm production in the region above $c\bar{c}$ threshold.

\subsection{Theoretical Predictions}

Soon after the discovery of the $D$ meson relative production ratios were predicted by counting available spin states \cite{RujulaGeorgiandGlashow} for the possible reactions:
\begin{itemize}
\item {$1^{-}\rightarrow{0^-0^-}$, as in
      $e^+e^-\rightarrow\gamma^*\rightarrow{D\bar{D}}$}
\item {$1^{-}\rightarrow{0^-1^-}$, as in
      $e^+e^-\rightarrow\gamma^*\rightarrow{D\bar{D}^*}$}
\item {$1^{-}\rightarrow{1^-1^-}$, as in
      $e^+e^-\rightarrow\gamma^*\rightarrow{D^*\bar{D}^*}$}
\end{itemize}
\CONT This argument gives thefollowing ratios:
\begin{equation}
	\sigma_{D\bar{D}} : \sigma_{D^*\bar{D}} : \sigma_{D^*\bar{D}^*} = 1 : 4 : 7
\end{equation}
\CONT This naive expectation disagrees with the experimentally observed ratios obtained by the Mark I experiment \cite{DStarMeson} at 4028 MeV:
\begin{equation}
	\sigma_{D\bar{D}} : \sigma_{D^*\bar{D}} : \sigma_{D^*\bar{D}^*} = 0.2\pm0.1 : 4.0\pm0.8 : 128\pm40,
\end{equation}
\CONT where the $p^3$ phase space factor has been removed. The severe
disagreement between what is expected and what is experimentally observed suggests
that major additional effects are present.  

\begin{figure}[!htbp]
\begin{center}
\hspace{2.5pt}
\includegraphics[width=14.5cm]{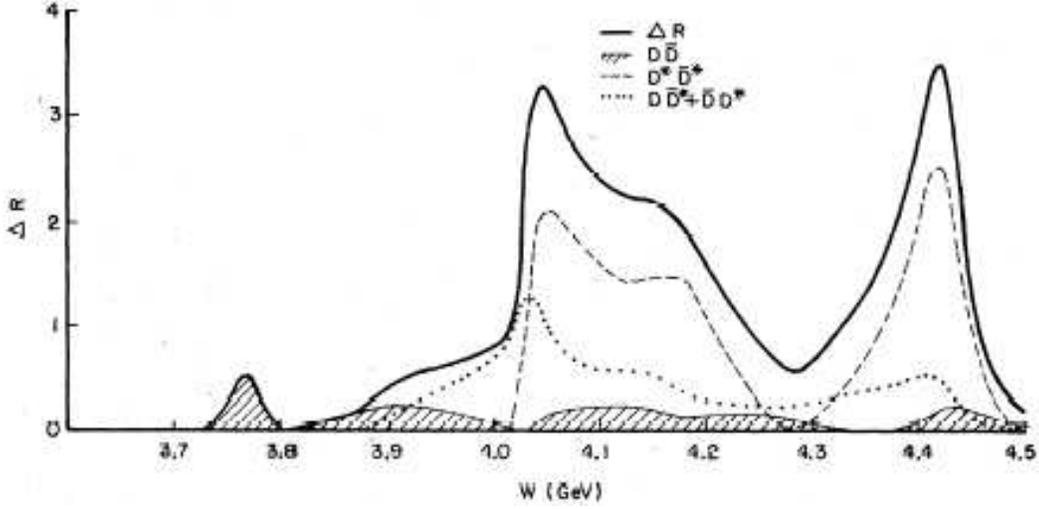}
\caption{The charm contribution to $R$ as calculated in the coupled
channel model of Eichten \etal \protect \cite{Eichten} for $D^{(*)}\bar{D}^{(*)}$ events.}
\vspace{0.2cm}
\label{fig:Eichten_d}
\end{center}
\end{figure}

\begin{figure}[!htbp]
\begin{center}
\hspace{2.5pt}
\includegraphics[width=14.5cm]{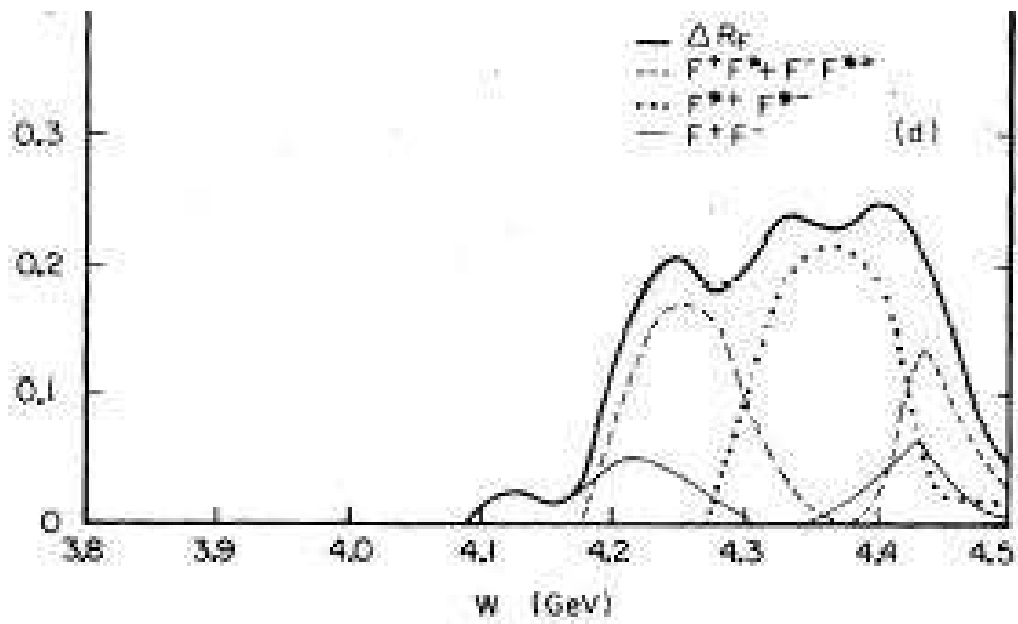}
\caption{The $D_{s}$ contribution to R as calculated in the coupled
channel model Eichten \etal \protect \cite{Eichten} for
$D^{(*)}\bar{D}^{(*)}$ of events. In the
figure $D_{s}$ is called $F$.}
\vspace{0.2cm}
\label{fig:Eichten_ds}
\end{center}
\end{figure}

A serious theoretical calculation of the charm cross sections was first attempted
by Eichten \etal in 1980 \cite{Eichten} with a coupled-channel potential model.
  Their predictions for the excess $\Delta{R}$ above
$uds$ production are shown in \FIGS \ref{fig:Eichten_d} and 
\ref{fig:Eichten_ds}.  According to these predictions the large enhancement 
in the cross section at $4030$~MeV is dominated by $D^*\bar{D}$ and
$D^*\bar{D}^*$ production. A detailed investigation of this energy region could definitively confirm or refute this prediction.   

The mass of the $D_{s}$ meson used in Eichten's 1980 prediction is
incorrect, as indicated by use of the older notation $F$ for that particle.  
To update Eichten's prediction, it is therefore necessary to
shift the cross sections downward in energy by $144$~MeV and $100$~MeV for 
$D_{s}^{+}D_{s}^{-}$ and $D_{s}^{*+}D_{s}^{-}$, respectively.  After this 
correction, Eichten predicts that the largest $D_{s}$ yield should be at
$\sim4160$~MeV, where the dominant source is $D_{s}^{+}D^{*-}_{s}$ events.  

More recently, T. Barnes \cite{Barnes} has presented calculations, using the phenomenological $^3P_{0}$ model \cite{Micu}, at 
4040 and 4159~MeV, which are summarized in \TAB \ref{tab:Barnes}. 
\begin{table}[!htbp]
\begin{center}
\caption{Partial widths, in units of MeV, of charm production as predicted by Barnes for
two center of mass energies.}
\vspace{0.2cm}
\label{tab:Barnes}
\begin{tabular}{|c|c|c|c|c|c|c|c|}
\hline
{Center-of-Mass}& {\(D\bar{D}\)} & {\(D^{*}\bar{D}\)}&
{\(D^{*}\bar{D}^{*}\)}& {\(D_{s}^{+}{D_{s}^{-}}\)}&
{\(D^{*+}_{s}{D^{-}_{s}}\)}& {SUM}& {Exp.} \\
Energy&&&&&&&
\\ \hline
4040 MeV & 0.1 & 33 & 33 & 7.8 & - & 74 & 52\(\pm\)10
\\ \hline
4159 MeV& 16 & 0.4 & 35 & 8.0 & 14 & 74 & 78\(\pm\)20
\\ \hline
\end{tabular}
\end{center}
\end{table}
Of these two energy points, Barnes predicts that $4159$~MeV is the better 
place for $D_{s}$ physics, with a total cross section for $D_{s}$ production 
that is three times larger than that at $4040$~MeV. In addition, he finds the
enhancement in $R$ at $\sim$4~GeV to be
due to an equal mixture of $D^{*}\bar{D}$ and $D^{*}\bar{D}^{*}$
events. While the experimentally measured summed rate differs by
$\sim2\sigma$ from Barnes's prediction at $4040$ MeV, the precision is
insufficient for a definitive conclusion.  Precise measurements of the
partial widths with CLEO-c would be decisive in testing this model.

\chapter{Experimental Apparatus}
\section{CESR - The Cornell Electron Storage Ring}

\begin{figure}[!htbp]
\begin{center}
\hspace{2.5pt}
\includegraphics[width=14.5cm]{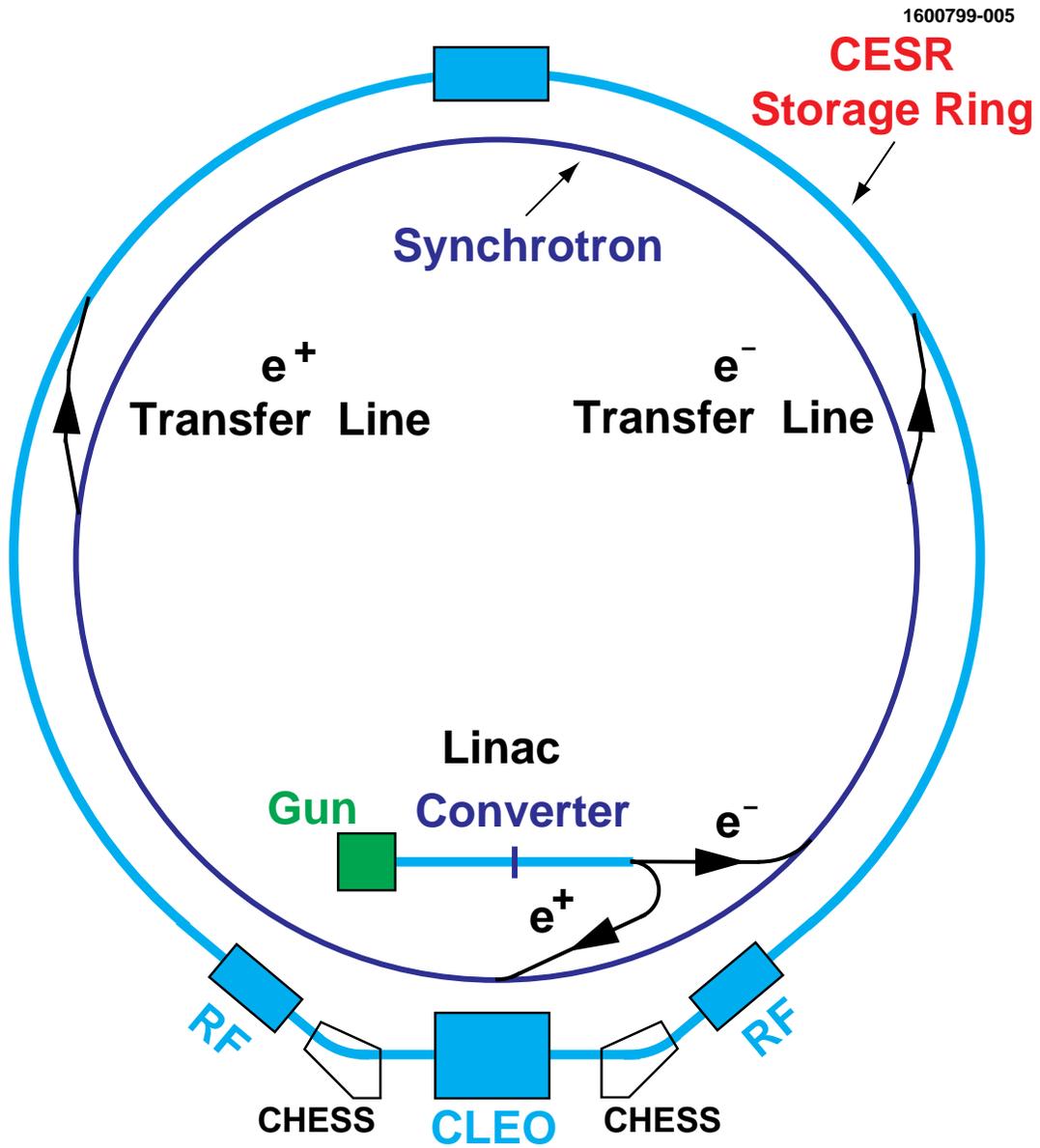}
\caption{The CESR accelerator facility.  The main components of the
facility are the linac, the synchrotron, and the storage ring.  The
linac converter is a piece of tungsten used to induce an electromagnetic
shower so as to produce positrons.  CHESS refers to the Cornell High
Energy Synchrotron Source.}
\vspace{0.2cm}
\label{fig:CESR}
\end{center}
\end{figure}

The Cornell Electron Storage Ring, or CESR, is
located in central New York on the Cornell University campus.  As shown in \FIG \ref{fig:CESR}, CESR
consists of three basic parts: a linear accelerator (linac), a
synchrotron, and the storage ring. The storage ring and
synchrotron are housed in a circular tunnel which has a diameter of 244 meters.  The
linac is located in the inner part of the ring.  The CLEO-c detector is located at
the south end of the tunnel.  

Electrons for acceleration are created by boiling them off a heating
filament.  Once the energy is sufficient for them to escape from the
surface, they are collected and bunched together for acceleration. The
linac consists of oscillating electric fields which accelerate the
electrons down the length of the linac.  By the end, the electrons have an
energy of approximately 200 MeV.

Positrons are created by inserting a tungsten target, a converter,
halfway down the linac.  The target is used to generate
electromagnetic showers consisting of electrons, positrons, and
photons.  The positrons in the shower are selected out and accelerated
the rest of the way down the length of the linac.  By the end of the acceleration, the
positrons reach an energy of 200 MeV.

The bunches of the electrons and positrons from the linac are
injected separately and in opposite directions into the
synchrotron. The synchrotron consists of a series of dipole magnets
and four three-meter-long linear accelerating cavities.  The dipole magnets steer the electrons and positrons around the ring while
accelerating cavities increase the particles' energies to
about 2 GeV.  As the energy of the particles increases the
magnetic fields of the dipoles are increased to keep the particles in the ring.  Once
accelerated to 2 GeV, the electrons and positrons are transferred to
the storage ring.

The storage ring consists of a series of a dipole magnets which steer
the electrons and positrons around the ring, in addition to
quadrupole and sextupole magnets which focus the beams.  As the
particles traverse the ring they lose energy due to synchrotron
radiation.  The energy is replaced by superconducting radio-frequency
(RF) cavities which operate at a frequency of 500 MHz. These RF
cavities are similar to those used in the linac and synchrotron, except
that they do not significantly accelerate the beams, but primarily replace the radiated energy. 

\begin{figure}[!htbp]
\begin{center}
\hspace{2.5pt}
\includegraphics[width=14.5cm]{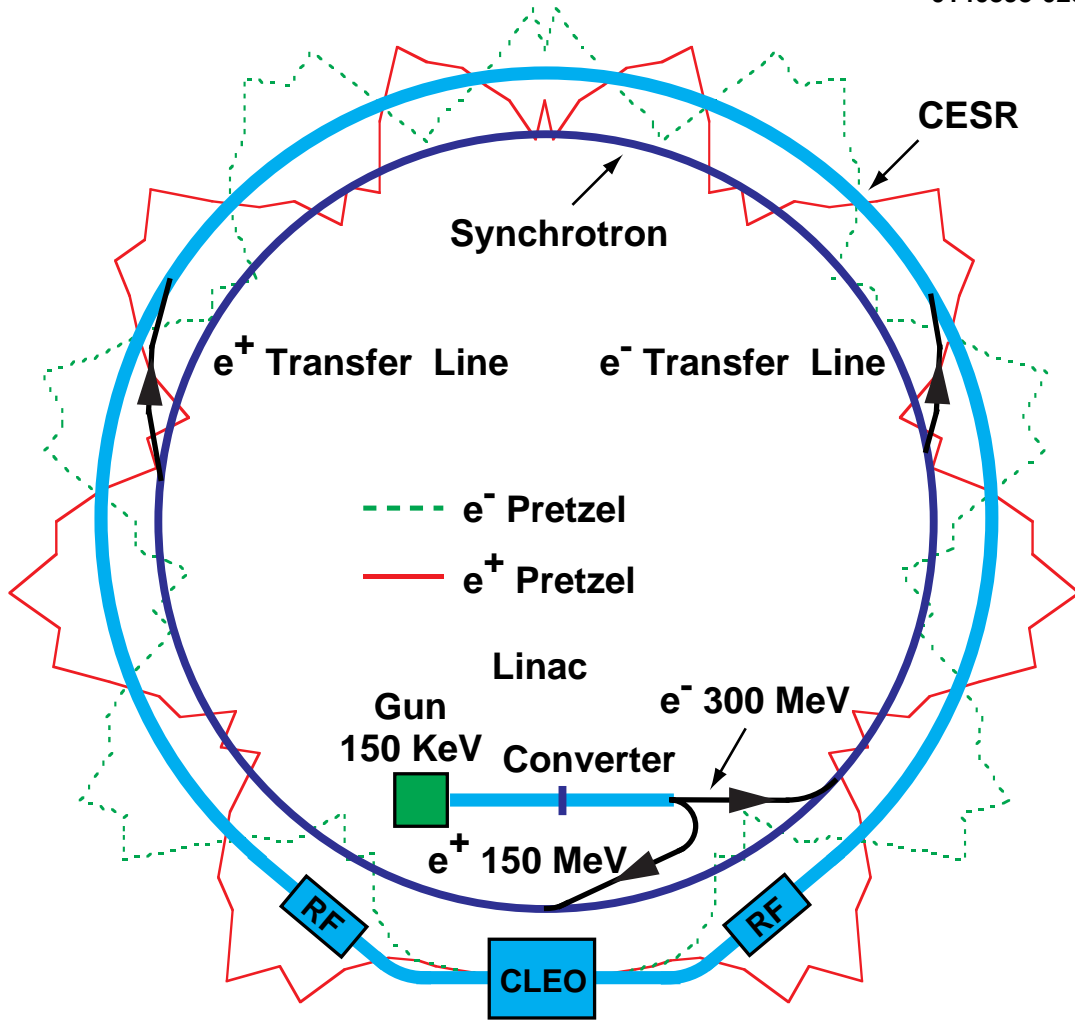}
\caption{Schematic of CESR showning the ``pretzel'' orbits.  The pretzel orbits
are used to separate the electron and positron beams at parasitic crossing
locations.}
\vspace{0.2cm}
\label{fig:pretzel}
\end{center}
\end{figure}

In current running conditions, CESR operates with nine bunch trains
each for the electrons and positrons.  Each train consists of as many as
five bunches.  To avoid unwanted interactions with the two counter-rotating beams four electrostatic horizontal separators are used.  These
separators set up what are known as pretzel orbits and ensure that
the electrons and positrons miss each other when they pass through the unwanted intersecting
locations, sometimes referred to a parasitic crossing. A picture showing an exaggerated view of the pretzel orbits is shown in \FIG
\ref{fig:pretzel}.  At the interaction point, which is
surrounded by the CLEO-c detector, the two beams are steered into each
other.  However, the two beams do not collide head-on, but rather at a
small crossing angle of 2.5 mrad ($\approx 1.7^{o}$).  This allows for
bunch-by-bunch interactions of the electron and positron trains.

Between 1979 and 2003, CESR operated at the $\Upsilon(4S)$ resonance, which corresponds to a
center-of-mass energy of 10.6 GeV or beam energy of 5.3 GeV.  The CLEO-c
charm program is carried out at center-of-mass energies between 3 and 5 GeV, which required major changes to
CESR.  The rate of synchrotron
radiation, energy radiated by the beams due to acceleration by the bending magnets,
is proportional to $E^4$ \cite{yellowbook}.  The decrease in the amount of synchrotron
radiation affects storage ring performance through two important beam parameters. 
One is the damping time with which perturbations in beam orbits caused by injection and other transitions decay away. In particular, a
particle that is off the ideal orbit because of a larger energy radiates slightly
more energy, whereas a particle with a lower energy radiates slightly less.
As a whole, the energy spread among the particles becomes reduced
which shows up as a damping of the oscillations \cite{Weidemann}.  In the original CESR design the radius was chosen to ensure adequate damping through this mechanism at a 5-GeV beam energy. Since the amount of
radiation is dependent upon the energy of the corresponding particles
in the beam, with a reduction in beam energy to $\sim{2}$ GeV, the damping time becomes too long for effective operation. The other parameter is the horizontal beam size, or
horizontal emittance, which measures the spread of particles in the bend plane.
The horizontal beam size,
which results from the betatron oscillation in addition to the rate of quantum
fluctuations in the synchrotron radiation, decreases with decreasing beam
energy thereby limiting the particle density per bunch
\cite{Weidemann}. 

These effects limited the
achievable collision rate (luminosity) of CESR at the lower center-of-mass energies of
CLEO-c to unacceptable levels. To increase the luminosity CESR accelerator physicists proposed
to increase the amount of radiation through
the use of wiggler magnets \cite{yellowbook}.

A wiggler magnet is a series of dipole magnets with high
magnetic fields.  Each successive dipole has its direction of magnetic
field flipped.  Therefore, when a particle passes, it will oscillate, which results in emission of additional synchrotron
radiation without changing the overall path of the
particle around the ring.  As a result, the damping time is decreased
while the horizontal beam size is increased, thereby increasing the luminosity.

CESR has installed twelve superconducting wiggler magnets for low-energy running.  Each
wiggler consists of eight dipole magnets with a maximum field strength
of 2.1 Tesla  \cite{yellowbook}.  These wigglers decrease the damping time by a
factor of 10, while increasing the beam emittance by a factor of 4 to 8
\cite{DRice}.  In addition, the energy resolution has increased, as compared to no
wigglers, by a factor of 4 to $\sigma_{E}/E = 8.6\times10^{-4}$ \cite{DRice}.

The center-of-mass (CM) energy of the collision is an important
quantity needed to put the cross section measurements into context.
In order to determine the CM energy of the colliding electrons and
positrons, one needs to know the energy of the beams.
The energy of the beams, to first order, can be determined by the
following \cite{CBX0070}: 
\begin{equation}
\label{beam_energy}
	E_{o} = \frac{ec}{2\pi}\sum_{i}|B_{i}|\Delta\theta_{i}\rho_{i},
\end{equation}
where $c$ is the speed of light in a vacuum and $e$ is the charge of an
electron.  The summation in \EQ \ref{beam_energy} is over all dipole
magnets in the ring, where $B_{i}$, $\theta_{i}$, and $\rho_i$ are the
magnitude of the magnetic field, the bending angle, and bending radius of
curvature of the $i$-th dipole magnet, respectively.  The result of \EQ
\ref{beam_energy} needs to be corrected for shifts in RF accelerating cavities,
for the currents of the the focusing and steering magnets, and for the
horizontal separators. The total uncertainty in the CM energy is of
order 1 MeV.

Another quantity needed in determining the cross sections is the
luminosity, which quantifies the rate
of $e^+e^-$ collisions.  The number of events expected for a particular process is given by
\begin{equation}
\label{Lum}
	N_{i} = \sigma_{i}\int{\cal{L}}dt,
\end{equation}
where $\cal{L}$ is the instantaneous luminosity and $\sigma_{i}$ is
the cross section for that process. The integral of the
instantaneous luminosity, $\int{\cal{L}}dt={\rm{L}}$, is the quantity
needed for this analysis and is referred to as the integrated
luminosity or just luminosity.
In CLEO-c, three final states are used to
obtain the luminosity.  The processes $e^+e^-\rightarrow{e^+e^-}$, $\mu^+\mu^-$, and
$\gamma\gamma$ are used since their cross sections are precisely determined by
QED.  Each of the three final states relies on different
components of the detector, with different systematic effects \cite{CBX0510}. The three individual
results are combined using a weighted average to obtain the
luminosity needed for this analysis.

\section{The CLEO-c Detector}

The electron and positron beams are steered into each other at the
central location known as the interaction point (IP) of the CLEO-c
detector.  CLEO-c is a general-purpose detector designed to identify and measure
relatively long-lived charged and neutral particles.  The charged
particles detected are the electron, muon, pion, kaon, and
proton.  While some neutral particles are easy to identify,
like the photon, others like the neutrinos are impossible.

\begin{figure}[p]
\begin{center}
\hspace{2.5pt}
\includegraphics[width=14.5cm]{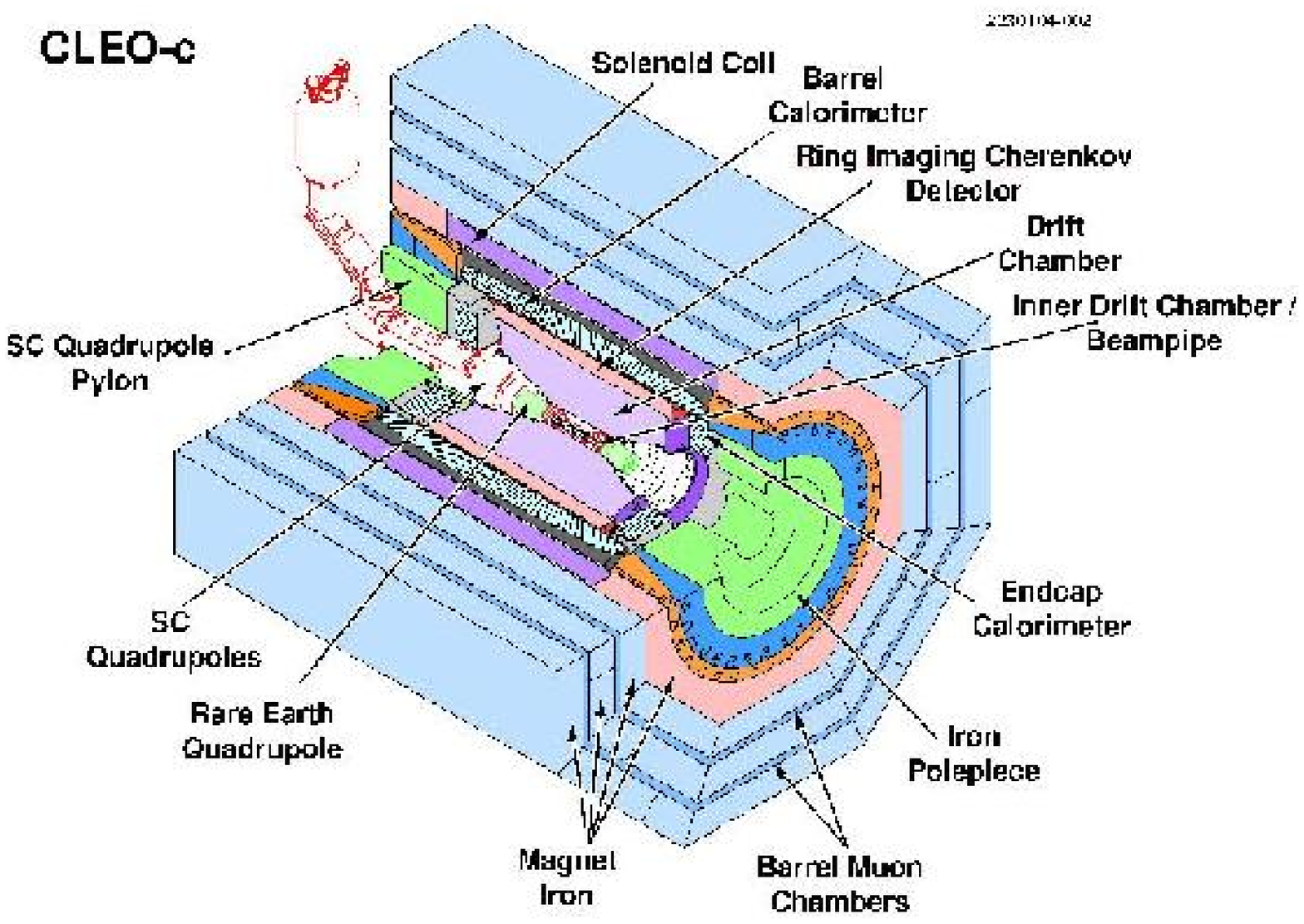}
\caption{The CLEO-c detector.}
\vspace{0.2cm}
\label{fig:Detector_2}
\end{center}
\end{figure}
\begin{figure}[p]
\begin{center}
\hspace{2.5pt}
\includegraphics[width=15cm]{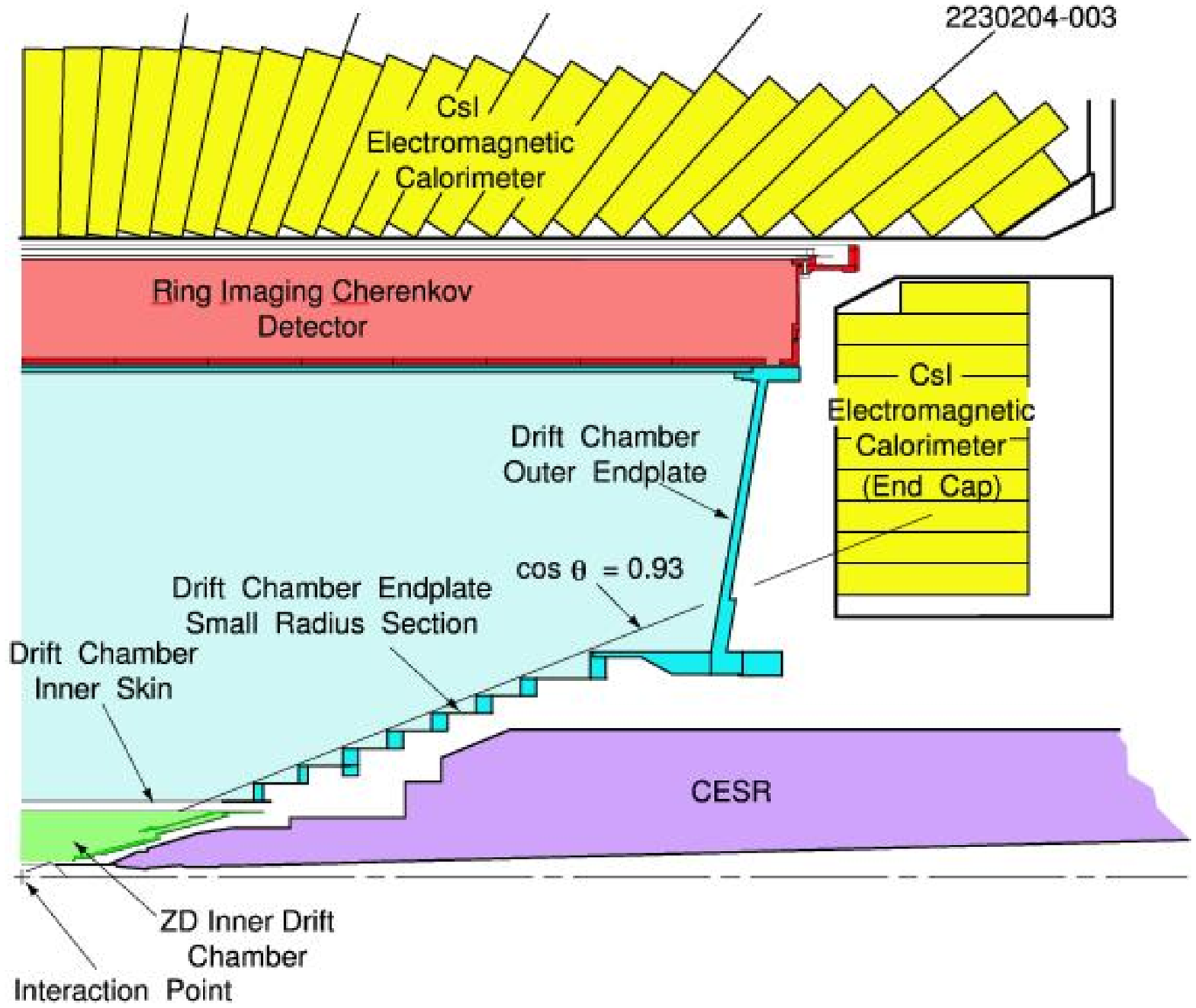}
\caption{A slice in the r-z plane of the CLEO-c detector.}
\vspace{0.2cm}
\label{fig:Detector_1}
\end{center}
\end{figure}

The remainder of this chapter is dedicated to describing the
detection systems that make up the CLEO-c detector.  As shown schematically in \FIGS \ref{fig:Detector_2} and
\ref{fig:Detector_1}, CLEO-c is a cylindrically symmetric
detector aligned along the beam line, the z-axis, that covers 93\% of the solid
angle.  Starting from the IP and moving out, the detector is composed
of an inner drift chamber \cite{yellowbook}, an outer or main drift
chamber \cite{yellowbook,NIMA478_142}, a Ring-Imaging Cherenkov detector
\cite{yellowbook,NIMA554_147,NIMA441_374}, denoted by the acronym
RICH, a crystal calorimeter \cite{yellowbook,NIMA320_66}, and finally
a muon detector \cite{yellowbook,NIMA320_66,NIMA320_114}. All systems,
except for the muon detector, are within a uniform 1-Tesla magnetic
field, produced by a superconducting solenoid, which is aligned with the beam axis.

\pagebreak
\subsection{Inner Drift Chamber}

The inner drift chamber (ZD) is the innermost detector of CLEO-c, located
right outside the beam pipe.  The inner drift chamber replaced the
silicon vertex detector of CLEO III because multiple scattering degrades the momentum resolution for
soft tracks, which are more prevalent at low center-of-mass energies. Therefore, minimizing the material is crucial.
The goal of the detector is to detect charged particles with
$|\cos\theta|<0.93$, where $\theta$ is defined as the angle of the
particle with respect to the beam pipe, the z-axis. It consists of
helium-propane-gas-filled volume segmented into 300 cells (half-cell
size of 5 mm), where
each cell consists of a sense wire held at +1900 V.  These cells are
surrounded by field wires, held at ground, which neighboring
cells share. When a charged particle travels through a cell the gas
is ionized.  The resulting ionized electrons travel away
from the field wire and toward the sense wire.  Since the electric
field increases close to the sense wire, the primary electrons will ionize other atoms
in the gas, thereby creating an avalanche of electrons at the sense
wire.  The time of the resulting electric pulse seen on the sense
wire, which is synchronized with the bunch crossing, is converted
using the drift velocity of the gas to a distance of closest approach
to the sense wire. This information can then be used to map out the
trajectories of the charged particles through the drift chamber. 

\subsection{Outer Drift Chamber}

Located directly outside the inner drift chamber is the main drift
chamber, sometimes referred to as the outer drift chamber (DR).  The main
drift chamber has a similar design to the ZD, except that it is larger
in size both overall, with 47 layers of field and sense sires as compared to 6 for the
ZD,  and in the size of the cell, with a half-cell
size of 7 mm as compared to 5 mm. In addition, the field wires are held at a
higher potential, +2100 V rather than +1900 V.

\begin{figure}[!hp]
\begin{center}
\hspace{2.5pt}
\includegraphics[width=14.5cm]{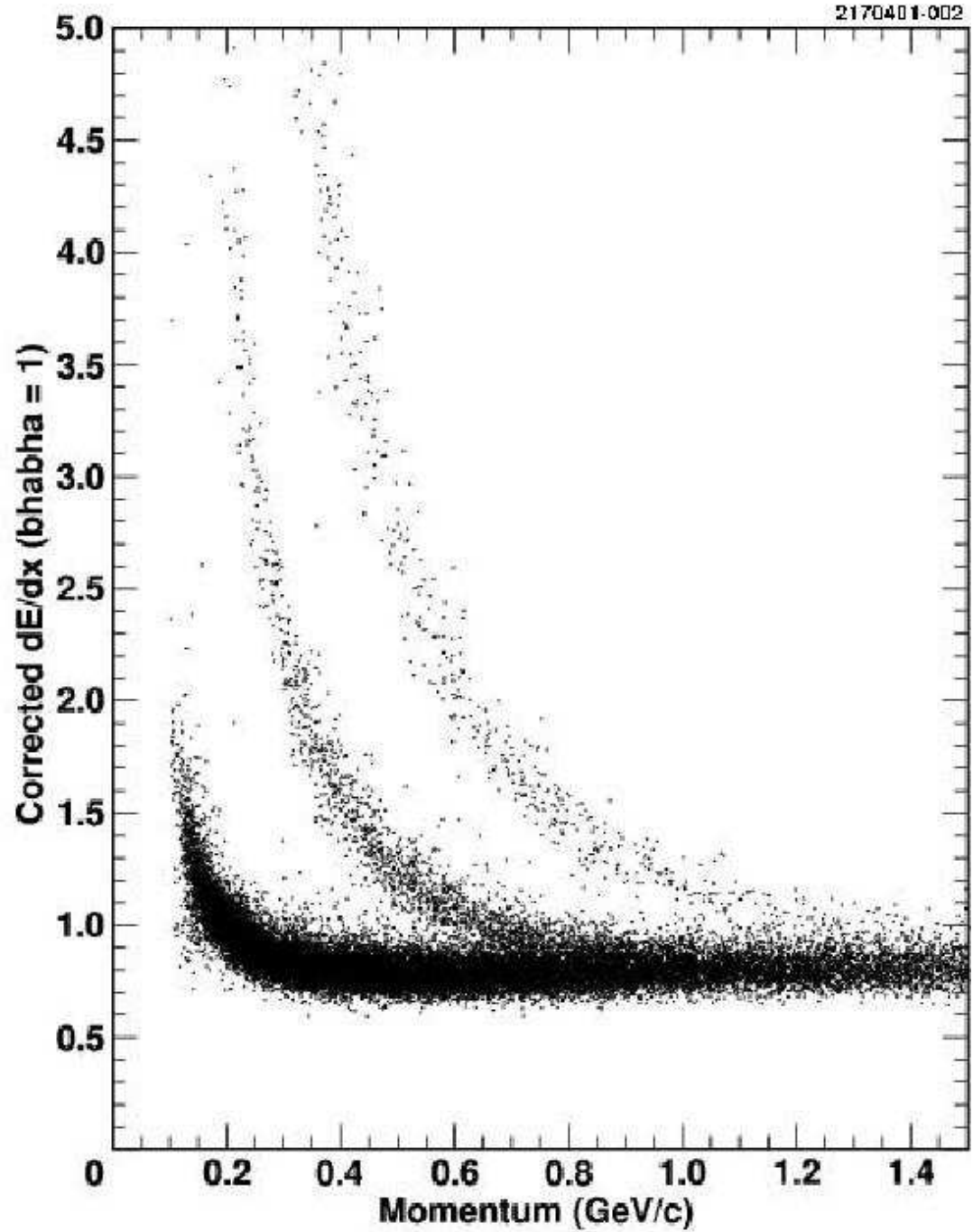}
\caption{$dE/dx$ as a function of momentum for a large population of charged particle measured in CLEO-c.  The $\pi,~K$, and $p$
bands, from left to right, are clearly seen.}
\vspace{0.2cm}
\label{fig:dedx}
\end{center}
\end{figure}

The energy lost by a charged particle in the drift chamber is used
in identifying what particle traversed the volume.  The energy lost
per unit length, $dE/dx$, is related to the particle's velocity.
By constructing a $\chi^2$-like variable, the consistency of the
actual energy lost per unit length with what is expected for different
particle hypotheses can be assessed:
\begin{equation}
X_{i} = \frac{dE/dx^{measured} - dE/dx^{expected}_{i}}{\sigma},
\end{equation}
where $i$ = $e,~\mu,~\pi,~K,$ or $p$. The quantity $\sigma$ is the uncertainty in the
measurement, usually approximately 6\%. \FIG  \ref{fig:dedx} shows
the measured
$dE/dx$ as a function of particle momentum for a large number of charged particles detected in CLEO-c.  The figure clearly shows
good $\pi-K$ separation for momenta below 500 MeV.  At momenta
above 500 MeV, $dE/dx$ is quite limited and the additional separation power of the RICH is needed.

As mentioned earlier, the ZD and DR are contained
inside a 1-Telsa magnetic field oriented along the beam.  This field causes charged particles to travel in helical paths
as they travel through the detector volume.  In other words, the
particle's path is circular in the x-y plane and moves with constant
velocity along the z-axis.  Pattern recognition computer programs group the hits into tracks and fitting programs are used to obtain the parameters \cite{CBX9620}. The CLEO-c fitter is an implementation of the Billoir or Kalman algorithm, which incorporates the expected energy loss of a particular particle to optimize the determination of its momentum and trajectory. At 1 GeV$/c$ the
charged-particle momentum resolution is approximately 0.6\%.

\subsection{RICH - Ring Imaging Cherenkov Detector}

Directly outside the main drift chamber is the Ring Imaging Cherenkov
Detector, commonly referred to as the RICH.  Radiation is emitted when
a charged particle's velocity is greater than the speed of light in the medium through
which the particle is traveling; this radiation is known as Cherenkov
radiation. The radiation is emitted in a cone with a characteristic
opening angle known as the Cherenkov angle. The Cherenkov angle is related to the velocity of the
particle by 
\begin{equation}
\label{eq:RICH}
\cos\theta = \frac{1}{\beta{n}},
\end{equation}
\CONT where $\beta$ is the velocity in units of $c$ and $n$ is the
index of refraction of the medium. Using $\beta=\frac{p}{E}$, in
addition to $E^2=m^2+p^2$, \EQ \ref{eq:RICH} can be rewritten in terms
of the particle's mass and momentum as follows:
\begin{equation}
\cos\theta = \frac{1}{n}\sqrt{1+\frac{m^2}{p^2}}.
\end{equation}

\CONT This shows that one can identify the type of particle by using its
momentum from the fitted track and the observed Cherenkov
angle.

\begin{figure}[!hp]
\begin{center}
\hspace{2.5pt}
\includegraphics[width=14.5cm]{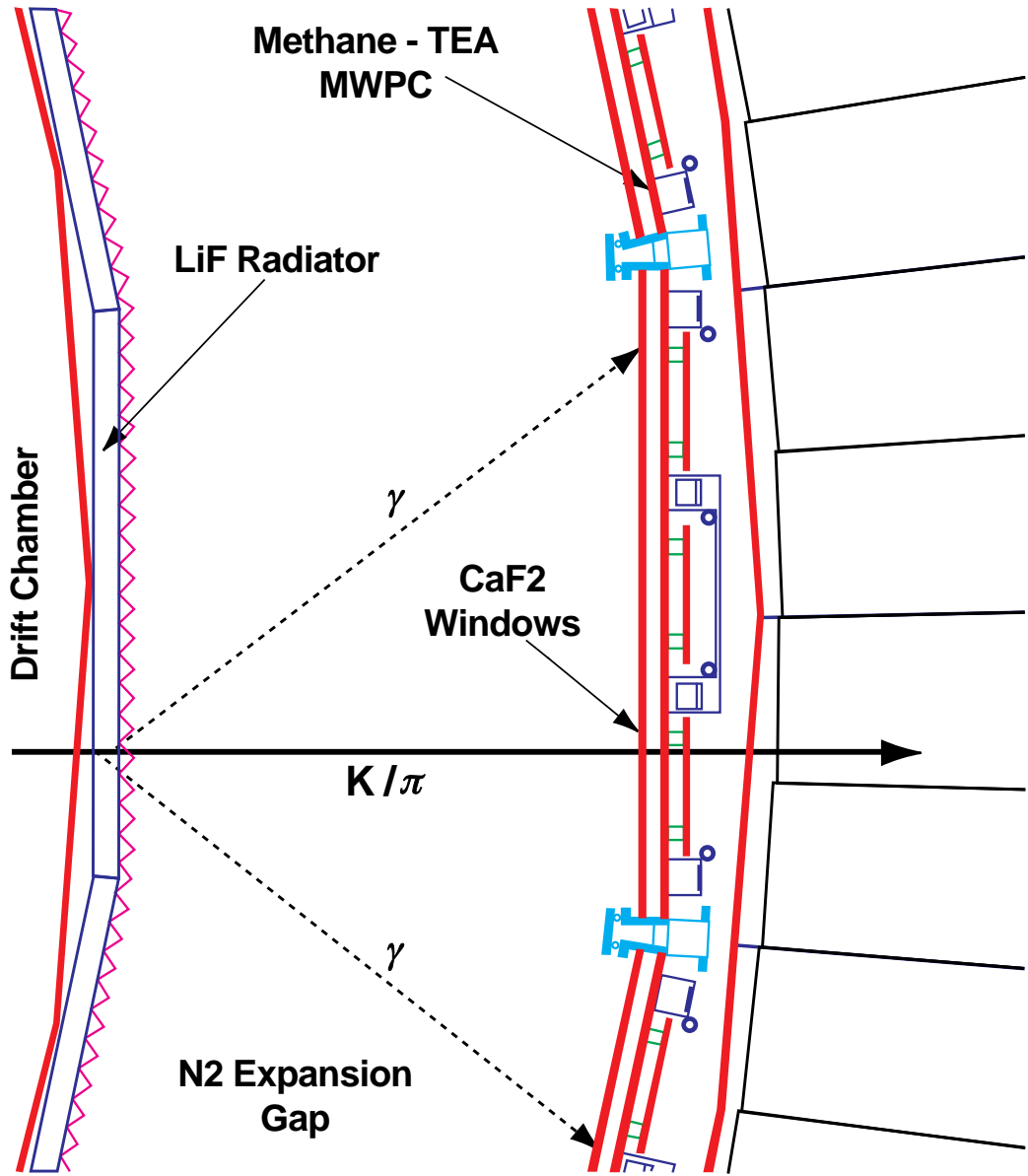}
\caption{A section of the CLEO-c RICH detector shown in $r-\phi$ cross section.}
\vspace{0.2cm}
\label{fig:RICH_drawing}
\end{center}
\end{figure}

\begin{figure}[!hp]
\begin{center}
\hspace{2.5pt}
\includegraphics[width=14.5cm]{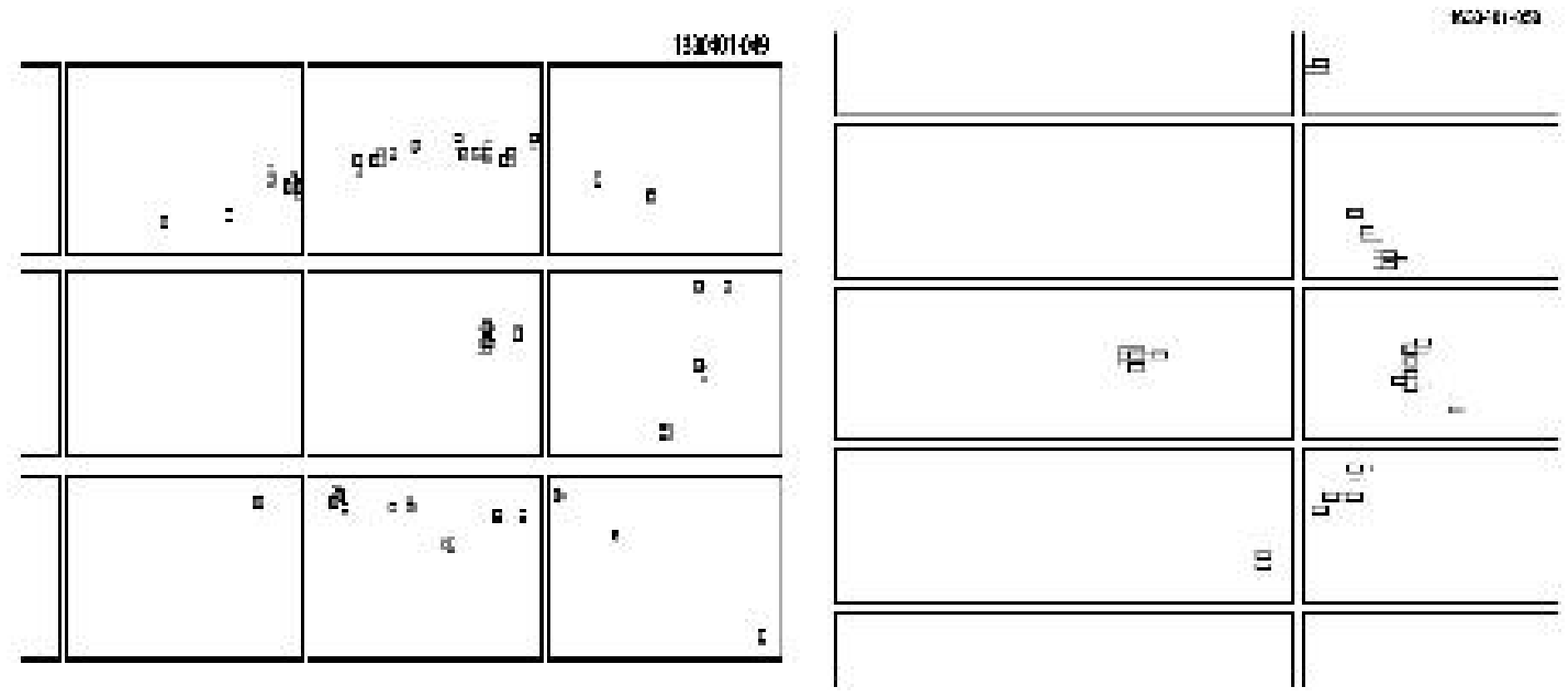}
\caption{Cherenkov rings produced by a charged track for the flat
radiator (right) and sawtooth (left).  The central hits of each image
are due to the charge track passing through the wire chamber.  The
other hits are due to the Chernkov photons that are produced.  Half of the
image of the flat radiator is missing because it was trapped in the
radiator by total internal reflection.  The sawtooth imagine is
distorted by refraction.}
\vspace{0.2cm}
\label{fig:RINGS}
\end{center}
\end{figure}

The RICH detector is shown schematically in \FIG
\ref{fig:RICH_drawing}.  It covers approximately 83\% of the full 4$\pi$ solid angle.
Cherenkov photons are produced when a charged particle passes
through entrance windows fabricated from LiF crystals, as shown in \FIG
\ref{fig:RICH_drawing}.  There are a total of 14 rows, or rings, of these
radiator crystals.  All but the central four have flat surfaces, while the remainder have a ``sawtooth'' surface to reduce loss of photons by total internal reflection.  The photons, which have a typical wavelength $\lambda=150$~nm, travel through an expansion volume
filled with nitrogen gas which is effectively transparent. After traveling through the expansion volume they pass through
a CaF$_2$ window and into a multi-wire proportional chamber (MWPC) filled
with a methane-TEA (tri-ethyl amine) mixture. Here photoelectrons are created which are
collected in the same manner as described above for the drift
chambers.  Examples of the Cherenkov rings are shown in \FIG \ref{fig:RINGS}.

Using Cherenkov photon images, like those in \FIG
\ref{fig:RINGS}, one can construct the likelihood for a particular
particle hypothesis.  A $\chi^2$-like variable for identification can be constructed
to discriminate between two different particle hypotheses:

\begin{equation}
\chi_{i}^2 - \chi_{j}^2 = -2\ln{L_{i}} + 2\ln{L_{j}},
\end{equation}

\begin{figure}[!p]
\begin{center}
\hspace{2.5pt}
\includegraphics[width=14.5cm]{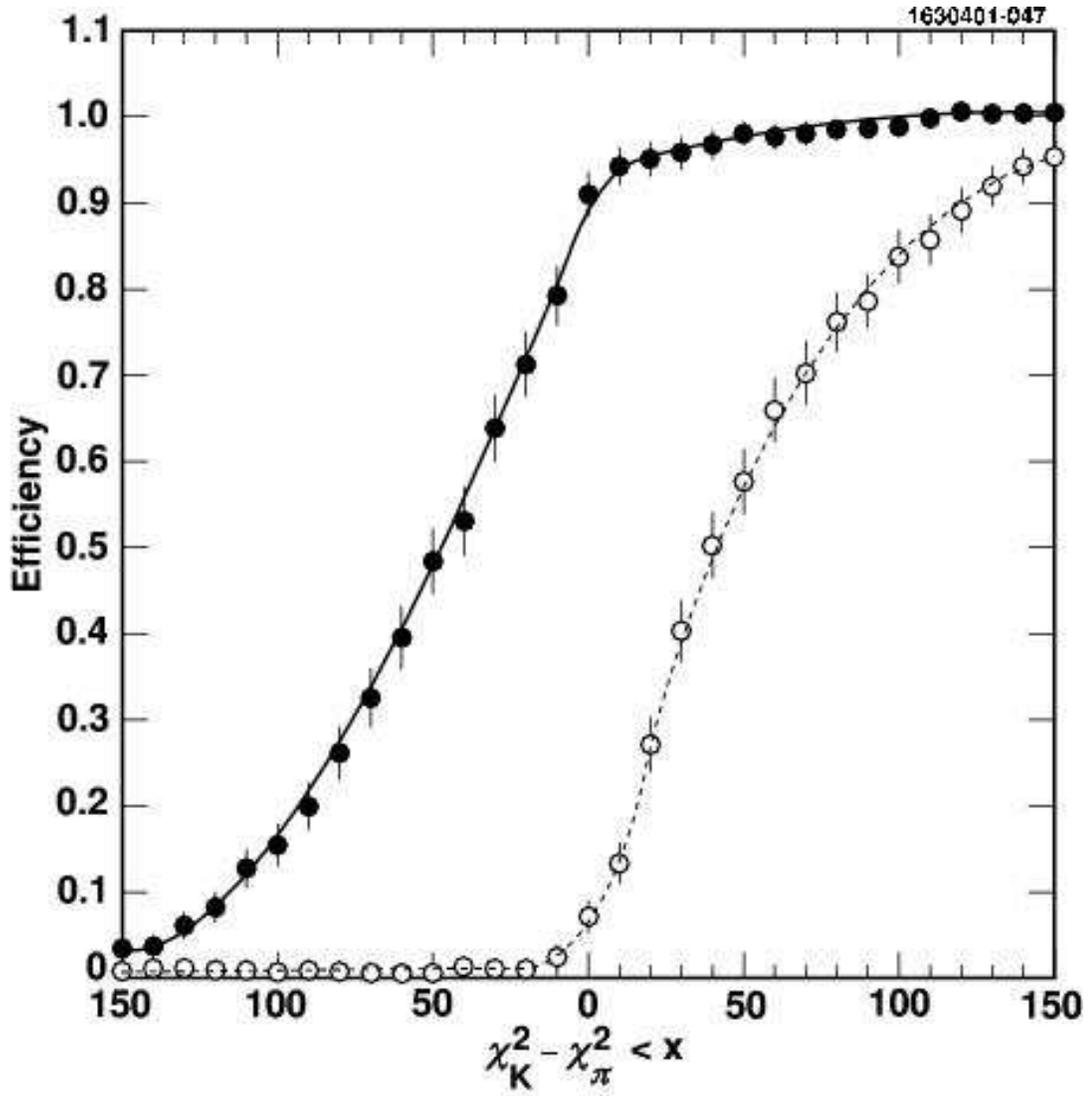}
\caption{Kaon identification efficiency (solid points) and pion fake
rate (open circles) as a function of various cuts on $\chi^2_{K} -
\chi^2_{\pi}$. Momentum is restricted to be above 700 MeV/$c$.}
\vspace{0.2cm}
\label{fig:RICH_pik}
\end{center}
\end{figure}

\begin{figure}[!p]
\begin{center}
\hspace{2.5pt}
\includegraphics[width=14.5cm]{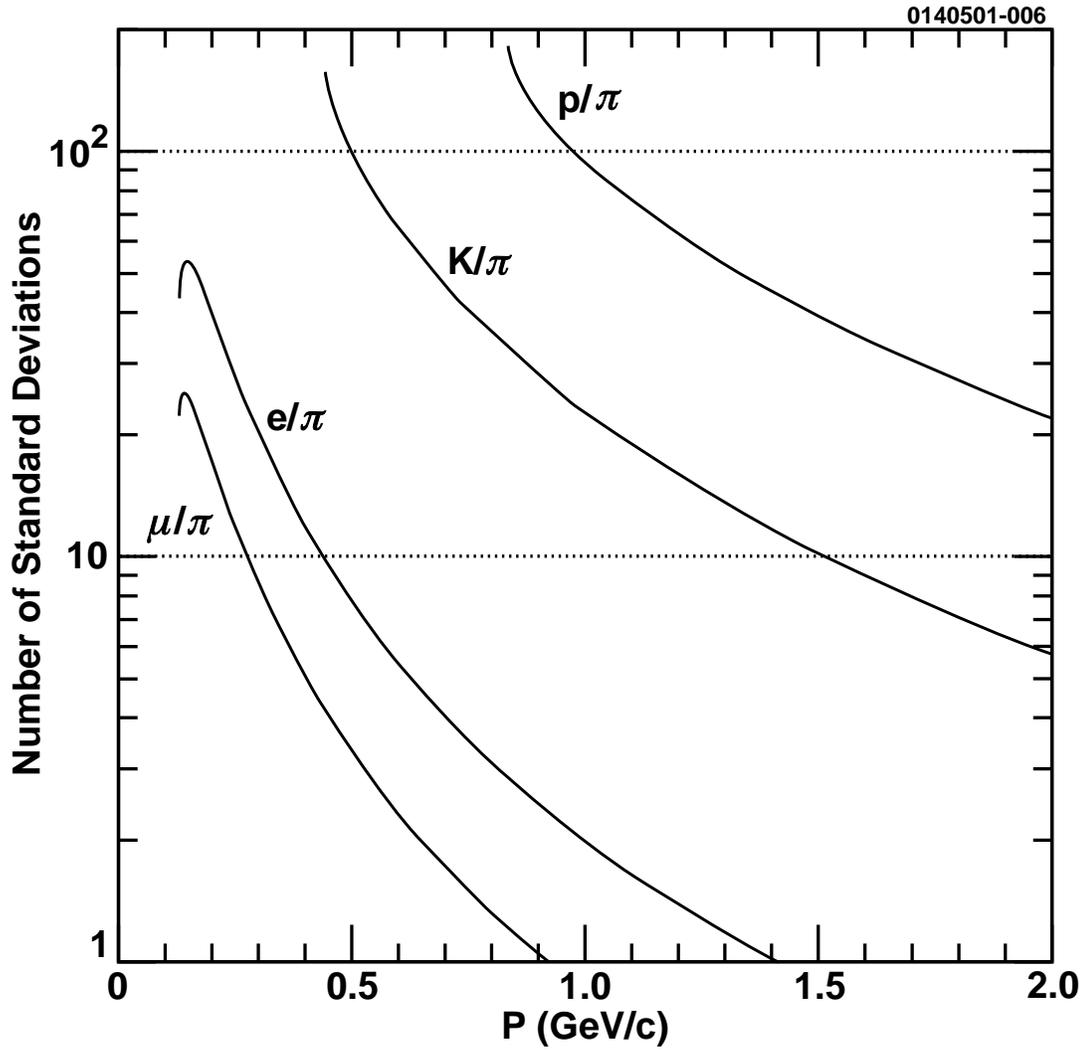}
\caption{Particle separation as a function of
momentum for different particle hypotheses in the RICH detector. Both
particles are restricted to be above their respective
threshold, where the threshold is determined by the index of refraction
of the LiF radiator, $n=1.4$.}
\vspace{0.2cm}
\label{fig:RICH_all}
\end{center}
\end{figure}

\CONT where $L_{i}$ and $L_{j}$ are the likelihoods for particle hypotheses
$i$ and $j$, respectively.  An illustration of the power of the RICH
detector is shown in \FIG \ref{fig:RICH_pik}.  Requiring
$\chi^2_{K} - \chi^2_{\pi}<0$, that is that the particle is more kaon-like
than pion-like, and that momentum is greater
than 700 MeV$/c$, leads to a kaon identification efficiency of 92\% with a pion-fake rate
of 8\%.  \FIG \ref{fig:RICH_all} shows particle separation as a function of
momentum for different particle hypotheses above their respective
thresholds, where the threshold is determined by the index of refraction
of the LiF radiator, $n=1.4$.

\pagebreak
%%%%%%%%%%%%%%%%%%%%%%%%%%%%%%%%%%%%%%%%%%%%%%%%%%%%%%%%%%%%%%%%%%%%
\subsection{Calorimetry}
%%%%%%%%%%%%%%%%%%%%%%%%%%%%%%%%%%%%%%%%%%%%%%%%%%%%%%%%%%%%%%%%%%%%
Located just outside the RICH and just inside CLEO-c's superconducting
magnet is the electromagnetic crystal calorimeter (CC).  The CC
consists of about 7,800 thallium-doped cesium iodide crystals.
About 80\% of the crystals are arranged in a projective geometry in the barrel region, defined by $|\cos\theta|<0.8$.  The remainder are in two end-caps, covering $0.85<|\cos\theta|<0.93$.  The transition regions between the barrel and end-caps provide substandard performance and are generally not used.

When charged particles or photons enter these highly dense crystals they interact and lose energy through various mechanisms: ionization,
bremsstrahlung, pair conversion, and nuclear interactions.
Electromagnetic interactions with these high-Z nuclei are very
effective in stopping electrons and photons.  Hadrons, on the
other hand, lose energy less quickly in the CC electromagnetically, and their hadronic showers extend
into the steel of the flux return of the superconducting solenoid.  Muons and noninteracting hadrons, which deposit only a small fraction of
their energy inside the calorimeter, and are referred to as minimum
ionizing particles (MIPS). While hadrons generally are absorbed in the steel, the muons continue to lose energy only by ionization and travel through the magnet and into the
proportional chambers that comprise the muon detector.

\begin{figure}[!hp]
\begin{center}
\hspace{2.5pt}
\includegraphics[width=14.5cm]{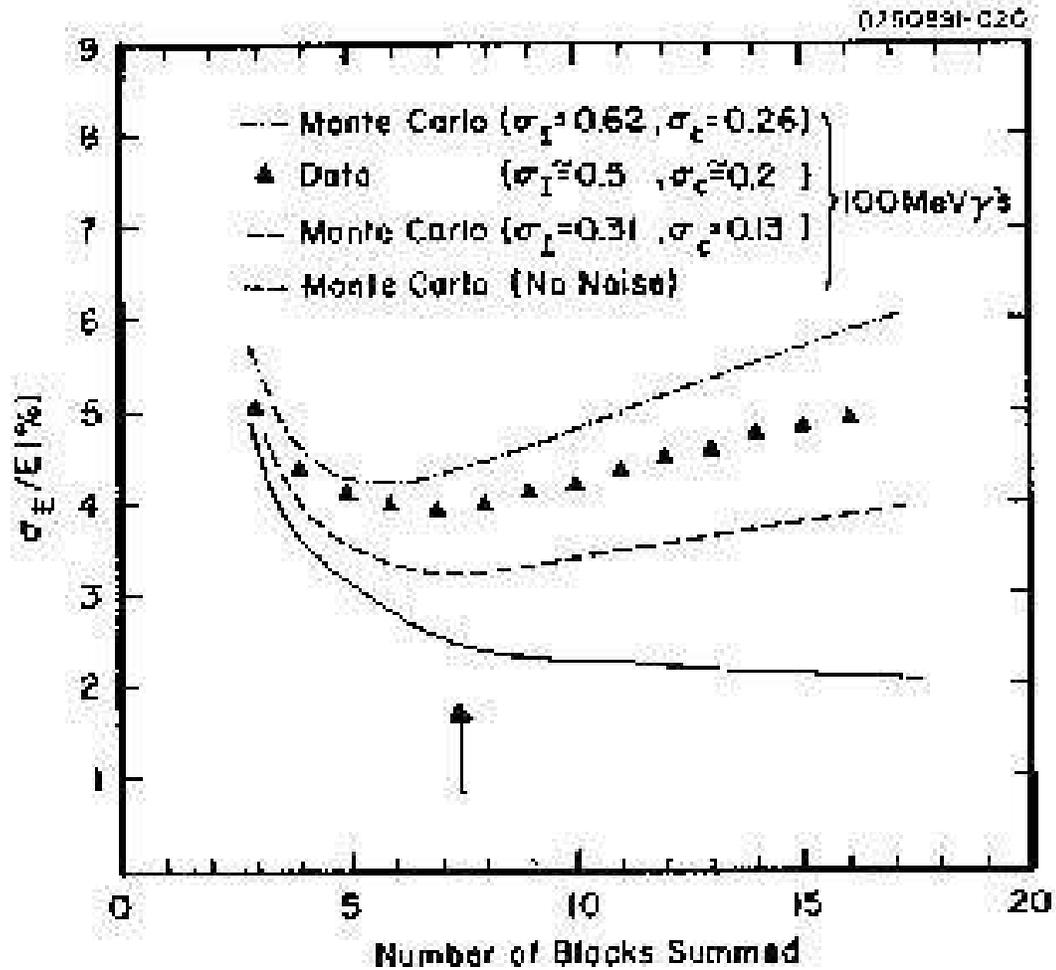}
\caption{Shower energy resolution as a function of the blocks used in
the shower reconstruction for the CLEO II detector.  The barrel calorimeter
has not changed since the installation of CLEO II.  The smooth lines are
from Monte-Carlo (MC) simulation of the 100 MeV photons.  The points are
experimental data from the 100 MeV transition photon in
$\Upsilon(3S)\rightarrow\gamma\chi_{bJ}(2P)$.  The arrow indicates the
number of crystals used in the shower reconstruction of 100 MeV photons.}
\vspace{0.2cm}
\label{fig:CC_res}
\end{center}
\end{figure}

Electrons and photons lose energy through the successive generation of
bremsstrahlung photons and $e^+e^-$ production, together known as an
electromagnetic shower.  These showers produce numerous low-energy
electrons which are then captured by the thallium atoms.  The photons emitted by
the de-excitation of thallium, $\lambda=560$~nm, are invisible to the rest of the
crystal. This means, they can propagate through the rest of the 30
cm-long crystal and be collected by the photo-diodes mounted on the
back of the crystal. The energy resolution is about 4.0\% at 100 MeV and 2.2\% at 1
GeV.  The resolution of the total shower energy is increased when more
than one crystal is used in the reconstruction. This can be seen in
\FIG \ref{fig:CC_res}.  The center of the shower is then
determined by an energy-weighted average of the blocks used in the
sum.  The number of crystals used is logarithmic, and ranges
from 4 at 25 MeV to 13 at 1 GeV.

\begin{figure}[!hp]
\begin{center}
\hspace{2.5pt}
\includegraphics[width=14.5cm]{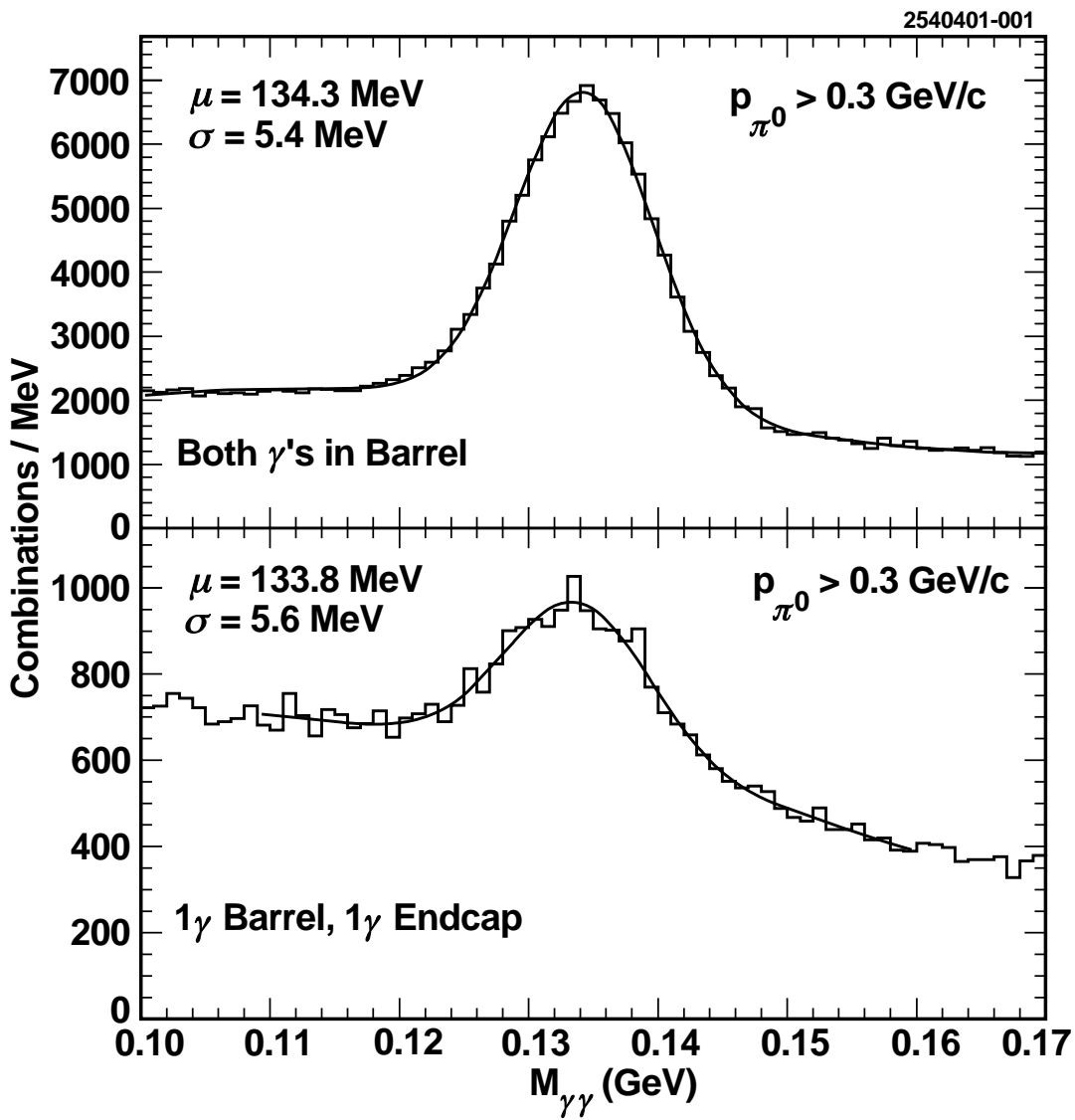}
\caption{$M(\gamma\gamma)$ resolution in CLEO III.}
\vspace{0.2cm}
\label{fig:CC_res2}
\end{center}
\end{figure}
\pagebreak
%%%%%%%%%%%%%%%%%%%%%%%%%%%%%%%%%%%%%%%%%%%%%%%%%%%%%%%%%%%%%%%%%%%%%
\subsection{Magnetic Field}
%%%%%%%%%%%%%%%%%%%%%%%%%%%%%%%%%%%%%%%%%%%%%%%%%%%%%%%%%%%%%%%%%%%%%
CLEO-c's 1-T magnetic field is provided by a large liquid-helium-cooled superconducting
solenoid with a diameter of 3 m and length of 3.5 m. The resulting field produced is uniform to
$\pm 0.02\%$ over the entire tracking volume.  The iron
flux return of the magnet is used in muon detection as an absorber,
which is described next.

%%%%%%%%%%%%%%%%%%%%%%%%%%%%%%%%%%%%%%%%%%%%%%%%%%%%%%%%%%%%%%%%%%%%%
\subsection{Muon Detection}
\begin{figure}[!hp]
\begin{center}
\hspace{2.5pt}
\includegraphics[width=14.5cm]{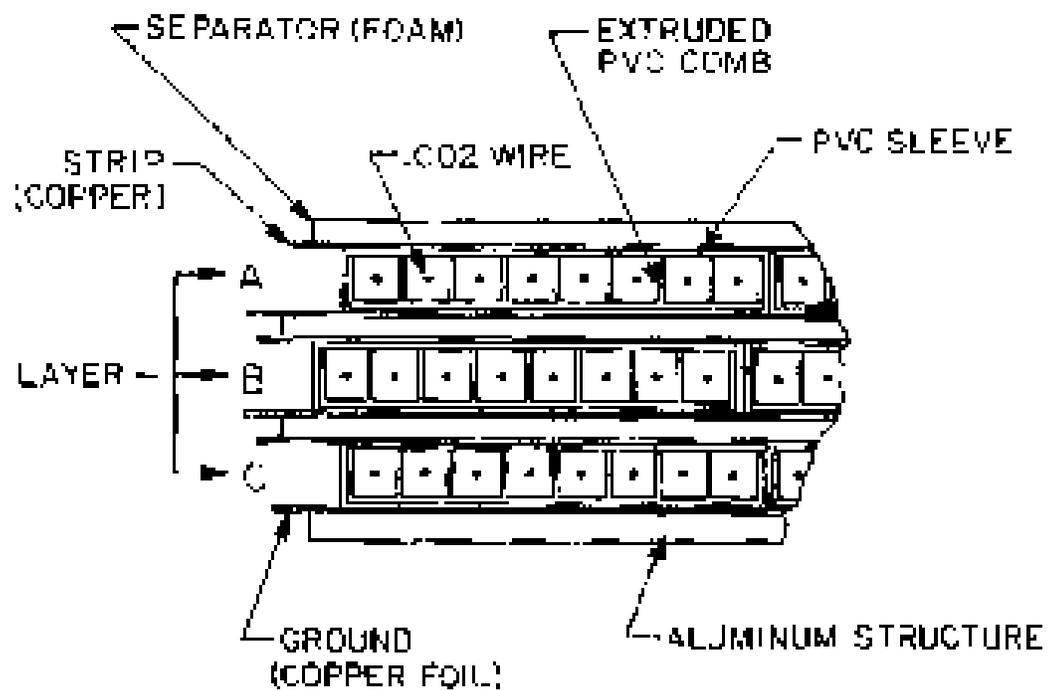}
\caption{Cross section of the muon chamber.  It consists of three
layers of 8-cell proportional counters.  The middle counter layer
is offset by one half a cell width to improve acceptance.}
\vspace{0.2cm}
\label{fig:muon_chamber}
\end{center}
\end{figure}

%%%%%%%%%%%%%%%%%%%%%%%%%%%%%%%%%%%%%%%%%%%%%%%%%%%%%%%%%%%%%%%%%%%%%
The muon detector exploits the fact that muons do not participate in
the strong interaction.  Since muons are much heavier than
electrons, they lose energy much more slowly as they travel through
material.  Therefore, muons can penetrate much greater depths of
material. The detector consists of three layers of gas-filled,
wire-proportional tracking chambers in between 36 cm iron absorbers
surrounding the detector (see \FIG \ref{fig:muon_chamber}).

The muon detector's layers provides information on a particle traveling
different interaction lengths.  The interaction
length is the average distance a charged hadron has to travel before
having an interaction.  The three layers of the detector are located at approximately 
3, 5, and 7 interaction lengths.  Information from this detector
is useful for identifying muons above about 1 GeV.  Its applicability to the studies with CLEO-c reported in this thesis is limited.

%%%%%%%%%%%%%%%%%%%%%%%%%%%%%%%%%%%%%%%%%%%%%%%%%%%%%%%%%%%%%%%%%%%%%%%%%%
\subsection{Data Acquisition}
%%%%%%%%%%%%%%%%%%%%%%%%%%%%%%%%%%%%%%%%%%%%%%%%%%%%%%%%%%%%%%%%%%%%%%%%%%
During the CLEO-c scan, bunch crossings happened at a rate on the order
of 1 MHz. However, the actual rate of interesting physics events
was much smaller, on the order of 1 Hz.

The CLEO-c detector includes hundreds of thousands of sensitive components
and associated electronic channels.  At every crossing each one of
these components can deliver a signal representing the traversal of a particle produced in the $e^+e^-$ annihilation, although in typical events only a small fraction of these events have valid data. In general data are read out locally and ``sparsified'' to eliminate uninteresting channels.  The nontrivial information is read out to on-line computers in tens of microseconds.
During the data-gathering process, the detector can not acquire any new events.
Because of this, it is essential to record only those events that
contain interesting physics.  This amount of ``dead-time'' is defined
as the time between the trigger signal and the end of the digitization
process.  The maximum readout-induced dead-time is $<3\%$ \cite{yellowbook}.

\begin{table}[!htbp]
\begin{center}
\caption{Definitions of the trigger lines.}
\vspace{0.2cm}
\label{tab:triggers}
\begin{tabular}{c|c}
\hline
{Name}& {Definition} \\ \hline
{Hadronic}& {$(N_{axial}>1)\&(N_{CB low}>0)$} \\ \hline
{Muon Pair}& {two back-to-back stereo tracks} \\ \hline
{Barrel Bhabha}& {back-to-back high showers in CB} \\ \hline
{End-cap Bhabha}& {back-to-back high showers in CE} \\ \hline
{Electron track}& {$(N_{axial}>0)\&(N_{CB med}>0)$} \\ \hline
{Tau}& {$(N_{stereo}>1)\&(N_{CB low}>0)$} \\ \hline
{Two Track}& {$(N_{axial}>1)$} \\ \hline
{Random}& {random 1 kHz source} \\ \hline
\end{tabular}
\end{center}
\end{table}

\begin{figure}[!hp]
\begin{center}
\hspace{2.5pt}
\includegraphics[width=14.5cm]{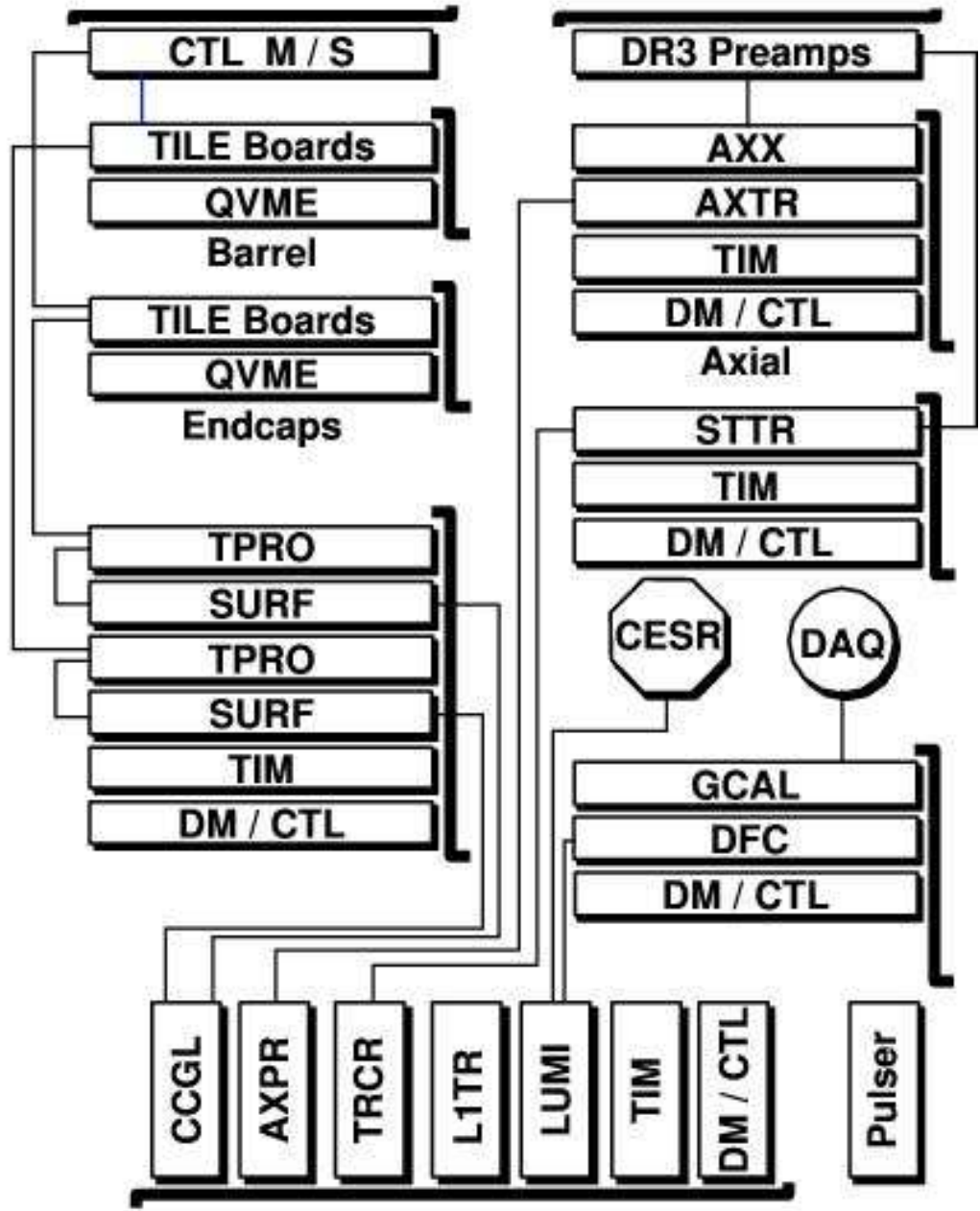}
\caption{Overview of the trigger system.}
\vspace{0.2cm}
\label{fig:Trigger}
\end{center}
\end{figure}

The selection of these interesting physics events is achieved with
a multi-layered trigger system
\cite{yellowbook,IEEE48552,IEEE48547,IEEE48562}. A schematic view of the
CLEO-c trigger system is shown in \FIG \ref{fig:Trigger}.  Currently, there
are eight trigger lines used, listed in \TAB \ref{tab:triggers}.  Data
from the DR and the CC are received
and processed in separate VME crates and yield basic trigger
parameters. These parameters are tracks counts, or multiplicity,
topology in the main drift chamber, and the number of showers and
topology in the calorimeter.  The information from both systems is
correlated by a global trigger which generates a ``pass'' signal every
time a valid trigger condition is satisfied.  The trigger system consists of two
tracking triggers, one using information provided by the axial wires
and the other using the stereo wires of the DR, a
CC trigger, and a decision and gating global trigger system.

\begin{figure}[!hp]
\begin{center}
\hspace{2.5pt}
\includegraphics[width=14.5cm]{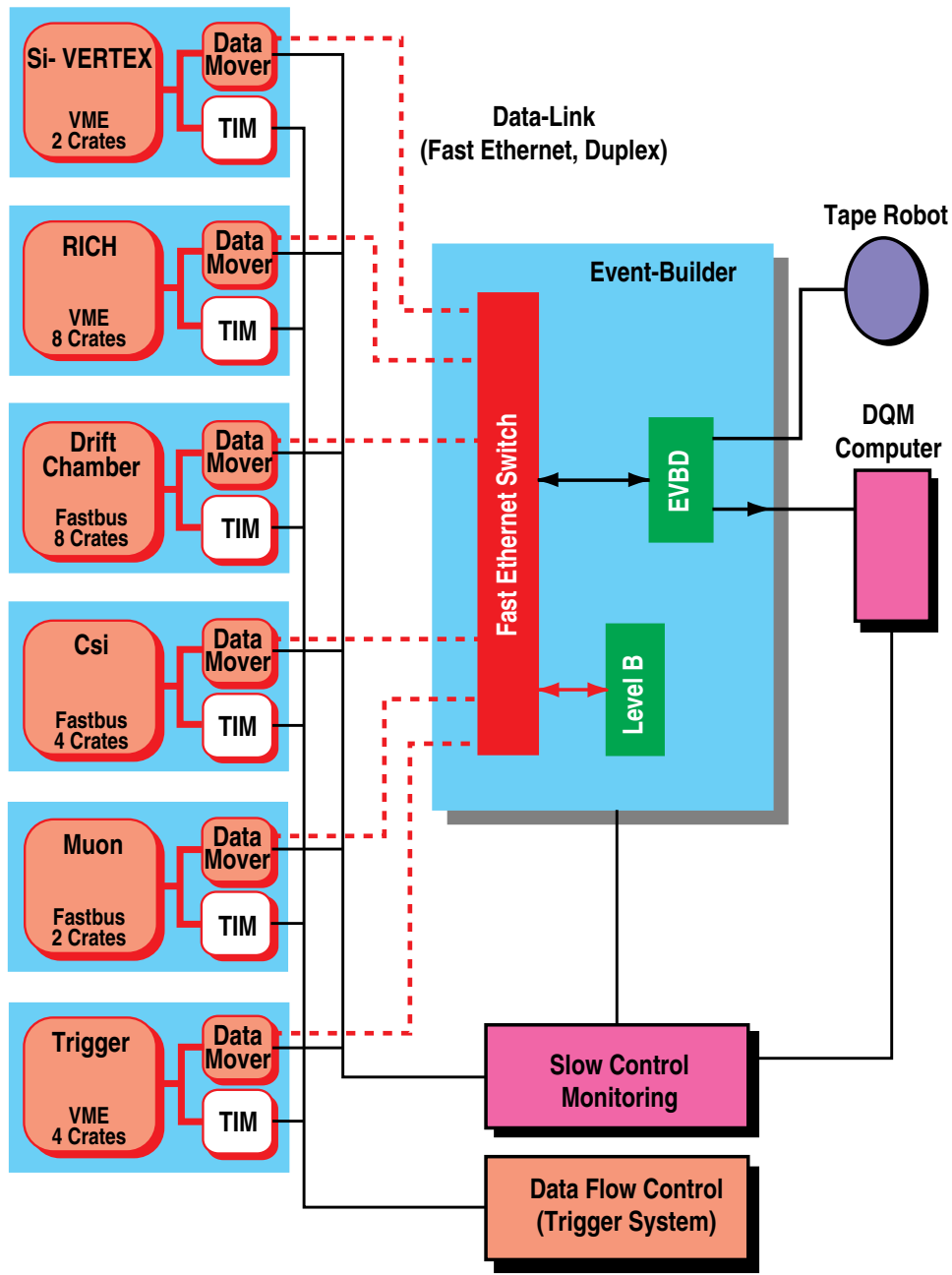}
\caption{Block diagram of the CLEO III data acquisition system (DAQ). The
only difference in the between CLEO III and CLEO-c detectors, and in
turn the DAQ system is the replacement of the silicon vertex detector
with the inner drift chamber. }
\vspace{0.2cm}
\label{fig:DAQ}
\end{center}
\end{figure}

The data acquisition system (DAQ) consists of two equally important
parts.\footnote{The review of the data acquisition system is based
upon Reference \cite{yellowbook}.}  The DAQ system is responsible for
the transfer of the data from the front end electronics to the mass
storage device, while the slow control monitors environmental conditions, the quality of the data,
and the status of detector components.

For each acceptable trigger from the CLEO-c detector, about 400,000
channels are digitized. The front-end electronics provide the data
conversion in parallel with a local buffer and waits for an
asynchronous readout by the DAQ.  A dedicated module, the Data-Mover, in each
front-end crate, assures transfer times of the data are below 500
$\mu{s}$, in addition to providing another buffer.  The Data-Mover
moves the data to the Event-Builder which reconstructs the accepted
events for transfer to mass storage.  Also, a fraction of
reconstructed events are analyzed on-the-fly by the CLEO monitoring
system, commonly referred to as {\tt{Online-Pass1}}, to quickly discover
problems and check the quality.

In addition to {\tt{Online-Pass1}}, there is an {\tt{Offline-Pass1}}, which is
commonly referred to as {\tt{Caliper}} (which stands for CALIbration and
PERformance monitoring).  {\tt{Caliper}}, unlike {\tt{Online-Pass1}}, can be run
over all the data in an efficient manner by applying harsh cuts to get at
the interesting physics events for data quality and monitoring.  As a
result, {\tt{Caliper}} was used extensively during the scan running used for the analysis in this thesis.  

A block diagram of the DAQ for the CLEO III detector is
shown in \FIG \ref{fig:DAQ}.  The only difference between the CLEO III
and the CLEO-c detector, and that of the DAQ, is the replacement of
the silicon vertex detector within the inner drift chamber.

\chapter{CLEO-c $D_{s}$ Scan}
%%%%%%%%%%%%%%%%%%%%%%%%%%%%%%%%%%%%%%%%%%%%%%%%%%%%%%%%%%%%%%%%%%%%%%%%%%
%

The CLEO-c project description (Yellow Book) \cite{yellowbook} includes 
measurements of $D_{s}$ branching fractions and other properties among 
the principal goals of the program. It makes a specific proposal that a
scan run be undertaken to determine the running energy above 
$D_{s}^{+}D_{s}^{-}$ threshold that will maximize the sensitivity for 
$D_s$ physics.  It also acknowledges that such a data sample would include
non-strange charmed mesons in the form of $D\bar{D}$, $D^{*}\bar{D}$ and
$D^*\bar{D}^*$ events produced in different quantum states from those
at the $\psi(3770)$.  It is on this scan data that this thesis is based.

\section{Data and Monte Carlo Samples}
\label{sec:samples}
\begin{figure}[ht]
\begin{center}
\hspace{2.5pt}
\includegraphics[width=14.5cm]{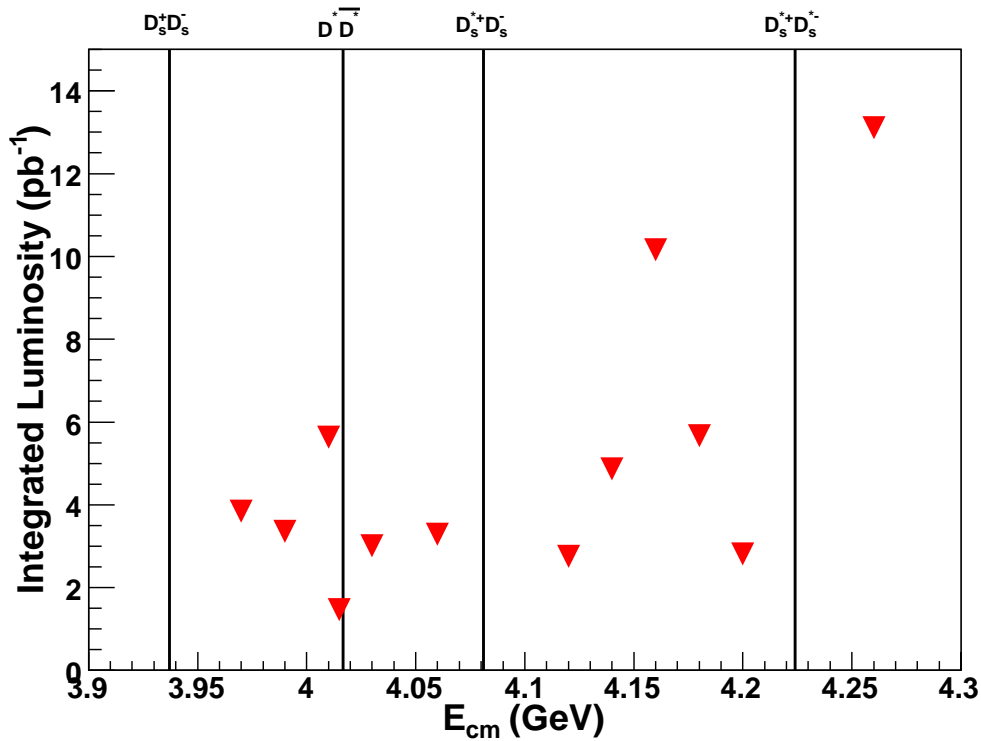}
\caption{Integrated luminosity as a function of energy: The two
largest integrated luminosity points correspond to 4160 and
4260~MeV. A point corresponding to 4170~MeV is left off, but has a
luminosity of about 178~pb${^{-1}}$}
\vspace{0.2cm}
\label{fig:Lum}
\end{center}
\end{figure}

%%%%%%%%%%%%%%%%%%%%%%%%%%%%%%%%%%%%%%%%%%%%%%%%%%%%%%%%%%%%%%%%%%%%%%%%%%
%
The scan run was designed to provide maximum information in the available running period of August-October, 2005.
The objective at each energy point was a measurement of the cross sections
for all accessible final states consisting of a pair of charmed mesons.  At 
the highest energy the possibilities include all of the following: 
$D\bar{D}$, $D^{*}\bar{D}$, $D^{*}\bar{D}^{*}$, $D_{s}\bar{D}_{s}$, 
$D_{s}^{*}\bar{D}_{s}$, and $D_{s}^{*}\bar{D}_{s}^{*}$, where the first four include both charged and neutral states.  The original plan 
for the scan included ten energy points, with an integrated luminosity target 
for each point of $\sim 5$pb$^{-1}$.  It was recognized that specific 
energies might reveal themselves as unpromising with less than this 
luminosity.  The plan was therefore designed to be flexible, with the 
option of adding or revisiting points to the extent that the data suggested
and time allowed.  In the end two points were added to the original list, 
including one at 4260~MeV to investigate the $Y(4260)$.  The center-of-mass
energies and integrated luminosities for the twelve scan points are given
in \TAB \ref{lum_table} and \FIG \ref{fig:Lum}.  An additional point
was added as a result of the scan and corresponds to the location that
maximizes the $D_{s}$ yield.  This point, 4170~MeV, is added to this
analysis and its larger data sample was essential in understanding the nature of charm
production throughtout this energy region.

\begin{table}[!htbp]
\begin{center}
\caption{Center-of-mass energies and integrated luminosities for all
data points in the CLEO-c $D_s$ scan.}
\label{lum_table}
\vspace{0.2cm}
\begin{tabular}{|c|c|}\hline
\raisebox{0pt}[12pt][6pt]{$\sqrt{s}$ (MeV)} &
\raisebox{0pt}[12pt][6pt]{${\int\cal{L}}dt$ (nb$^{-1}$)}
\\
\hline
\raisebox{0pt}[12pt][6pt]{$3970$} &
\raisebox{0pt}[12pt][6pt]{$3854.30 \pm 8.48$} \\ \hline
\raisebox{0pt}[12pt][6pt]{$3990$} &
\raisebox{0pt}[12pt][6pt]{$3356.45 \pm 7.73$} \\ \hline
\raisebox{0pt}[12pt][6pt]{$4010$} &
\raisebox{0pt}[12pt][6pt]{$5625.65 \pm 10.20$} \\ \hline
\raisebox{0pt}[12pt][6pt]{$4015$} &
\raisebox{0pt}[12pt][6pt]{$1470.35 \pm 5.32$} \\ \hline
\raisebox{0pt}[12pt][6pt]{$4030$} &
\raisebox{0pt}[12pt][6pt]{$3005.55 \pm 7.48$} \\ \hline
\raisebox{0pt}[12pt][6pt]{$4060$} &
\raisebox{0pt}[12pt][6pt]{$3285.65 \pm 7.79$} \\ \hline
\raisebox{0pt}[12pt][6pt]{$4120$} &
\raisebox{0pt}[12pt][6pt]{$2759.20 \pm 7.29$} \\ \hline
\raisebox{0pt}[12pt][6pt]{$4140$} &
\raisebox{0pt}[12pt][6pt]{$4871.85 \pm 9.77$} \\ \hline
\raisebox{0pt}[12pt][6pt]{$4160$} &
\raisebox{0pt}[12pt][6pt]{$10155.40 \pm 13.96$} \\ \hline
\raisebox{0pt}[12pt][6pt]{$4160$} &
\raisebox{0pt}[12pt][6pt]{$178942.15 \pm 59.37$} \\ \hline
\raisebox{0pt}[12pt][6pt]{$4180$} &
\raisebox{0pt}[12pt][6pt]{$5666.90 \pm 10.51$} \\ \hline
\raisebox{0pt}[12pt][6pt]{$4200$} &
\raisebox{0pt}[12pt][6pt]{$2809.90 \pm 7.59$} \\ \hline
\raisebox{0pt}[12pt][6pt]{$4260$} &
\raisebox{0pt}[12pt][6pt]{$13107.60 \pm 16.45$} \\ \hline
\end{tabular}
\end{center}
\end{table}

Numerous Monte Carlo (MC) samples have been generated in the development of the
procedures and the determination of efficiencies and backgrounds for this 
analysis.  The goal in specifying these MC samples was not to reproduce
reality precisely, but to include all relevant 
final states in sufficient quantity to develop selection criteria and assess 
the potential for cross-feed backgrounds.  Thirteen 100~pb$^{-1}$ samples (one 
for each center-of-mass energy), both signal and continuum, were generated.  
At each energy the signal MC sample includes all kinematically allowed 
$D_{(s)}^{(*)}\bar{D}_{(s)}^{(*)}$ final states.  The continuum
samples included only the $uds$ background. 
All MC samples for the scan were generated on a system of $\sim50$ computers operated by the Minnesota CLEO-c group (``MN MC farm''). A 
breakdown of the signal MC samples is shown in \TAB \ref{tab:MC1}.  Additional
samples, described later, were subsequently produced to aid in assessing the
potential contributions of ``multi-body'' production.
\begin{table}[!htbp]
\label{tab:MC1}
\vspace{0.2cm}
\caption{Composition of the $100$~pb$^{-1}$ signal MC
samples. The assumed total charm cross section was $11$~nb for
$E_{\rm{cm}}=4015-4200$~Mev and $5$~nb for $E_{\rm{cm}}=3970-4015$ and $4260$~MeV.}
\vspace{0.2cm}

\begin{center}
\begin{tabular}{|c|c|c|c|c|c|}\hline
Event&3970&4015&4020-4080& 4080-4200&
4260\\ \hline
{\(D^{0}\bar{D^{0}}\)}& \(11.6\%\) & \(6.9\%\) & \(6.9\%\) &
\(6.6\%\) & \(6.3\%\) \\ \hline
{\(D^{+}{D^{-}}\)}& \(11.6\%\) & \(6.9\%\) & \(6.9\%\) &
\(6.6\%\) & \(6.3\%\) \\ \hline
{\(D^{*0}\bar{D^{0}}\)} & \(35\%\) & \(13.7\%\) & \(13.7\%\) &
\(13.2\%\) & \(12.6\%\) \\ \hline
{\(D^{*+}{D^{-}}\)} & \(35\%\) & \(13.7\%\) & \(13.7\%\) &
\(13.2\%\) & \(12.6\%\) \\ \hline
{\(D^{*0}\bar{D^{*0}}\)} & \(-\) & \(54.9\%\) & \(27.4\%\) &
\(26.4\%\) & \(25.1\%\) \\ \hline
{\(D^{*+}{D^{*-}}\)} &  \(-\) & \(-\) & \(27.4\%\) &
\(26.4\%\) & \(25.1\%\) \\ \hline
{\(D_{s}^{+}D_{s}^{-}\)}& \(7\%\) & \(4\%\) & \(4\%\) &
\(2.5\%\) & \(1\%\) \\ \hline
{\(D_{s}^{*+}D_{s}^{-}\)}& \(-\) & \(-\) & \(-\) &
\(5\%\) & \(2\%\) \\ \hline
{\(D_{s}^{*+}D_{s}^{*-}\)} & \(-\) & \(-\) & \(-\) &
\(-\) & \(9\%\) \\ \hline
\end{tabular}
\end{center}
\end{table}
%

%%%%%%%%%%%%%%%%%%%%%%%%%%%%%%%%%%%%%%%%%%%%%%%%%%%%%%%%%%%%%%%%%%%%%%%%%%
%       SECTION 3
%%%%%%%%%%%%%%%%%%%%%%%%%%%%%%%%%%%%%%%%%%%%%%%%%%%%%%%%%%%%%%%%%%%%%%%%%%
\section{Decay Modes and Reconstruction}
\label{sec:recon}
%%%%%%%%%%%%%%%%%%%%%%%%%%%%%%%%%%%%%%%%%%%%%%%%%%%%%%%%%%%%%%%%%%%%%%%%%%
%
Eight $D_s$ modes are used to measure
the $D_s$ production at each energy point. These modes are
listed in \TAB \ref{Dsmode_table}. In addition to these $D_{s}$
decays, several $D^{0}$ and $D^{+}$ decay modes, given in  \TAB
\ref{Dmode_table}, were used to investigate the amount of $D\bar{D}$,
$D^{*}\bar{D}$, and $D^{*}\bar{D}^{*}$ produced at each energy point.

\begin{table}[!tb]
\begin{center}
\caption{The $D_{s}$ branching fractions including the updated
branching fractions, in $10^{-2}$.}
\vspace{0.2cm}
\label{Dsmode_table}
\begin{tabular}{|c|c|}
\hline
{Modes}& {Branching Fraction}
\\ \hline
\(\phi\pi^{+}\), 10~MeV cut on the Invariant
\(\phi\rightarrow{K}^{+}K^{-}\) Mass \cite{CLEO_DsBF}&\(1.98\pm0.15\)
\\ \hline
\(K^{*0}K^{+},K^{*0}\rightarrow{K^{-}\pi^{-}}\)~\cite{pdg}&\(2.2\pm0.6\)
\\ \hline
\(\eta\pi^{+},\eta\rightarrow{\gamma\gamma}\)~\cite{pdg,CLEO_DsBF}&\(0.58\pm0.07\)
\\ \hline
\(\eta\rho^{+},\eta\rightarrow{\gamma\gamma},\rho^{+}\rightarrow{\pi^{+}\pi^{0}}\)~\cite{pdg}&\(4.3\pm1.2\)
\\ \hline
\(\eta^{'}\pi^{+},\eta^{'}\rightarrow{\pi^{+}\pi^{-}\eta},\eta\rightarrow{\gamma\gamma}\)~\cite{pdg,CLEO_DsBF}&\(0.7\pm0.1\)
\\ \hline
\(\eta^{'}\rho^{+},\eta^{'}\rightarrow{\pi^{+}\pi^{-}\eta},\eta\rightarrow{\gamma\gamma},\rho^{+}\rightarrow{\pi^{+}\pi^{0}}\)~\cite{pdg}&\(1.8\pm0.5\)
\\ \hline
\(\phi\rho^{+},\phi\rightarrow{K^{+}K^{-}},\rho^{+}\rightarrow{\pi^{+}\pi^{0}}\)~\cite{pdg}&\(3.4\pm1.2\)
\\ \hline
\(K_{s}K^{+},K_{s}\rightarrow{\pi^{+}\pi^{-}}\)~\cite{pdg,CLEO_DsBF}&\(1.0\pm0.07\)
\\ \hline
\end{tabular}
\end{center}
\end{table}

\begin{table}[!htbp]
\begin{center}
\caption{Decay modes used to select $D^{0}$ and $D^{+}$ in the
scan data.  The branching fractions are from a published CLEO-c analysis \cite{CLEO_DHad_PRL,BFcbx}.}
\vspace{0.2cm}
\label{Dmode_table}
\begin{tabular}{|c|c|}
\hline
{Modes}& {Branching Fraction}
\\ \hline
\(D^{0}\) decay mode &
\\ \hline
 \(K^{-}\pi^{+}\)&\(3.91\pm0.12\%\)
\\ \hline
 \(K^{-}\pi^{+}\pi^{0}\)&\(14.94\pm0.56\%\)
\\ \hline
 \(K^{-}\pi^{+}\pi^{+}\pi^{-}\)&\(8.29\pm0.36\%\)
\\ \hline
\(D^{+}\) decay mode &
\\ \hline
\(K^{-}\pi^{+}\pi^{+}\)&\(9.52\pm0.37\%\)
\\ \hline
\(K^{-}\pi^{+}\pi^{+}\pi^{0}\)&\(6.04\pm0.28\%\)
\\ \hline
\(K_{s}\pi^{+}\)&\(1.55\pm0.08\%\)
\\ \hline
 \(K_{s}\pi^{+}\pi^{0}\)&\(7.17\pm0.43\%\)
\\ \hline
\(K_{s}\pi^{+}\pi^{-}\pi^{+}\)&\(3.2\pm0.19\%\)
\\ \hline
\(K^{+}K^{-}\pi^{+}\)&\(0.97\pm0.06\%\)
\\ \hline
\end{tabular}
\end{center}
\end{table}

In selecting these decays the CLEO-c standard {\tt DTAG}
\cite{DTAGcbx} code was used with the following modifications to
the usual criteria:
\begin{itemize}
\item{The $\frac{dE}{dX}$ requirements for charged pions and kaons were relaxed
from $3\sigma$ to $3.5\sigma$.}
\item{The $K_{s}$ mass requirement was tightened from $4.5\sigma$ to $3\sigma$.}
\item{The $\Delta E$ cut for tag selection was relaxed from $|\Delta E|< 0.1$~GeV to $|\Delta E|<0.5$~GeV}.
\item{The $M_{\rm{bc}}$ cut for tag selection was relaxed from $1.83<M_{\rm{bc}}<2.0$~GeV to $1.7<M_{\rm{bc}}<2.14$~GeV}.
\end{itemize}
\CONT In addition to the above, the following cuts on intermediate-particle masses relative to nominal values were 
applied for the $D_{s}$ modes:
\begin{itemize}
\item {\(|M_{\rho}-M^{PDG}_{\rho}|\leq150\){~MeV}}
\item {\(|M_{K^{*}}-M^{PDG}_{K^{*}}|\leq75\){~MeV}}
\item {\(|M_{\phi}-M^{PDG}_{\phi}|\leq10\){~MeV}}
\item {\(|M_{\eta'}-M^{PDG}_{\eta'}|\leq10\){~MeV}}
\end{itemize}

\begin{figure}[ht]
\begin{center}
\hspace{2.5pt}
\includegraphics[width=14.5cm]{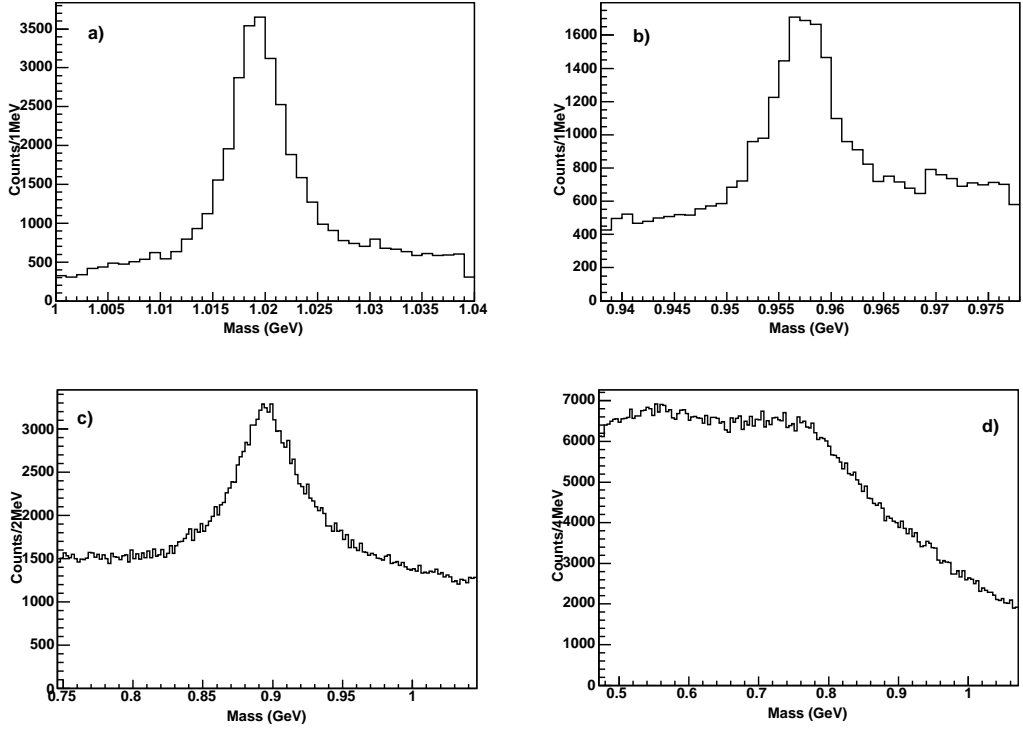}
\caption{Invariant-mass distributions for the intermediate states
involved in $D_{s}$ reconstruction.  The plots are for 4160~MeV MC:  a$)$
$\phi\rightarrow{K^{+}{K^{-}}}$,  b$)$
$\eta^{'}\rightarrow{\pi^{+}{\pi^{-}}\eta}$,
c$)$ $K^{*}\rightarrow{K^{-}{\pi^{+}}}$, and d$)$
$\rho^{+}\rightarrow{\pi^{+}{\pi^{0}}}$.
}
\vspace{0.2cm}
\label{fig:Inter_mass}
\end{center}
\end{figure}

\CONT MC invariant-mass plots for the selection of intermediate 
states used in the reconstruction of $D_{s}$ are shown in 
\FIG \ref{fig:Inter_mass}.
%%%%%%%%%%%%%%%%%%%%%%%%%%%%%%%%%%%%%%%%%%%%%%%%%%%%%%%%%%%%%%%%%%%%%%%%%%
%       SECTION 5
%%%%%%%%%%%%%%%%%%%%%%%%%%%%%%%%%%%%%%%%%%%%%%%%%%%%%%%%%%%%%%%%%%%%%%%%%%
\section{Event Selection and Kinematics}
\label{sec:eventsel}
%%%%%%%%%%%%%%%%%%%%%%%%%%%%%%%%%%%%%%%%%%%%%%%%%%%%%%%%%%%%%%%%%%%%%%%%%%
%
We begin by assuming that charm production in the threshold region is 
dominated by final states with two charmed mesons and no other particles: 
$e^+e^-\rightarrow{D_{(s)}^{(*)}\bar{D}_{(s)}^{(*)}}$.  For these
events, the energy and momentum of the primary charmed mesons 
are well defined. In general, for $e^{+}e^{-}\rightarrow{X}{Y}$ in the
center-of-mass frame we have
\begin{equation}
        E_{X} = \frac{s + M_{X}^{2} - M_{Y}^{2}}{2\sqrt{s}}
\label{eq:ED}	
\end{equation}
\CONT and
\begin{equation}
        |\vec{P_{X}}| = |\vec{P_{Y}}| = |\vec{P}| = \sqrt{E_{X}^{2} - M_{X}^{2}},
\label{eq:PD}
\end{equation}
\CONT where $s=4E_{\rm{beam}}^{2}$.  The energy and momentum of one of the
two charmed mesons is therefore sufficient to assign an event to one of the
possible two-body processes.  In practice, we use familiar forms of these
variables for this classification: the candidate's beam-constrained mass 
($M_{\rm{bc}} \equiv \sqrt{E_{\rm beam}^{2}-|\vec{P}|^{2}}~$) and its
energy deficit relative to the beam ($\Delta E \equiv E_{\rm{beam}}- E_{D}$).

As the center-of-mass energy increases above $D\bar{D}$ and $D_{s}^{+}{D}_{s}^{-}$
thresholds, it becomes possible to produce the ``starred''
states, $D^{*0}$, $D^{*+}$ and $D_{s}^{*+}$. These are not fully reconstructed 
in this analysis, since momenta and energies are sufficient to identify the
origin of the reconstructed $D_{(s)}$. Reconstructed $D$ and $D_{s}$
candidates from these sources do not have a
well-defined momentum, since they are daughters of starred parents and
exhibit Doppler broadening.  This Doppler broadening manifests itself
through smeared distributions in both energy and momentum. Some properties of the
intermediate starred states are summarized in \TAB \ref{DStarmode_table}.

\begin{table}[!htbp]
\begin{center}
\caption{$D^{*0}$, $D^{*+}$, and $D_{s}^{*+}$ decay modes \cite{pdg}}
\label{DStarmode_table}
\vspace{0.2cm}
\begin{tabular}{|c|c|}
\hline
{Modes}& {Branching Fraction}
\\ \hline
\(D^{*0}\) decays mode &
\\ \hline
 \(D^{0}\pi^{0}\)&\(61.9\pm2.9\%\)
\\ \hline
 \(D^{0}\gamma\)&\(38.1\pm2.9\%\)
\\ \hline
\(D^{*+}\) decays mode &
\\ \hline
\(D^{0}\pi^{+}\)&\(67.7\pm0.5\%\)
\\ \hline
\(D^{+}\pi^{0}\)&\(30.7\pm0.5\%\)
\\ \hline
\(D^{+}\gamma\)&\(1.6\pm0.4\%\)
\\ \hline
\(D_{s}^{*+}\) decays mode &
\\ \hline
\(D_{s}^{+}\gamma\)&\(94.2\pm2.5\%\)
\\ \hline
\(D_{s}^{+}\pi^{0}\)&\(5.8\pm2.5\%\)
\\ \hline
\end{tabular}
\end{center}
\end{table}

\begin{figure}[!htbp]
\begin{center}
\hspace{2.5pt}
\includegraphics[width=14.5cm]{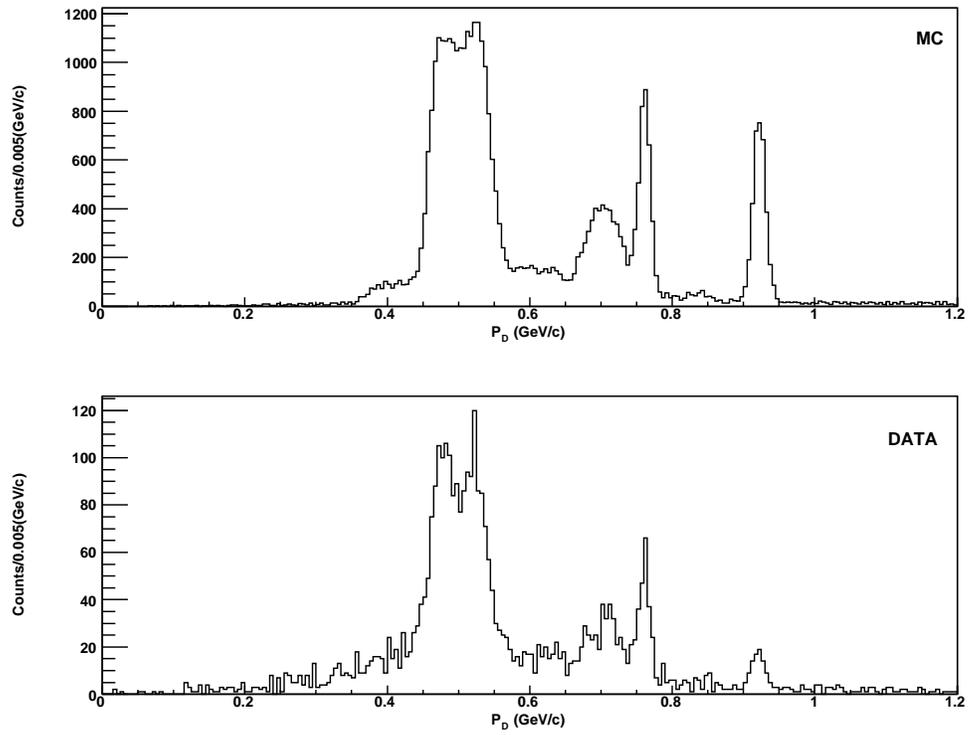}
\caption{The momentum spectrum at 4160~MeV for
$D^{0}\rightarrow{K^{-}}{\pi^{+}}$ candidates within 15~MeV of the
nominal mass. The top plot is from the MC described in the text
and the bottom plot is from the $10.16$~pb${^{-1}}$ of data at this energy. The
three distinct concentrations of entries near 0.95, 0.73 and 
0.5~GeV/$c$ correspond to $D\bar{D}$, $D^{*}\bar{D}$,
and $D^{*}\bar{D}^{*}$ production, respectively.}
\vspace{0.2cm}
\label{fig:D0_mom_spectrum}
\end{center}
\end{figure}

\begin{figure}[!htbp]
\begin{center}
\hspace{2.5pt}
\includegraphics[width=14.5cm]{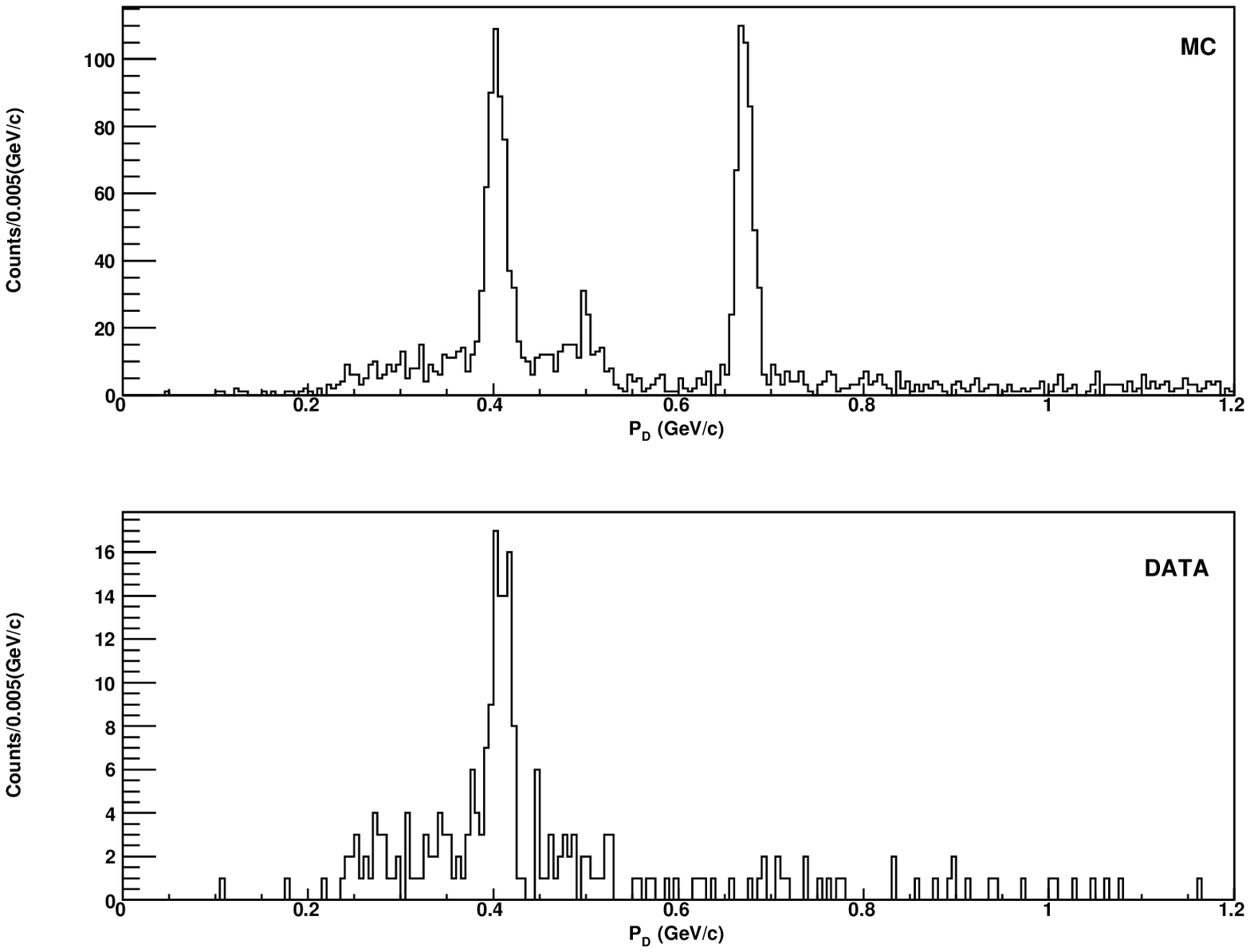}
\caption{The momentum spectrum at 4160~MeV for
$D_{s}^{+}\rightarrow{\phi}{\pi^{+}}$ candidates within 15~MeV of the
nominal mass. The top plot is from the generic MC discussed in the text
and the bottom plot is from the $10.16$~pb${^{-1}}$ of data at this energy.  The
peak at $\sim0.675$~MeV/$c$ is missing from the data, giving advanced
notice that the cross section for ${D_{s}^{+}}{D_{s}^{-}}$ at this energy 
is consistent with zero.
}
\vspace{0.2cm}
\label{fig:Ds_mom_spectrum}
\end{center}
\end{figure}

To illustrate the separation of events we show the momentum spectrum of $D^{0}$ 
candidates within 15~MeV of the nominal mass in \FIG \ref{fig:D0_mom_spectrum} for 
the center-of-mass energy 4160~MeV. The top plot in \FIG \ref{fig:D0_mom_spectrum} is
from the MC described in Sect.~\ref{sec:samples} and the bottom plot is from the 10.16~pb${^{-1}}$ of
data. There are three distinct concentrations of entries near 0.95, 0.73 and 0.5~GeV/$c$,
corresponding to $D\bar{D}$, $D^{*}\bar{D}$, and $D^{*}\bar{D}^{*}$ production,
respectively.  Similar distributions for $D_{s}$ candidates are shown
in \FIG \ref{fig:Ds_mom_spectrum}. In this case there are two distinct peaks in the MC
at 0.675 and 0.4~GeV/$c$, corresponding to $D_{s}^{+}{D_{s}^{-}}$ and 
$D_{s}^{*+}{D_{s}^{-}}$, respectively, which are the only two accessible final states
at this energy.  Only the peak at 0.4~GeV/$c$ is visible in data, demonstrating that the 
cross section for $D_{s}^{+}{D_{s}^{-}}$ at this energy is consistent with zero.  At 
center-of-mass energies above $D_{s}^{*+}D_{s}^{*-}$ threshold, such as 4260~MeV (not shown), 
there are three accessible final states, so three peaks in momentum are possible.

\begin{figure}[!htbp]
\begin{center}
\hspace{2.5pt}
\includegraphics[width=14.5cm]{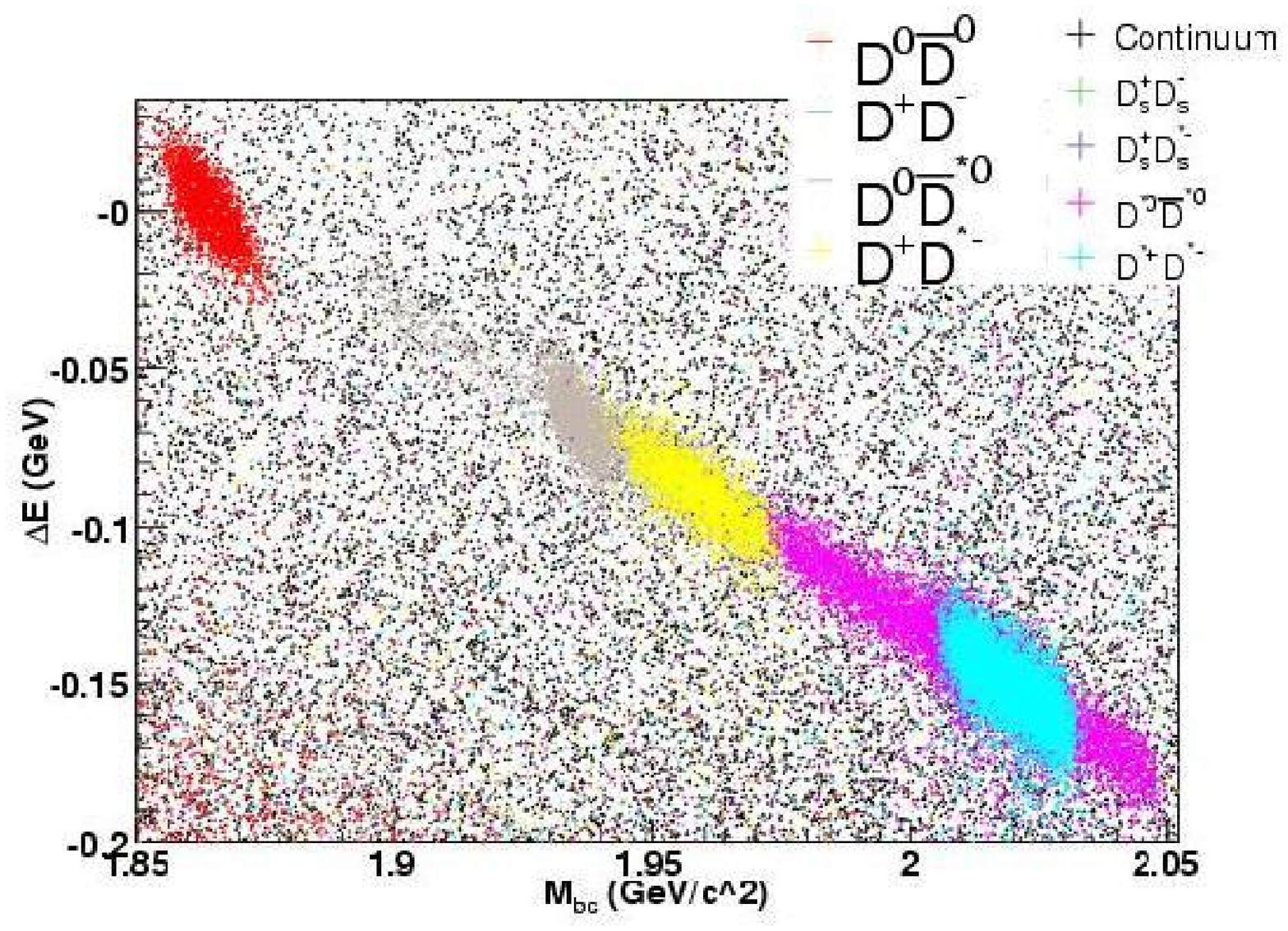}
\caption{$\Delta E$ vs. $M_{\rm{bc}}$ for $D^{0}\rightarrow{K^{-}\pi^{+}}$ at 4160 MeV using the 100 pb$^{-1}$ MC sample.  The plot 
illustrates the clear separation that is achieved for this choice of variables.}
\vspace{0.2cm}
\label{fig:D0kp_de_mbc}
\end{center}
\end{figure}

\begin{figure}[!htbp]
\begin{center}
\hspace{2.5pt}
\includegraphics[width=14.5cm]{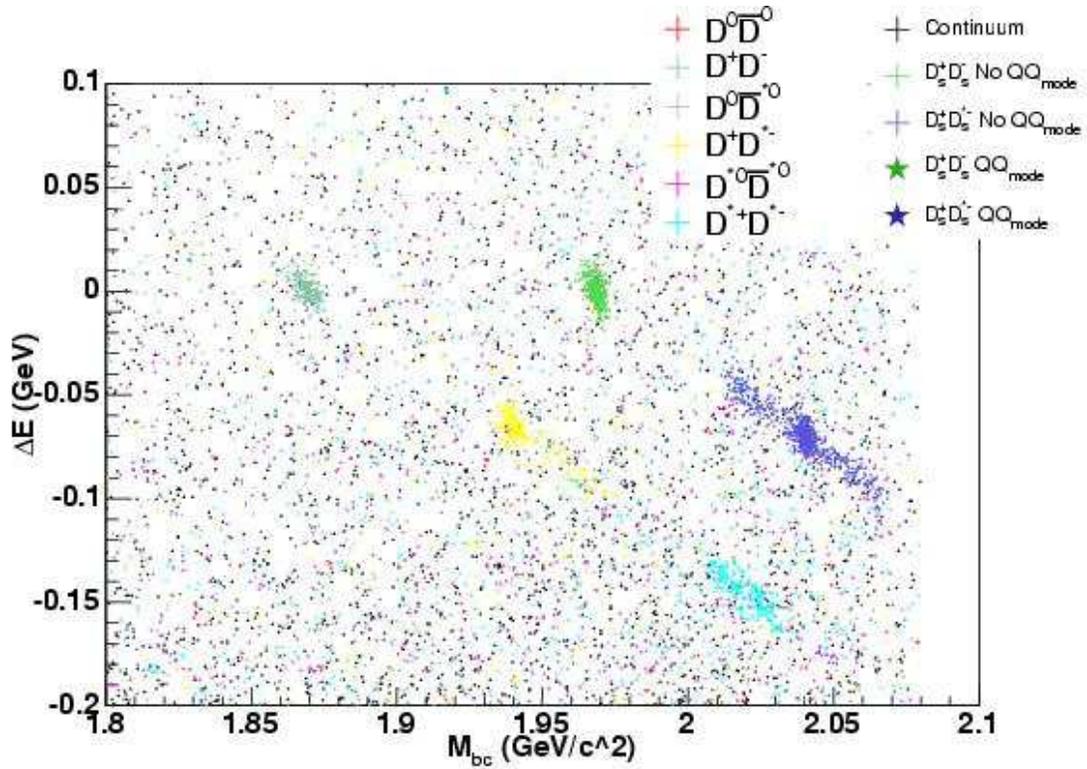}
\caption{$\Delta E$ vs. $M_{\rm{bc}}$ for $D_{s}^{+}\rightarrow{\phi}{\pi^{+}}$ at 4160 MeV using the 100 pb$^{-1}$ MC sample.  The plot 
illustrates the clear separation that is achieved by the choice of variables not just for $D_s$ but also for $D^{+}$ since 
the branching ratio for $D^{+}\rightarrow{\phi}{\pi^{+}}$ is nonzero.}
\vspace{0.2cm}
\label{fig:Ds_de_mbc}
\end{center}
\end{figure}

For this analysis we need to separate clearly and measure the cross
sections for the nine event types:  $D^{0}\bar{D}^{0}$,
$D^{*0}\bar{D}^{0}$, $D^{*0}\bar{D}^{*0}$, $D^{+}{D}^{-}$, $D^{*+}{D}^{-}$,
$D^{*+}{D}^{*-}$, $D_{s}^{+}{D}_{s}^{-}$, $D_{s}^{*+}{D}_{s}^{-}$,
and $D_{s}^{*+}{D}_{s}^{*-}$.  We want to choose variables that
are as orthogonal as possible for the purpose of this separation.  For
$D^{0}\bar{D}^{0}$, $D^{+}D^{-}$ and $D_{s}^{+}D_{s}^{-}$, the variables used
were the $D$ candidate's energy ($\Delta E$) and its
momentum (in the form of $M_{\rm{bc}}$).  
The separation of $D$-meson candidates at 4160 MeV into the possible event types 
is illustrated in the $\Delta E$ vs. $M_{\rm{bc}}$ plot in \FIG \ref{fig:D0kp_de_mbc}. 
The corresponding plot for $D_s$ is given in \FIG \ref{fig:Ds_de_mbc}.  The quantitative 
task is to count the events and determine the cross section for each event category while 
controlling contributions from backgrounds and cross-feed from other two-charm final states.  

\begin{figure}[!htbp]
\begin{center}
\hspace{2.5pt}
\includegraphics[width=14.5cm]{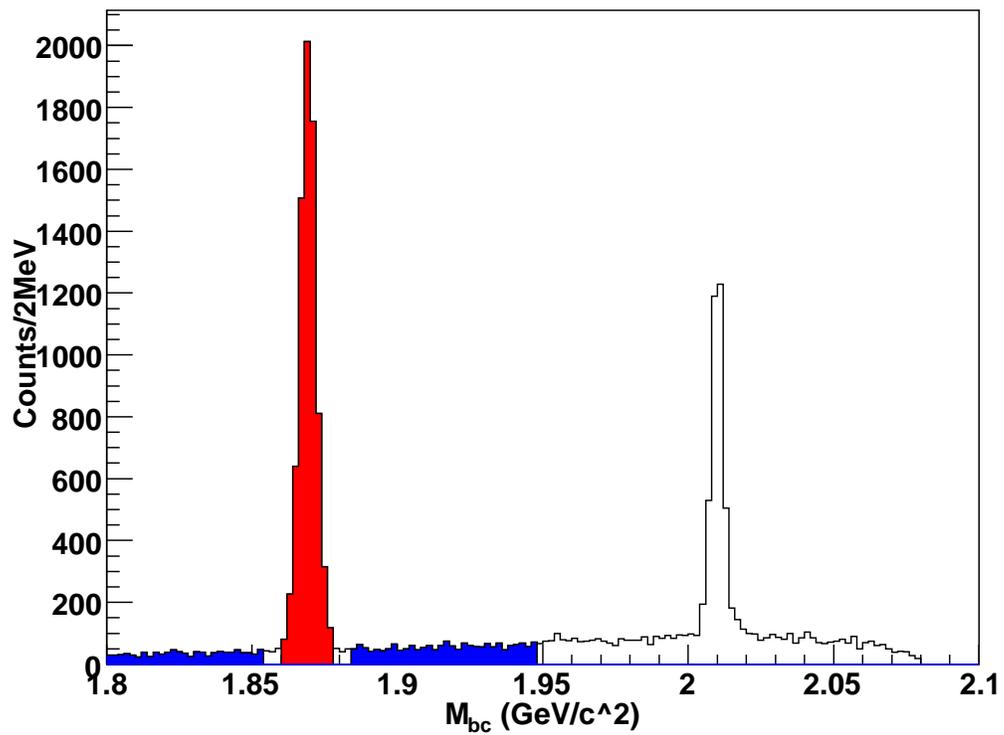}
\caption{Plot of $M_{\rm{bc}}$ for $D^+\rightarrow{K^{-}\pi^+\pi^+}$ in $D^{+}D^{-}$ events in MC at 
4160~MeV.  The red indicates the signal region and the blue the sideband.}
\vspace{0.2cm}
\label{fig:Dp_mbc_SB}
\end{center}
\end{figure}

\begin{figure}[!htbp]
\begin{center}
\hspace{2.5pt}
\includegraphics[width=14.5cm]{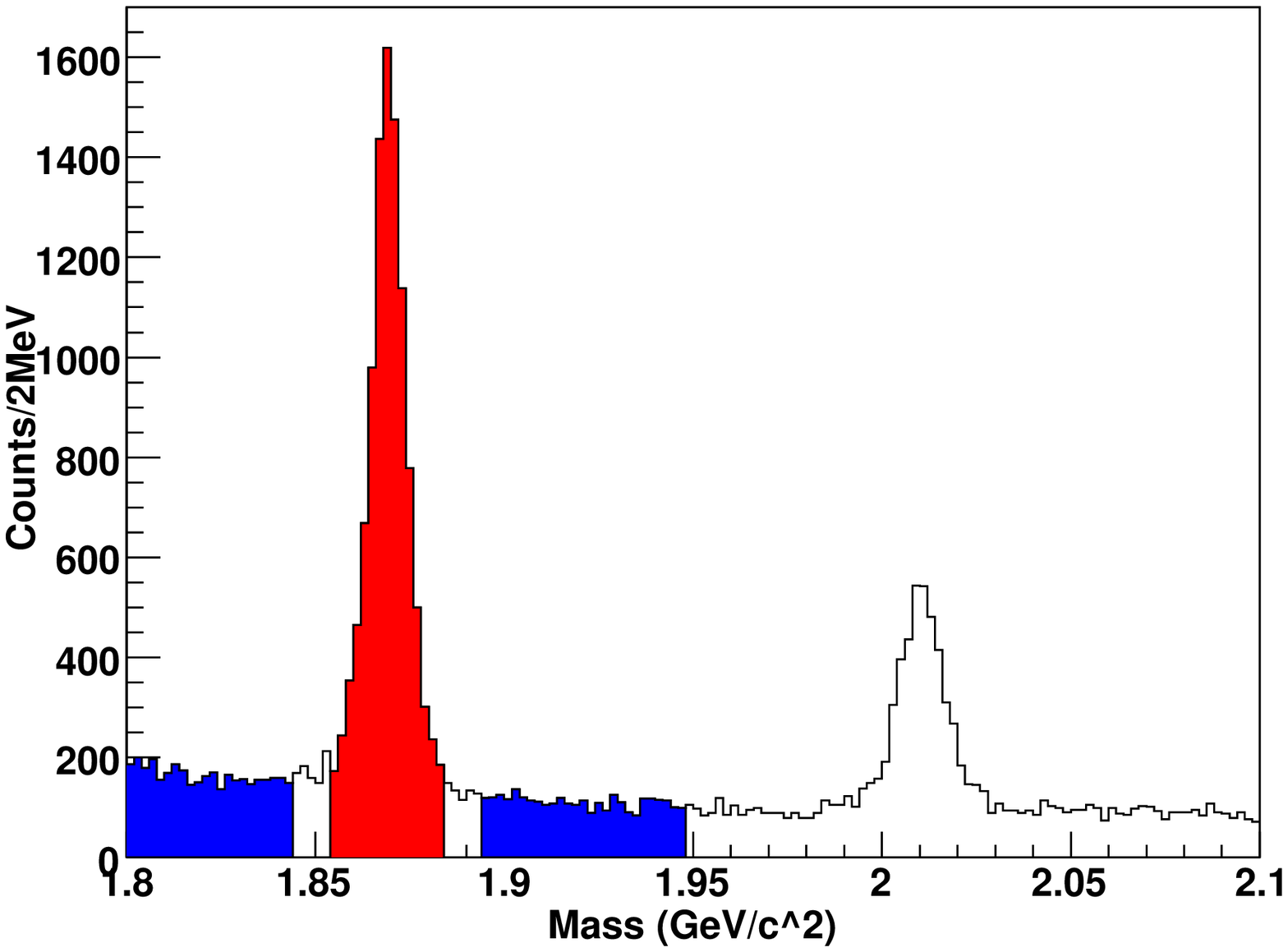}
\caption{Plot of the invariant mass for $D^+\rightarrow K^{-}\pi^+\pi^+$ in $D^{*+}D^{*-}$ events
in MC at 4160 MeV.  The red indicates the signal region and the blue the sideband.}
\vspace{0.2cm}
\label{fig:DpSS_mass_SB}
\end{center}
\end{figure}

To obtain the number of signal candidates for each event type, a signal region must first be 
defined.  The signal region in $M_{\rm{bc}}$ for $D^{0}\bar{D}^{0}$, $D^{+}D^{-}$ and 
$D_{s}^{+}D_{s}^{-}$ is $\pm 9$~MeV around the respective particle masses.  For the other event
types ($D^{*0}\bar{D}^{0}$, $D^{*0}\bar{D}^{*0}$, $D^{*+}{D}^{-}$, $D^{*+}{D}^{*-}$, 
$D_{s}^{*+}{D}_{s}^{-}$, and $D_{s}^{*+}{D}_{s}^{*-}$) the requirement is $\pm 15$~MeV in invariant
mass.  To estimate combinatoric and other backgrounds, a sideband is defined on either side of the 
signal region.  In every case the sidebands are spaced from the nominal particle mass by 5~$\sigma$.  
The sizes vary from mode to mode because of differing resolutions and the need to exclude
potential peaking contributions (such as decay modes that are common between $D^+$ and $D_s^+$). 
These sidebands are all chosen to be significantly larger than the signal region to minimize the 
statistical uncertainty of the background subtraction.  MC is used to determine the sideband
normalization, which is defined as the total sideband yield divided by the MC-tagged background 
contribution in the signal region.  In almost all cases the normalization given by MC
is consistent with the ratio of the sizes of the signal and sideband regions.  The background 
procedure is illustrated with one example for $M_{\rm{bc}}$ 
($D^+ \rightarrow K^- \pi^+ \pi^+$ in $D^{+} D^{-}$) and one for invariant mass 
($D^+ \rightarrow K^- \pi^+ \pi^+$ in $D^{*+} D^{*-}$) in \FIGS \ref{fig:Dp_mbc_SB} and 
\ref{fig:DpSS_mass_SB}, respectively.

\begin{figure}[!htbp]
\begin{center}
\hspace{2.5pt}
\includegraphics[width=14.5cm]{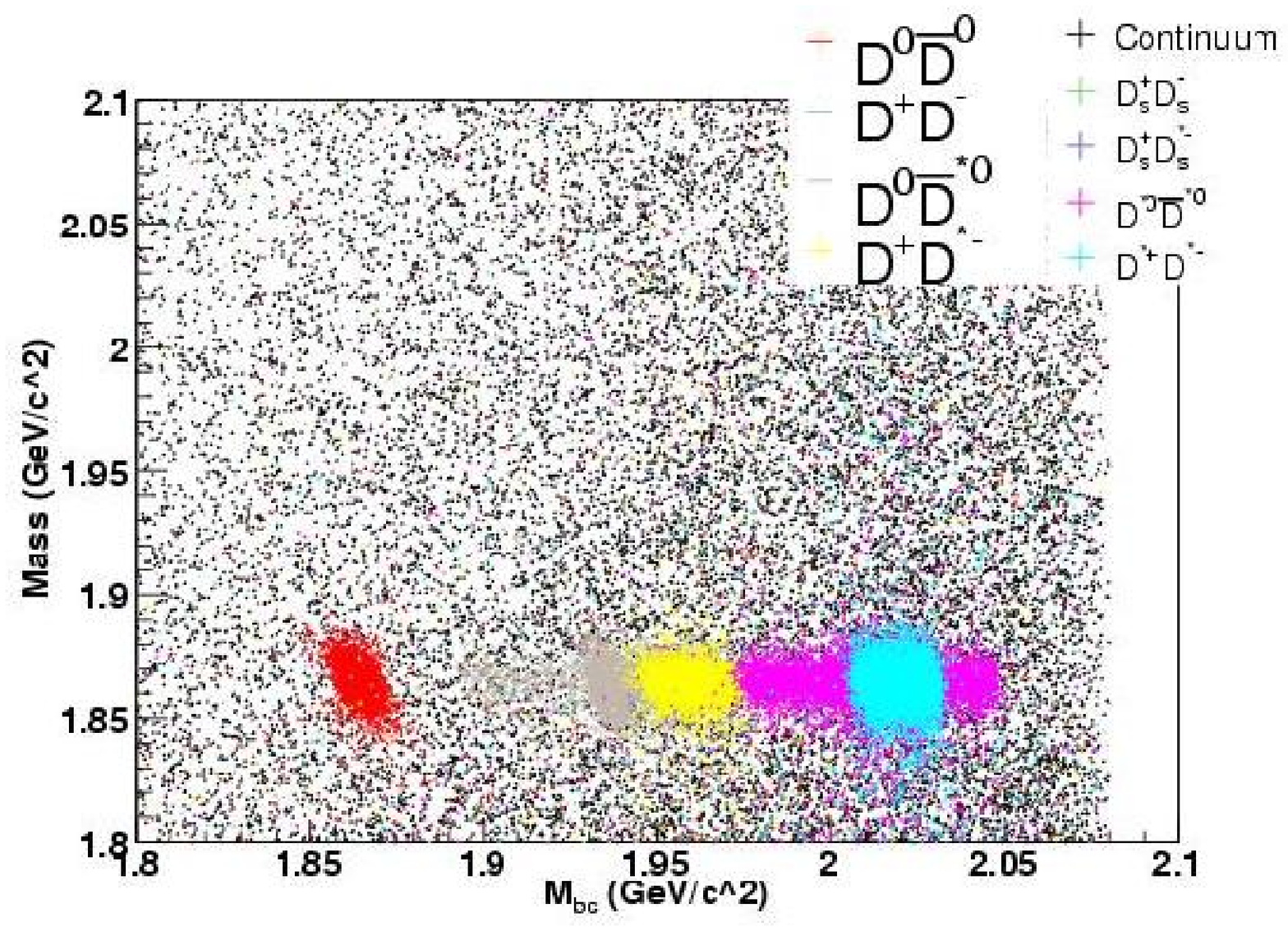}
\caption{Invariant mass vs. $M_{\rm{bc}}$ for $D^{0}\rightarrow{K^{-}\pi^{+}}$ at 4160 MeV using the 100 pb$^{-1}$ MC sample.  The plot 
illustrates the clear separation that is achieved with this choice of variables.}
\vspace{0.2cm}
\label{fig:D0kp_mass_mbc}
\end{center}
\end{figure}

\begin{figure}[!htbp]
\begin{center}
\hspace{2.5pt}
\includegraphics[width=14.5cm]{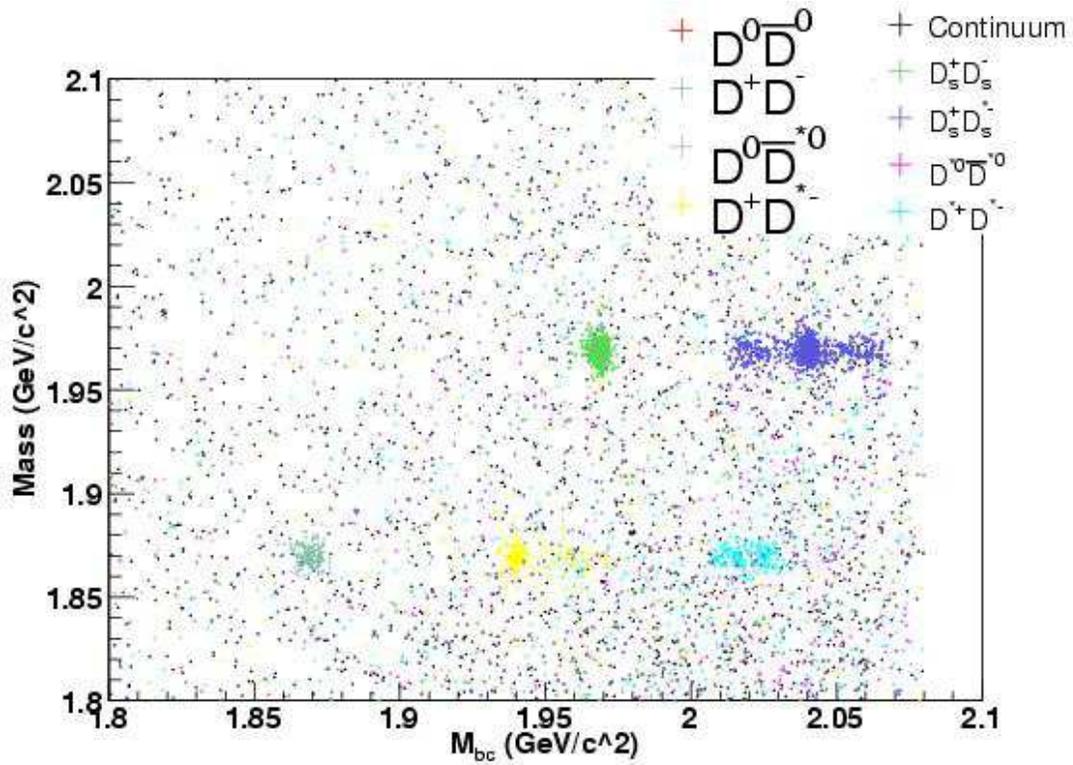}
\caption{Invariant mass vs. $M_{\rm{bc}}$ for $D_{s}^{+}\rightarrow{\phi}{\pi^{+}}$ at 4160 MeV using the 100 pb$^{-1}$ MC sample.  
The plot illustrates the clear separation that is achieved with this choice of variables, not 
just for $D_s$ but also for $D^{+}$, since the branching ratio for 
$D^{+}\rightarrow{\phi}{\pi^{+}}$ is nonzero.}
\vspace{0.2cm}
\label{fig:Ds_mass_mbc}
\end{center}
\end{figure}

For the other event types (i.e. those involving one or two ``starred'' charmed mesons)
the variables used for the separation were $M_{\rm{bc}}$ (candidate momentum)
and invariant mass, which is a combination of momentum and energy. 
Regardless of the origin of a $D$ candidate, the invariant mass
peaks at the $D$ mass.  Since invariant mass does not
differentiate between event types, $M_{\rm{bc}}$ provides all of the event-type separation
for these events.  The separation can be seen in \FIGS \ref{fig:D0kp_mass_mbc} and 
\ref{fig:Ds_mass_mbc}.

The difference in variable choices between unstarred and starred events is
a matter of convenience.  We use the same procedure that has been used with
great success in the CLEO-c analysis of $\psi(3770)\rightarrow{D\bar{D}}$
in Refs.~\cite{CLEO_DHad_PRL,BFcbx}.  $M_{\rm{bc}}$ has a practical advantage over the raw 
momentum in that $M_{\rm{bc}}$ changes much more slowly with beam energy.  
For a given center-of-mass energy, the expected values of $M_{\rm{bc}}$ and 
$\Delta E$ in $e^{+}e^{-}\rightarrow{X}{Y}$ are given by 
\begin{equation}
	M_{\rm{bc}}=\sqrt{\frac{1}{2}(M_{X}^{2}+M_{Y}^{2}-\frac{(M_{X}^{2}-M_{Y}^{2})^{2}}{2s})},
\end{equation}
\CONT and
\begin{equation}
        \Delta{E}=-\frac{1}{2}\frac{M_{X}^{2}-M_{Y}^{2}}{\sqrt{s}}.
\end{equation}
As the center-of-mass energy increases from 3970 to 4180~MeV, $M_{\rm{bc}}$ for the $D$ 
in $D\bar{D}^{*}$ changes by $\sim0.01\%$, while the momentum changes by $\sim80\%$.  

As discussed above, the method for
$D^{0}\bar{D}^{0}$, $D^{+}D^{-}$ and $D_{s}^{+}D_{s}^{-}$ is to
cut on $\Delta E$ ($\Delta E < 15$~MeV) and use the $M_{\rm{bc}}$ 
distribution to determine the yield (\FIG \ref{fig:Dp_mbc_SB}).  
For all other event types the method is to cut on $M_{\rm{bc}}$ and use the
invariant-mass distribution to determine the yield
(\FIG \ref{fig:DpSS_mass_SB}).  The cut on $M_{\rm{bc}}$ was
determined by kinematics, since $M_{\rm{bc}}$ is center-of-mass energy
dependent in
addition to being dependent on the nature of the decay of the starred
state.  In order to choose the cut range we assumed that
$D^{*0}$ and $D_{s}^{*+}$ decay $100\%$ of the time by
$D^{*0}\rightarrow\gamma{D^{0}}$ and
$D_{s}^{*+}\rightarrow\gamma{D_{s}^{+}}$, respectively, since the pion
transitions 
will fall in the cut window for the $\gamma$ decay.
We assumed that $D^{*+}$ decays only to $D^+\pi^0$ and $D^0\pi^+$, 
since the branching ratio for $D^{*+}\rightarrow{D^{+}}\gamma$
is only $1.6\%$.
The equation that determines the $M_{\rm{bc}}$ cut window for
$D^{*}\bar{D}$ and $D^{*}\bar{D}^{*}$ with
$D^{*}\rightarrow{D}{X}$ is
\begin{equation}
        M_{\rm{bc}}^{\rm{Lab}}=\sqrt{E_{\rm{beam}}^{2}-M_{D}^{2} - (\gamma{E_{D}}\pm\beta\gamma{P_{D}})^{2}},
\end{equation}
\CONT with $\gamma=\frac{E_{D^{*}}}{M_{D^{*}}}$,
$\beta\gamma=\frac{P_{D^{*}}}{M_{D^{*}}}$ and $E_{D} =
\frac{M_{D^{*}}^{2} + M_{X}^{2} - M_{D}^{2}}{2M_{D^{*}}}$, where
$M_{D},E_{D}$ and ${P_{D}}$ are the mass, energy, and momentum of the
daughter $D$ in the rest frame of the $D^{*}$, and $E_{D^{*}}$ and
$P_{D^{*}}$ are determined by \EQS \ref{eq:ED} and \ref{eq:PD},
respectively. For $D^{*0}$ and $D_{s}^{*+}$, $M_{X}=M_{\gamma}=0$, since only the
$\gamma$ decay is considered in the calculation and for
$D^{*+}$, $M_{X}=M_{\pi^0}=135$~MeV.  After calculating the maximum
and minimum values using the above equation and assumptions,
we expand the interval by 5~MeV on each end to account for resolution effects.

An important point is that there is a certain energy at
which an overlap between the $D^{0}\bar{D}^{*0}$ and $D^{*0}\bar{D}^{*0}$ will
occur.  This is only a problem for $D^{0}\bar{D}^{*0}$ and $D^{*0}\bar{D}^{*0}$,
and not $D^{-}D^{*+}$ and $D^{*+}D^{*-}$, because only the ${\pi}$
decay was considered. This means these event types will have a
significantly smaller cut window in $M_{\rm{bc}}$ as compared to neutral
$D$ events.  If this occurs, the high $M_{\rm{bc}}$ cut of
$D^{0}\bar{D}^{*0}$ will be set equal to the low $M_{\rm{bc}}$ cut of
$D^{*0}\bar{D}^{*0}$ so that events cannot pass both
event-type cuts and be double-counted. This is done because
$D^{*0}\bar{D}^{*0}$ will have twice as many $D$ mesons populating this
area, assuming equal rates, therefore there will be less
contamination from $D^{0}\bar{D}^{*0}$ in $D^{*0}\bar{D}^{*0}$ by this
method as compared to its inverse.

\begin{table}[!htbp]
\begin{center}
\caption{Selection criteria, in units of ${\rm{M_{bc}}}$, for measuring the cross sections for
$D^{*}\bar{D}$ and $D^{*}\bar{D}^{*}$.  A 15~MeV cut on
$\Delta E$ was made when selecting $D\bar{D}$.}
\label{D0Dp_cut_table}
\vspace{0.2cm}
%\small{
\begin{tabular}{|c|c|c|c|c|}\hline
$E_{\rm{cm}}$~(MeV) & $D^{0}\bar{D}^{*0}$~($\frac{\rm{GeV}}{c^2}$) & $D^{*0}\bar{D}^{*0}$~($\frac{\rm{GeV}}{c^2}$) &
$D^{+}{D}^{*-}$~($\frac{\rm{GeV}}{c^2}$) & $D^{*+}{D}^{*-}$~($\frac{\rm{GeV}}{c^2}$) \\ \hline
3970&$(1.902,1.972)$&$-$&$(1.933,1.958)$&$-$ \\ \hline
3990&$(1.901,1.976)$&$-$&$(1.933,1.961)$&$-$ \\ \hline
4010&$(1.899,1.98)$&$-$&$(1.934,1.963)$&$-$ \\ \hline
4015&$(1.899,1.981)$&$(1.994,2.01)$&$(1.934,1.963)$&$-$ \\ \hline
4030&$(1.898,1.984)$&$(1.987,2.02)$&$(1.934,1.965)$&$(2.003,2.018)$ \\ \hline
4060&$(1.896,1.98)$&$(1.98,2.03)$&$(1.935,1.968)$&$(2.002,2.022)$ \\ \hline
4120&$(1.893,1.974)$&$(1.974,2.044)$&$(1.935,1.974)$&$(2.003,2.029)$ \\ \hline
4140&$(1.892,1.972)$&$(1.972,2.048)$&$(1.935,1.976)$&$(2.004,2.032)$ \\ \hline
4160&$(1.891,1.97)$&$(1.97,2.052)$&$(1.935,1.978)$&$(2.004,2.034)$ \\
\hline
4170&$(1.89,1.97)$&$(1.97,2.054)$&$(1.935,1.979)$&$(2.005,2.035)$ \\
\hline
4180&$(1.89,1.969)$&$(1.969,2.056)$&$(1.935,1.98)$&$(2.005,2.036)$ \\ \hline
4200&$(1.889,1.968)$&$(1.968,2.06)$&$(1.935,1.982)$&$(2.006,2.038)$ \\ \hline
4260&$(1.887,1.965)$&$(1.965,2.071)$&$(1.935,1.988)$&$(2.008,2.044)$ \\ \hline
\end{tabular}%}
\end{center}
\end{table}

\begin{table}[!htbp]
\begin{center}
\caption{Selection criteria, in units of ${\rm{M_{bc}}}$, for measuring the cross section for
$D_{s}^{*}\bar{D}_{s}$ and $D^{*}_{s}\bar{D}^{*}_{s}$.  A 15~MeV cut on
$\Delta E$ was made when selecting $D_{s}\bar{D}_{s}$.}
\label{Ds_cut_table}
\vspace{0.2cm}
\begin{tabular}{|c|c|c|}\hline
{E$_{\rm{cm}}$~MeV}& {$D^{+}_{s}{D}^{*-}_{s}~\frac{\rm{GeV}}{\rm{c^{2}}}$}& {$D^{*+}_{s}{D}^{*-}_{s}~\frac{\rm{GeV}}{\rm{c^{2}}}$} \\ \hline
3970&$-$&$-$ \\ \hline
3990&$-$&$-$ \\ \hline
4010&$-$&$-$ \\ \hline
4015&$-$&$-$ \\ \hline
4030&$-$&$-$ \\ \hline
4060&$-$&$-$ \\ \hline
4120&$(2.015,2.06)$&$-$ \\ \hline
4140&$(2.011,2.067)$&$-$ \\ \hline
4160&$(2.009,2.071)$&$-$ \\ \hline
4170&$(2.008,2.073)$&$-$ \\ \hline
4180&$(2.007,2.076)$&$-$ \\ \hline
4200&$(2.005,2.08)$&$-$ \\ \hline
4260&$(2.001,2.087)$&$(2.087,2.132)$ \\ \hline
\end{tabular}
\end{center}
\end{table}

\chapter{Cross Section Calculation}
\section{Determination of Cross Sections}
\label{sec:sigmas}
%%%%%%%%%%%%%%%%%%%%%%%%%%%%%%%%%%%%%%%%%%%%%%%%%%%%%%%%%%%%%%%%%%%%%%%%%%
%
\subsection{Exclusive Cross Sections}

The cross section for $D_{(s)}$ from any of the nine possible
event types can be computed with the following equation:
\begin{equation}
         \sigma_{D_{(s)}^{(*)}\bar{D}_{(s)}^{(*)}}(D_{(s)})=
\frac{N(\rm{signal})} {\BR{\cal{L}}\epsilon},
\end{equation}
\CONT where $N(\rm{signal})$ is the number of signal events, $\BR$ is the 
branching ratio for the particular $D$ decay being used 
(\TAB \ref{Dsmode_table} \cite{pdg,CLEO_DsBF} and \TAB \ref{Dmode_table} 
\cite{CLEO_DHad_PRL,BFcbx}), $\cal{L}$ is the integrated luminosity,
and $\epsilon$ is the detection efficiency (determined with MC).

$N(\rm{signal})$ is obtained by counting candidates in the signal region and
subtracting sideband-estimated backgrounds with normalizations determined
by MC.  This procedure and the definitions of the signal and sideband 
regions for each mode appear in Sect.~\ref{sec:eventsel}.

The cross sections for the $D\bar{D}$, $D^{*}\bar{D}$ and
$D^{*}\bar{D}^{*}$ production modes are calculated as follows:
\begin{equation}
	\sigma(D^{*+}D^{*-}) =
	\frac{\sigma_{D^{*}\bar{D}^{*}}(D^{+})}{2(1-\BR(D^{*+}\rightarrow{D^{0}\pi^{+}}))},
\label{eqDSDS}
\end{equation}
\begin{equation}
	\sigma(D^{*0}\bar{D}^{*0}) = \frac{1}{2}(\sigma_{D^{*}\bar{D}^{*}}(D^{0})-2\BR(D^{*+}\rightarrow{D^{0}\pi^{+}})\sigma(D^{*+}D^{*-})),
\end{equation}
\begin{equation}
	\sigma(D^{*+}D^{-}) = \frac{\sigma_{D^{*}\bar{D}}(D^{+})}{2-\BR(D^{*+}\rightarrow{D^{0}\pi^{+}})},
\end{equation}
\begin{equation}
	\sigma(D^{*0}\bar{D}^{0}) =
	\frac{1}{2}(\sigma_{D^{*}\bar{D}}(D^{0})-\BR(D^{*+}\rightarrow{D^{0}\pi^{+}})\sigma(D^{*+}D^{-})),
\end{equation}
\begin{equation}
	\sigma(D^{+}D^{-}) = \frac{\sigma_{D\bar{D}}(D^{+})}{2},
\end{equation}
\CONT and
\begin{equation}
	\sigma(D\bar{D}) = \frac{\sigma_{D\bar{D}}(D^{0})}{2},
\label{eqDD}
\end{equation}
\CONT where $\sigma_{D^{*}\bar{D}^{*}}(D^{+})$ is the cross section of
$D^{+}$ produced in $D^{*}\bar{D}^{*}$ events,
$\sigma_{D^{*}\bar{D}^{*}}(D^{0})$ is the cross section of $D^{0}$
produced in $D^{*}\bar{D}^{*}$ events, $\sigma_{D^{*}\bar{D}}(D^{+})$ is the cross
section of $D^{+}$ produced in $D^{*}\bar{D}$ events,
$\sigma_{D^{*}\bar{D}}(D^{0})$ is the cross section of $D^{0}$ produced in
$D^{*}\bar{D}$ events, $\sigma_{D\bar{D}}(D^{+})$ is the cross section of
$D^{+}$ produced in $D\bar{D}$ events, and $\sigma_{D\bar{D}}(D^{0})$ is the
cross section of $D^{0}$ produced in $D\bar{D}$ events.

To determine $\sigma(D^0)$ and $\sigma(D^+)$, weighted averages of the 
three $D^{0}$ modes and five $D^{+}$ modes are calculated, with weights 
defined as $\frac{1}{\sigma_{\sigma(D)_i}^2}$. 

For $D_{s}\bar{D}_{s}$, $D_{s}^{*}\bar{D}_{s}$ and 
$D_{s}^{*}\bar{D}_{s}^{*}$, a weighted sum technique is used to 
combine the eight $D_s$ decay modes and obtain the cross sections.  The 
weights that minimize the error are given by the following:
\begin{equation}
           w_{i} = \frac{\frac{1}{f_i^{2}N_{i}}}{\sum_{i=0}^{8}\frac{1}{f_i^{2}N_{i}}},
\label{eq:wt}	
\end{equation}

\CONT where $f_i=\frac{\sigma_{i}}{N_{i}}$, $N_{i}$ is the yield,
$\sigma_{i}$ is the uncertainty on the yield for mode $i$.  It is perhaps
counterintuitive that for modes with equal precision the weight is inversely 
proportional to the yield, thereby suppressing the weight of modes with 
high yields.  This feature guarantees that the weighting of different 
modes in the cross section measurement is determined by the precision
of the yield measurement rather than its magnitude.  (The conclusion can be 
easily verified by considering a ``toy'' example of two modes, each with 
10\% precision and yields of 100 and 1000, respectively.  Eq.~\ref{eq:wt}
properly assigns roughly equal weighting and achieves an uncertainty of 
$\sim 7\%$, rather than the 10\% that would follow if the ``bigger'' 
mode were allowed to dominate.)

The weights were determined by MC for each of the eight modes for each
of the three possible event types, $D_{s}\bar{D}_{s}$, $D_{s}^{*}\bar{D}_{s}$,
$D_{s}^{*}\bar{D}_{s}^{*}$.  The weights were then averaged across
these possible event types and used in
calculating the cross sections. The equation for
determining either the $D_{s}\bar{D}_{s}$, $D_{s}^{*}\bar{D}_{s}$,
$D_{s}^{*}\bar{D}_{s}^{*}$ cross sections is as follows:
\begin{equation}
           \sigma(D_{s}^{(*)+}D_{s}^{(*)-}) = \frac{(\sum^{8}_{i=1}N_{i}^{DATA}w_{i})}{\epsilon{\cal{L}}},	
\end{equation}
\CONT where $\epsilon$ is determined with MC by the following:
\begin{equation}
           \epsilon = \frac{(\sum_{i=1}^{8}N^{MC}_{i}w_{i})}{N_{generated}},	
\end{equation}
\CONT where the $8$ in the summation refers to the eight $D_{s}$ modes
that were used during the scan.

\subsection{Efficiencies for Exclusive Selection}

\label{sec:effs}

Efficiencies for exclusive selection of all accessible event types and 
center-of-mass energies were determined by analyzing the MC samples described 
in Sect.~\ref{sec:samples}.  The efficiency for detecting a particular decay mode is 
defined as follows:
\begin{equation}
	\epsilon = \frac{N^{\rm{MC}}_{\rm{detected}}}{N^{\rm{MC}}_{\rm{generated}}},
\end{equation}
\CONT where the error is determined by binomial statistics:
\begin{equation}
         \sigma_{\epsilon} = \sqrt{\frac{\epsilon(1-\epsilon)}{N^{\rm{MC}}_{\rm{generated}}}}.	
\end{equation}
\CONT $N^{\rm{MC}}_{\rm{detected}}$ is obtained by applying the same selection 
and background-correction procedures to MC as are used to determine the signal yields 
in data.  The efficiencies for each of the twelve energies and event
types are listed in \TABS \ref{ExclusiveD0_eff_table} ($D^0$), 
\ref{ExclusiveDp_eff_table} ($D^+$) and \ref{ExclusiveDs_eff_table} ($D_s^+$ - note that 
this multi-page table appears on P.193 after the References).

\begin{table}[htbp]
\begin{center}
\caption{Efficiencies (units of $10^{-2}$) at each scan-energy point for 
selection of $D^{0}\rightarrow{K^{-}}\pi^{+}$ decays in the three exclusive
event types.}
\label{ExclusiveD0_eff_table}
\vspace{0.2cm}
\begin{tabular}{|c|c|c|c|}\hline
{E$_{cm}$ (MeV)}& {$D\bar{D}$}& {$D^{*}\bar{D}$}& {$D^{*}\bar{D}^{*}$} \\ \hline
$3970$ & $62.42\pm0.72$& $61.78\pm0.36$ & $ -\pm-$ \\ \hline 
$3990$ & $63.04\pm0.72$& $62.37\pm0.36$ & $ -\pm-$ \\ \hline 
$4010$ & $61.75\pm0.73$& $62.83\pm0.36$ & $ -\pm-$ \\ \hline 
$4015$ & $62.29\pm0.59$& $63.31\pm0.36$ & $64.23\pm0.24$ \\ \hline
$4030$ & $63.75\pm0.63$& $62.68\pm0.39$ & $64.60\pm0.24$ \\ \hline
$4060$ & $60.17\pm0.64$& $61.24\pm0.39$ & $64.42\pm0.24$ \\ \hline
$4120$ & $58.97\pm0.66$& $59.75\pm0.40$ & $64.23\pm0.25$ \\ \hline
$4140$ & $59.49\pm0.47$& $57.90\pm0.29$ & $64.84\pm0.17$ \\ \hline
$4160$ & $57.94\pm0.66$& $58.12\pm0.40$ & $64.80\pm0.25$ \\ \hline
$4170$ & $58.12\pm0.66$& $56.53\pm0.40$ & $65.23\pm0.25$ \\ \hline
$4180$ & $59.30\pm0.66$& $54.72\pm0.41$ & $66.00\pm0.25$ \\ \hline
$4200$ & $59.06\pm0.66$& $52.52\pm0.41$ & $66.88\pm0.24$ \\ \hline
$4260$ & $56.29\pm1.01$& $41.36\pm0.61$ & $70.02\pm0.36$ \\ \hline
\end{tabular}
\end{center}
\end{table}

\begin{table}[!htbp]
\begin{center}
\caption{Efficiencies (units of $10^{-2}$) at each scan-energy point for 
selection of $D^{+}\rightarrow{K^{-}}\pi^{+}\pi^{+}$ decays in the three exclusive
event types.}
\label{ExclusiveDp_eff_table}
\vspace{0.2cm}
\begin{tabular}{|c|c|c|c|}\hline
{E$_{cm}$~MeV}& {$D\bar{D}$}& {$D^{*}\bar{D}$}& {$D^{*}\bar{D}^{*}$} \\ \hline
$3970$ & $56.07\pm0.49$& $52.82\pm0.35$ & $ -\pm-$ \\ \hline 
$3990$ & $55.33\pm0.48$& $52.78\pm0.35$ & $ -\pm-$ \\ \hline 
$4010$ & $55.59\pm0.49$& $53.13\pm0.35$ & $ -\pm-$ \\ \hline 
$4015$ & $53.60\pm0.39$& $53.37\pm0.34$ & $ -\pm-$ \\ \hline 
$4030$ & $56.79\pm0.43$& $52.33\pm0.37$ & $53.60\pm0.38$ \\ \hline
$4060$ & $53.93\pm0.43$& $52.53\pm0.37$ & $53.12\pm0.38$ \\ \hline
$4120$ & $54.42\pm0.44$& $52.55\pm0.38$ & $51.64\pm0.38$ \\ \hline
$4140$ & $53.83\pm0.31$& $51.18\pm0.27$ & $52.47\pm0.27$ \\ \hline
$4160$ & $54.12\pm0.44$& $51.10\pm0.38$ & $50.98\pm0.38$ \\ \hline
$4170$ & $54.15\pm0.44$& $51.15\pm0.38$ & $51.31\pm0.38$ \\ \hline
$4180$ & $54.25\pm0.44$& $51.39\pm0.38$ & $52.46\pm0.38$ \\ \hline
$4200$ & $53.38\pm0.44$& $51.37\pm0.38$ & $52.47\pm0.38$ \\ \hline
$4260$ & $54.51\pm0.66$& $49.15\pm0.58$ & $51.29\pm0.59$ \\ \hline

\end{tabular}
\end{center}
\end{table}

\subsection{Cross Section Results}

\begin{figure}[!htbp]
\begin{center}
\hspace{2.5pt}
\includegraphics[width=11cm, angle=270]{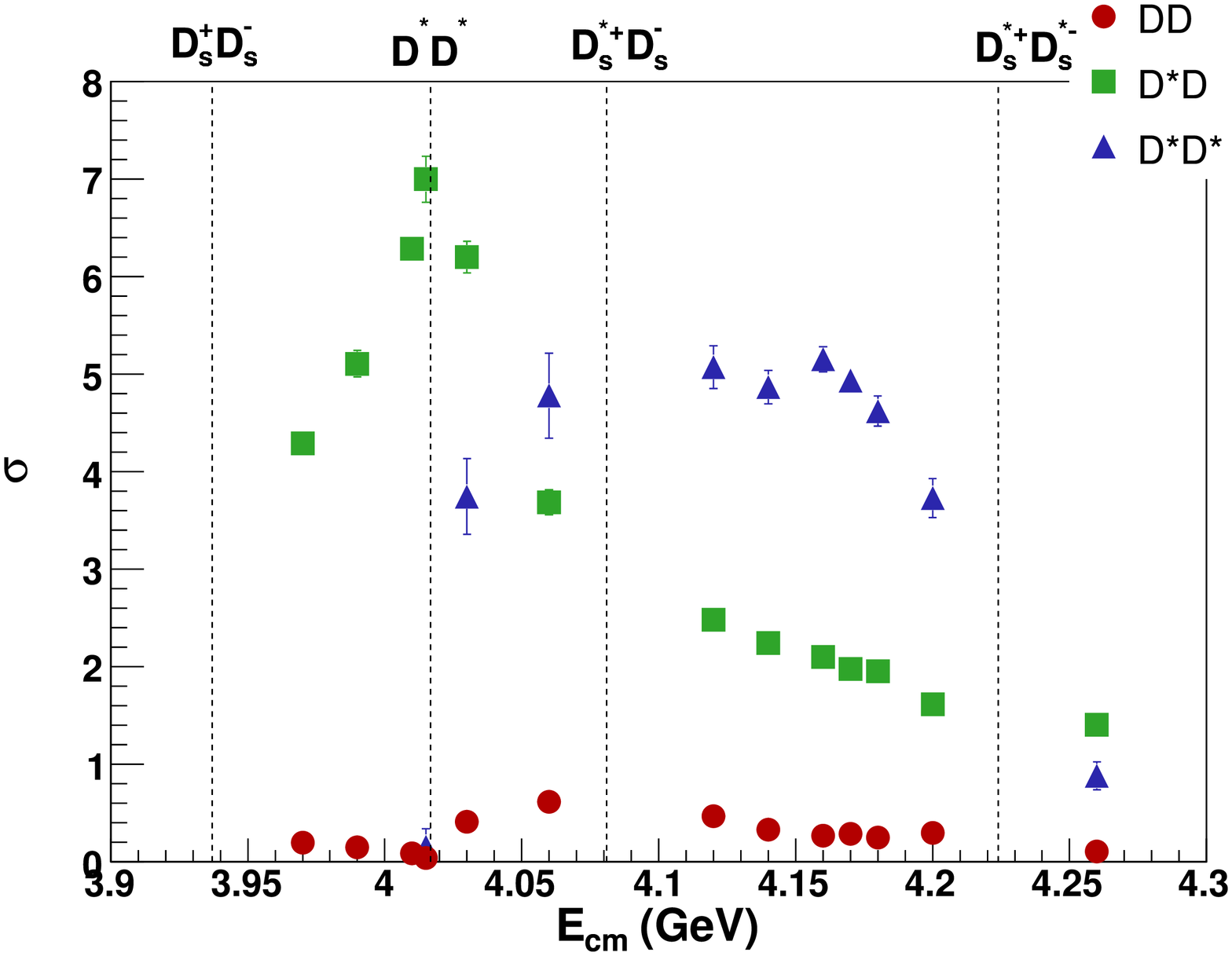}
\caption{Observed cross sections for $e^+e^-\rightarrow{D\bar{D}}$, $D^{*}\bar{D}$ and
$D^{*}\bar{D}^{*}$ as a function of center-of-mass energy. Errors are just statistical.}
\vspace{0.2cm}
\label{fig:XS_scan_DD}
\end{center}
\end{figure}

\begin{figure}[!htbp]
\begin{center}
\hspace{2.5pt}
\includegraphics[width=11cm, angle=270]{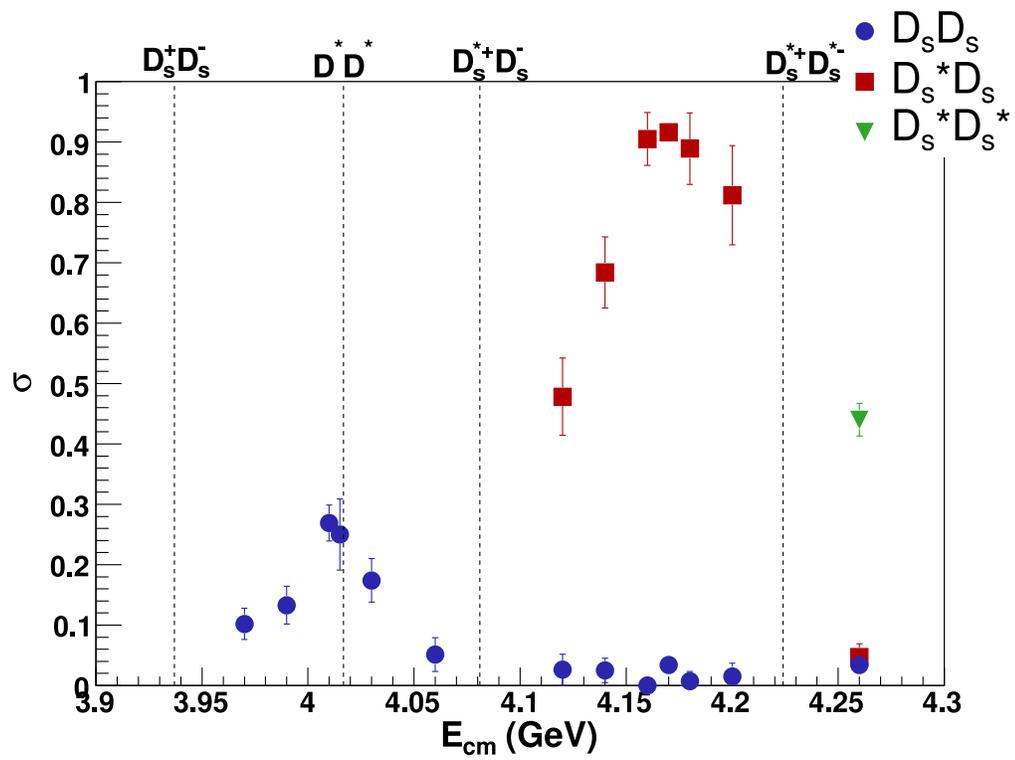}
\caption{Observed cross sections for $e^+e^-\rightarrow{D_{s}\bar{D}_{s}}$, $D_{s}^{*}\bar{D}_{s}$ and
$D_{s}^{*}\bar{D}_{s}^{*}$ as a function of center-of-mass energy from
the scan. Errors are just statistical.}
\vspace{0.2cm}
\label{fig:XS_scan_Ds}
\end{center}
\end{figure}

Following the procedure laid out in the previous sections, the
production cross
sections were measured at thirteen center-of-mass energies.
The results are shown in \TABS \ref{XS_results_1} and
\ref{XS_results_2} and \FIGS
\ref{fig:XS_scan_DD} and \ref{fig:XS_scan_Ds}.

\CONT These cross sections were then used in \EQS \ref{eqDSDS}-\ref{eqDD} to obtain the results 
shown in \TABS \ref{XS_results_1} and \ref{XS_results_2}.

The weights and efficiencies are shown in \TABS
\ref{totalEffDs_table} and \ref{weight_table}.
By summing the individual exclusive cross
sections, one arrives at the total observed charm cross section
which is shown, along with the exclusive cross sections, in \TAB \ref{XS_results_1} and
\ref{XS_results_2}.
\begin{table}[!htbp]
\begin{center}
\caption{The efficiencies in units of $10^{-2}$ for detecting $D_{s}\bar{D}_{s}$, $D_{s}^{*}\bar{D}_{s}$, and 
$D_{s}^{*}\bar{D}_{s}^{*}$ at each energy point in the CLEO-c scan.}
\label{totalEffDs_table}
\vspace{0.2cm}
\begin{tabular}{|c|c|c|c|}\hline
{E$_{cm}$ MeV}& {$\epsilon({D_{s}^{+}D_{s}^{-}})$}& {$\epsilon(D_{s}^{*+}{D_{s}^{-}})$}& {$\epsilon({D^{*+}_{s}D^{*-}_{s}})$}
\\ \hline
$3970$&$1.14 \pm 0.02$&$-\pm-$&$-\pm-$ \\ \hline
$3990$&$1.11 \pm 0.02$&$-\pm-$&$-\pm-$ \\ \hline
$4010$&$1.11 \pm 0.02$&$-\pm-$&$-\pm-$ \\ \hline
$4015$&$1.13 \pm 0.02$&$-\pm-$&$-\pm-$ \\ \hline
$4030$&$1.20 \pm 0.02$&$-\pm-$&$-\pm-$ \\ \hline
$4060$&$1.09 \pm 0.02$&$-\pm-$&$-\pm-$ \\ \hline
$4100$&$1.12 \pm 0.03$&$1.00 \pm 0.02$&$-\pm-$ \\ \hline
$4120$&$1.04 \pm 0.03$&$1.04 \pm 0.02$&$-\pm-$ \\ \hline
$4140$&$1.04 \pm 0.02$&$1.06 \pm 0.02$&$-\pm-$ \\ \hline
$4160$&$1.09 \pm 0.03$&$1.06 \pm 0.02$&$-\pm-$ \\ \hline
$4170$&$1.09\pm 0.03 $&$1.06 \pm 0.02$&$-\pm-$ \\ \hline
$4180$&$1.09 \pm 0.03$&$1.10 \pm 0.03$&$-\pm-$ \\ \hline 
$4200$&$1.09 \pm 0.03$&$1.07 \pm 0.03$&$-\pm-$ \\ \hline
$4260$&$1.09 \pm 0.03$&$1.10 \pm 0.03$&$1.05\pm0.03$ \\ \hline
\end{tabular}
\end{center}
\end{table}
\begin{table}[!htbp]
\begin{center}
\caption{For $D_{s}\bar{D}_{s}$, $D_{s}^{*}\bar{D}_{s}$, and 
$D_{s}^{*}\bar{D}_{s}^{*}$, a weighted sum technique that minimize
the error is used to determine the cross sections. The weights
for each mode are shown below. The mode with the largest weight is
$\phi\pi^+$, which is by far the cleanest of all the modes.}
\label{weight_table}
\vspace{0.2cm}
\begin{tabular}{|c|c|}\hline
{Mode}& {Weight}
\\ \hline
$K_{s}K^{+}$&$0.14$ \\ \hline
$\eta\pi^{+}$&$0.06$ \\ \hline
$\phi\pi^{+}$&$0.23$ \\ \hline
$K^{*}K^{+}$&$0.12$ \\ \hline
$\eta\rho^{+}$&$0.05$ \\ \hline
$\eta^{'}\pi^{+}$&$0.16$ \\ \hline
$\eta^{'}\rho^{+}$&$0.10$ \\ \hline
$\phi\rho^{+}$&$0.13$ \\ \hline
\end{tabular}
\end{center}
\end{table}

%%%%%%%%%%%%%%%%%%%%%%%%%%%%%%%%%%%%%%%%%%%%%%%%%%%%%%%%%%%%%%%%%%%%
\subsection{Two Other Methods for Measuring the Total Charm Cross Section}
%%%%%%%%%%%%%%%%%%%%%%%%%%%%%%%%%%%%%%%%%%%%%%%%%%%%%%%%%%%%%%%%%%%%
\subsubsection{Inclusive $D$ Method}
%%%%%%%%%%%%%%%%%%%%%%%%%%%%%%%%%%%%%%%%%%%%%%%%%%%%%%%%%%%%%%%%%%%%

In addition to measuring the separate cross sections for all 
expected charm event types, one can perform inclusive measurements to
obtain the total observed charm cross section.

As for the exclusive measurements, the efficiencies for inclusively selecting events 
with charmed mesons are determined with the MC samples described in 
Sect.~\ref{sec:samples}.  They are given in \TAB \ref{Inclusive_eff_table}.
\begin{table}[!htbp]
\begin{center}
\caption{Efficiencies at each scan-energy point for inclusive selection of
$D^{0}\rightarrow{K^{-}}\pi^{+}$, $D^{+}\rightarrow{K^{-}}\pi^{+}\pi^{+}$,
and $D_{s}^{+}\rightarrow{K^{-}}K^{+}\pi^{+}$ (units of $10^{-2}$).}
\label{Inclusive_eff_table}
\vspace{0.2cm}
\begin{tabular}{|c|c|c|c|}\hline
{E$_{cm}$~MeV}& {$\epsilon(D^{0}\rightarrow{K^{-}}\pi^{+})$}& {$\epsilon(D^{+}\rightarrow{K^{-}}\pi^{+}\pi^{+})$}& {$\epsilon(D_{s}^{+}\rightarrow{K^{-}}K^{+}\pi^{+})$} \\ \hline
$3970$ & $63.26\pm0.32$& $53.07\pm0.28$& $54.31\pm0.90$ \\ \hline
$3990$ & $61.99\pm0.33$& $52.29\pm0.28$& $53.04\pm0.90$ \\ \hline
$4010$ & $62.52\pm0.32$& $53.57\pm0.28$& $53.62\pm0.90$ \\ \hline
$4015$ & $63.79\pm0.19$& $51.91\pm0.26$& $52.10\pm0.74$ \\ \hline
$4030$ & $64.52\pm0.20$& $53.47\pm0.23$& $52.87\pm0.81$ \\ \hline
$4060$ & $63.24\pm0.20$& $51.95\pm0.23$& $48.83\pm0.80$ \\ \hline
$4120$ & $62.49\pm0.20$& $52.33\pm0.23$& $50.90\pm0.59$ \\ \hline
$4140$ & $62.61\pm0.14$& $51.79\pm0.16$& $49.84\pm0.42$ \\ \hline
$4160$ & $62.54\pm0.20$& $51.17\pm0.23$& $52.75\pm0.59$ \\ \hline
$4170$ & $62.26\pm0.20$& $51.39\pm0.23$& $52.93\pm0.59$ \\ \hline
$4180$ & $62.32\pm0.20$& $51.95\pm0.23$& $53.67\pm0.59$ \\ \hline
$4200$ & $63.13\pm0.20$& $52.00\pm0.23$& $51.63\pm0.59$ \\ \hline
$4260$ & $60.82\pm0.31$& $51.35\pm0.35$& $53.74\pm0.69$ \\ \hline
\end{tabular}
\end{center}
\end{table}

\begin{figure}[!htbp]
\begin{center}
\hspace{2.5pt}
\includegraphics[width=14.5cm]{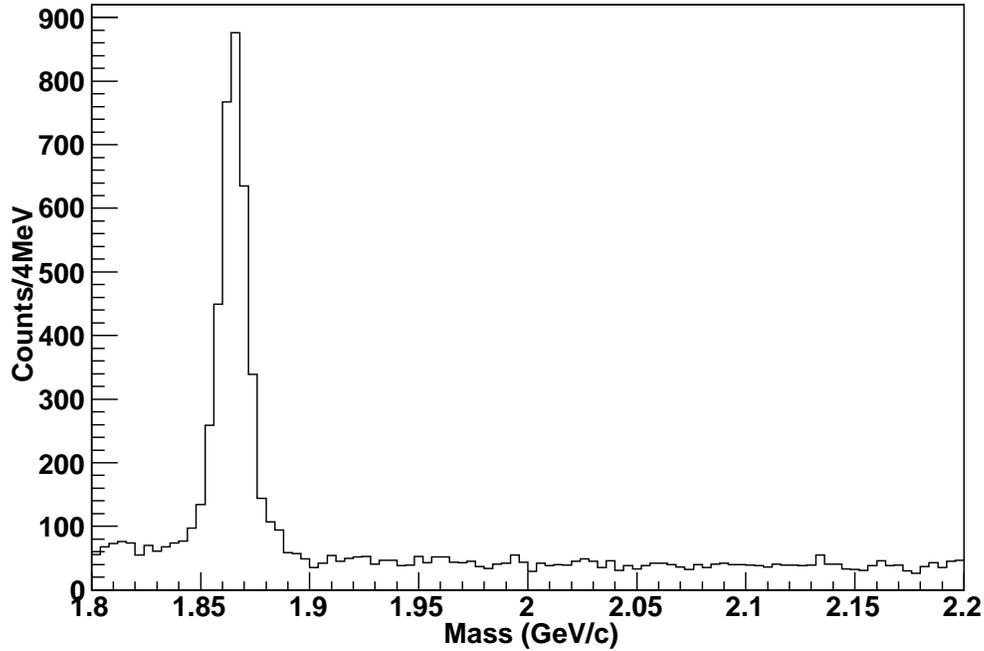}
\caption{The invariant mass of $D^{0}\rightarrow{K^{-}}{\pi^{+}}$ in data at $E_{cm}=4160$~MeV.}
\vspace{0.2cm}
\label{fig:D0_IM}
\end{center}

\end{figure}

\begin{figure}[!htbp]
\begin{center}
\hspace{2.5pt}
\includegraphics[width=14.5cm]{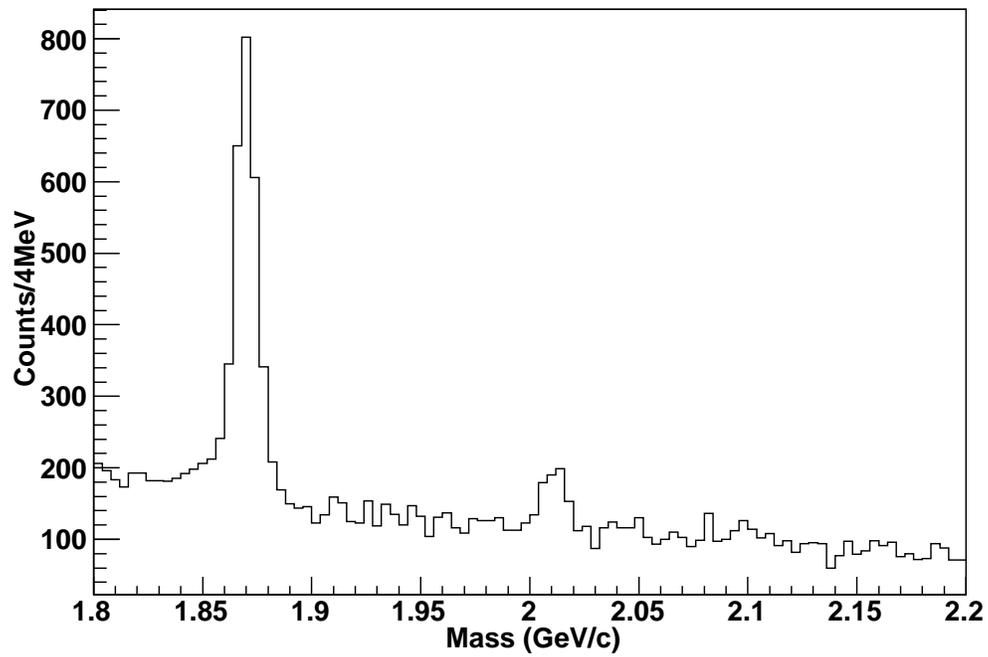}
\caption{The invariant mass of
$D^{+}\rightarrow{K^{-}}{\pi^{+}}{\pi^{+}}$ in data at $E_{cm}=4160$~MeV. The
peak at $\sim2$~GeV is fully reconstructed
$D^{*+}\rightarrow{D^{0}\pi^{+}}$, $D^{0}\rightarrow{K^{-}\pi^{+}}$.}
\vspace{0.2cm}
\label{fig:Dp_IM}
\end{center}
\end{figure}

\begin{figure}[!htbp]
\begin{center}
\hspace{2.5pt}
\includegraphics[width=14.5cm]{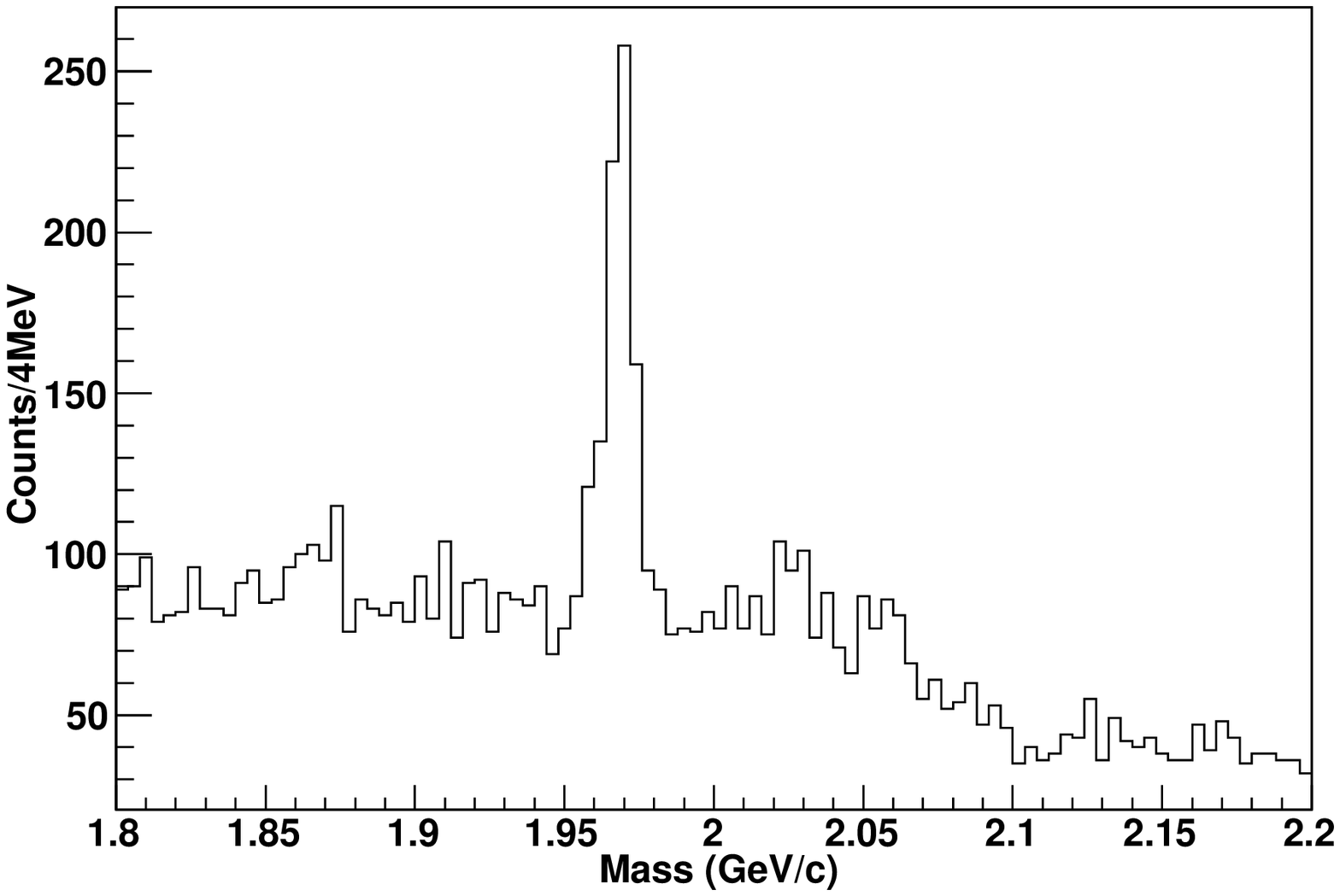}
\caption{The invariant mass of $D_{s}^{+}\rightarrow{K^{+}}{K^{-}}{\pi^{+}}$ in data at $E_{cm}=4160$~MeV.}
\vspace{0.2cm}
\label{fig:Ds_IM}
\end{center}
\end{figure}

For $D^{0}$ and $D^{+}$, the event-type requirements on
$|\Delta E|$ and $M_{\rm{bc}}$ are lifted and the
invariant mass is used to extract the yields. 
The inclusive $D^{0}\rightarrow{K^{-}\pi^{+}}$ and
$D^{+}\rightarrow{K^{-}\pi^{+}\pi^{+}}$ invariant-mass spectra are
shown for the 4160 MeV data sample in \FIGS \ref{fig:D0_IM} and \ref{fig:Dp_IM}.
For $D_{s}$ the event-type requirements are preserved because of the
need to suppress the background for the high-yield mode
$K^{-}K^{+}\pi^{+}$.  At energies above $4100$~MeV, for all
candidates that pass the selection requirements for $D_{s}^{+}D_{s}^{-}$,
$D_{s}^{*+}D_{s}^{-}$ and $D_{s}^{*+}D_{s}^{*-}$ (the last only for 
$4260$~MeV), the invariant mass
is used to determine the inclusive yield.
At energies below $4100$~MeV, $M_{\rm{bc}}$ is
used to determine the yield since these energies are below
$D_{s}^{*+}D_{s}^{-}$ and $D_{s}^{*+}D_{s}^{*-}$ thresholds.
The inclusive $D_{s}^{+}\rightarrow{K^{-}K^{+}\pi^{+}}$ invariant
mass in 4160 MeV data is shown in \FIG \ref{fig:Ds_IM}.
Each histogram is fitted to a function that includes a Gaussian signal
and an appropriate background function.  For 
$D^{0}\rightarrow{K^{-}\pi^{+}}$, $D^{+}\rightarrow{K^{-}\pi^{+}\pi^{+}}$ 
and $D_{s}^{+}\rightarrow{K^{-}K^{+}\pi^{+}}$ above $4100$~MeV,
the background function is a second-order polynomial.
For $D_{s}^{+}\rightarrow{K^{-}K^{+}\pi^{+}}$ below
$4100$~MeV, the background function is an Argus function \cite{Argus}.
The results of the fits are shown in \TAB \ref{InFit_table}.

\begin{table}[!htbp]
\begin{center}
\caption{Yields from inclusive fits for ${D^{0}}$, ${D^{+}}$ and ${D^{+}_{s}}$.}
\label{InFit_table}
\vspace{0.2cm}
\begin{tabular}{|c|c|c|c|}\hline
{E$_{cm}$ MeV}& {${(D^{0})}$}& {${(D^{+})}$}& {${(D^{+}_{s})}$}
\\ \hline
3970 & $ 595.52 \pm 28.25 $ & $ 636.42 \pm 32.33 $ & $ 35.82 \pm 14.47 $ \\ \hline
3990 & $ 643.91 \pm 29.35 $ & $ 600.57 \pm 30.36 $ & $ 34.04 \pm 7.50 $ \\ \hline
4010 & $ 1280.07 \pm 40.56 $ & $ 1256.24 \pm 43.55 $ & $ 69.50 \pm 11.70 $ \\ \hline
4015 & $ 370.27 \pm 21.35 $ & $ 346.65 \pm 22.61 $ & $ 16.19 \pm 5.58 $ \\ \hline
4030 & $ 1192.53 \pm 37.29 $ & $ 844.87 \pm 36.18 $ & $ 43.42 \pm 9.14 $ \\ \hline
4060 & $ 1191.66 \pm 37.68 $ & $ 859.62 \pm 36.77 $ & $ 0.00 \pm 0.00 $ \\ \hline
4120 & $ 881.87 \pm 32.89 $ & $ 576.56 \pm 30.62 $ & $ 77.30 \pm 14.50 $ \\ \hline
4140 & $ 1512.51 \pm 43.45 $ & $ 1006.17 \pm 42.40 $ & $ 181.21 \pm 21.29 $ \\ \hline
4160 & $ 3130.11 \pm 62.79 $ & $ 2028.33 \pm 59.92 $ & $ 448.12 \pm
32.75 $ \\ \hline
4170 & $ 54958 \pm 262.702 $ & $ 34050.8 \pm 247.879 $ & $ 8379.97\pm146.051 $ \\ \hline

4180 & $ 1668.05 \pm 46.10 $ & $ 1088.37 \pm 44.12 $ & $ 243.47 \pm 23.80 $ \\ \hline
4200 & $ 727.11 \pm 30.61 $ & $ 445.36 \pm 27.59 $ & $ 97.80 \pm 15.19 $ \\ \hline
4260 & $ 1691.56 \pm 51.13 $ & $ 1370.82 \pm 52.60 $ & $ 330.49 \pm 34.93 $ \\ \hline
\end{tabular}
\end{center}
\end{table}

The observed inclusive cross section can be determined as follows:
\begin{equation}
         \sigma(e^+e^-\rightarrow{D_{(s)}}X)=
                        \frac{N(\rm{signal})}
                        {\BR{\cal{L}}\epsilon},
\end{equation}
\CONT where $N(\rm{signal})$ is the signal yield from the inclusive 
fit (\TAB \ref{InFit_table}), $\epsilon$ is the efficiency for detecting 
the signal (\TAB \ref{Inclusive_eff_table}), $\BR$ is the branching ratio for the 
particular $D$ decay in question, and $\cal{L}$ is the integrated luminosity.  
The cross section times branching ratio for $D^{0}$, $D^{+}$ and $D_{s}$ and
the cross sections determined from these with the branching ratios in
Ref.~\cite{CLEO_DHad_PRL} and Ref.~\cite{pdg} are shown in \TABS \ref{InXSBF_table}
and \ref{InXS_table}, respectively.

\begin{table}[!htbp]
\begin{center}
\caption{The cross section, in nb, times branching ratio of ${D^{0}}$, ${D^{+}}$ and ${D^{+}_{s}}$.}
\label{InXSBF_table}
\vspace{0.2cm}
\begin{tabular}{|c|c|c|c|}\hline
{E$_{cm}$}& {$\sigma\cdot{\BR}(D^{0}\rightarrow{K^{-}\pi^{+}})$}& {$\sigma\cdot{\BR}(D^{+}\rightarrow{K^{-}\pi^{+}\pi^{+}})$}& {$\sigma\cdot{\BR}(D^{+}_{s}\rightarrow{K^{-}K^{+}\pi^{+}})$}
\\ \hline
3970 &$ 0.244\pm0.012 $&$0.311\pm 0.016$&$0.017\pm 0.007$
\\ \hline
3990 &$ 0.309\pm0.014 $&$0.342\pm 0.017$&$0.019\pm 0.004$
\\ \hline
4010 &$ 0.364\pm0.012 $&$0.417\pm 0.014$&$0.023\pm 0.004$
\\ \hline
4015 &$ 0.395\pm0.023 $&$0.454\pm 0.030$&$0.021\pm 0.007$
\\ \hline
4030 &$ 0.615\pm0.019 $&$0.526\pm 0.023$&$0.027\pm 0.006$
\\ \hline
4060 &$ 0.574\pm0.018 $&$0.504\pm 0.022$&$0.000\pm 0.000$
\\ \hline
4120 &$ 0.511\pm0.019 $&$0.399\pm 0.021$&$0.055\pm 0.010$
\\ \hline
4140 &$ 0.496\pm0.014 $&$0.399\pm 0.017$&$0.075\pm 0.009$
\\ \hline
4160 &$ 0.493\pm0.010 $&$0.390\pm 0.012$&$0.084\pm 0.006$
\\ \hline
4170 &$ 0.493\pm0.002$&$0.370 \pm 0.003$&$0.088\pm 0.002$
\\ \hline

4180 &$ 0.472\pm0.013 $&$0.370\pm 0.015$&$0.080\pm 0.008$
\\ \hline
4200 &$ 0.410\pm0.017 $&$0.305\pm 0.019$&$0.067\pm 0.010$
\\ \hline
4260 &$ 0.212\pm0.006 $&$0.204\pm 0.008$&$0.048\pm 0.005$
\\ \hline
\end{tabular}
\end{center}
\end{table}

\begin{table}[!htbp]
\begin{center}
\caption{The inclusive cross section of ${D^{0}}$, ${D^{+}}$, and ${D^{+}_{s}}$}
\label{InXS_table}
\vspace{0.2cm}
\begin{tabular}{|c|c|c|c|}\hline
{E$_{cm}$ MeV}& {$\sigma_{(D^{0})}$~nb}& {$\sigma_{(D^{+})}$~nb}& {$\sigma_{(D^{+}_{s})}$~nb}
\\ \hline
3970 & $6.25\pm0.30$ & $3.27\pm0.17$ & $0.31\pm 0.13$
\\ \hline
3990 &$ 7.92\pm0.36 $&$3.59\pm 0.18$&$0.34\pm 0.08$
\\ \hline
4010 &$ 9.31\pm0.30 $&$4.38\pm 0.15$&$0.41\pm 0.07$
\\ \hline
4015 &$ 10.10\pm0.58 $&$4.77\pm 0.31$&$0.38\pm 0.13$
\\ \hline
4030 &$ 15.73\pm0.49 $&$5.52\pm 0.24$&$0.49\pm 0.10$
\\ \hline
4060 &$ 14.67\pm0.46 $&$5.29\pm 0.23$&$-\pm -$
\\ \hline
4120 &$ 13.08\pm0.49 $&$4.19\pm 0.22$&$0.99\pm 0.19$
\\ \hline
4140 &$ 12.68\pm0.36 $&$4.19\pm 0.18$&$1.34\pm 0.16$
\\ \hline
4160 &$ 12.60\pm0.25 $&$4.10\pm 0.12$&$1.50\pm 0.11$
\\ \hline
4170 &$ 12.51\pm0.06 $&$3.90\pm0.03$&$1.59\pm0.03$
\\ \hline
4180 &$ 12.08\pm0.33 $&$3.88\pm 0.16$&$1.44\pm 0.14$
\\ \hline
4200 &$ 10.48\pm0.44 $&$3.20\pm 0.20$&$1.21\pm 0.19$
\\ \hline
4260 &$ 5.43\pm0.16 $&$2.14\pm 0.08$&$0.86\pm 0.09$
\\ \hline
\end{tabular}
\end{center}
\end{table}

From the inclusive cross sections, the total charm cross section can be
obtained. Since all $D$ mesons are produced in pairs the total
cross section for production of charm events is given by
\begin{equation}
         \sigma(e^+e^-\rightarrow{D\bar{D}}X)= \frac{\sigma_{D^{0}} +
         \sigma_{D^{+}} + \sigma_{D^{+}_{s}}}{2}
\end{equation}

\CONT at each energy point.  These results are given in \TAB \ref{total_InXS_table}.

\begin{table}[!htbp]
\begin{center}
\caption{The total charm cross section as obtained by the inclusive method}
\label{total_InXS_table}
\vspace{0.2cm}
\begin{tabular}{|c|c|}\hline
{E$_{cm}$ MeV}& {$\sigma(e^+e^-\rightarrow{D\bar{D}X})$~nb}
\\ \hline
3970 &$4.910\pm 0.183$ \\ \hline
3990 &$5.926\pm 0.219$ \\ \hline
4010 &$7.050\pm 0.180$ \\ \hline
4015 &$7.623\pm 0.360$ \\ \hline
4030 &$10.870\pm 0.294$ \\ \hline
4060 &$9.979\pm 0.258$ \\ \hline
4120 &$9.132\pm 0.281$ \\ \hline
4140 &$9.105\pm 0.212$ \\ \hline
4160 &$9.103\pm 0.147$ \\ \hline
4170 &$9.094\pm 0.066$ \\ \hline
4180 &$8.701\pm 0.193$ \\ \hline
4200 &$7.448\pm 0.250$ \\ \hline
4260 &$4.212\pm 0.093$ \\ \hline
\end{tabular}
\end{center}
\end{table}

%%%%%%%%%%%%%%%%%%%%%%%%%%%%%%%%%%%%%%%%%%%%%%%%%%%%%%%%%%%%%%%%%%%
\subsubsection{Hadron-Counting Method}
%%%%%%%%%%%%%%%%%%%%%%%%%%%%%%%%%%%%%%%%%%%%%%%%%%%%%%%%%%%%%%%%%%%
Still another method and cross-check can be done.  This method
involves counting the multihadronic events in the data at all thirteen
energy points and using data collected below $c {\bar c}$ threshold 
at \(E_{\rm{cm}}=3671\)~MeV to subtract $uds$ continuum production.
Except for one difference to be discussed 
later, this method is the same as that used by
CLEO-c \cite{Hajime,Hajime_update,Hajime_PRL} to determine the 
cross section of \(e^+e^-\rightarrow\psi(3770)\rightarrow{hadrons}\) at
\(E_{\rm{cm}}=3770\)~MeV.  The Standard Hadron ({\tt SHAD}) cuts as
discussed in Refs. \cite{Hajime,Hajime_PRL} are used.

This method first starts by calculating the number of hadronic
continuum events at \(E_{\rm{cm}}=3671\)~MeV:
\begin{equation}
\label{eq:Hajime_cont}
           N(3671) = (N_{{\rm{obs}}} - N_{\psi(2S)} -
           N_{J/\psi}\cdot\epsilon_{J/\psi}- N_{ee}\cdot\epsilon_{ee}-
           N_{\mu\mu}\cdot\epsilon_{\mu\mu} - N_{\tau\tau}\cdot\epsilon_{\tau\tau})/\epsilon_{{\rm{cont}}},
\end{equation}
\CONT where the $N$'s are the numbers of events of different types as determined 
from data or by calculating ${\cal{L}}\cdot\sigma$, and the $\epsilon$'s are
the efficiencies for passing the {\tt SHAD} cuts of the same event types.  The
hadronic efficiency for events from the $uds$ continuum is $\epsilon_{cont}$.  The
quantity \(N_{\psi(2S)}\) in \EQ \ref{eq:Hajime_cont} is the contribution due to 
the Breit-Wigner tail of \(\psi(2S)\).  It is estimated from the data collected at
$E_{\rm{cm}}=3671$ and 3686~MeV as follows:
\begin{equation}
\label{eq:NPsi2S}
           N(\psi(2S))_{3671} = \frac{N(\pi^+\pi^-l^+l^-)_{3671}}
	{N(\pi^+\pi^-l^+l^-)_{3686}}\cdot{N(\psi(2S))_{3686}},
\end{equation}
\CONT where ${N(\psi(2S))_{3686}}$ is the number of hadronic events in $\psi(2S)$ decays at 
\(E_{\rm{cm}}=3686\)~MeV. The values used for $N(\pi^+\pi^-l^+l^-)_{3686}$ and 
$N(\pi^+\pi^-l^+l^-)_{3671}$ in \EQ \ref{eq:Hajime_cont} were obtained by CLEO-c \cite{BrianandHanna} and give a scale factor of 
$\frac{N(\pi^+\pi^-l^+l^-)_{3671}}{N(\pi^+\pi^-l^+l^-)_{3686}} = \frac{221}{30462} = 0.0073$.
The number of hadronic events in $\psi(2S)$ decays at \(E_{\rm{cm}}=3686\)~MeV is determined by
\begin{equation}
\label{eq:RawPsi2S}
           N(\psi(2S))_{3686} = N(3868)_{\rm{obs}} - S\cdot{N_{{\rm{obs}}}(3671)},
\end{equation}
\CONT where the scale factor, 
$S=\frac{{\cal{L}}_{3686}}{{\cal{L}}_{3671}}\cdot(\frac{3671}{3686})^{2}=0.139$.  In \EQ 
\ref{eq:RawPsi2S} we neglect the contamination of $\psi(2S)$ in the off-resonance data 
and QED events, which is small compared to the large number of $\psi(2S)$ decays present in 
the sample.

After determining \(N(3671)_{\rm{obs}}\), it can be used in determining the number
of hadronic events at each scan point as follows:
\begin{eqnarray}
\label{eq:Hajime_scan}
           N(X) = (N_{{\rm{obs}}}(X) - S\cdot{N_{{\rm{obs}}}(3671)} -
           N_{\psi(2S)}\cdot\epsilon_{\psi(2S)}
	   - N_{\psi(3770)}\cdot\epsilon_{\psi(3770)} - \nonumber\\ 
           N_{J/\psi}\cdot\epsilon_{J/\psi}- N_{ee}\cdot\epsilon_{ee}-
           N_{\mu\mu}\cdot\epsilon_{\mu\mu} - N_{\tau\tau}\cdot\epsilon_{\tau\tau}),
\end{eqnarray}

\CONT where X stands for the energy point in question (3970, 3990,
4010, 4015, 4030, 4060, 4120, 4140, 4160, 4170, 4180, 4200, and 4260~MeV),
and \(S\) is the scale factor given by
$\frac{{\cal{L}}_{X}}{{\cal{L}}_{3671}}\cdot(\frac{3671~{\rm{MeV}}}{X~{\rm{MeV}}})^{2}$.

In determining \(N(3671)\) CLEO-c's method was followed
exactly; that is the data collected at the \(\psi(2S)\) resonance was used
in determining \(N_{\psi(2S)}\) in \EQ \ref{eq:Hajime_cont}.
However, in regards to $N(X)$
we used the calculated production cross section in determining the
amount of \(\psi(2S)\), in addition to the amount of
\(J/\psi\) and \(\psi(3770)\) present at each energy point. The calculated production cross section is a convolution of
a $\delta$-function-approximated Breit-Wigner and an ISR kernel:
\begin{equation}
\sigma(e^+e^-\rightarrow\gamma{X}) = \frac{12\pi^2\Gamma_{ee}}{s{\rm{M_X}}}
\cdot f(s,x),
\end{equation}
\CONT where the ISR kernel f(x,s) is defined in \EQ 28 of Ref. \cite{KF} and reproduced 
here as \EQ \ref{eq:KFEQ} in Sect.~\ref{sec:radcor}, with 
$x=\frac{(s-{\rm{M_X}}^2)}{s}$ and ${\rm{M_X}}$ referring to the mass of 
the \(\psi(2S)\), \(J/\psi\) or \(\psi(3770)\).  The
calculated production cross sections for \(\psi(2S)\), \(J/\psi\), and
$\psi(3770)$ are shown in \TAB \ref{prod_JPsiPsi2S_HCXS}. It should be
noted that we could use the results from our data samples, the number of
$e^+e^-\rightarrow{\gamma\psi(2S)}\rightarrow\gamma\pi^+\pi^-J/\psi$ events
at each of the scan energies, to improve this result in the future.
Also, in calculating $N(X)$ it was assumed that effects due to interference 
between $e^+e^-\rightarrow\psi(2S)\rightarrow\gamma^*\rightarrow{hadrons}$ and
the continuum were negligible at all the scan energies, and so were only
included in the calculation of \(N(3671)\) using the method described
in Ref. \cite{Hajime_update,Hajime_PRL}.
The numbers determined and used in this method are shown for
all energy points in \TABS \ref{HadCount_breakdown_off}
and \ref{HadCount_breakdown}.
\begin{table}[!htbp]
\begin{center}
\caption{Calculated production cross sections, in nb, for $J/\psi$, 
$\psi(2S)$, and $\psi(3770)$ following the procedure discussed in the text.}
\label{prod_JPsiPsi2S_HCXS}
\vspace{0.2cm}
\begin{tabular}{|c|c|c|c|}\hline
{E$_{cm}$ (MeV)}& {$\sigma(e^+e^-\rightarrow{\gamma{J/\psi}})$}& {$\sigma(e^+e^-\rightarrow{\gamma\psi(2S)})$}& {$\sigma(e^+e^-\rightarrow{\gamma\psi(3770)})$}
\\ \hline
3970 & $ 0.70 $ & $ 0.92 $&$ 0.13 $ \\ \hline
3990 & $ 0.68 $ & $ 0.85 $&$ 0.12 $ \\ \hline
4010 & $ 0.66 $ & $ 0.79 $&$ 0.11 $ \\ \hline
4015 & $ 0.66 $ & $ 0.78 $&$ 0.11 $ \\ \hline
4030 & $ 0.64 $ & $ 0.74 $&$ 0.10 $ \\ \hline
4060 & $ 0.62 $ & $ 0.68 $&$ 0.09 $ \\ \hline
4120 & $ 0.57 $ & $ 0.57 $&$ 0.07 $ \\ \hline
4140 & $ 0.56 $ & $ 0.54 $&$ 0.07 $ \\ \hline
4160 & $ 0.54 $ & $ 0.52 $&$ 0.06 $ \\ \hline
4170 & $ 0.54 $ & $ 0.51 $&$ 0.06 $ \\ \hline

4180 & $ 0.53 $ & $ 0.49 $&$ 0.06 $ \\ \hline
4200 & $ 0.52 $ & $ 0.47 $&$ 0.06 $ \\ \hline
4260 & $ 0.48 $ & $ 0.41 $&$ 0.05 $ \\ \hline
\end{tabular}
\end{center}
\end{table}

Wide-angle Bhabha events were generated with {\tt BHLUMI} \cite{Bhabha} using a cut off angle of $21.57^o$; 
$\mu$-pairs and $\tau$-pairs were generated with {\tt FPAIR} \cite{Fpair} and
{\tt KORALB} \cite{Koralb}, respectively.  These calculations provide the production 
cross sections needed for \EQS \ref{eq:Hajime_scan} and \ref{eq:Hajime_scan}
and were used to produce MC samples that were used to determine the {\tt SHAD} 
selection efficiencies.  The QED production cross sections are shown in 
\TAB \ref{prod_XS_HCXS} and their efficiencies are given in \TAB \ref{eff_XS_HCXS}. 

\begin{table}[!htbp]
\begin{center}
\caption{Calculated QED production cross sections at twelve energy points.}
\label{prod_XS_HCXS}
\vspace{0.2cm}
\begin{tabular}{|c|c|c|c|}\hline
{E$_{cm}$ (MeV)}& {$e^+e^-(\theta_{min} = 21.57^o)$~(nb)}& {$\mu^+\mu^-$~(nb)}& {$\tau^+\tau^-$~(nb)}
\\ \hline
3670 & 448.2 & 8.11& 2.1 \\ \hline
3970 & 383.47 & 6.99& 3.32 \\ \hline
3990 & 379.59 & 6.98& 3.38 \\ \hline
4010 & 376.39 & 6.83& 3.4 \\ \hline
4015 & 374.01 & 6.86& 3.4 \\ \hline
4030 & 372.34 & 6.81& 3.44 \\ \hline
4060 & 366.5 & 6.69& 3.45 \\ \hline
4120 & 355.95 & 6.5& 3.51 \\ \hline
4140 & 352.72 & 6.47& 3.53 \\ \hline
4160 & 348.78 & 6.4& 3.53 \\ \hline
4170 & 346.56 & 6.36& 3.54 \\ \hline
4180 & 344.77 & 6.33& 3.55 \\ \hline
4200 & 344.71 & 6.29& 3.56 \\ \hline
4260 & 332.4 & 6.17& 3.57 \\ \hline
\end{tabular}
\end{center}
\end{table}

\begin{table}[!htbp]
\begin{center}
\caption{Efficiencies (units of $10^{-2}$) for various event types to pas {\tt SHAD} 
hadronic event selection criteria.}
\label{eff_XS_HCXS}
\vspace{0.2cm}
\footnotesize{
\begin{tabular}{|c|c|c|c|c|c|c|c|} \hline
{E$_{cm}$}& {$e^+e^-$}& {$\mu^+\mu^-$}& {$\tau^+\tau^-$}& {$J/\psi$}& {$\psi(2S)$}& {$q\bar{q}$}& {$Signal$}
\\ \hline
3670 & $0.03\pm0.01$ & $0.12\pm0.02$ & $22.0\pm0.1$& $41.9\pm0.6$ & $68.3\pm0.1$& $59.0\pm0.2$ & $-\pm-$ \\ \hline
3970 & $0.02\pm0.01$ & $0.12\pm0.02$ & $22.8\pm0.8$& $39.3\pm0.3$ & $66.3\pm0.3$& $63.1\pm0.2$ & $80.9\pm0.1$ \\ \hline
3990 & $0.04\pm0.01$ & $0.12\pm0.02$ & $20.9\pm0.8$& $39.0\pm0.3$ & $65.5\pm0.3$& $62.8\pm0.2$ & $80.9\pm0.1$ \\ \hline
4010 & $0.03\pm0.01$ & $0.12\pm0.02$ & $21.6\pm0.8$& $38.5\pm0.3$ & $65.9\pm0.3$& $63.3\pm0.2$ & $81.0\pm0.1$ \\ \hline
4015 & $0.03\pm0.01$ & $0.12\pm0.02$ & $22.8\pm0.8$& $38.2\pm0.3$ & $65.6\pm0.3$& $63.3\pm0.2$ & $82.0\pm0.1$ \\ \hline
4030 & $0.05\pm0.01$ & $0.12\pm0.02$ & $21.8\pm0.8$& $38.1\pm0.3$ & $65.7\pm0.3$& $63.6\pm0.2$ & $81.9\pm0.1$ \\ \hline
4060 & $0.02\pm0.01$ & $0.12\pm0.02$ & $22.8\pm0.8$& $39.2\pm0.3$ & $65.7\pm0.3$& $63.7\pm0.2$ & $82.7\pm0.1$ \\ \hline
4120 & $0.05\pm0.01$ & $0.12\pm0.02$ & $22.8\pm0.8$& $35.1\pm0.3$ & $65.3\pm0.3$& $64.1\pm0.2$ & $82.6\pm0.1$ \\ \hline
4140 & $0.04\pm0.01$ & $0.12\pm0.02$ & $20.9\pm0.8$& $35.0\pm0.3$ & $64.8\pm0.3$& $64.5\pm0.2$ & $82.8\pm0.1$ \\ \hline
4160 & $0.04\pm0.01$ & $0.12\pm0.02$ & $21.7\pm0.8$& $34.2\pm0.3$ &
$65.7\pm0.3$& $64.8\pm0.2$ & $82.8\pm0.1$ \\ \hline
4170 & $0.04\pm0.01$ & $0.12\pm0.02$ & $21.9\pm0.8$& $34.3\pm0.3$ &$65.4\pm0.3$& $64.9\pm0.2$ & $82.6\pm0.1$ \\ \hline
4180 & $0.04\pm0.01$ & $0.12\pm0.02$ & $23.6\pm0.8$& $33.6\pm0.3$ & $65.0\pm0.3$& $65.1\pm0.2$ & $82.7\pm0.1$ \\ \hline
4200 & $0.03\pm0.01$ & $0.12\pm0.02$ & $21.6\pm0.8$& $32.2\pm0.3$ & $64.5\pm0.3$& $65.1\pm0.2$ & $82.8\pm0.1$ \\ \hline
4260 & $0.05\pm0.01$ & $0.12\pm0.02$ & $21.9\pm0.8$& $28.8\pm0.3$ & $64.7\pm0.3$& $65.7\pm0.2$ & $82.7\pm0.1$ \\ \hline
\end{tabular}}
\end{center}
\end{table}

\begin{figure}[!htbp]
\begin{center}
\hspace{2.5pt}
\includegraphics[width=14.5cm]{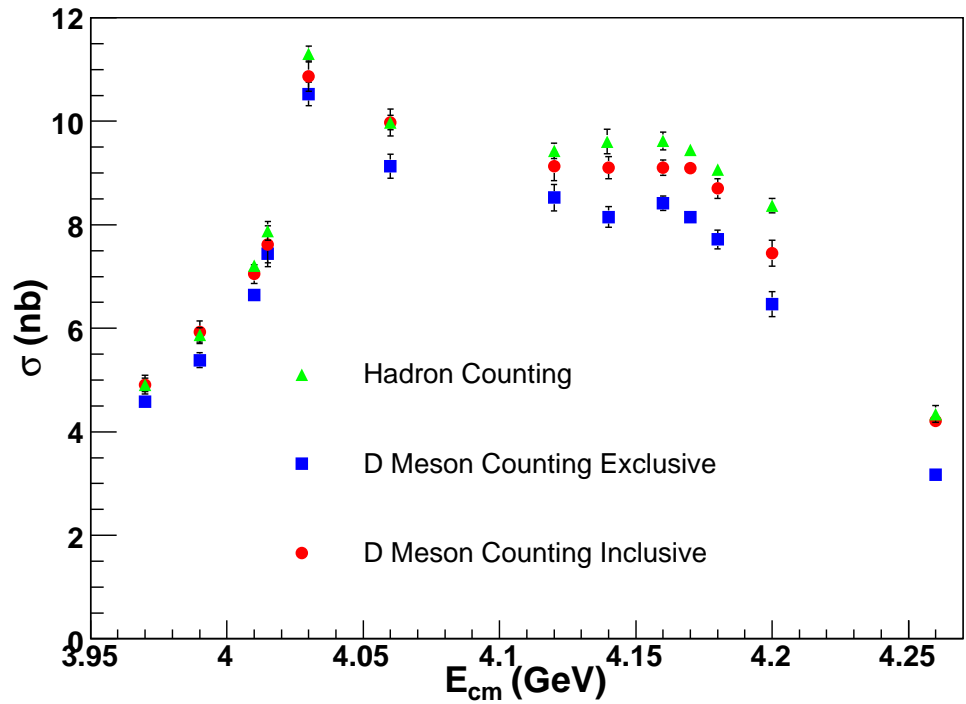}
\caption{Plot of the total charm cross section as calculated in each of the
three methods described in the text.  The reason
for the discrepancy between the inclusive and exclusive method is due
to the presence of multi-body production, which is described in the text. Only statistical errors are shown.}
\vspace{0.2cm}
\label{all_plot}
\end{center}
\end{figure}

Once the number of supposed pure charm decays \(N(X)\) has been
obtained, the total charm cross section can be calculated 
at each energy point as follows:
\begin{equation}
           \sigma_{c}(X) = \frac{N(X)}{{\cal{L}}_{X}\epsilon_{X}}.
\end{equation}
\CONT
The results using this hadron-counting method are shown in \TAB
\ref{total_HCXS_table}.  A comparison of all methods is shown in \FIG \ref{all_plot}.

\begin{table}[!htbp]
\begin{center}
\caption{The total charm cross section as obtained by the Hadron
Counting Method. Only statistical errors are shown.}
\label{total_HCXS_table}
\vspace{0.2cm}
\begin{tabular}{|c|c|}\hline
{E$_{cm}$ MeV}& {$\sigma(e^+e^-\rightarrow{Charm})$~nb}
\\ \hline
$3970$ & $4.91\pm0.13$\\ \hline
$3990$ & $5.87\pm0.14$\\ \hline
$4010$ & $7.21\pm0.12$\\ \hline
$4015$ & $7.88\pm0.18$\\ \hline
$4030$ & $11.30\pm0.15$\\ \hline
$4060$ & $9.98\pm0.14$\\ \hline
$4120$ & $9.43\pm0.15$\\ \hline
$4140$ & $9.58\pm0.12$\\ \hline
$4160$ & $9.62\pm0.09$\\ \hline
$4170$ & $9.44\pm0.09$\\ \hline
$4180$ & $9.07\pm0.12$\\ \hline
$4200$ & $8.37\pm0.14$\\ \hline
$4260$ & $4.34\pm0.08$\\ \hline

\end{tabular}
\end{center}
\end{table}

\chapter{Momentum Spectrum Analysis}
%%%%%%%%%%%%%%%%%%%%%%%%%%%%%%%%%%%%%%%%%%%%%%%%%%%%%%%%%%%%%%%%%%%%%%%%
\section{Multi-Body, Initial-State Radiation, and \\Momentum-Spectrum Fits}
%%%%%%%%%%%%%%%%%%%%%%%%%%%%%%%%%%%%%%%%%%%%%%%%%%%%%%%%%%%%%%%%%%%%%%%%
Throughout this analysis we have assumed that all charm production is
through two-body events, that is
\(e^{+}e^{-}\rightarrow{D_{(s)}^{(*)}\bar{D}_{(s)}^{(*)}}\).
With sufficient energy, however, there is no reason that 
final states like \(e^{+}e^{-}\rightarrow{D\bar{D}^{(*)}\pi}\) or any 
other energetically allowed combination with extra pions should not
be produced.  From here on,
these types of events are referred to as the multi-body production or
just multi-body. For production of non-strange $D$ states, there is no
{{\it a priori}} expectation for the amount of multi-body. For $D_{s}$
we expect it to be small, if not zero, since $D_{s}\bar{D}_{s}\pi$
violates isospin. In both cases it is appropriate to examine our data
for evidence of such multi-body processes.

The first step is to determine whether multi-body exists.  Assuming
that it does, we then need to develop and apply procedures to
determine its composition: \(D\bar{D}\pi\), \(D^{*}\bar{D}\pi\), or \(D^{*}\bar{D}\pi\pi\).

\begin{figure}[!htbp]
\begin{center}
\hspace{2.5pt}
\includegraphics[width=14.5cm]{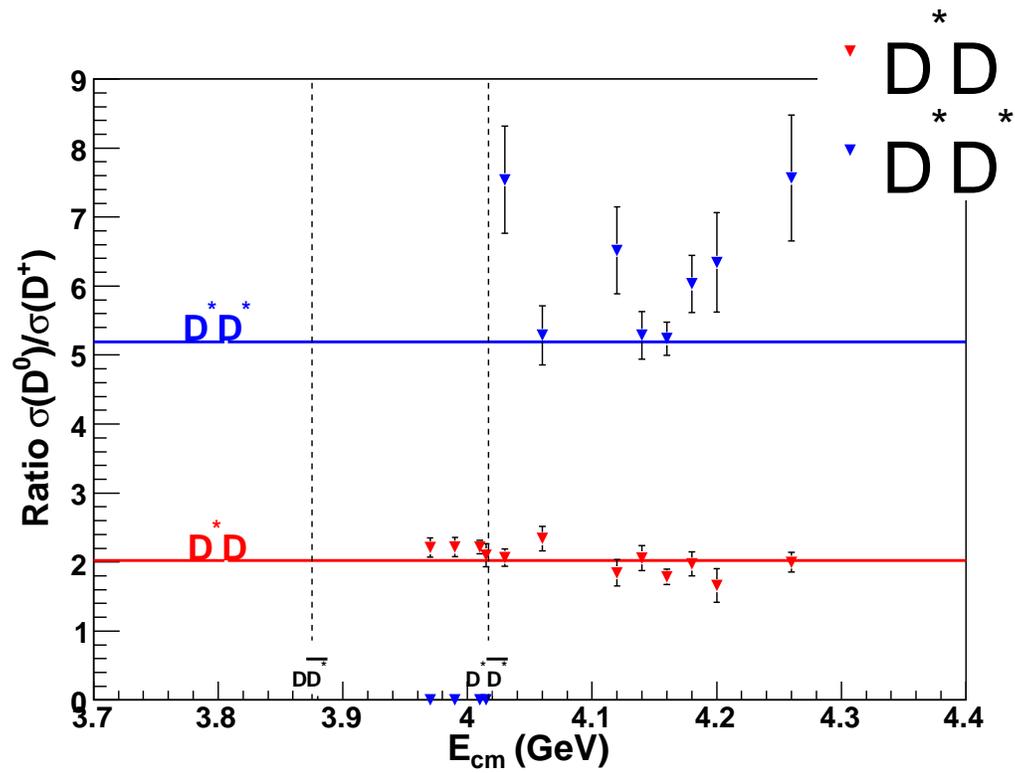}
\caption{The measured ratio of $D^{0}$ to $D^{+}$ as a function of energy.}
\vspace{0.2cm}
\label{fig:D0Dp_ratio}
\end{center}
\end{figure}

To determine if multi-body exists we can apply tests
of consistency between our measurements and the expectations for pure
$D_{(s)}^{(*)}\bar{D}_{(s)}^{(*)}$ states.  One observable is the ratio of 
\(\frac{D^{0}}{D^{+}}\) as a function of energy,
which is shown in \FIG \ref{fig:D0Dp_ratio}.  The bold horizontal
lines are the predicted ratios as determined by the decays of the
\(D^{*}\) using the information in \TAB \ref{DStarmode_table}.
It is evident that the observed ratio deviates from that expected for
\(D^{*}\bar{D}^{*}\) events.  The only candidate explanation for
this observation is multi-body production.  The cut window in 
$M_{\rm{bc}}$ is different for neutral and charged \(D^{*}\). The 
charged window is quite a bit smaller than the neutral, since the $\gamma$ decay is excluded when
determining the cut window. Therefore, the cuts can select different
amounts of the multi-body, which leads to a result which is not
consistent with that expected based on known branching fractions.  The
fact that the \(D^{*}\bar{D}\) events have the correct ratio will help
in pin-pointing the make-up of the background. 

\begin{figure}[t]
\begin{center}
\hspace{2.5pt}
\includegraphics[width=14.5cm]{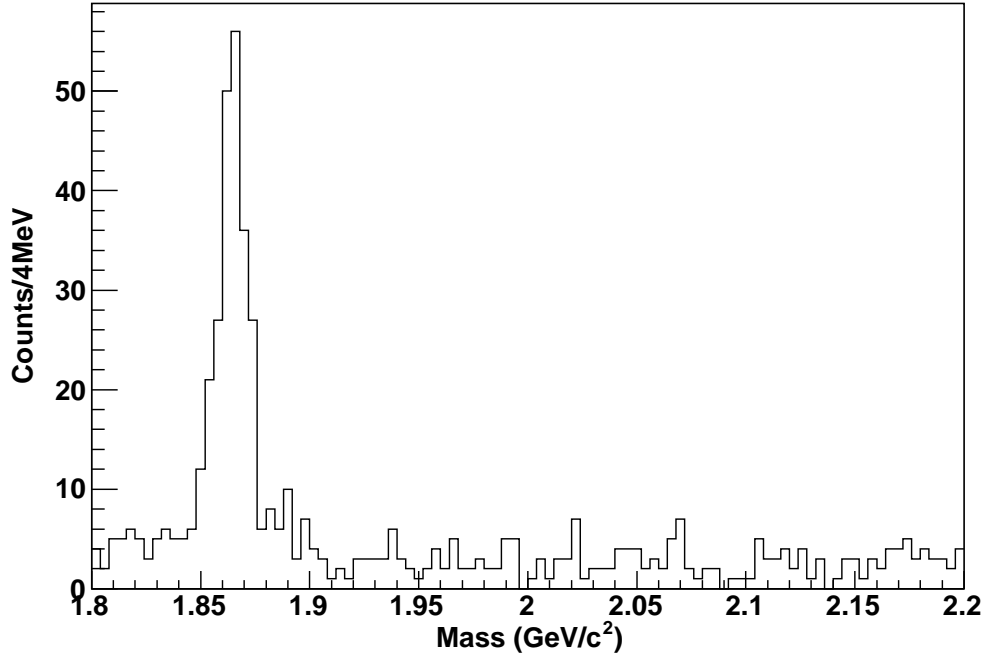}
\caption{The measured $D^{0}\rightarrow{K^-}\pi^+$ invariant mass distribution for
      $p_{D^{0}}<{300~{\rm{MeV/}}c}$.  If the assumption of pure
      two-body events is made, then no $D^{0}$ decays are
      kinematically allowed below $\sim500$~MeV/$c$ at $4260$~MeV. }
\vspace{0.2cm}
\label{fig:D0_4260_Momcut}
\end{center}
\end{figure}

The ratio of \(\frac{D^{0}}{D^{+}}\) gives a hint that there might be
multi-body background present at the energies of interest, but it is 
far from conclusive.  Another indication that points to a multi-body
background is the noticeable difference in the total charm
cross section between the exclusive and inclusive methods.  The
difference between these two methods as a function of energy can be
seen in \FIG \ref{all_plot}. The best way to prove the presence of multi-body is by looking for
\(D\) or \(D^{*}\) mesons in a kinematically forbidden region. That is, 
we look for \(D\) mesons in a momentum region where one would expect none 
under the assumption of pure two-body events. \FIG \ref{fig:D0_4260_Momcut} 
shows a plot of the invariant mass of \(D^{0}\rightarrow{K^-\pi^+}\) at the 
center-of-mass energy \(4260\)~MeV
for \(D^{0}\) candidates with momenta less than \(300\)~MeV/$c$.  Under the
assumption that only pure \(D^{*}\bar{D}^{*}\), \(D^{*}\bar{D}\), and \(D\bar{D}\) are
present, no \(D^{0}\) are allowed below
\(\sim500\)~MeV/$c$.  The figure shows a clearly defined peak at
the \(D^{0}\) mass, demonstrating the multi-body contribution.

\begin{figure}[!htbp]
\begin{center}
\hspace{2.5pt}
\includegraphics[width=14.5cm]{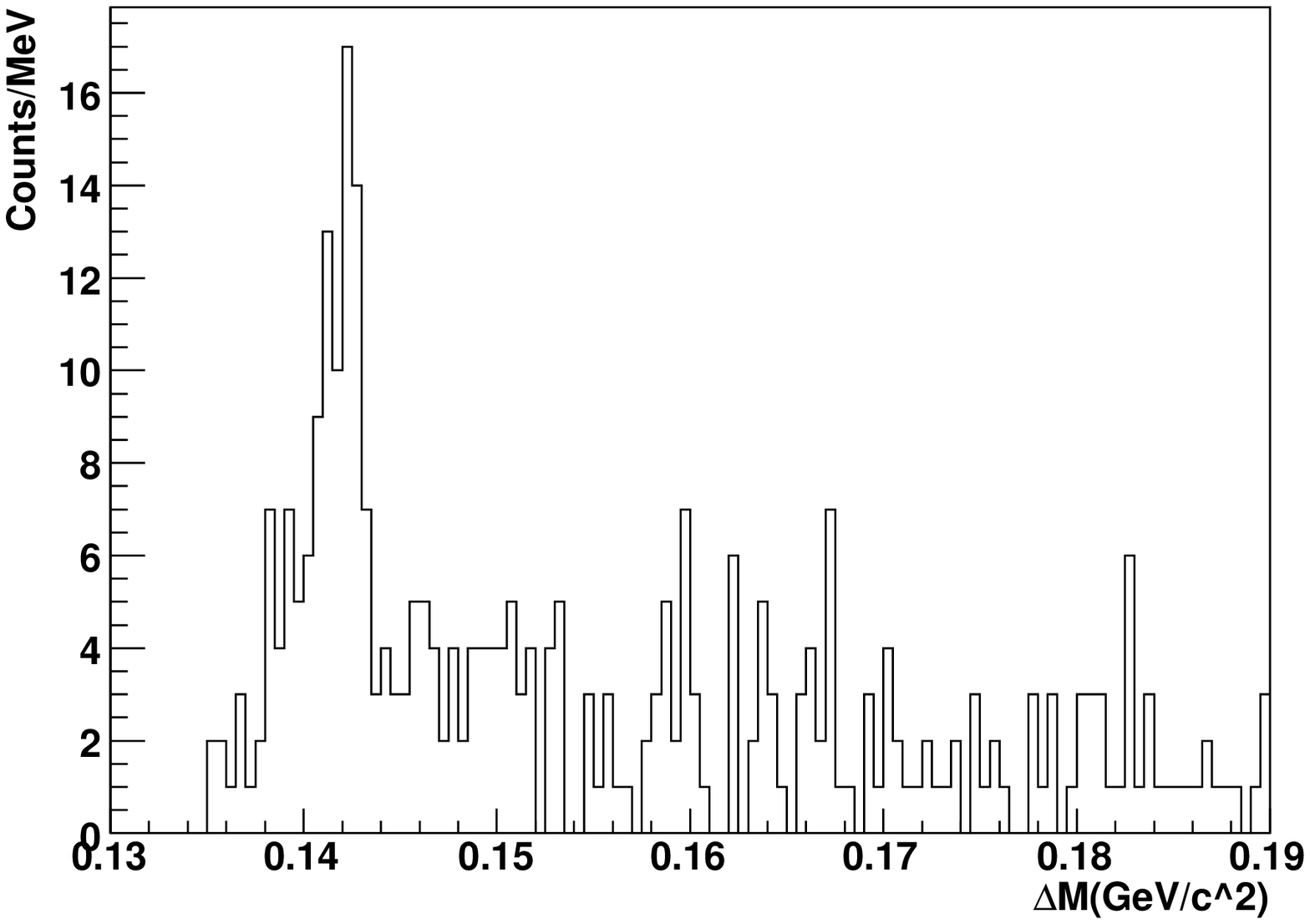}
\caption{The measured $D^{*0}\rightarrow{D^0}\pi^0$ $\Delta$M distribution for 
$D^{0}\rightarrow{K^-}\pi^+$ satisfying
      $p_{D^{*0}}<{500~{\rm{MeV/}}c}$.  If the assumption of pure
      two-body events is made, then no $D^{*0}$ decays are
      kinematically allowed below $\sim700$~MeV/$c$ at $4260$~MeV.}
\vspace{0.2cm}
\label{fig:DS0_4260_Momcut}
\end{center}
\end{figure}

This demonstrates that
charm is produced through more than the two-body event categories, but
it sheds no light on the composition.  Are the multi-body events $D\bar{D}\pi$, $D\bar{D}\pi\pi$,
$D^*\bar{D}\pi$, etc., or a combination of all possible types?  To help
answer this question a similar study to the one above, was performed
for $D^{*0}$. \FIG \ref{fig:DS0_4260_Momcut} shows a plot of the $\Delta$M spectrum for
$p_{D^{*0}}<{500~{\rm{MeV/}}c}$.  If the assumption of only pure two-body events is made, then no $D^{*0}$ decays are kinematically
allowed below $\sim700$~MeV/$c$. The figure clearly shows that there
exist multi-body events of the type $D^*\bar{D}\pi$ in the data,
evident in the well-defined peak located at
$\Delta{M}=M_{D^{*0}}-M_{D^{0}}=0.142~{\rm{MeV/}}c^2$.

Looking in the kinematically forbidden region has given clues to the
possible composition of the multi-body events, but still does not
provide a definitive and 
quantitative breakdown. One reason is that initial-state radiation
(ISR), can lead to $D$ mesons smeared
outside of the two-body kinematic regions.  A more
definitive test is to
reconstruct a $D^{*}$, add a charged or neutral $\pi$, and
look at the missing mass (recoil mass) of the event.  A peak in the
missing mass at the $D$ would clearly demonstrate the presence of $D^{*}\bar{D}\pi$ in the data.
We concentrate on multi-body events of the type
$D^{(*)0}D^{\pm}\pi^{\mp}$, which by isospin are twice as likely to occur as
$D^{(*)0}\bar{D^{0}}\pi^{0}$ and $D^{(*)+}\bar{D^{-}}\pi^{0}$. In addition to the factor
of two from isospin, the multi-body events with a charged pion
should be cleaner than those with a neutral pion, since the additional
pion will be soft, $P_{\pi}<{150}$~MeV/$c$. With this method the
observation of multi-body cannot be obscured by ISR, because the
presence of the radiative photon will prohibit peaks in the
missing-mass spectrum.

\begin{figure}[!htbp]
\begin{center}
\hspace{2.5pt}
\includegraphics[width=10.5cm, angle=270]{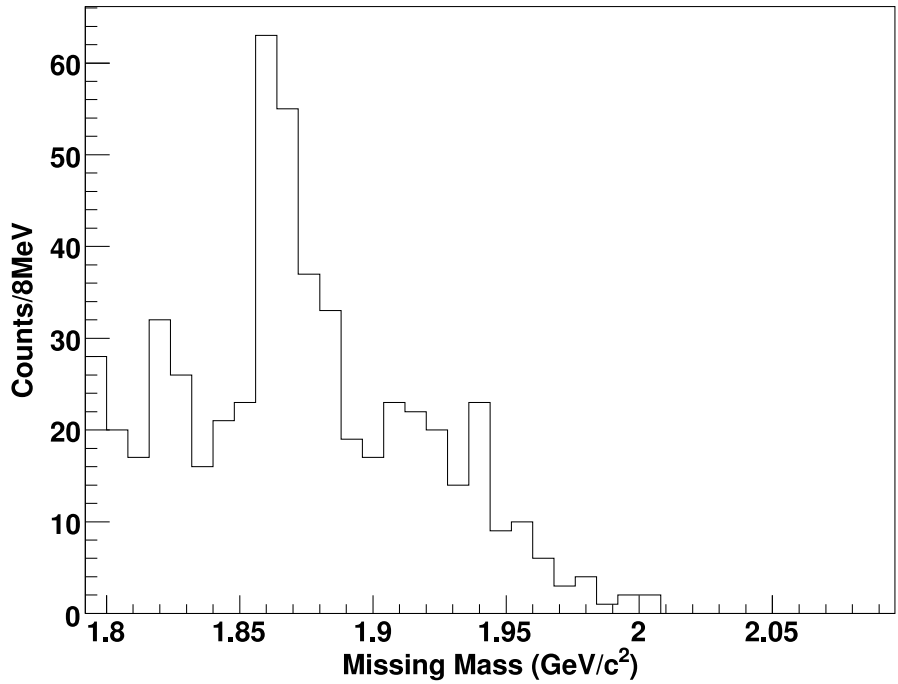}
\caption{The mass spectrum of $X$, in
$e^+e^-\rightarrow{D^{*\pm}\pi^{\mp}X}$, using the
$179~{\rm{pb}}^{-1}$ collected at $4170$~MeV. The peak at the $D$
mass shows conclusive evidence that events of the form
$D^{*}\bar{D}\pi$ are present in the data. A cut on the reconstructed $D^{*}$
momentum of 400 ${\rm{MeV}/}c$ was used to exclude two-body events.}
\vspace{0.2cm}
\label{fig:MultiBody_DSDPiPlus_4170}
\end{center}
\end{figure}

\begin{figure}[!htbp]
\begin{center}
\hspace{2.5pt}
\includegraphics[width=10.5cm, angle=270]{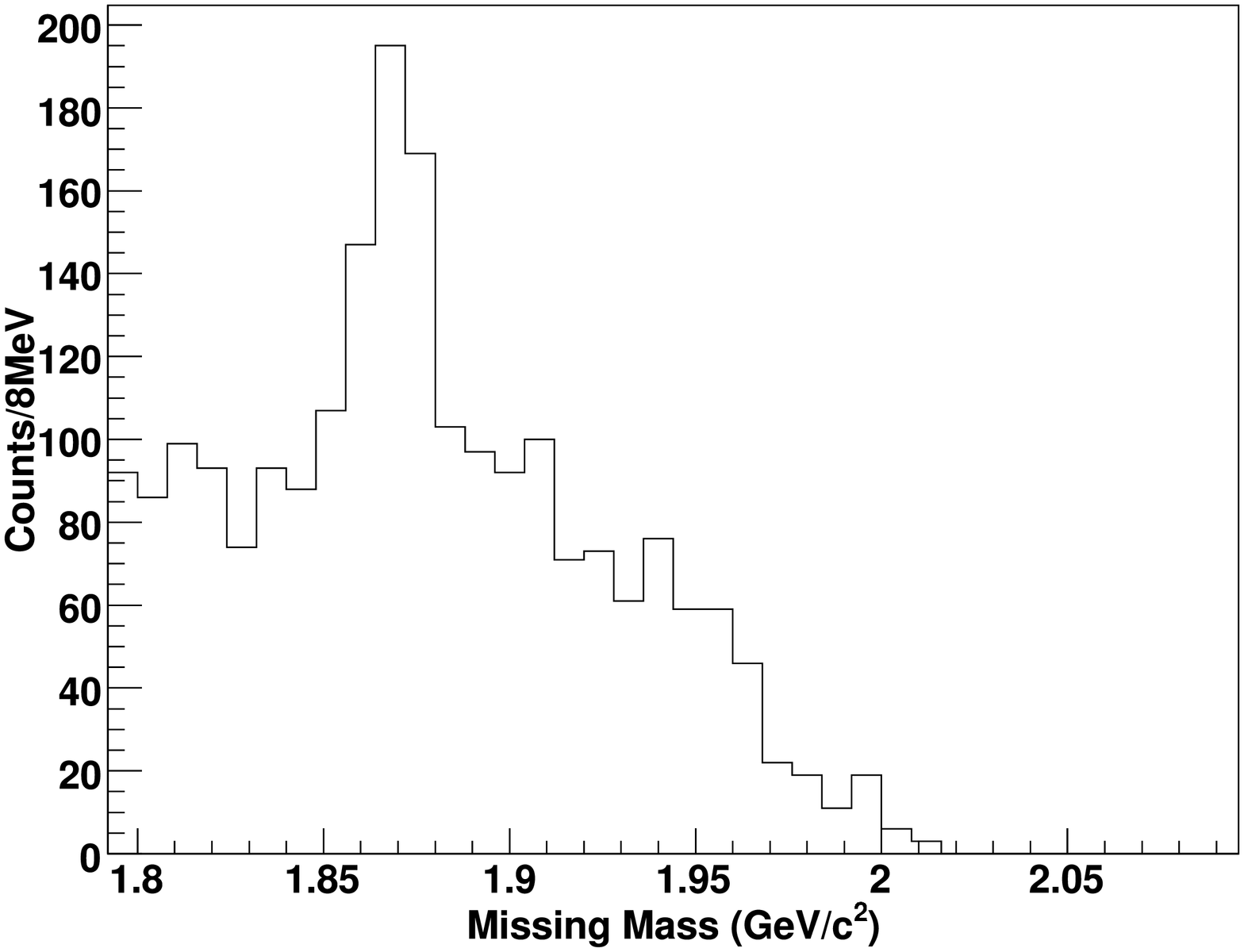}
\caption{The mass spectrum of $X$, in
$e^+e^-\rightarrow{D^{*0}\pi^{\pm}X}$, using the
$179~{\rm{pb}}^{-1}$ collected at $4170$~MeV. The peak at the $D$
mass shows conclusive evidence for production of
$D^{*}\bar{D}\pi$ events. A cut on the reconstructed $D^{*}$
momentum of 400 ${\rm{MeV}/}c$ was used to exclude two-body events. In
addition, the charge of the $D$-daughter kaon is used to obtain the
correct combination of the neutral $D^{*}$ and the charged pion.}
\vspace{0.2cm}
\label{fig:MultiBody_DSDPiPlus_4170_2}
\end{center}
\end{figure}

For this study we use the high-statistics data sample collected at 4170 MeV,
consisting of 179 pb$^{-1}$. Using $D^{*+}\rightarrow{D^+}\pi^0$ with
$D^{+}\rightarrow{K^-}\pi^+\pi^+$, and $D^{*0}\rightarrow{D^0}\pi^0$
with $D^{0}\rightarrow{K^-}\pi^+$, $D^{0}\rightarrow{K^-}\pi^+\pi^{0}$, or
$D^{0}\rightarrow{K^-}\pi^+\pi^{+}\pi^{-}$, in addition to {\tt{DTAG}}-like cuts for
the charged pion, we obtain the missing-mass spectra shown in \FIGS
\ref{fig:MultiBody_DSDPiPlus_4170} and
\ref{fig:MultiBody_DSDPiPlus_4170_2}.
\CONT \FIG \ref{fig:MultiBody_DSDPiPlus_4170} is the invariant mass spectrum of $X$ in
$e^+e^-\rightarrow{D^{*\pm}\pi^{\mp}X}$, while \FIG
\ref{fig:MultiBody_DSDPiPlus_4170_2} is for $X$ in
$e^+e^-\rightarrow{D^{*0}\pi^{\pm}X}$, where for the latter decay the
charge of the $D$-daughter kaon is used to obtain the
correct combination of the neutral $D^{*}$ and the charged pion.  For
both cases we define the signal region as a 6~MeV window centered on the previously measured
(PDG) mass difference.  A cut on the reconstructed $D^{*}$
momentum of 400 ${\rm{MeV}/}c$ was used to exclude two-body events.
These figures provide conclusive evidence that multi-body events of the
form $D^{*}\bar{D}\pi$ exist in the data.
We must now determine the amount of multi-body present to assess the effect on the two-body cross sections that were determined earlier.

\begin{figure}[!htbp]
\begin{center}
\hspace{2.5pt}
\includegraphics[width=14.5cm]{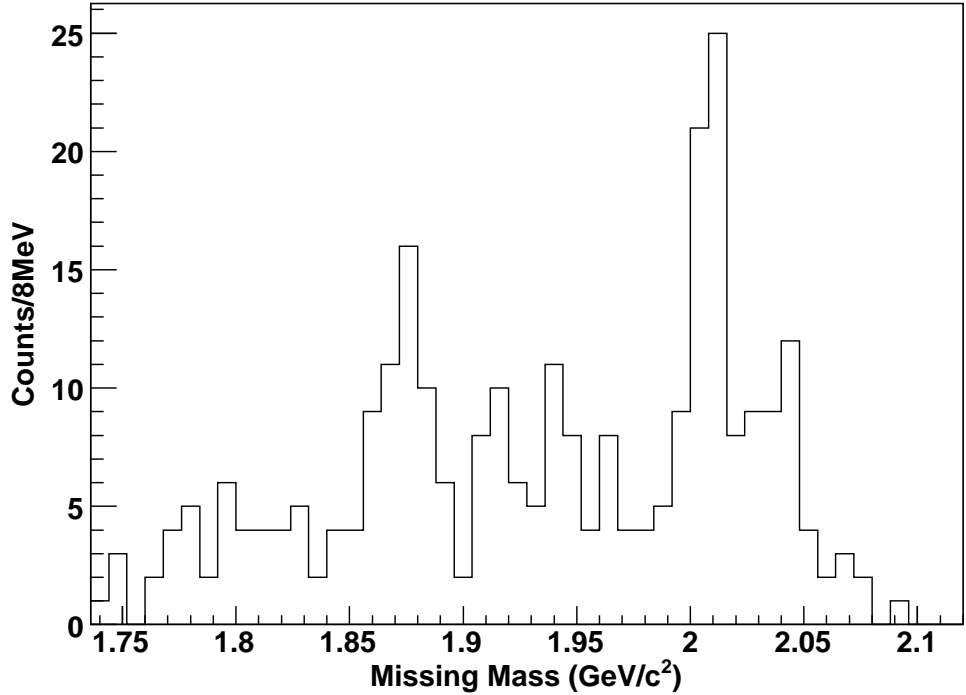}
\caption{Mass spectrum for $X$, in
$e^+e^-\rightarrow{D^{*0}\pi^{\mp}X}$, at 4260 MeV. The
charge of the daughter kaon is used to obtain the correct combination
of the neutral $D$ and the charged pion. The peak at the $D$ mass is
additional evidence of the existence of $D^*D\pi$, whereas the peak at
the $D^*$ mass is evidence of the production of $D^*D^*\pi$. A cut of
500 ${\rm{MeV}/}c$ was made on the reconstructed $D^{*}$ momentum to
select entries from the multi-body region.}
\vspace{0.2cm}
\label{fig:MultiBody_DSDPiPlus_4260}
\end{center}
\end{figure}

We performed a similar study with the data collected at 4260 MeV. The corresponding missing-mass
spectrum, only for $D^{*0}\rightarrow{D^{0}\pi^{0}}$ with
$D^{0}\rightarrow{K^-}\pi^+$, $D^{0}\rightarrow{K^-}\pi^+\pi^{0}$, or
$D^{0}\rightarrow{K^-}\pi^+\pi^{+}\pi^{-}$, is shown in \FIG
\ref{fig:MultiBody_DSDPiPlus_4260} for the $13.11~{\rm{pb}}^{-1}$
collected at $4260$~MeV.  A cut of 500 ${\rm{MeV}/}c$ was made on the
reconstructed $D^{*}$ momentum to select entries from the
multi-body region.  In addition, the charge of the $D$-daughter kaon
is used to obtain the correct combination of the neutral $D^{*}$ and
the charged pion. It is clear from the plot that $D^*\bar{D}^*\pi$
multi-body events exist, in addition to $D^*\bar{D}\pi$, at 4260
MeV. This is shown by the clear peaks at both the $D$ and
the $D^{*}$ masses in \FIG \ref{fig:MultiBody_DSDPiPlus_4260}.

\begin{figure}[!htbp]
\begin{center}
\hspace{2.5pt}
\includegraphics[width=10.5cm, angle=270]{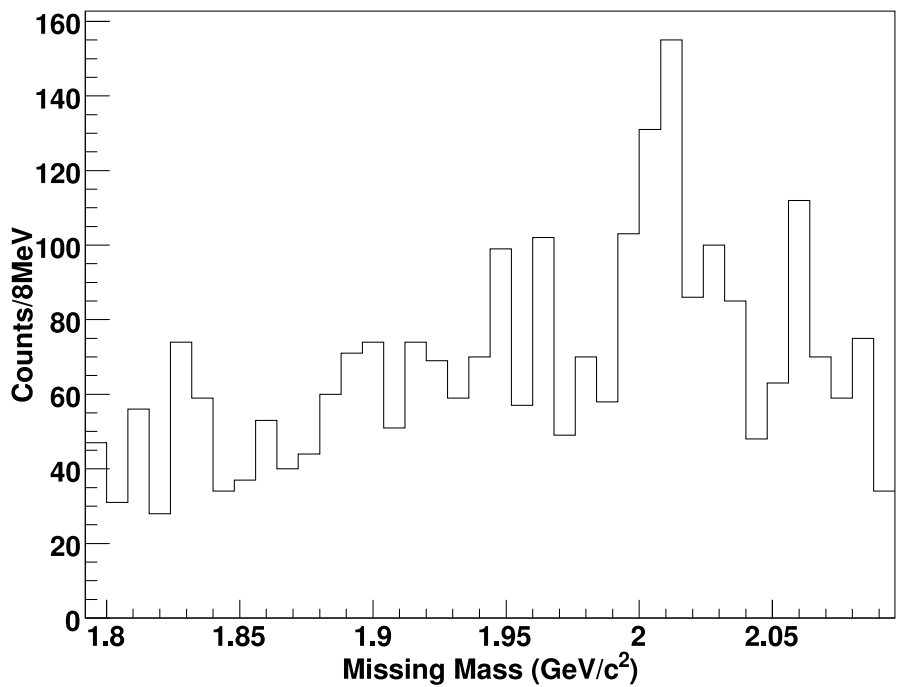}
\caption{The mass spectrum of $X$, in
$e^+e^-\rightarrow{D^{\pm}\pi^{\mp}X}$, using the
$179~{\rm{pb}}^{-1}$ collected at $4170$~MeV.  A cut on the
reconstructed $D$ momentum of 250 ${\rm{MeV}/}c$ was made to
exclude two-body events. The peak at the $D^{*}$
mass shows, again, conclusive evidence for events of the form
$D^{*}\bar{D}\pi$ being present in the data.  The lack of a peak at
the $D$ mass shows the lack of $D\bar{D}\pi$ type events in the data.}
\vspace{0.2cm}
\label{fig:MultiBody_DDPiPlus_4170}
\end{center}
\end{figure}

\begin{figure}[!htbp]
\begin{center}
\hspace{2.5pt}
\includegraphics[width=10.5cm, angle=270]{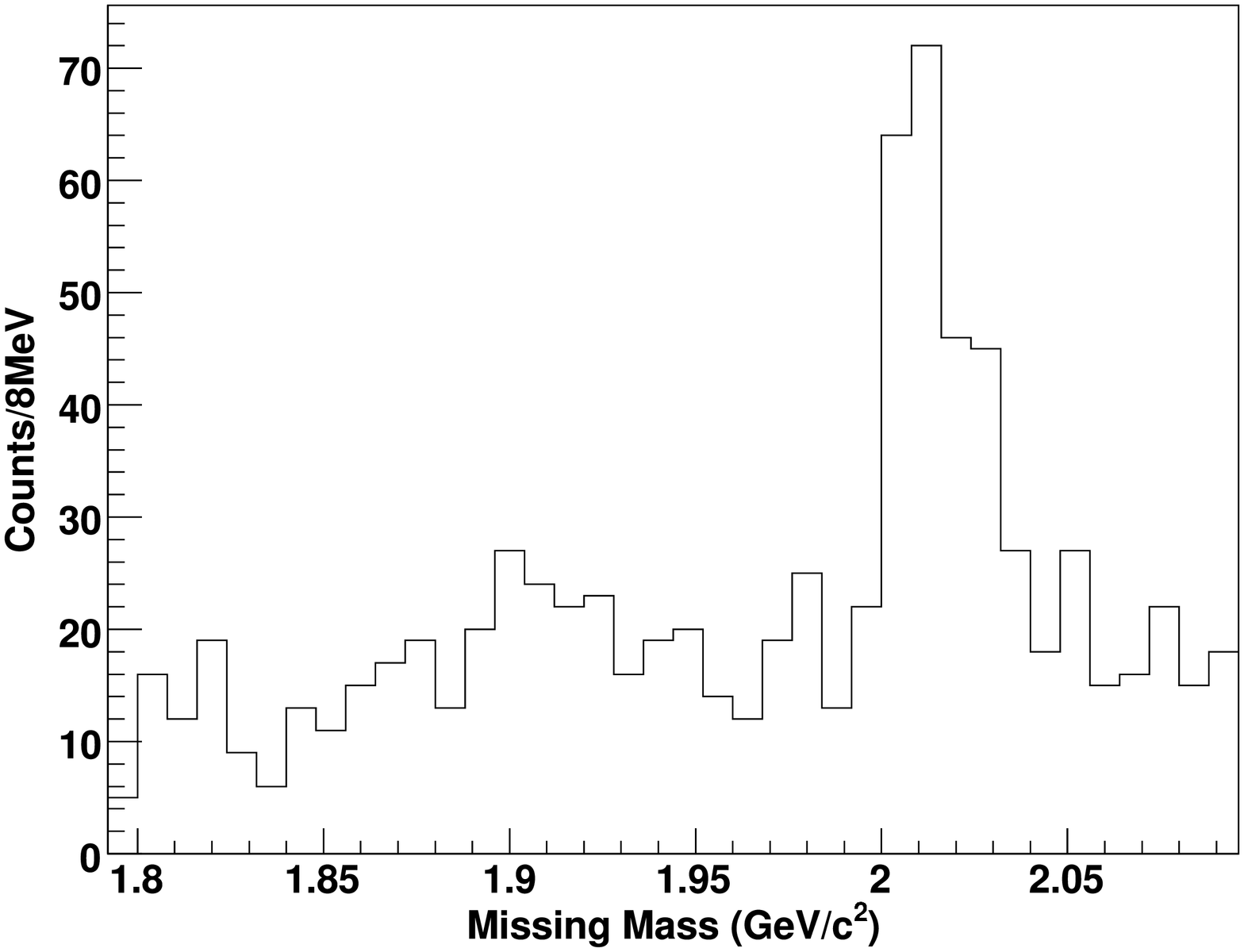}
\caption{The mass spectrum of $X$, in
$e^+e^-\rightarrow{D^{0}\pi^{\pm}X}$, using the
$179~{\rm{pb}}^{-1}$ collected at $4170$~MeV.  A cut on the
reconstructed $D$ momentum of 250 ${\rm{MeV}/}c$ was made to
exclude two-body events.  In
addition, the charge of the $D$-daughter kaon is used to obtain the
correct combination of the neutral $D^{*}$ and the charged pion. The peak at the $D^{*}$
mass shows further evidence for events of the form
$D^{*}\bar{D}\pi$.  The absence of a peak at
the $D$ mass indicates the lack of $D\bar{D}\pi$ events in the
data.}
\vspace{0.2cm}
\label{fig:MultiBody_DDPiPlus_4170_2}
\end{center}
\end{figure}

Having demonstrated unambiguously that events of the form $D^*\bar{D}\pi$ are present in
our energy region, it remains to be determined if
other types of events, like $D\bar{D}\pi$, also contribute. We performed a
similar study to the one above using a $D$ rather than a
$D^*$ to investigate this possibility in the 4170 MeV data sample.  Using only
$D^{0}\rightarrow{K^-}\pi^+$ and $D^{+}\rightarrow{K^-}\pi^+\pi^+$, in
addition to another charged pion, we obtained the missing-mass spectra
in \FIGS \ref{fig:MultiBody_DDPiPlus_4170} and
\ref{fig:MultiBody_DDPiPlus_4170_2}. \FIG \ref{fig:MultiBody_DDPiPlus_4170} shows the mass of
$X$ in $e^+e^-\rightarrow{D^{\pm}\pi^{\mp}X}$, while \FIG
\ref{fig:MultiBody_DDPiPlus_4170_2} gives the mass of $X$ in
$e^+e^-\rightarrow{D^{0}\pi^{\mp}X}$.
\CONT For the latter decay we use the
charge of the $D$-daughter kaon to obtain the
correct combination of the neutral $D$ and the charged pion.  In
both plots, the signal region is defined to be a 30-MeV window
centered on the known $D$ mass.  A cut on the
reconstructed $D$ momentum of 250 ${\rm{MeV}/}c$ was made to
exclude two-body events. In both
\FIGS \ref{fig:MultiBody_DDPiPlus_4170} and
\ref{fig:MultiBody_DDPiPlus_4170_2}, clear peaks at the
$D^{*}$ mass give further evidence for $D^{*}\bar{D}\pi$
multi-body events.  The absence of a peak at the $D$ mass
indicates that there is no evidence for $D\bar{D}\pi$ production,
although such a contribution may still be present in the data and
suppressed by the $D$-momentum cut.

Now that the components of the multi-body have been identified, a method to
measure its yield is needed so that we can correct the exclusive $D^*\bar{D}^*$ and
total exclusive charm cross sections for the portion of the
multi-body that was excluded.  To
accomplish this, we use a two-body MC representation of the various
exclusive channels and a spin-averaged phase-space model MC
representation of $D^*\bar{D}\pi$, and $D^*\bar{D}^*\pi$,
and fit the sideband-subtracted momentum spectrum for
$D^0\rightarrow{K^-}\pi^+$. 

\begin{figure}[!p]
\begin{center}
\hspace{2.5pt}
\includegraphics[width=10.5cm, angle=270]{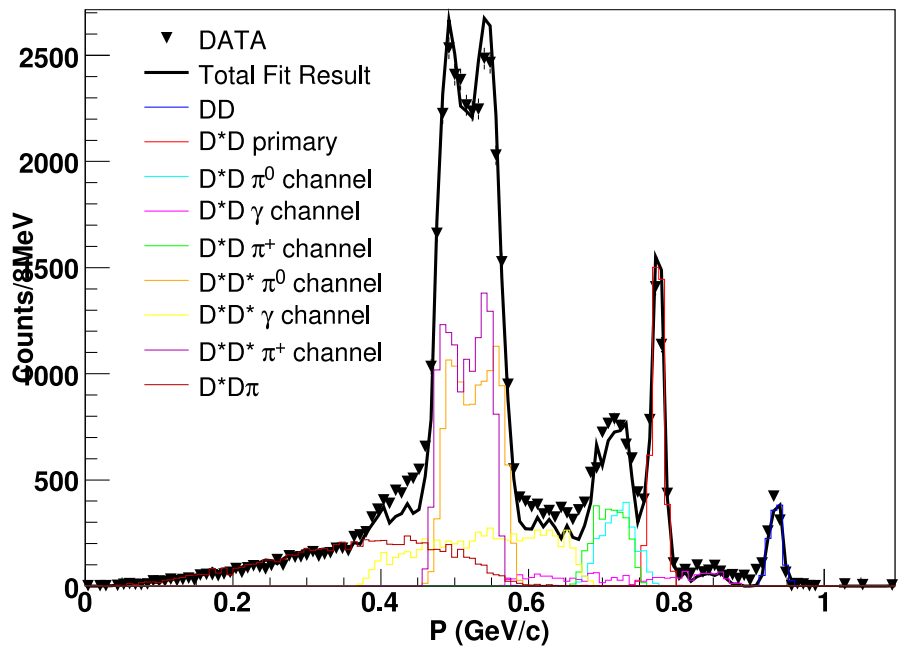}
\caption{Sideband-subtracted momentum spectrum for
$D^0\rightarrow{K^-}\pi^+$ at 4170~MeV. Data are shown as the points with error bars
which are fit to the MC (histograms). The fit uses a spin-averaged phase-space model MC representation of $D^*\bar{D}\pi$, shown
in dark red.  The fit seems to replicate the overall structure
reasonable well, but is clearly deficient in some regions.}
\vspace{0.2cm}
\label{fig:D0_mcfit_mom_4170}
\end{center}
\end{figure}

\CONT In the sideband-subtraction
method the signal region, in invariant mass, is defined to be
$\pm3\sigma$, where $\sigma$ is 5 MeV, about the PDG mass.  The
low-side and high-side sidebands start at 7.5 $\sigma$ away from the
peak and extend outward by 3 $\sigma$.  The
reconstructed $D$ momentum for each region is plotted and the resulting
histograms are subtracted. The fit made at 4170 MeV is
shown in \FIG \ref{fig:D0_mcfit_mom_4170}.  The fit seems to replicate
the overall structure reasonably well, but is clearly lacking in some regions. 

\begin{figure}[!htbp]
\begin{center}
\hspace{2.5pt}
\includegraphics[width=10.5cm, angle=270]{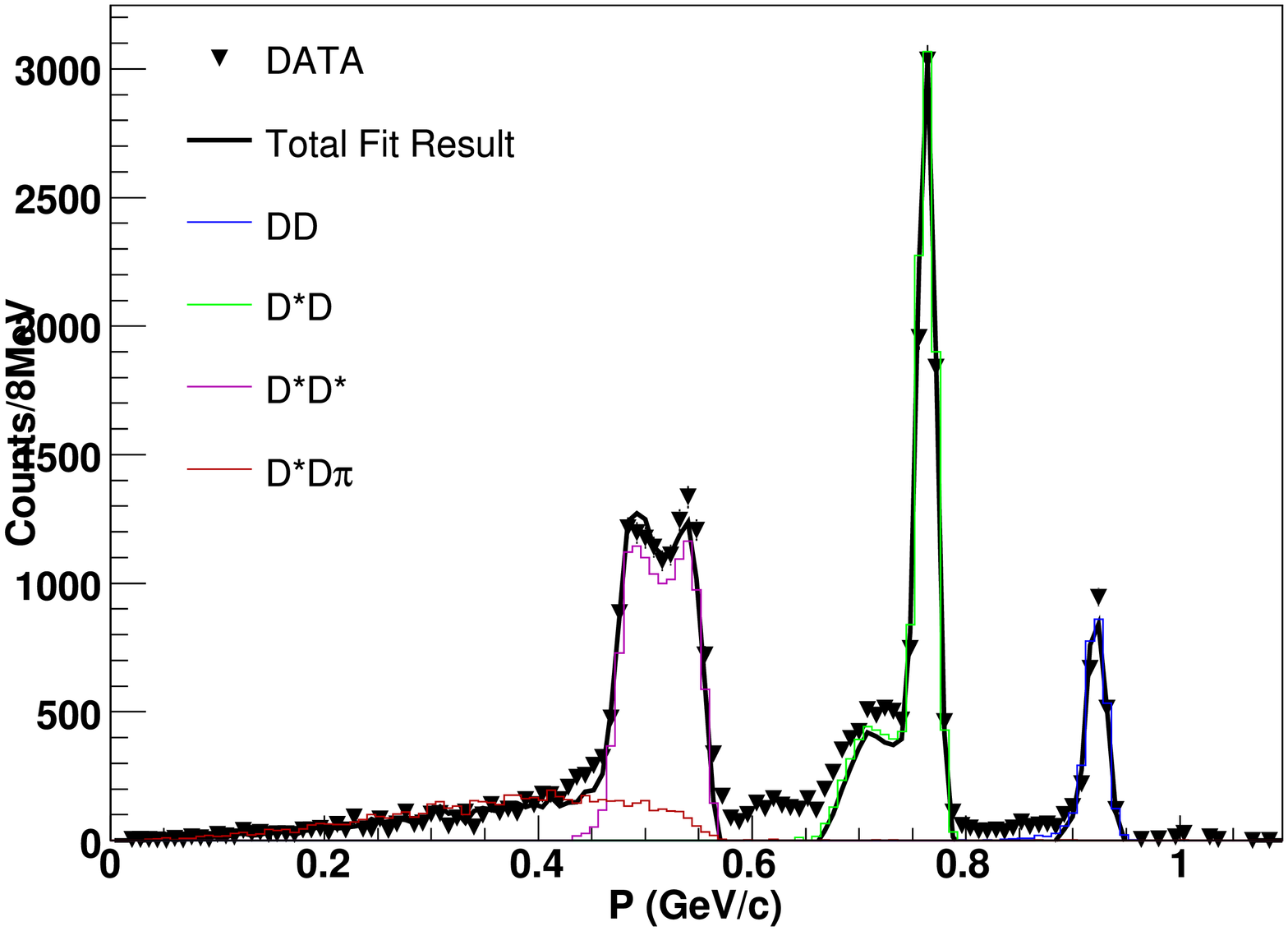}
\caption{Sideband-subtracted momentum spectrum for
$D^+\rightarrow{K^-}\pi^+\pi^+$ at 4170~MeV. Data are shown as points with error bars
which are fit to the MC (histograms). The fit uses a spin-averaged phase-space model MC
representation of $D^*\bar{D}\pi$, shown in dark red.  There is a large discrepancy between
what is expected and what is seen around 600-750
${\rm{MeV}}/c$.  This discrepancy suggests that $D\bar{D}\pi$
may be present at this energy.}
\vspace{0.2cm}
\label{fig:Dp_mcfit_mom_4170}
\end{center}
\end{figure}

Besides fitting the $D^0\rightarrow{K^-}\pi^+$ momentum spectrum, a fit to the
$D^+\rightarrow{K^-}\pi^+\pi^+$ can be a good cross-check and
test of the method.  \FIG \ref{fig:Dp_mcfit_mom_4170} shows this test
using all the assumptions made up to this point.  The fit is not as
good as the $D^0$, and it is clear that that something is missing.
The disagreement between the fit and the data is dramatic
between 600 and 750 ${\rm{MeV}}/c$.  

\begin{figure}[!htbp]
\begin{center}
\hspace{2.5pt}
\includegraphics[width=10.5cm, angle=270]{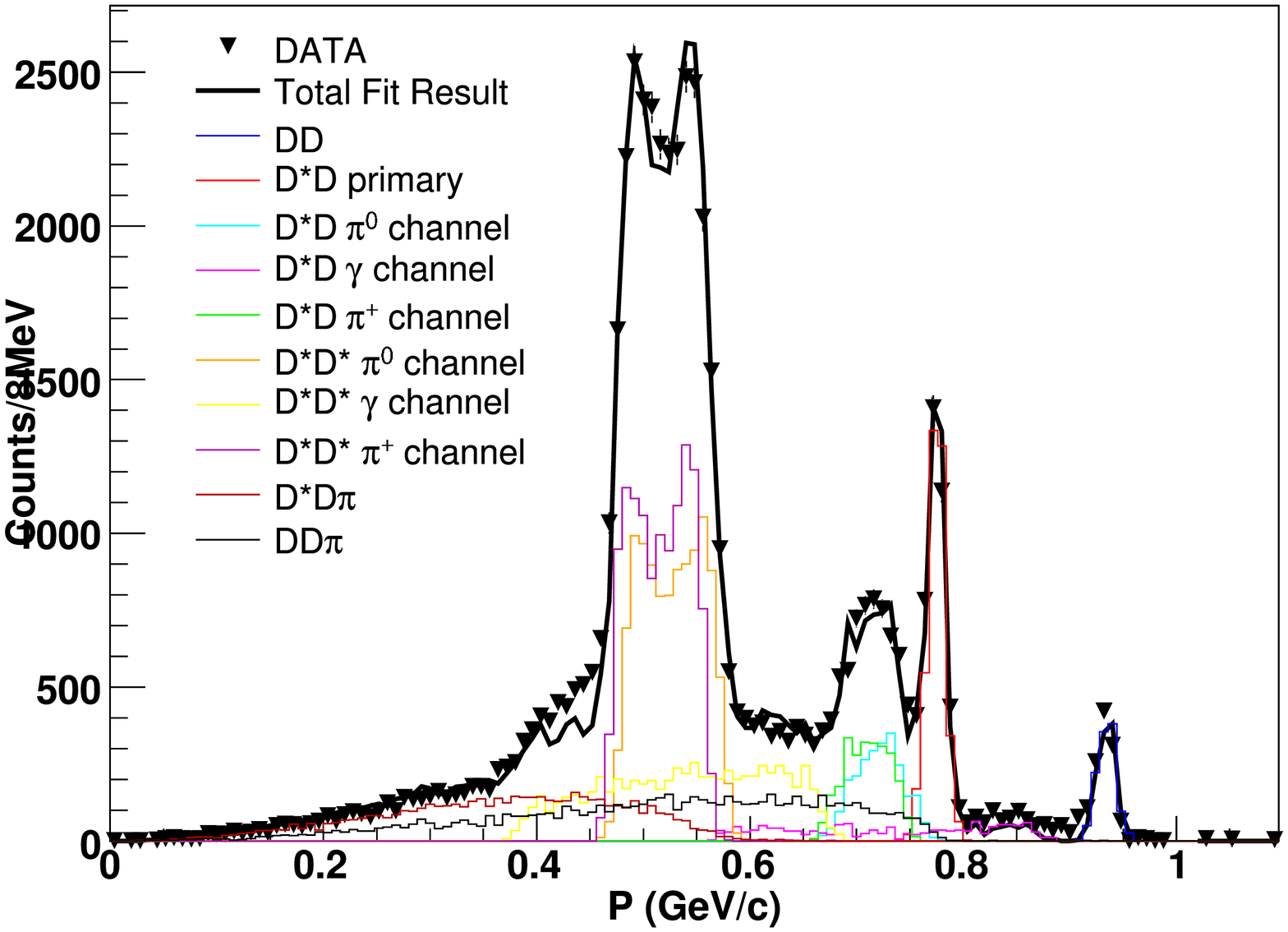}
\caption{Sideband-subtracted momentum spectrum for
$D^0\rightarrow{K^-}\pi^+$ at 4170~MeV. Data are shown as points with error bars
which are fit to the MC (histograms). The fit uses a spin-averaged phase-space model MC
representation for $D^*\bar{D}\pi$, shown in dark red, and
$D\bar{D}\pi$ is shown in black.}
\vspace{0.2cm}
\label{fig:D0_mcfit_mom_4170_ddpi}
\end{center}
\end{figure}

\begin{figure}[!htbp]
\begin{center}
\hspace{2.5pt}
\includegraphics[width=10.5cm, angle=270]{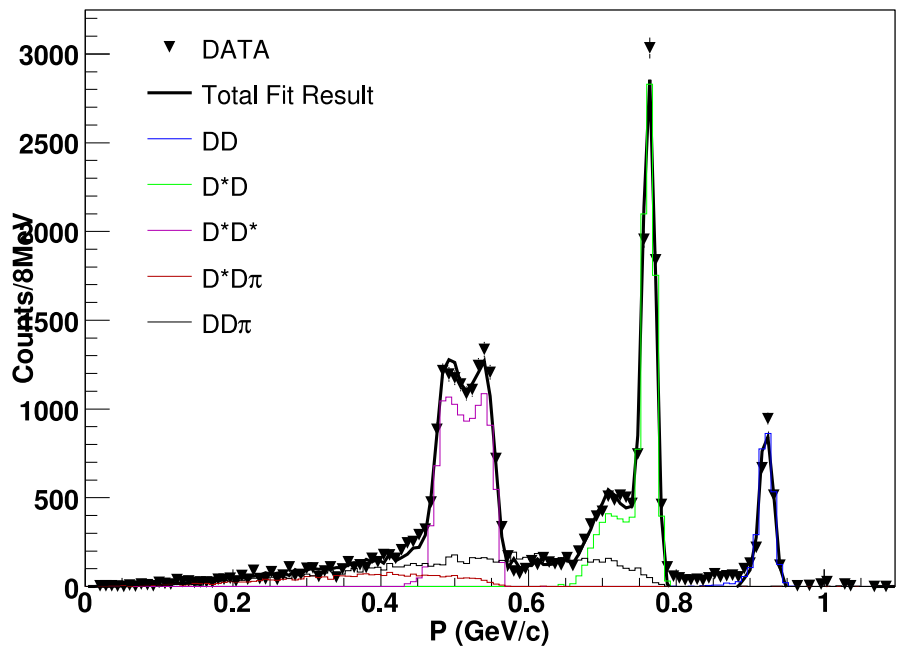}
\caption{Sideband-subtracted momentum spectrum for
$D^+\rightarrow{K^-}\pi^+\pi^+$ at 4170~MeV. Data are shown as points with error bars
which are fit to the MC (histograms). The fit uses a spin-averaged phase-space model MC
representation for $D^*\bar{D}\pi$, shown in dark red, and $D\bar{D}\pi$ is shown in black.}
\vspace{0.2cm}
\label{fig:Dp_mcfit_mom_4170_ddpi}
\end{center}
\end{figure}

We considered three possible explanations for this discrepancy:
(1) the branching fraction for the gamma decay channel is wrong,
(2) $D\bar{D}\pi$ is present in the data; and (3) something more fundamental is
missing.  The first possibility seems
unlikely, because enhancements are not seen in the region around
$D^{*+}D^{*-}$ outside that of the $\pi^0$ enhancement, like those
seen in the  $D^0\rightarrow{K^-}\pi^+$ spectrum.
The second possibility seems more likely since all the searches for
multi-body thus far only looked in momentum regions that limited the
allowable amount of $D\bar{D}\pi$.  If we take \FIG
\ref{fig:Dp_mcfit_mom_4170} as evidence for the presence of $D\bar{D}\pi$,
we can modify our fit procedure to allow a $D\bar{D}\pi$
contribution. This greatly improves the fit quality,
which can be seen in \FIG \ref{fig:D0_mcfit_mom_4170_ddpi} for
$D^0\rightarrow{K^-}\pi^+$ and \FIG \ref{fig:Dp_mcfit_mom_4170_ddpi} for
$D^+\rightarrow{K^-}\pi^+\pi^+$.

While the fits are improved, we find large discrepancies in the
$D^0/D^+$ ratio for $D^{*}\bar{D}\pi$ using the composition obtained
from the fits.  The ratio of $\sigma(D^{0})_{D^{*}\bar{D}\pi}$ to
$\sigma(D^{+})_{D^{*}\bar{D}\pi}$ should be 2, while the result from the
fit is about 5.  Also, the ratio of $\sigma(D^{0})_{D\bar{D}\pi}$ to
$\sigma(D^{+})_{D\bar{D}\pi}$ should be 1, but the result from the fit
is about 2.  So even though the fits look good, the corresponding
results are not consistent and further investigation is warranted.

\begin{figure}[!h]
\begin{center}
\hspace{2.5pt}
\includegraphics[width=14.5cm]{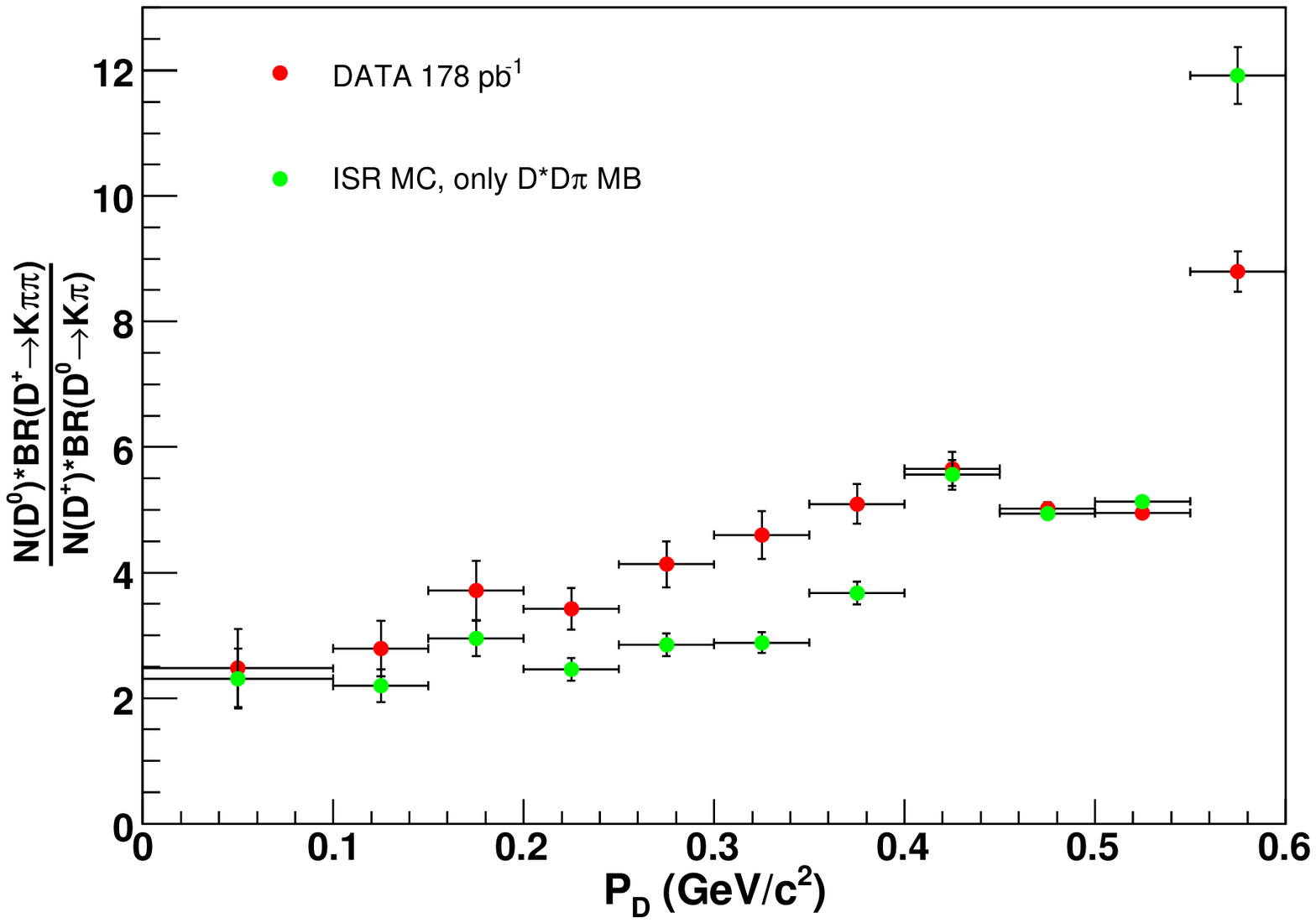}
\caption{Ratio of the number of $D^{0}$ to $D^{+}$ as a function of
reconstructed $D$ momentum at $E_{{\rm{cm}}}=4170$ MeV.  The
differences in the branching fractions were included in the ratio, while
the efficiencies were take to be constant over the momentum
range.  The MC includes only $D^*\bar{D}\pi$ multi-body production and uses a simple ISR model that assumes all
events are produced from the $\psi(4160)$ resonance, with a
width of 78 MeV.  The large discrepancy between the
data and the MC between 200 and 400 MeV/$c$ clearly indicates that
something is incorrect in our assumptions used to generate the MC sample.}
\vspace{0.2cm}
\label{fig:ratioD0Dp_function_energy}
\end{center}
\end{figure}

Using all the data collected at 4170 MeV, it is possible to break down
the ratio of $D^{0}$ to $D^{+}$ as a function of energy.  The ratio can be
seen in \FIG \ref{fig:ratioD0Dp_function_energy}.
\CONT Notice that the
ratio is not corrected for efficiency, which is found with MC to be constant across
the momentum range.  \FIG \ref{fig:ratioD0Dp_function_energy} shows the
ratio both for the 4170
MeV data set and for a MC sample that only assumes
$D^{*}\bar{D}\pi$ multi-body and implements a simple model for the
effects of ISR.  The ISR
model treats all charm production as coming from the $\psi(4160)$ resonance, which has a
width of 78 MeV \cite{pdg}.  There is a large discrepancy between the
MC and data in the momentum range 200-400 MeV/$c$.  This discrepancy would
be even worse if $D\bar{D}\pi$ were added to the MC.  While this
particular model does solve the problem of the discrepancies seen in the
momentum fits, it does suggest the need for improvements in our MC.

\begin{figure}[!h]
\begin{center}
\hspace{2.5pt}
\includegraphics[width=14.5cm]{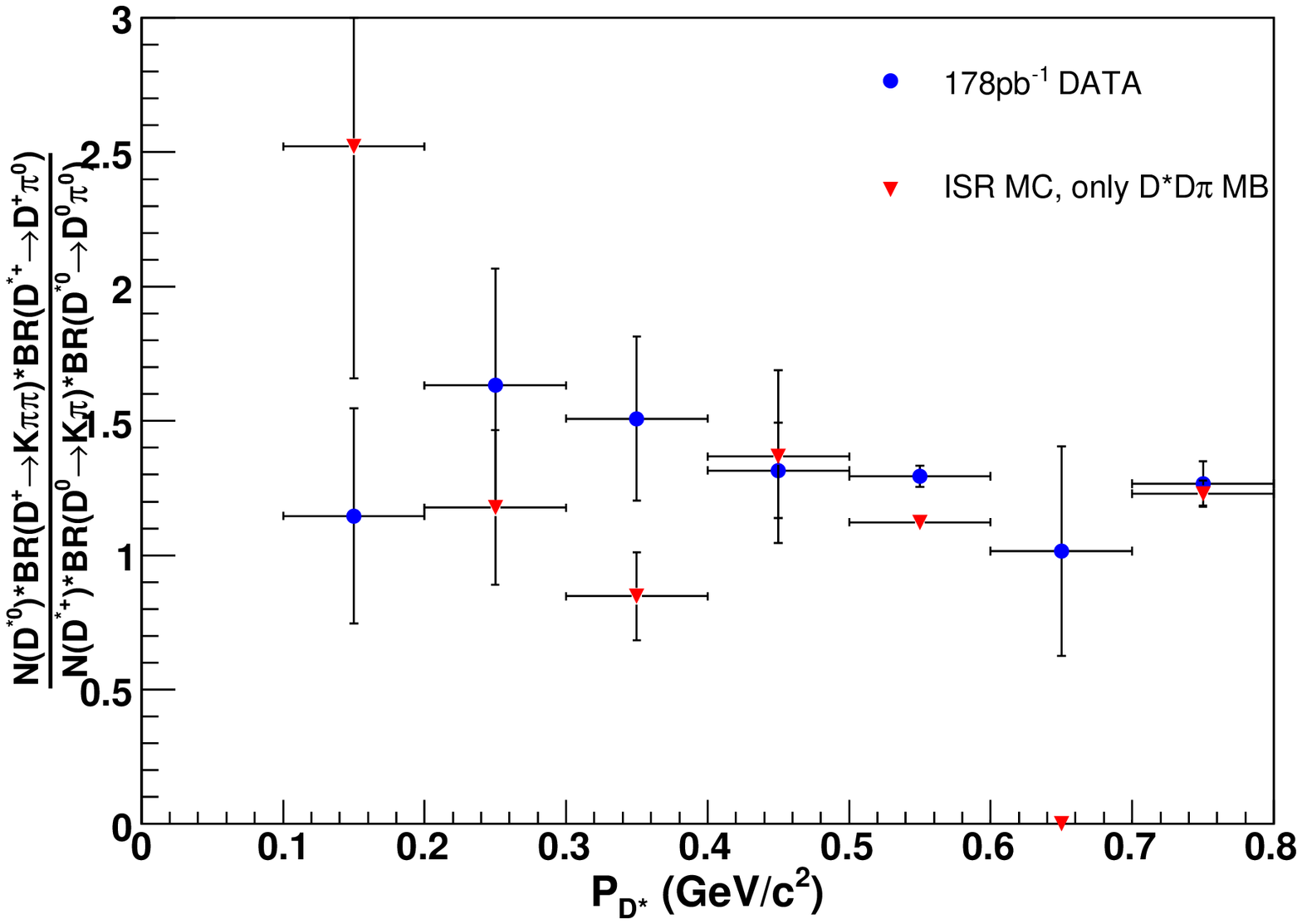}
\caption{Ratio of $D^{*0}$ to $D^{*+}$, both using the $\pi^0$
channel, with the subsequent $D$-meson decays
$D^0\rightarrow{K^-}\pi^+$ and $D^+\rightarrow{K^-}\pi^+\pi^+$,
respectively.  The ratio includes the branching fractions
but excludes the efficiencies, which are constant over the momentum
range.  The MC includes only $D^*\bar{D}\pi$ multi-body production
and uses a simple ISR model that assumes all events
are produced from the $\psi(4160)$ resonance, with a
width of 78 MeV.  The discrepancy at 650 MeV/$c$ indicates that ISR
is important and that it is incorrectly handled in our MC sample.}
\vspace{0.2cm}
\label{fig:ratioDS0DSp_function_energy}
\end{center}
\end{figure}

To investigate further, we similarly computed the ratio of $D^{*0}$ to $D^{*+}$, again
not correcting for efficiency, which is found to be constant across
the momentum range. The result is shown in \FIG
\ref{fig:ratioDS0DSp_function_energy}.
\CONT The same MC sample with a simple resonance
model is used.  The most surprising result can be seen at 650 MeV/$c$ in
the region between the peaks for the two-body events $D^*\bar{D}^*$ and $D\bar{D}^*$. In the limit of perfect resolution, there should be two delta
functions, one for $D^*$ in $D^*\bar{D}$ events and one $D^*$ in $D^*\bar{D}^*$
events, where the 650 MeV/$c$ bin falls right in the middle. This is true
even accounting for resolution, and therefore, the ratio should be
zero. That this ratio is zero for MC but not data demonstrates that
our simple ISR MC is incorrect and a better model is needed.

\begin{figure}[!htbp]
\begin{center}
\hspace{2.5pt}
\includegraphics[width=14.5cm]{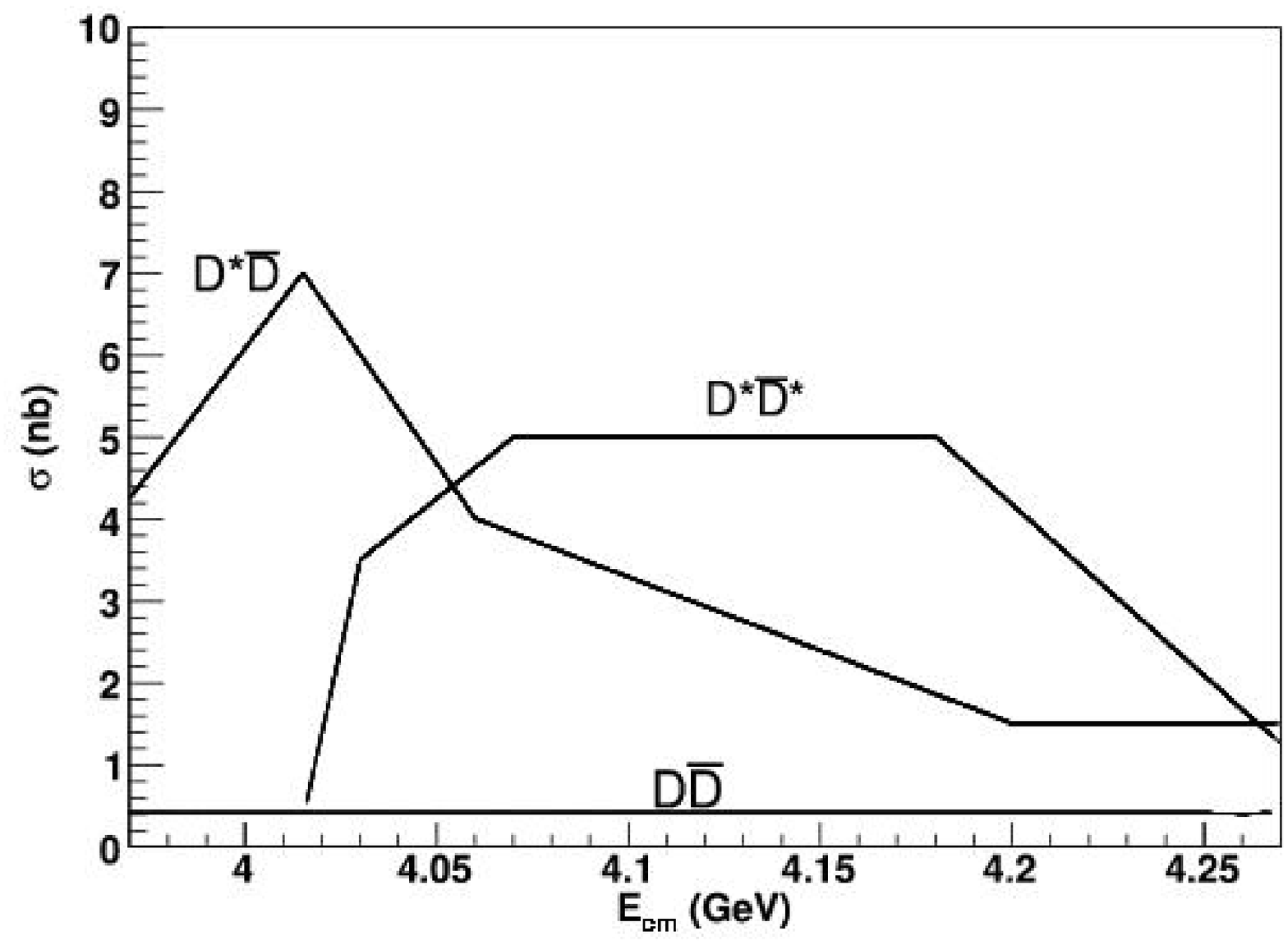}
\caption{The two-body cross sections used in {\tt{EVTGEN}} for
$D\bar{D}$, $D^*\bar{D}$, and $D^{*}\bar{D}^*$ for the MC
samples with improved treatment of ISR.  A simple implementation of the cross sections were made using
straight-line approximation for the real, Born-level, two-body, cross sections. The
main assumption is that the shape of the cross section will be
preserved after applying radiative corrections, that is the observed
cross section shape is the same as the Born-level cross section shape.}
\vspace{0.2cm}
\label{fig:EvtGen_DD}
\end{center}
\end{figure}

\begin{figure}[!htbp]
\begin{center}
\hspace{2.5pt}
\includegraphics[width=14.5cm]{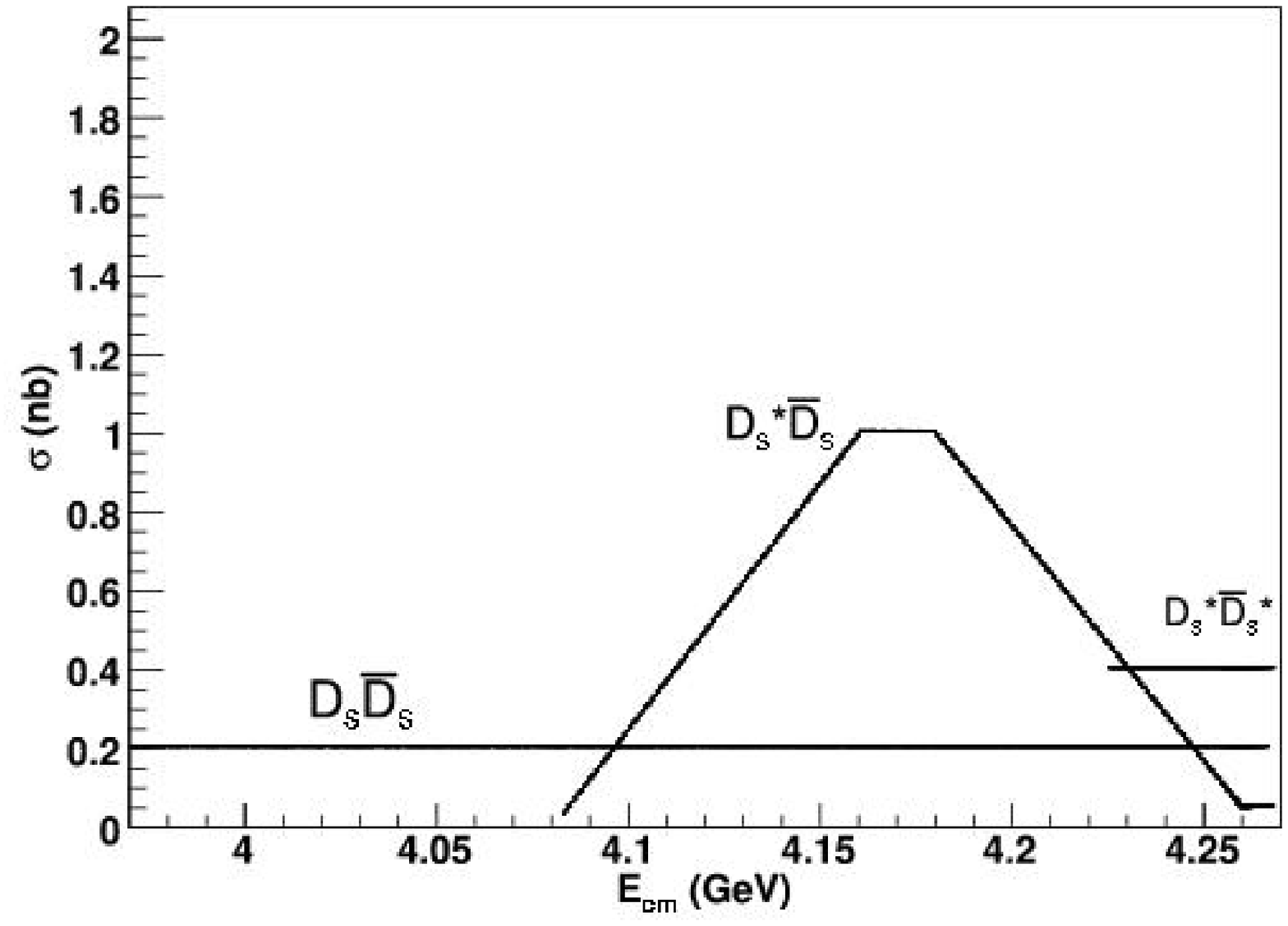}
\caption{The two-body cross sections used in {\tt{EVTGEN}} for
$D_{s}\bar{D}_{s}$, $D_{s}^*\bar{D}_{s}$, and $D_{s}^{*}\bar{D}_{s}^*$ for the MC
samples with improved treatment of ISR.  A simple implementation of the cross sections were made using
straight-line approximation for the real, Born-level, two-body cross sections. The
main assumption is that the shape of the cross section will be
preserved after applying radiative corrections, that is the observed
cross section shape is the same as the Born-level cross section shape.}
\vspace{0.2cm}
\label{fig:EvtGen_DsDs}
\end{center}
\end{figure}

\begin{figure}[!htbp]
\begin{center}
\hspace{2.5pt}
\includegraphics[width=14.5cm]{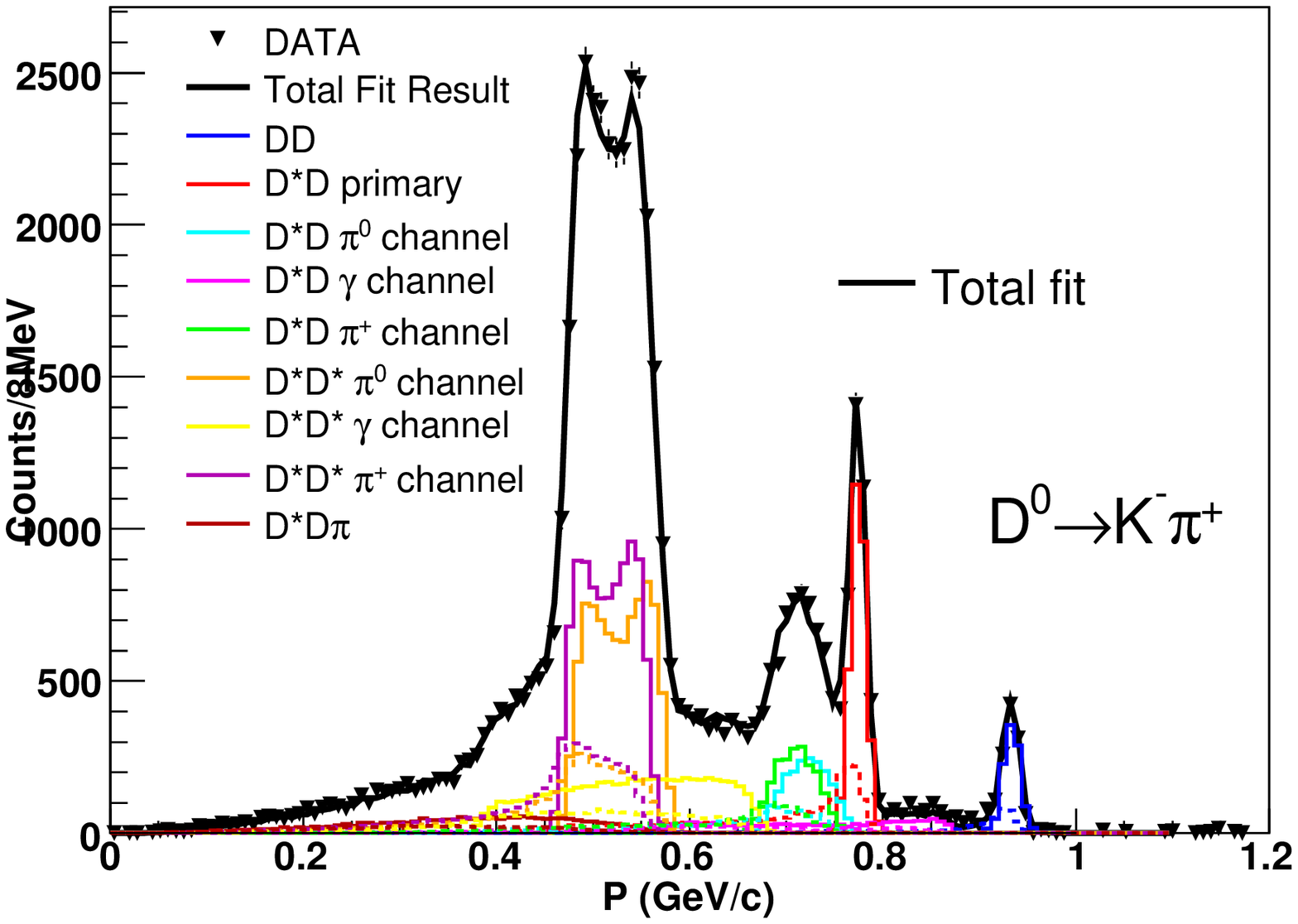}
\caption{Sideband-subtracted momentum spectrum for
$D^0\rightarrow{K^-}\pi^+$ at 4170~MeV. Data are shown as points with error bars
which are fit to the improved ISR MC (histograms). The fit
uses a spin-averaged phase-space model MC representation of $D^*\bar{D}\pi$, shown in
dark red.  The solid histograms correspond to $E_{\gamma}<1$
MeV (no ISR) while the dashed histograms correspond to $E_{\gamma}>1$
MeV (with ISR).  The total fit result is shown by the solid black line ($\chi^2/NDF=248/132$).}
\vspace{0.2cm}
\label{fig:Momentumfit_4170_D0_CBX_correct}
\end{center}
\end{figure}

\begin{figure}[!htbp]
\begin{center}
\hspace{2.5pt}
\includegraphics[width=14.5cm]{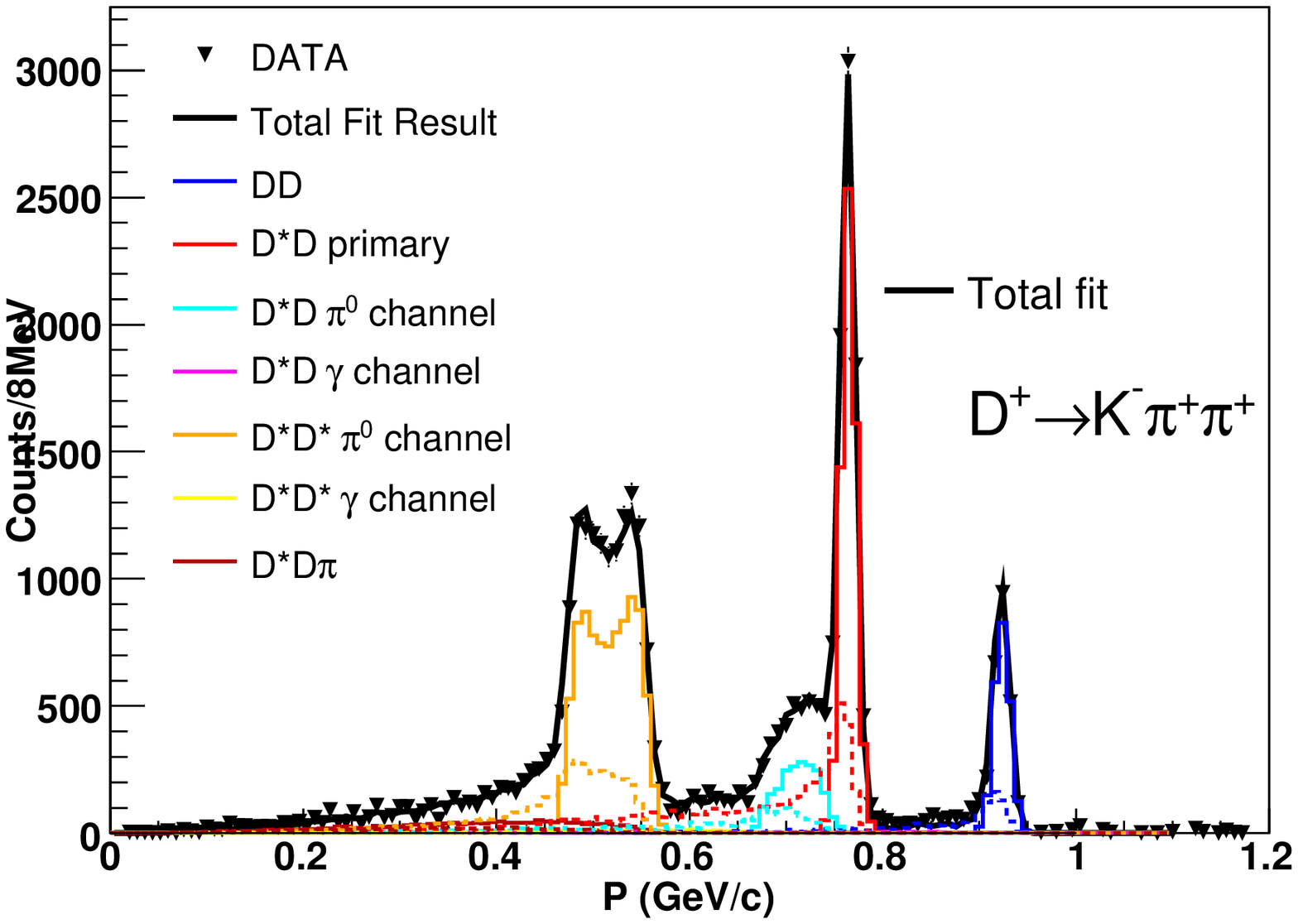}
\caption{Sideband-subtracted momentum spectrum for
$D^+\rightarrow{K^-}\pi^+\pi^+$ at 4170~MeV. Data are shown as points
with error bars which are fit to the improved ISR MC, the
histograms. The fit uses a spin-averaged phase-space model MC
representation of $D^*D\pi$, shown in dark red.  The solid
histograms correspond to $E_{\gamma}<1$ MeV (no ISR) while the dashed
histograms correspond to $E_{\gamma}>1$ MeV (with ISR).  The total fit result is
shown by the solid black line ($\chi^2/NDF=182/132$).}
\vspace{0.2cm}
\label{fig:Momentumfit_4170_Dp_CBX_correct}
\end{center}
\end{figure}

\begin{figure}[!htbp]
\begin{center}
\hspace{2.5pt}
\includegraphics[width=14.5cm]{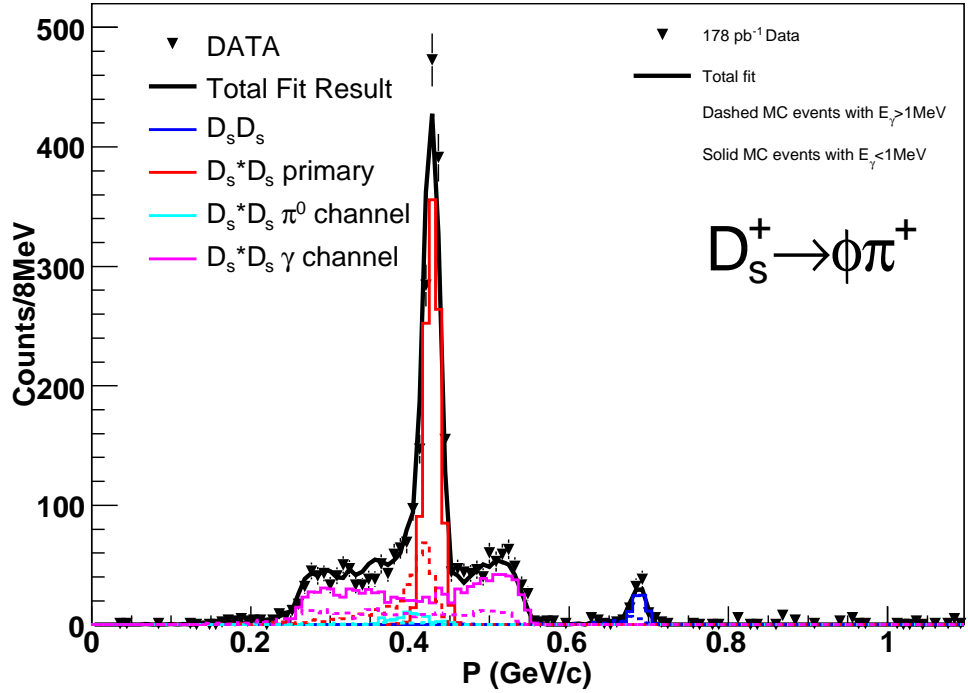}
\caption{Sideband-subtracted momentum spectrum for
$D_{s}^+\rightarrow\phi\pi^+$ at 4170~MeV. Data are shown as points with error bars
which are fit to the improved ISR MC (histograms).
The solid histograms correspond to $E_{\gamma}<1$
MeV (no ISR) while the dashed histograms correspond to $E_{\gamma}>1$
MeV (with ISR).  The total fit result is shown by the solid black line ($\chi^2/NDF=165/124$).}
\vspace{0.2cm}
\label{fig:Momentumfit_4170_Ds_CBX_correct}
\end{center}
\end{figure}

Our improved approach is to use {\tt{EVTGEN}}'s model of {\tt{EvtVPHOtoVISR}}
\cite{EVTGEN}, representing the cross sections for each event type of
two-body events (\FIGS \ref{fig:XS_scan_DD} and \ref{fig:XS_scan_Ds})
with simple linear interpolations between our measured data points, as is shown in
\FIGS \ref{fig:EvtGen_DD} and \ref{fig:EvtGen_DsDs}. For
multi-body constant cross sections are assumed.  Momentum fits to these
models give the
results for $D^{0}\rightarrow{K^-}\pi^+$ and
$D^{+}\rightarrow{K^-}\pi^+\pi^+$ shown in \FIG
\ref{fig:Momentumfit_4170_D0_CBX_correct} and \FIG
\ref{fig:Momentumfit_4170_Dp_CBX_correct}, respectively. In addition to fitting
$D^0$ and $D^+$, one can fit the $D_{s}$ momentum distribution using the same procedure.
This result, only for $D_{s}^+\rightarrow\phi\pi^+$, is given in
\FIG \ref{fig:Momentumfit_4170_Ds_CBX_correct}.

\begin{figure}[!p]
\begin{center}
\hspace{2.5pt}
\includegraphics[width=14.5cm]{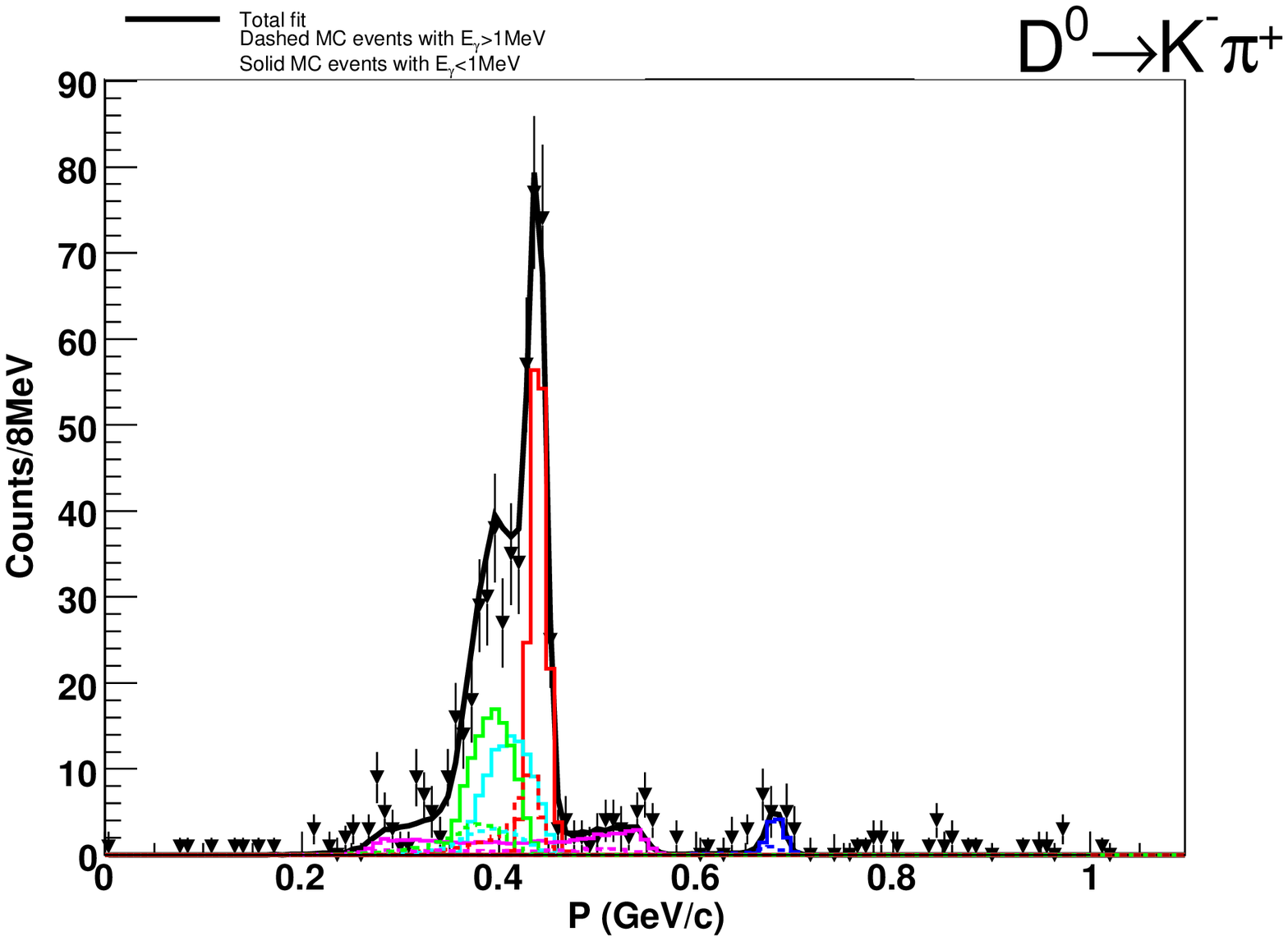}
\caption{Sideband-subtracted momentum spectrum for
$D^0\rightarrow{K^-}\pi^+$ at 3970 MeV.  The data is the points
with error bars and the histograms are MC.}
\vspace{0.2cm}
\label{fig:Mom_D0_3970}
\end{center}
\end{figure}

\begin{figure}[!p]
\begin{center}
\hspace{2.5pt}
\includegraphics[width=14.5cm]{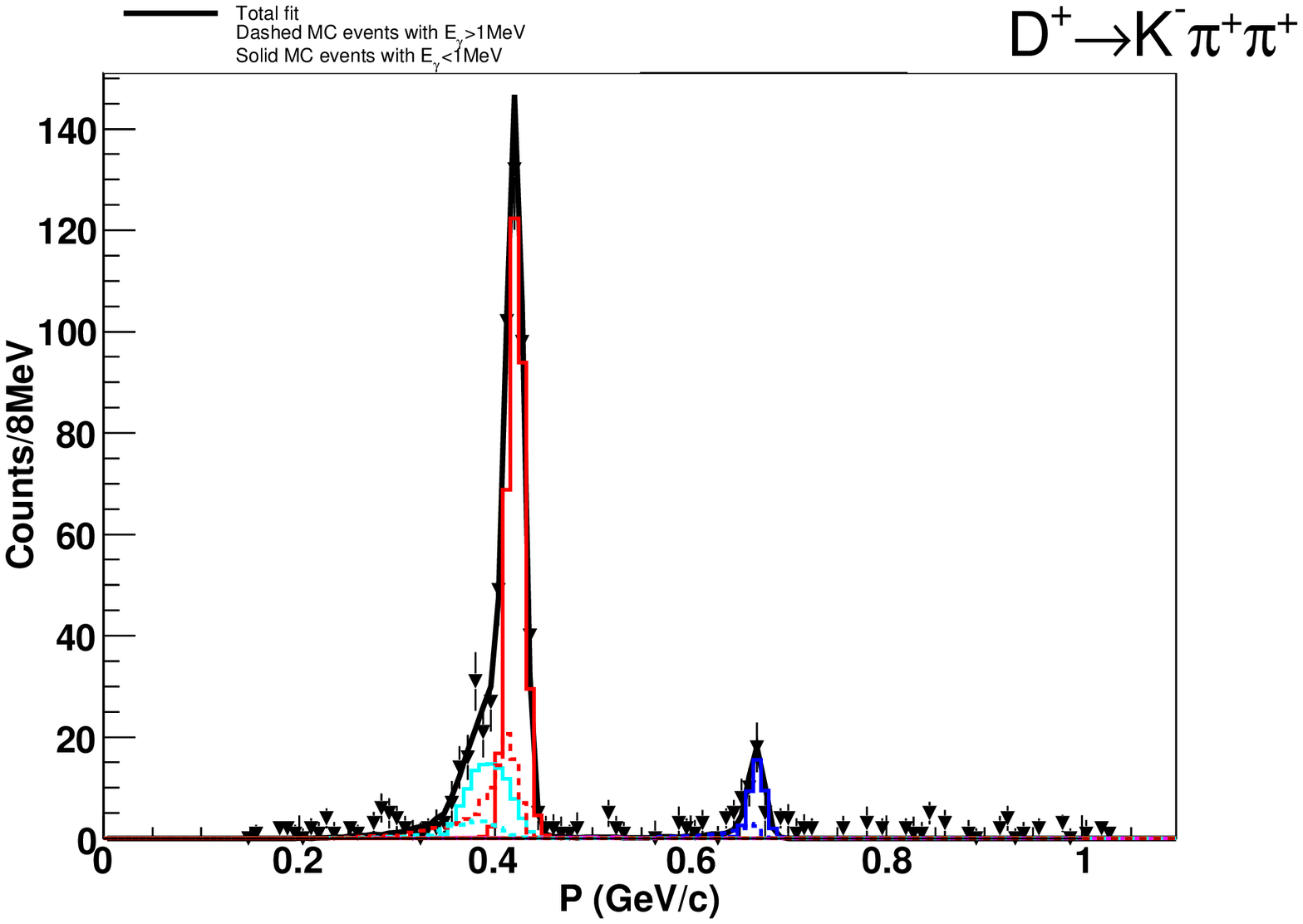}
\caption{Sideband-subtracted momentum spectrum for
$D^+\rightarrow{K^-}\pi^+\pi^+$ at 3970 MeV.  The data is the points
with error bars and the histograms are MC.}
\vspace{0.2cm}
\label{fig:Mom_Dp_3970}
\end{center}
\end{figure}

\begin{figure}[!p]
\begin{center}
\hspace{2.5pt}
\includegraphics[width=14.5cm]{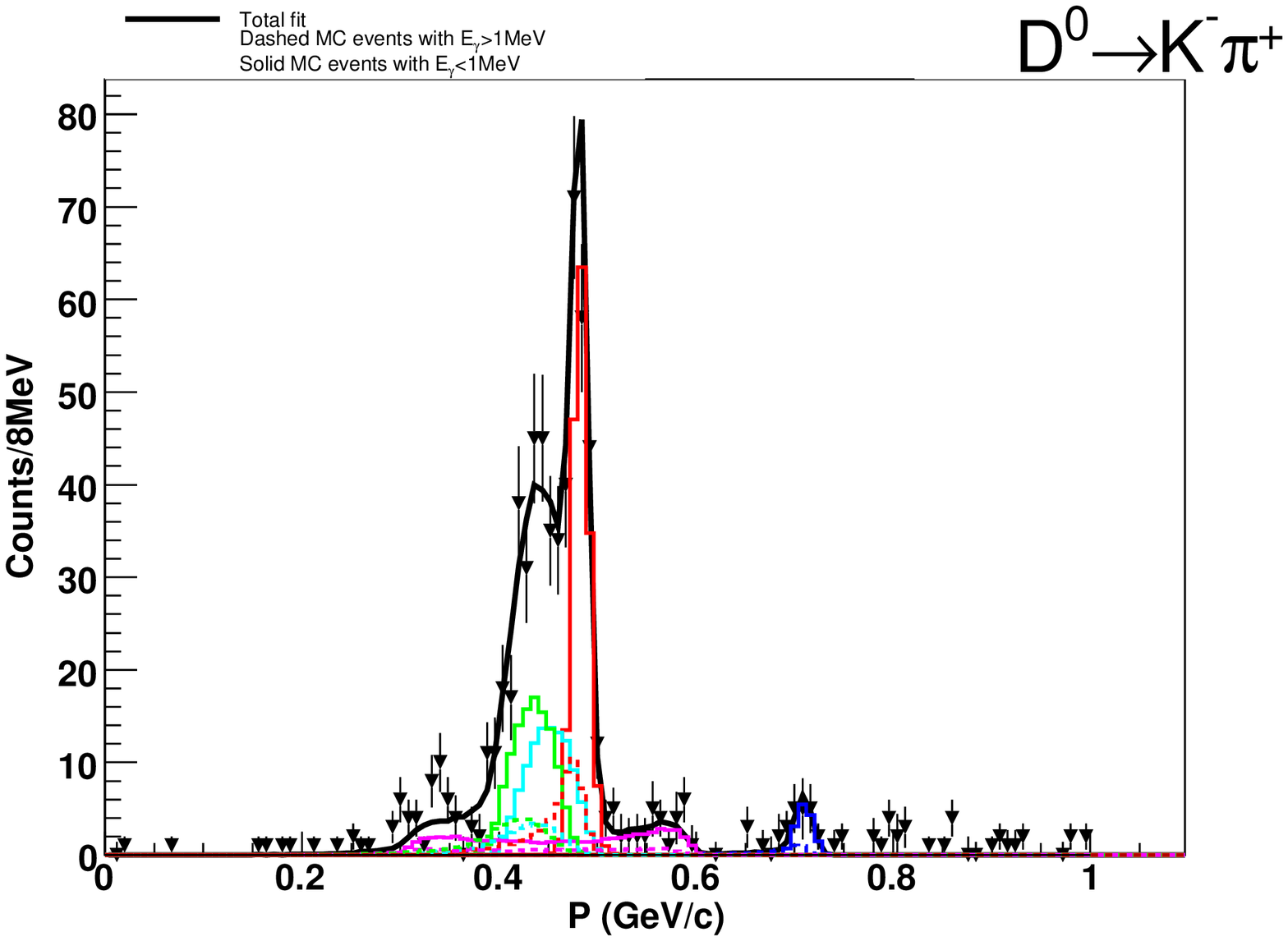}
\caption{Sideband-subtracted momentum spectrum for
$D^0\rightarrow{K^-}\pi^+$ at 3990 MeV.  The data is the points
with error bars and the histograms are MC.}
\vspace{0.2cm}
\label{fig:Mom_D0_3990}
\end{center}
\end{figure}

\begin{figure}[!p]
\begin{center}
\hspace{2.5pt}
\includegraphics[width=14.5cm]{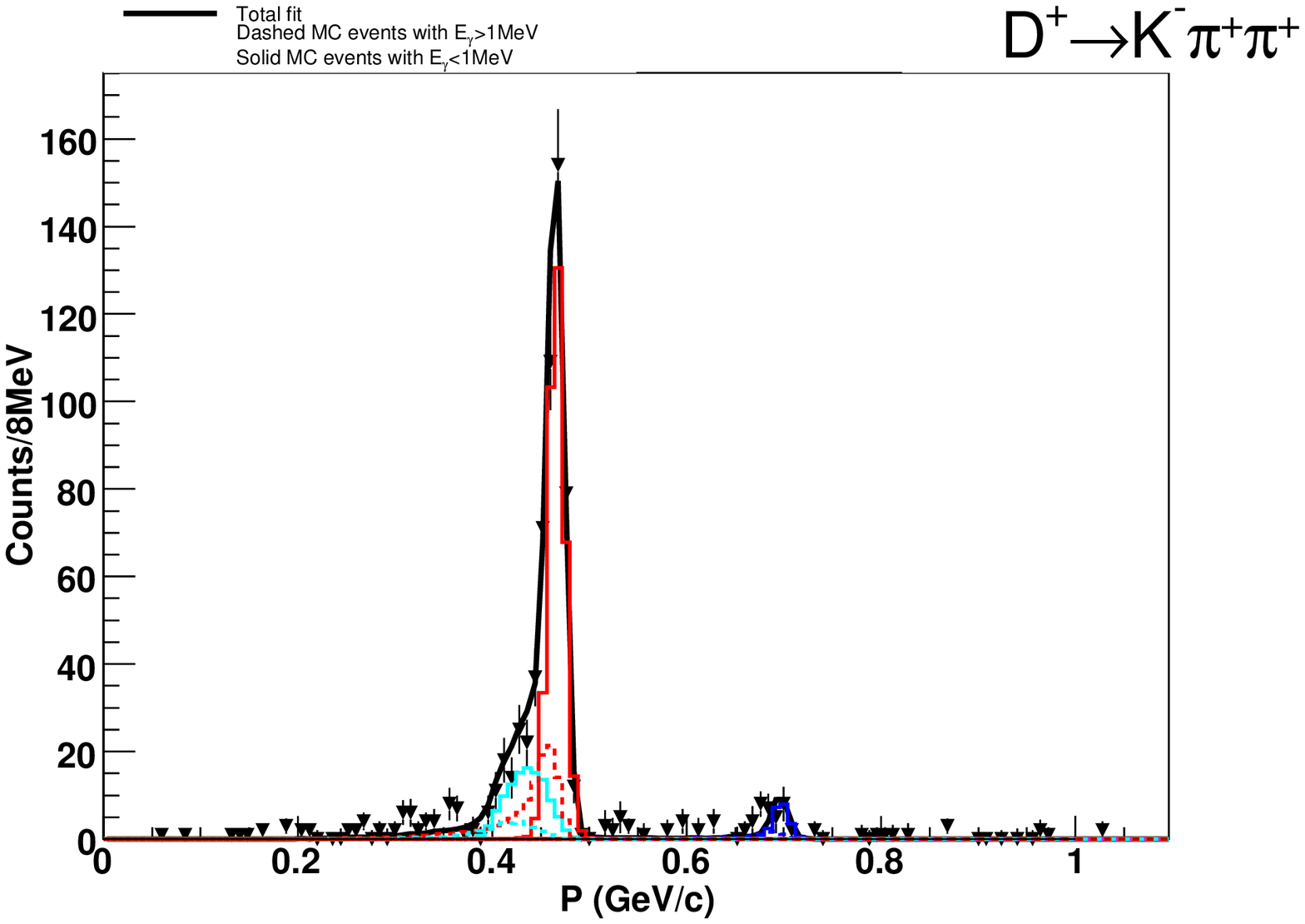}
\caption{Sideband-subtracted momentum spectrum for
$D^+\rightarrow{K^-}\pi^+\pi^+$ at 3990 MeV.  The data is the points
with error bars and the histograms are MC.}
\vspace{0.2cm}
\label{fig:Mom_Dp_3990}
\end{center}
\end{figure}

\begin{figure}[!p]
\begin{center}
\hspace{2.5pt}
\includegraphics[width=14.5cm]{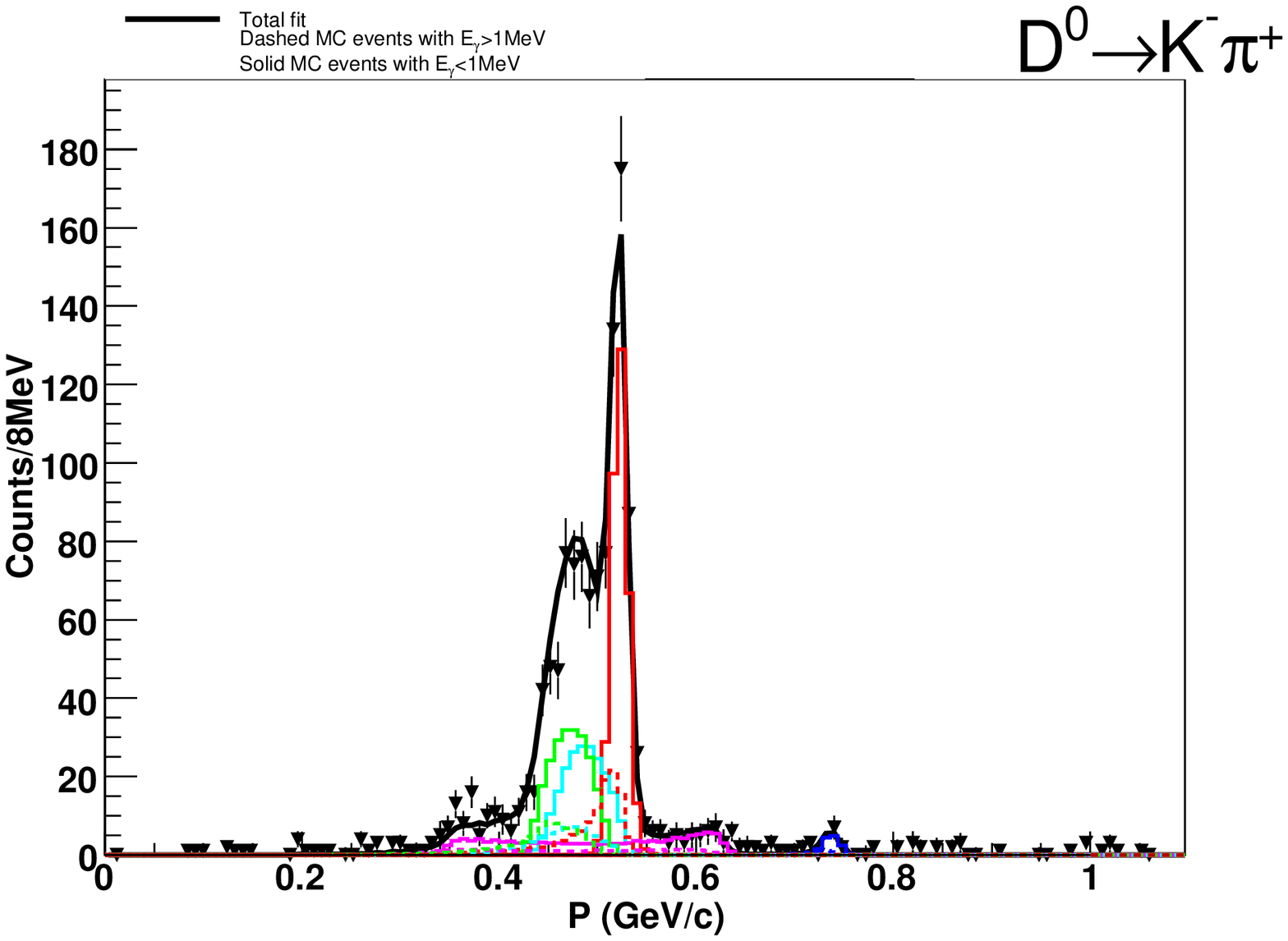}
\caption{Sideband-subtracted momentum spectrum for
$D^0\rightarrow{K^-}\pi^+$ at 4010 MeV.  The data is the points
with error bars and the histograms are MC.}
\vspace{0.2cm}
\label{fig:Mom_D0_4010}
\end{center}
\end{figure}

\begin{figure}[!p]
\begin{center}
\hspace{2.5pt}
\includegraphics[width=14.5cm]{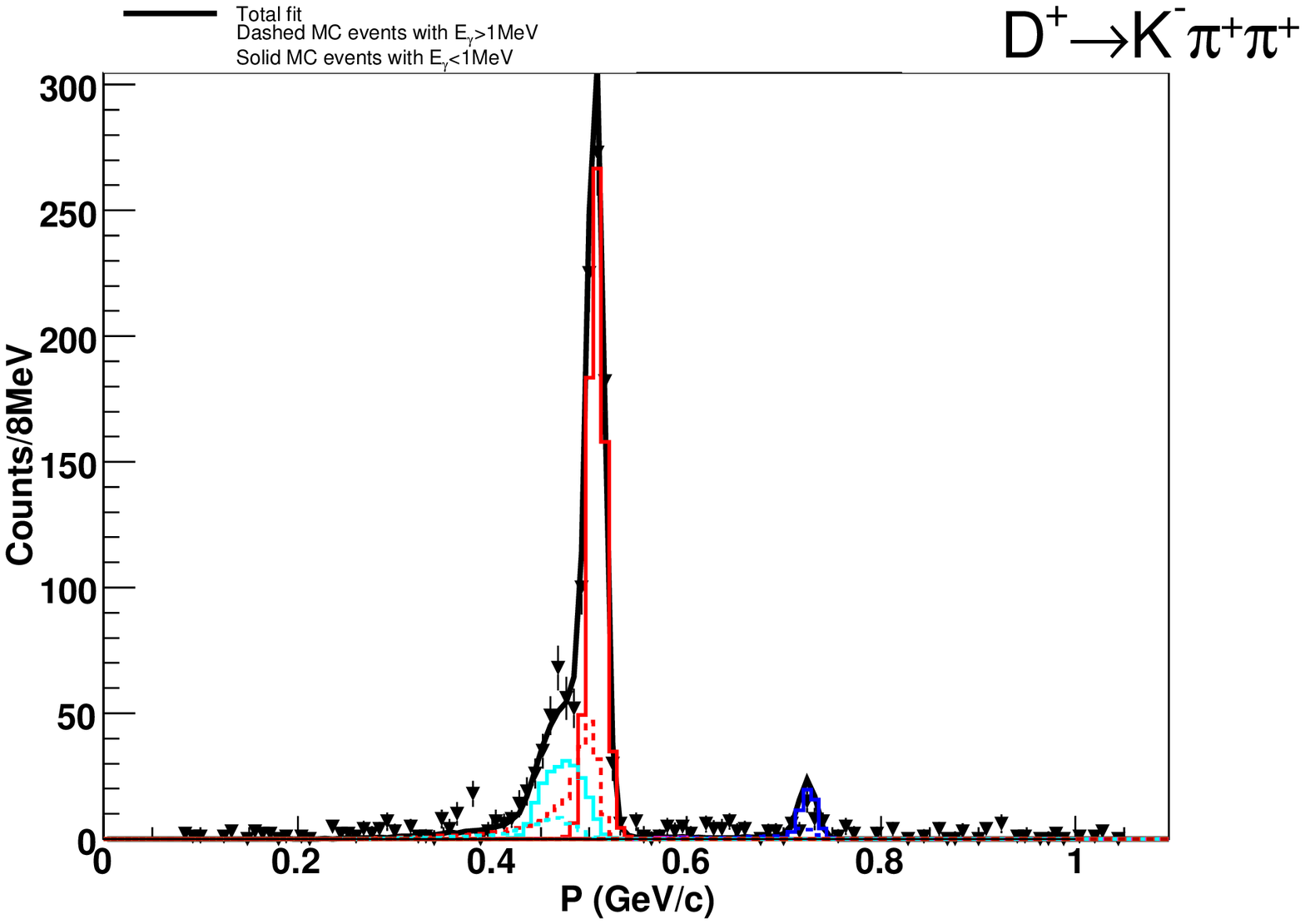}
\caption{Sideband-subtracted momentum spectrum for
$D^+\rightarrow{K^-}\pi^+\pi^+$ at 4010 MeV.  The data is the points
with error bars and the histograms are MC.}
\vspace{0.2cm}
\label{fig:Mom_Dp_4010}
\end{center}
\end{figure}

\begin{figure}[!p]
\begin{center}
\hspace{2.5pt}
\includegraphics[width=14.5cm]{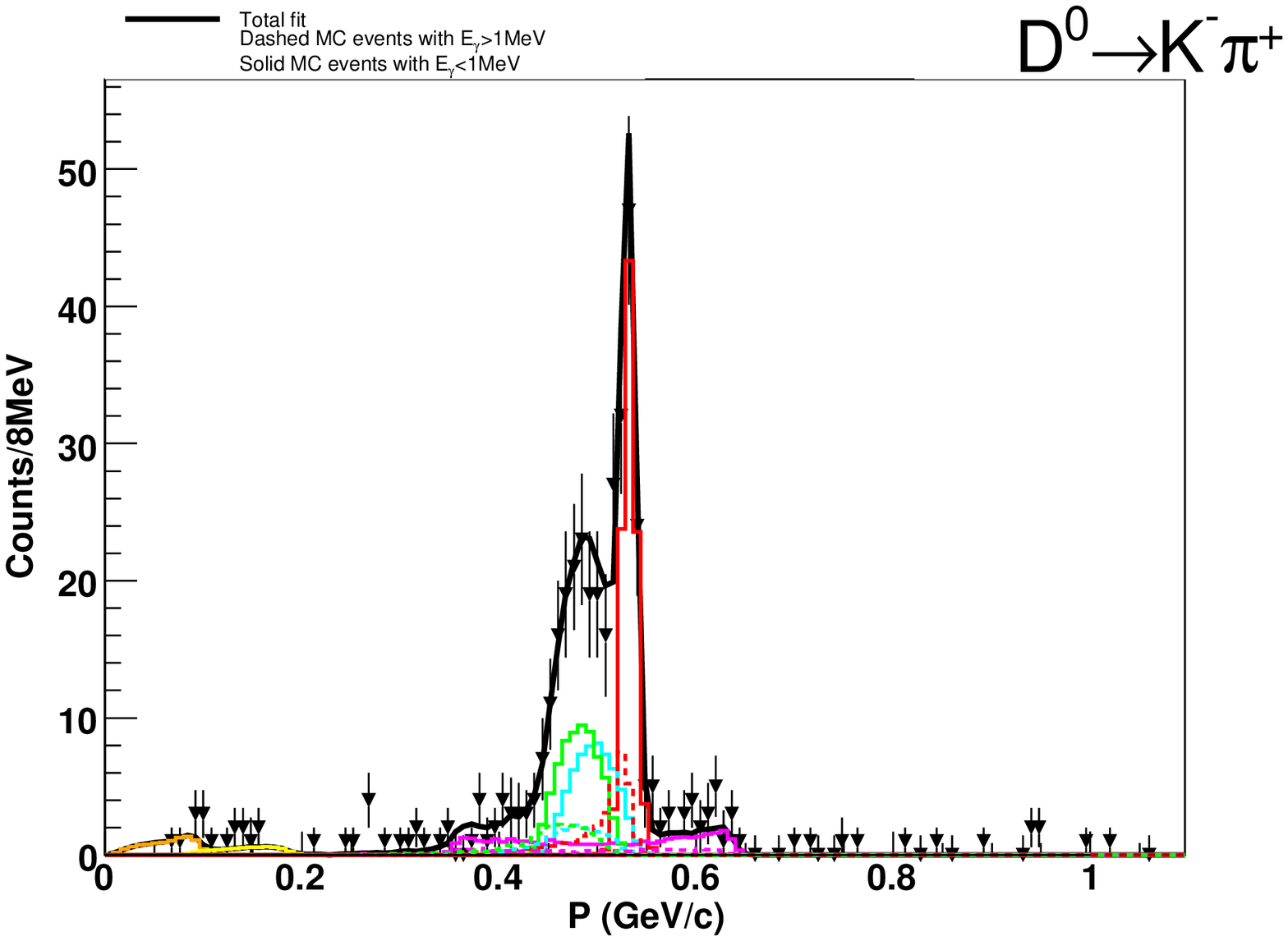}
\caption{Sideband-subtracted momentum spectrum for
$D^0\rightarrow{K^-}\pi^+$ at 4015 MeV.  The data is the points
with error bars and the histograms are MC.}
\vspace{0.2cm}
\label{fig:Mom_D0_4015}
\end{center}
\end{figure}

\begin{figure}[!p]
\begin{center}
\hspace{2.5pt}
\includegraphics[width=14.5cm]{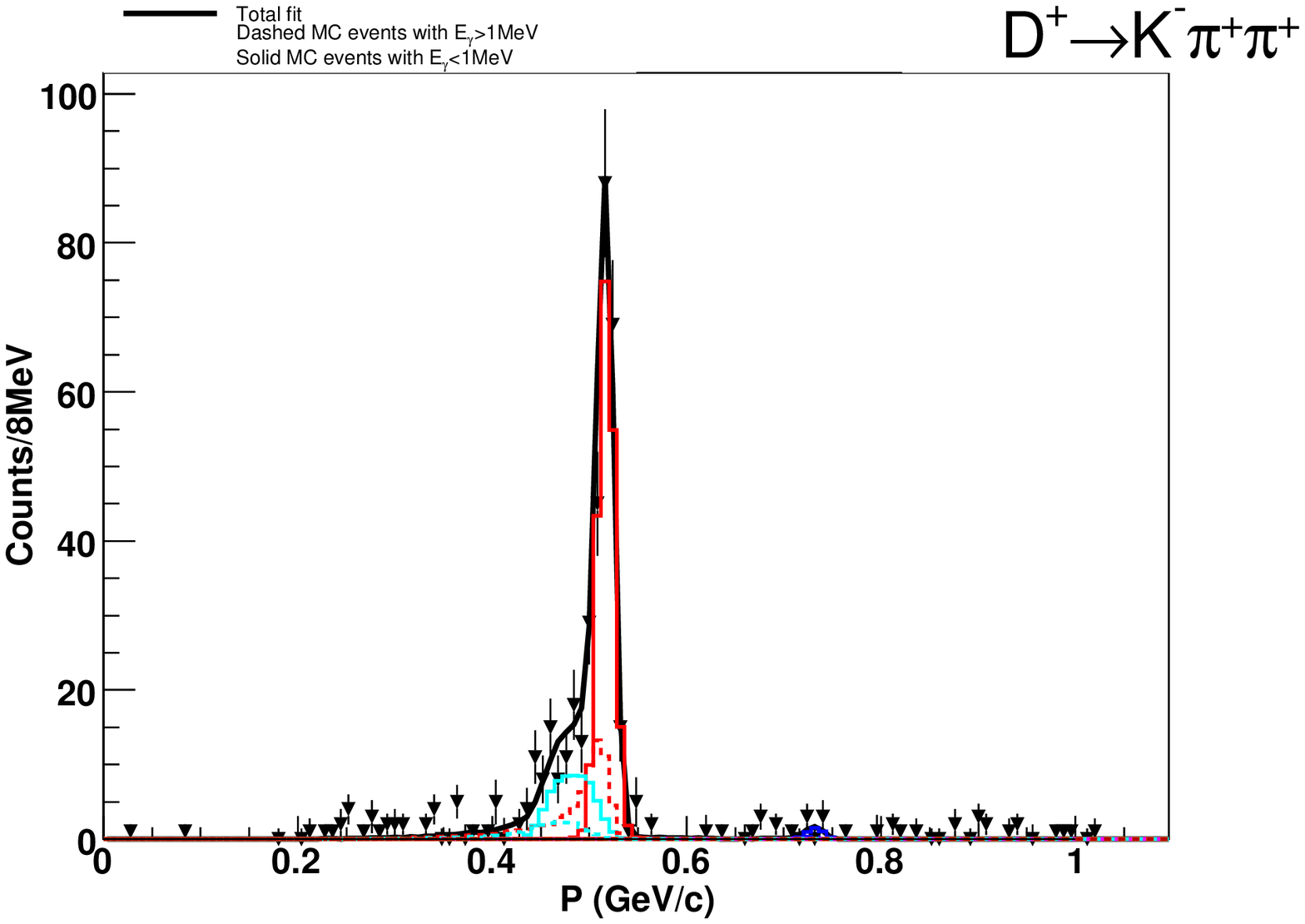}
\caption{Sideband-subtracted momentum spectrum for
$D^+\rightarrow{K^-}\pi^+\pi^+$ at 4015 MeV.  The data is the points
with error bars and the histograms are MC.}
\vspace{0.2cm}
\label{fig:Mom_Dp_4015}
\end{center}
\end{figure}

\begin{figure}[!p]
\begin{center}
\hspace{2.5pt}
\includegraphics[width=14.5cm]{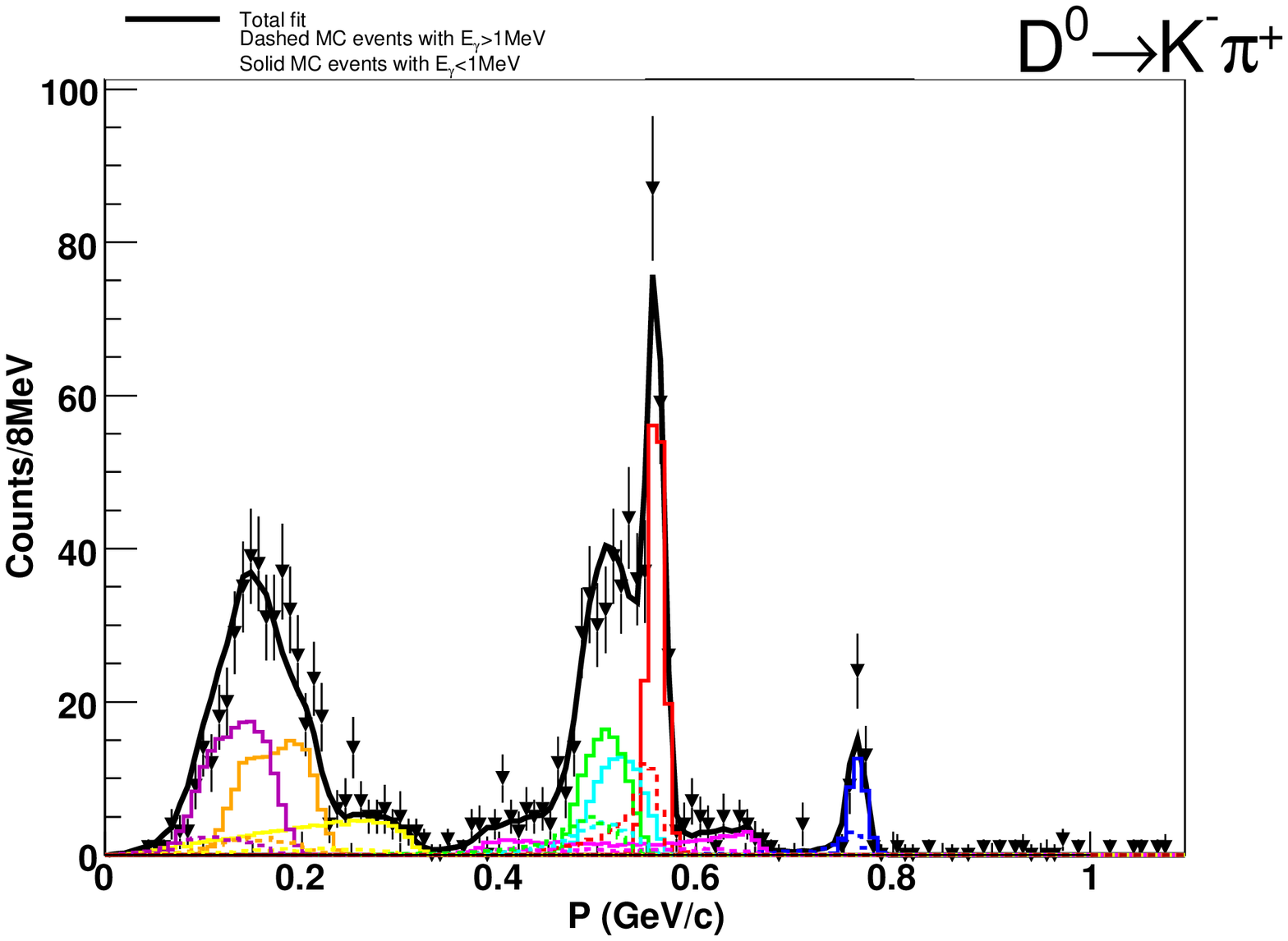}
\caption{Sideband-subtracted momentum spectrum for
$D^0\rightarrow{K^-}\pi^+$ at 4030 MeV.  The data is the points
with error bars and the histograms are MC.}
\vspace{0.2cm}
\label{fig:Mom_D0_4030}
\end{center}
\end{figure}

\begin{figure}[!p]
\begin{center}
\hspace{2.5pt}
\includegraphics[width=14.5cm]{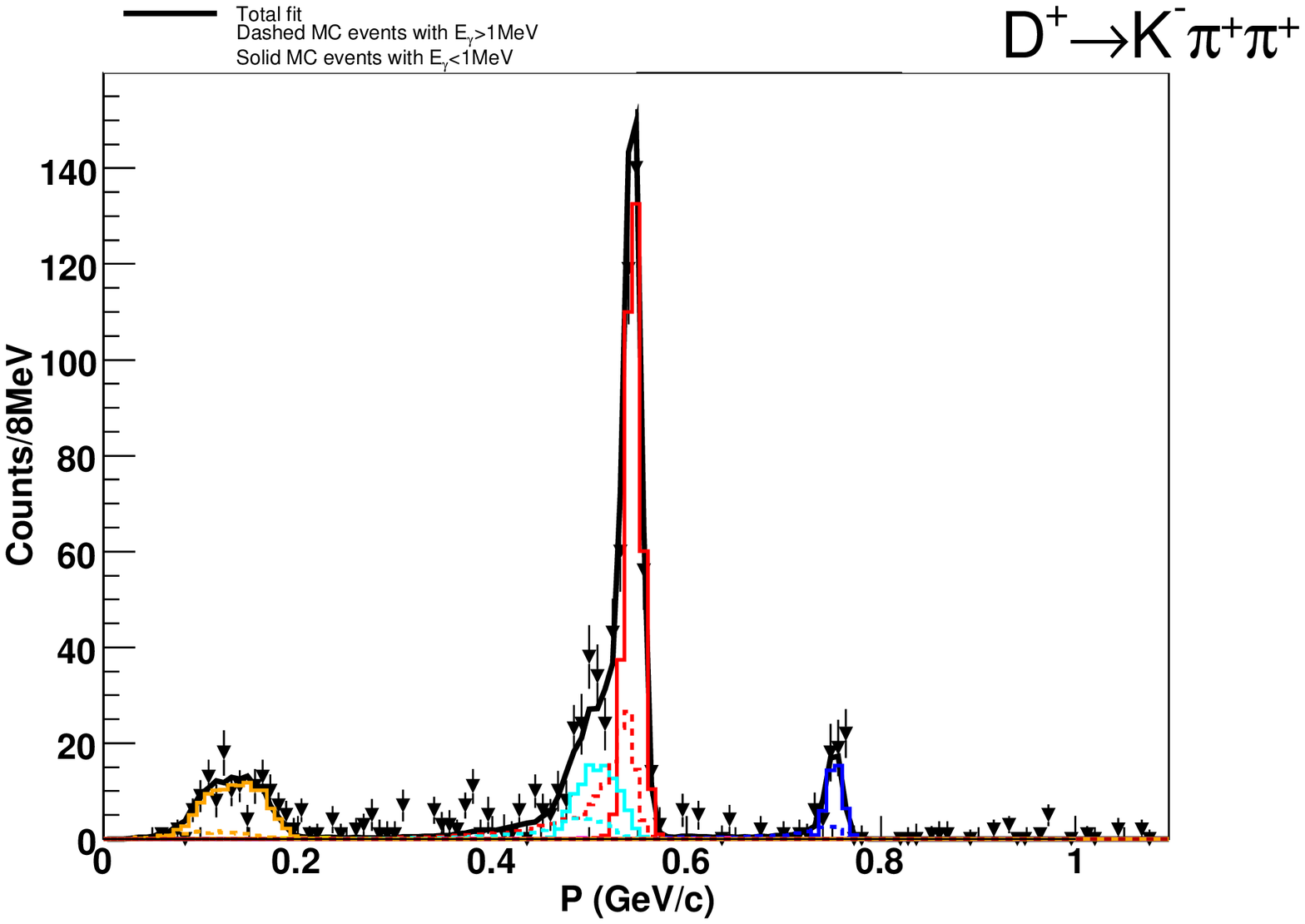}
\caption{Sideband-subtracted momentum spectrum for
$D^+\rightarrow{K^-}\pi^+\pi^+$ at 4030 MeV.  The data is the points
with error bars and the histograms are MC.}
\vspace{0.2cm}
\label{fig:Mom_Dp_4030}
\end{center}
\end{figure}

\begin{figure}[!p]
\begin{center}
\hspace{2.5pt}
\includegraphics[width=14.5cm]{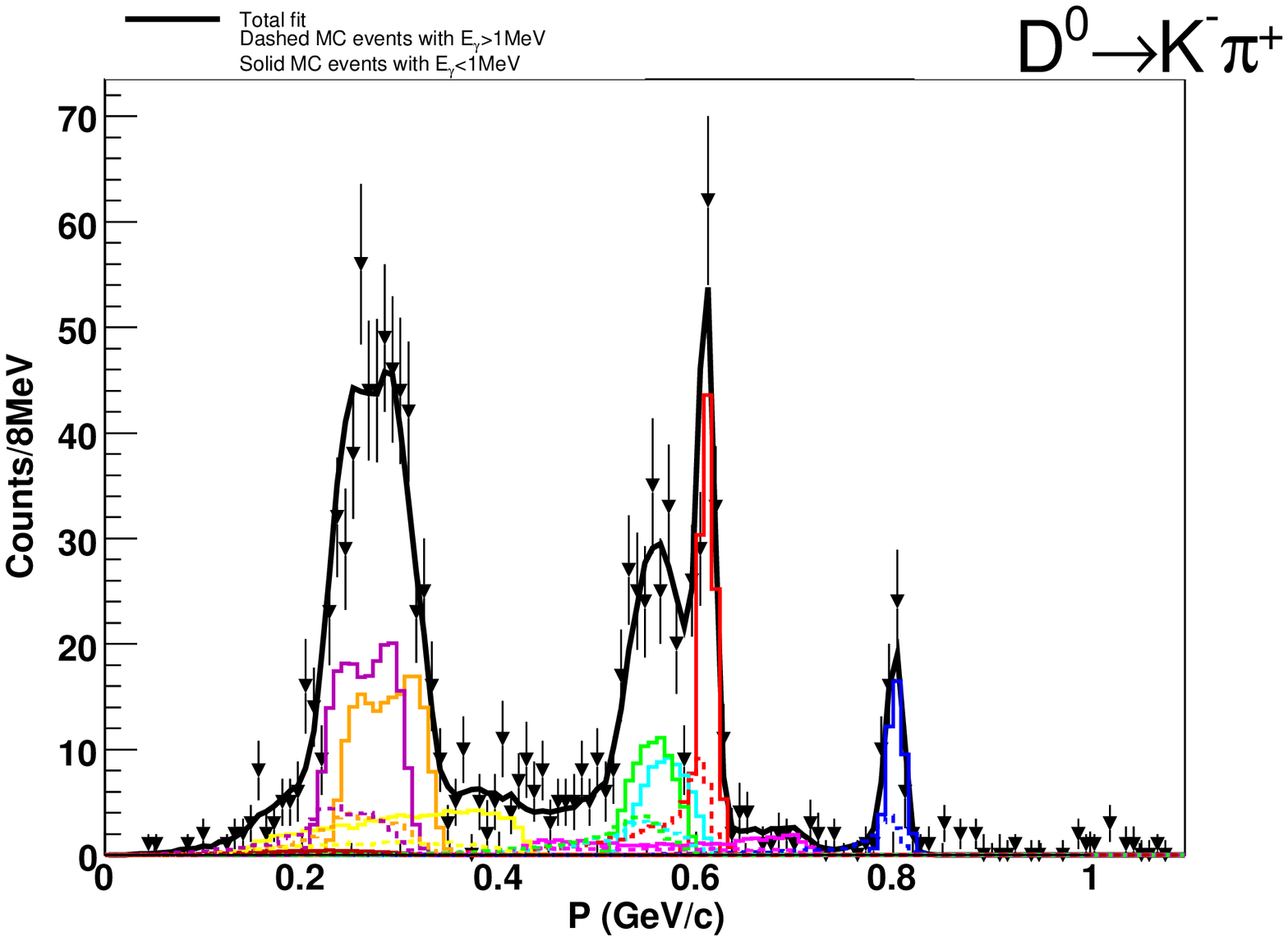}
\caption{Sideband-subtracted momentum spectrum for
$D^0\rightarrow{K^-}\pi^+$ at 4060 MeV.  The data is the points
with error bars and the histograms are MC.}
\vspace{0.2cm}
\label{fig:Mom_D0_4060}
\end{center}
\end{figure}

\begin{figure}[!p]
\begin{center}
\hspace{2.5pt}
\includegraphics[width=14.5cm]{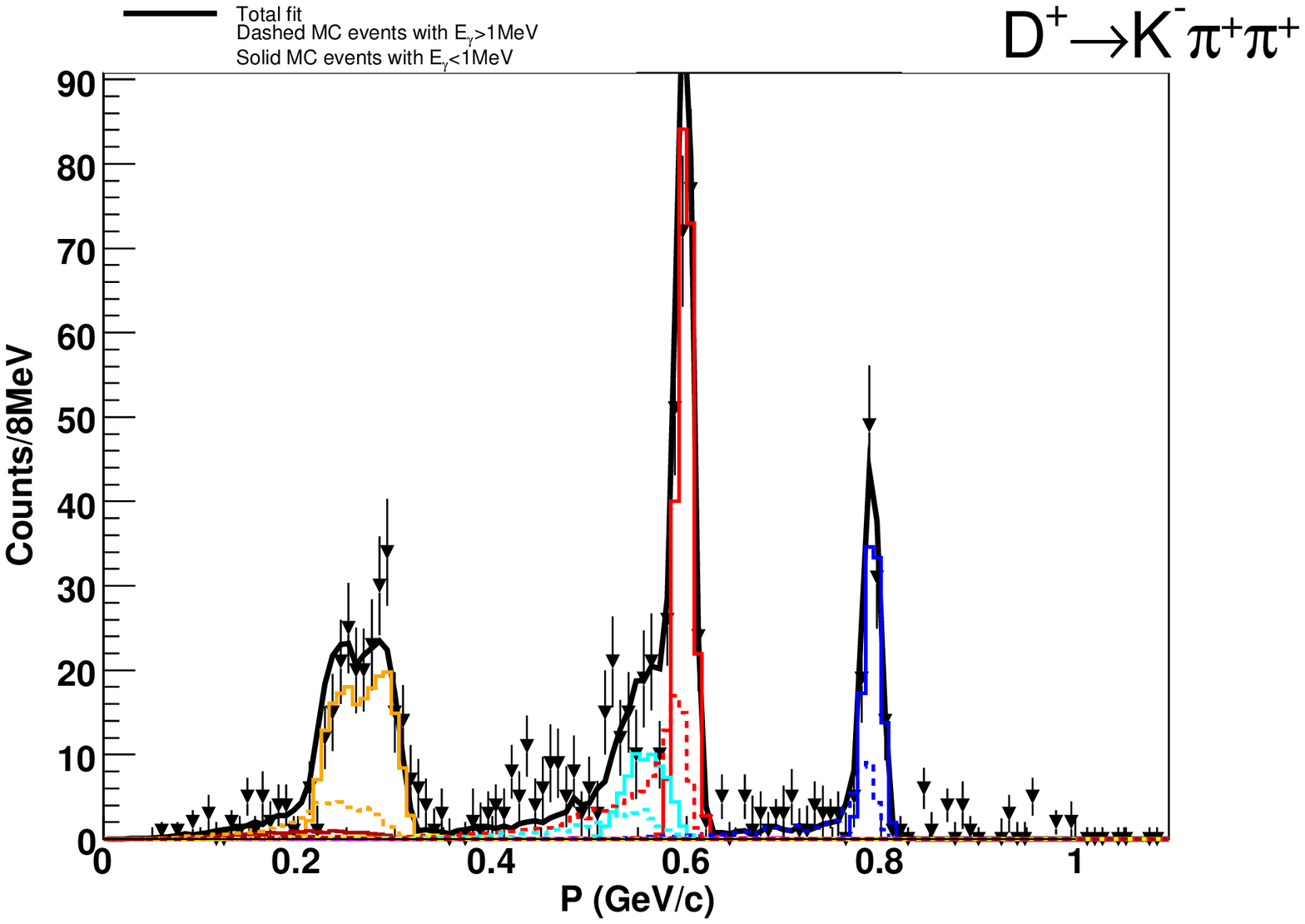}
\caption{Sideband-subtracted momentum spectrum for
$D^+\rightarrow{K^-}\pi^+\pi^+$ at 4060 MeV.  The data is the points
with error bars and the histograms are MC.}
\vspace{0.2cm}
\label{fig:Mom_Dp_4060}
\end{center}
\end{figure}

\begin{figure}[!p]
\begin{center}
\hspace{2.5pt}
\includegraphics[width=14.5cm]{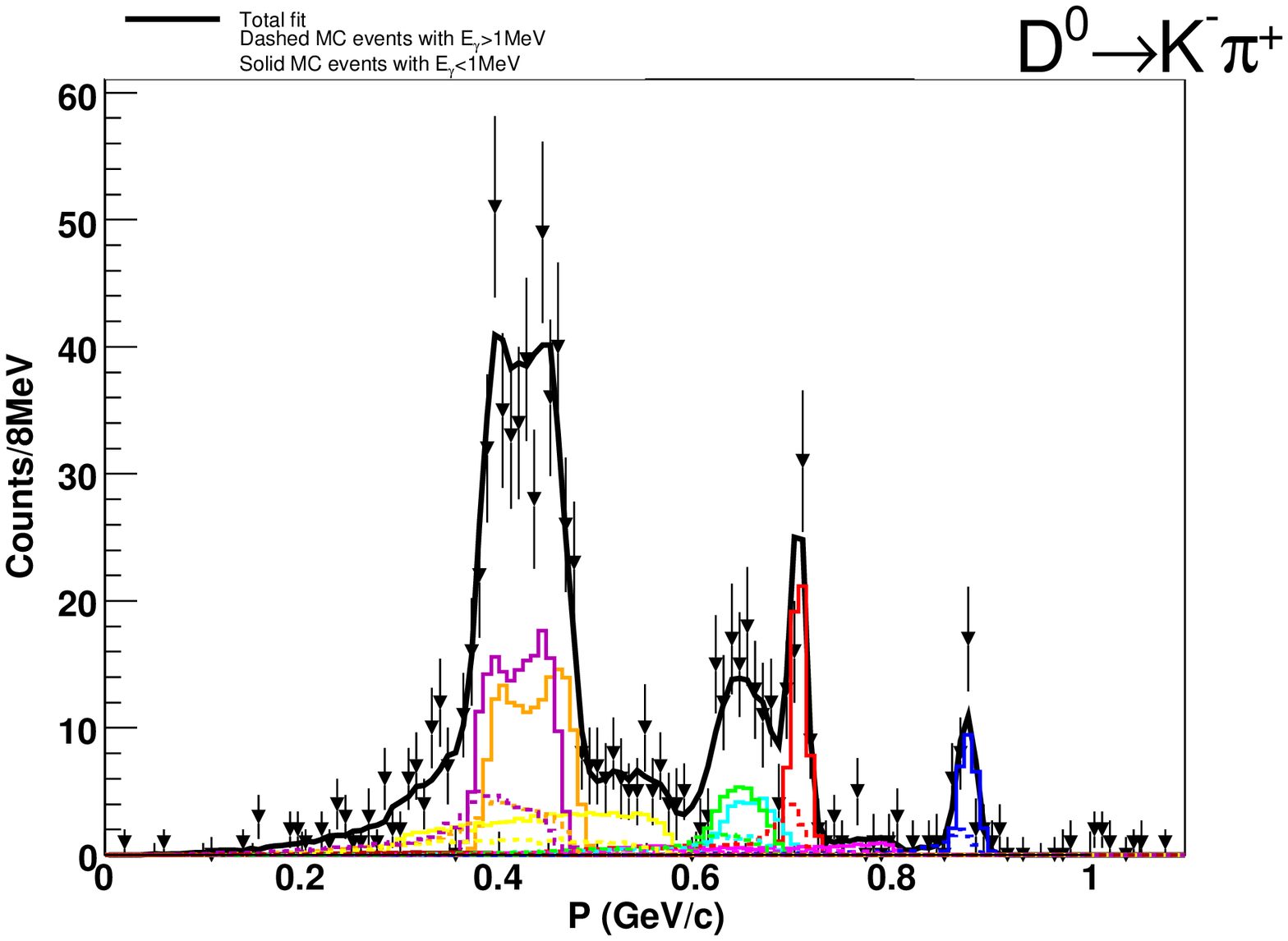}
\caption{Sideband-subtracted momentum spectrum for
$D^0\rightarrow{K^-}\pi^+$ at 4120 MeV.  The data is the points
with error bars and the histograms are MC.}
\vspace{0.2cm}
\label{fig:Mom_D0_4120}
\end{center}
\end{figure}

\begin{figure}[!p]
\begin{center}
\hspace{2.5pt}
\includegraphics[width=14.5cm]{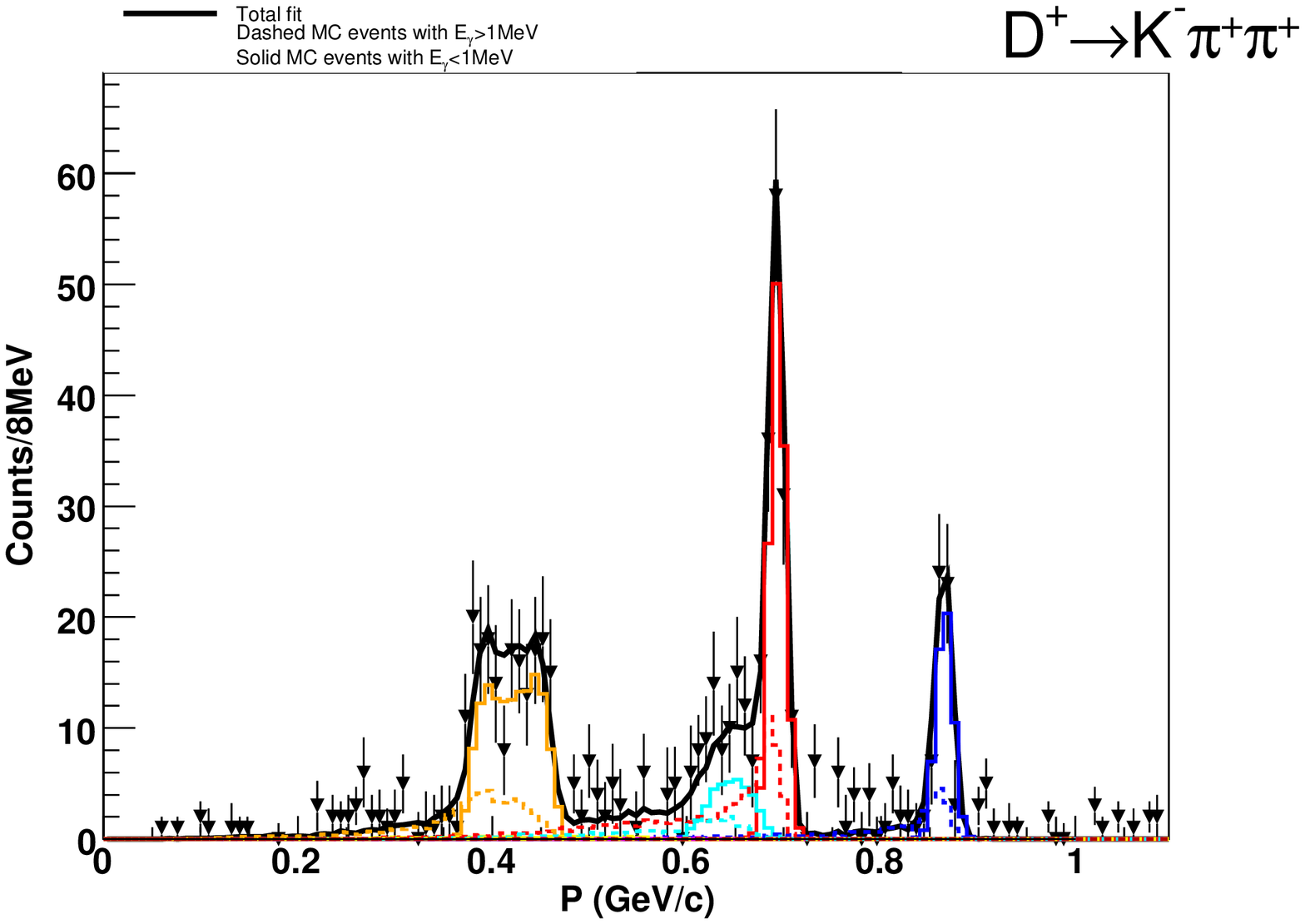}
\caption{Sideband-subtracted momentum spectrum for
$D^+\rightarrow{K^-}\pi^+\pi^+$ at 4120 MeV.  The data is the points
with error bars and the histograms are MC.}
\vspace{0.2cm}
\label{fig:Mom_Dp_4120}
\end{center}
\end{figure}

\begin{figure}[!p]
\begin{center}
\hspace{2.5pt}
\includegraphics[width=14.5cm]{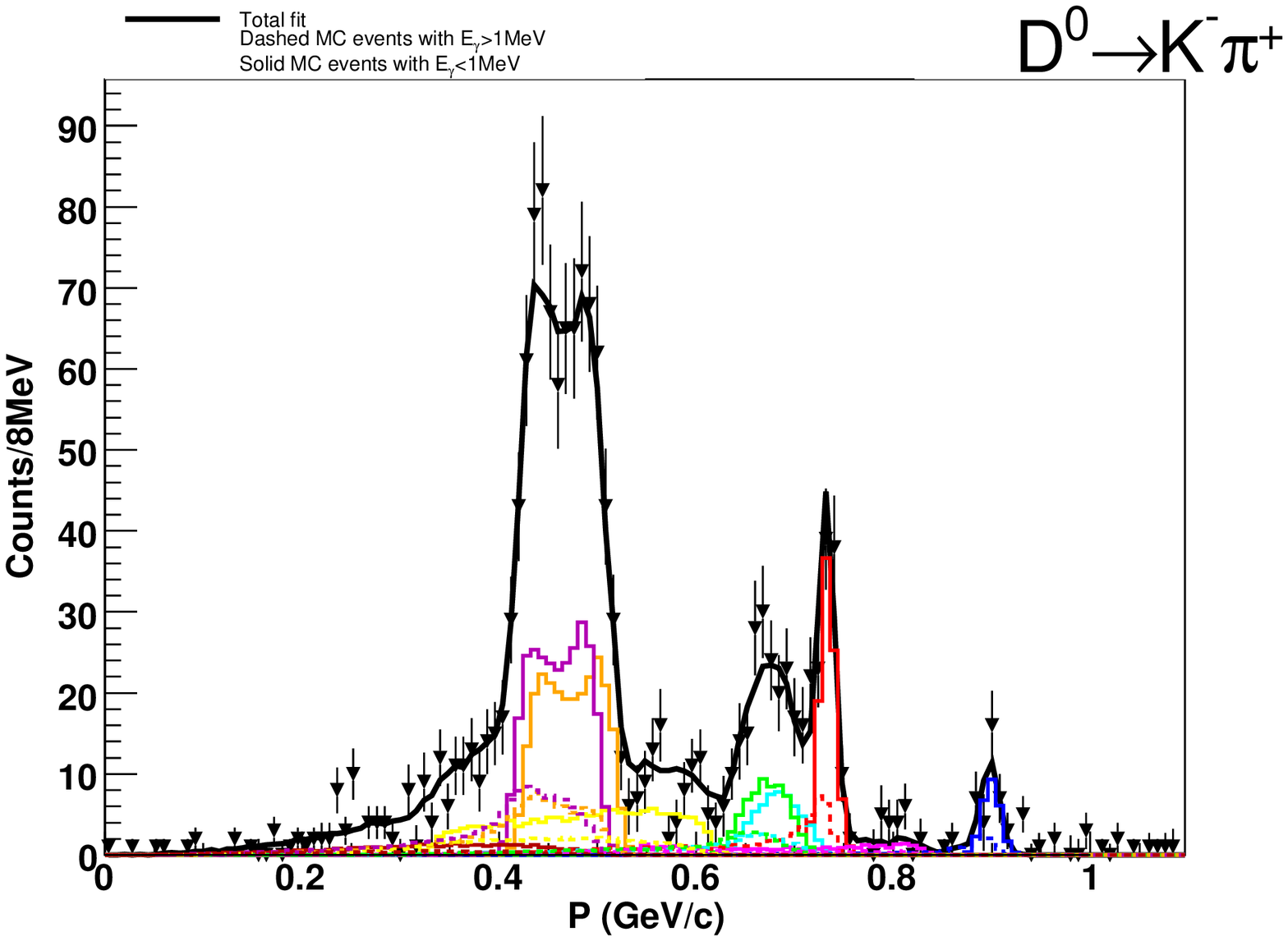}
\caption{Sideband-subtracted momentum spectrum for
$D^0\rightarrow{K^-}\pi^+$ at 4140 MeV.  The data is the points
with error bars and the histograms are MC.}
\vspace{0.2cm}
\label{fig:Mom_D0_4140}
\end{center}
\end{figure}

\begin{figure}[!p]
\begin{center}
\hspace{2.5pt}
\includegraphics[width=14.5cm]{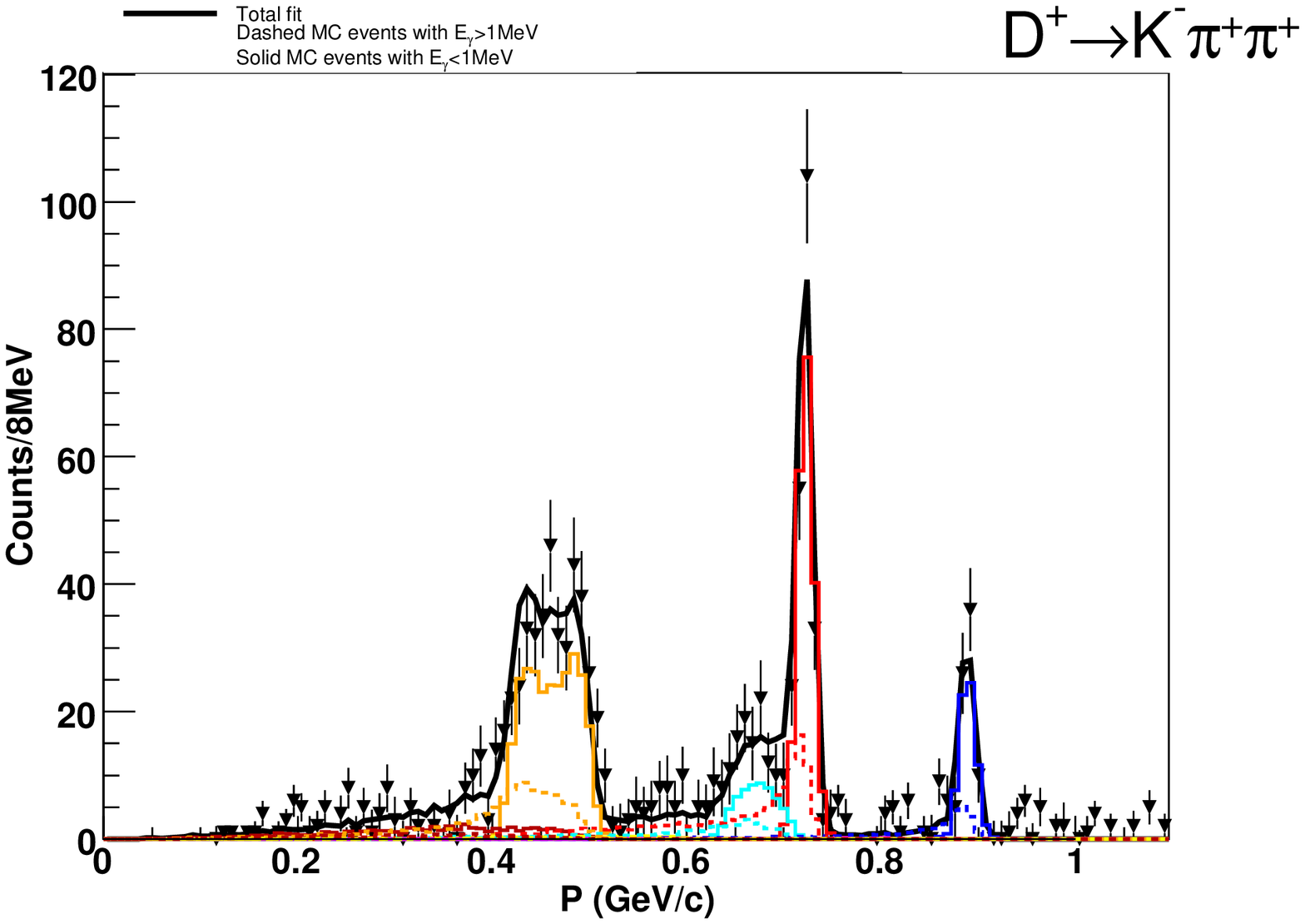}
\caption{Sideband-subtracted momentum spectrum for
$D^+\rightarrow{K^-}\pi^+\pi^+$ at 4140 MeV.  The data is the points
with error bars and the histograms are MC.}
\vspace{0.2cm}
\label{fig:Mom_Dp_4140}
\end{center}
\end{figure}

\begin{figure}[!p]
\begin{center}
\hspace{2.5pt}
\includegraphics[width=14.5cm]{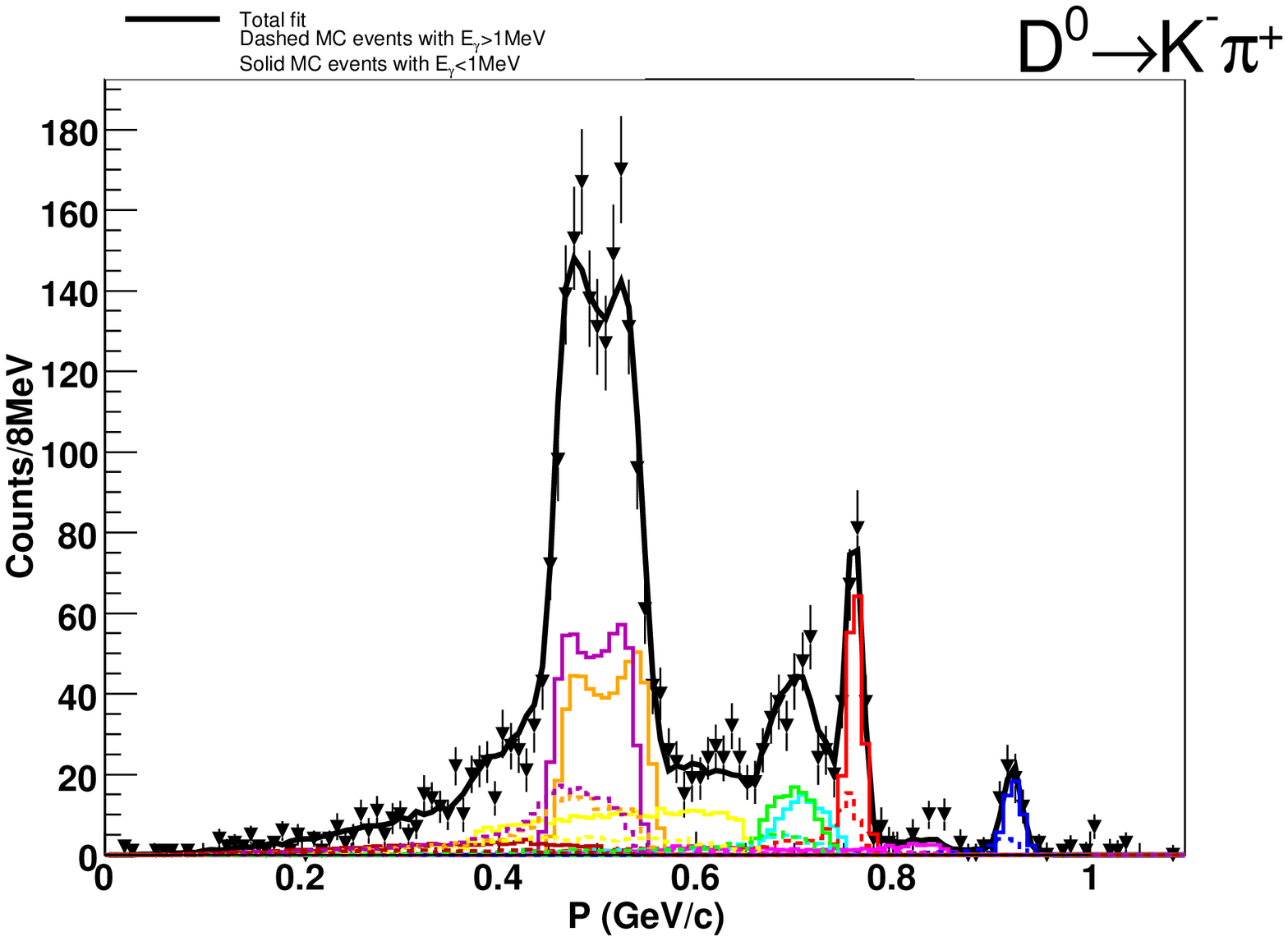}
\caption{Sideband-subtracted momentum spectrum for
$D^0\rightarrow{K^-}\pi^+$ at 4160 MeV.  The data is the points
with error bars and the histograms are MC.}
\vspace{0.2cm}
\label{fig:Mom_D0_4160}
\end{center}
\end{figure}

\begin{figure}[!p]
\begin{center}
\hspace{2.5pt}
\includegraphics[width=14.5cm]{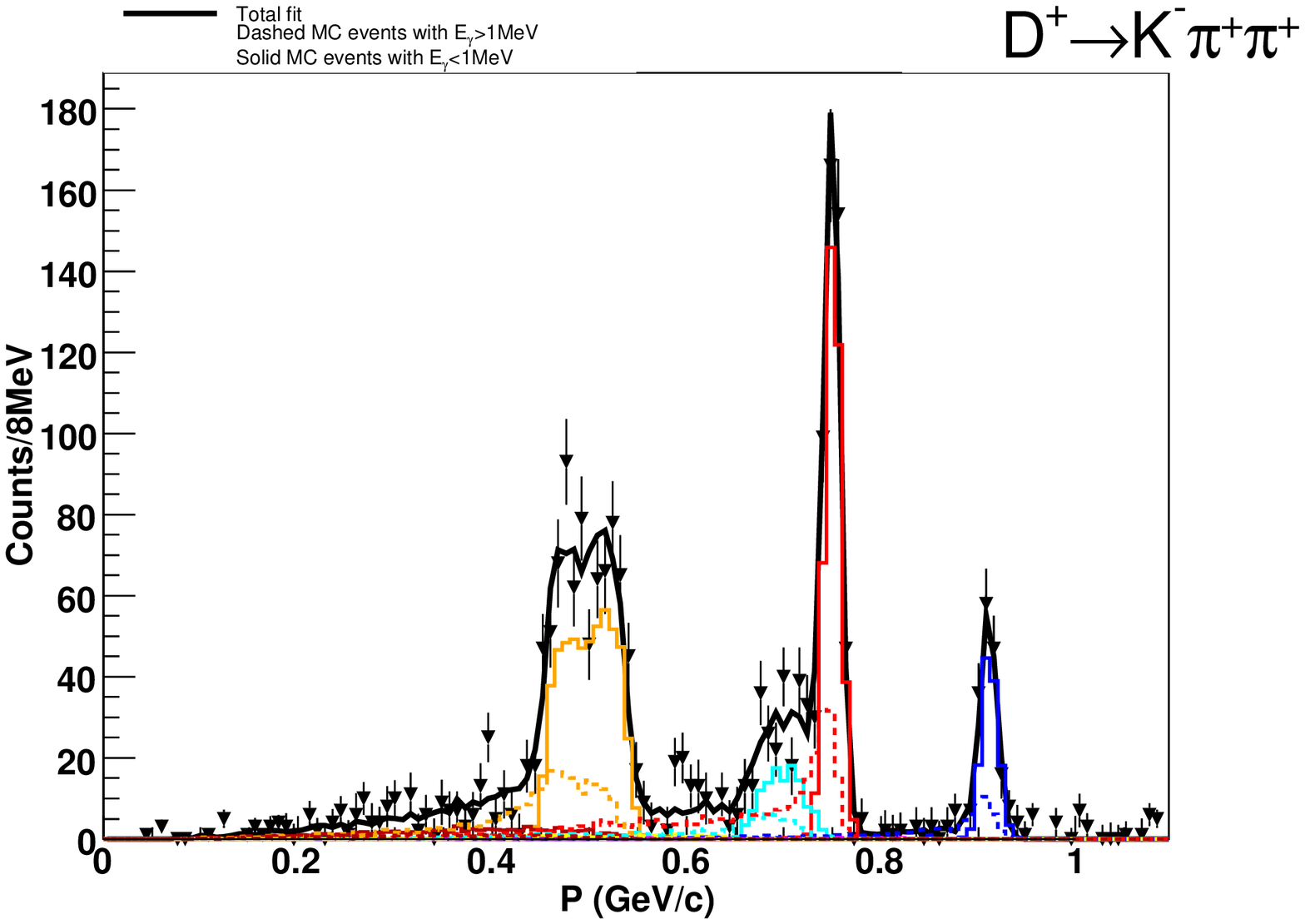}
\caption{Sideband-subtracted momentum spectrum for
$D^+\rightarrow{K^-}\pi^+\pi^+$ at 4160 MeV.  The data is the points
with error bars and the histograms are MC.}
\vspace{0.2cm}
\label{fig:Mom_Dp_4160}
\end{center}
\end{figure}

\begin{figure}[!p]
\begin{center}
\hspace{2.5pt}
\includegraphics[width=14.5cm]{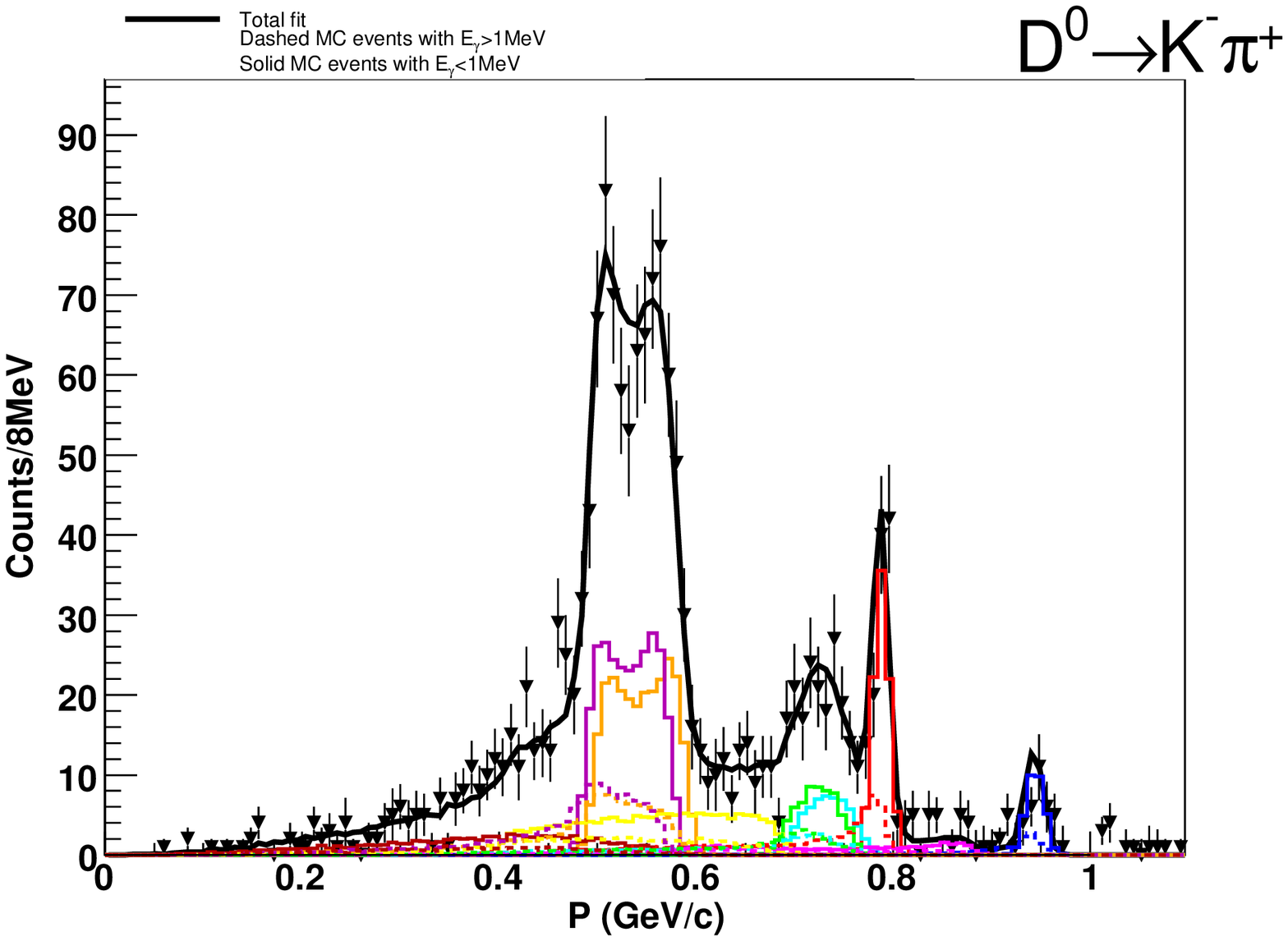}
\caption{Sideband-subtracted momentum spectrum for
$D^0\rightarrow{K^-}\pi^+$ at 4180 MeV.  The data is the points
with error bars and the histograms are MC.}
\vspace{0.2cm}
\label{fig:Mom_D0_4180}
\end{center}
\end{figure}

\begin{figure}[!p]
\begin{center}
\hspace{2.5pt}
\includegraphics[width=14.5cm]{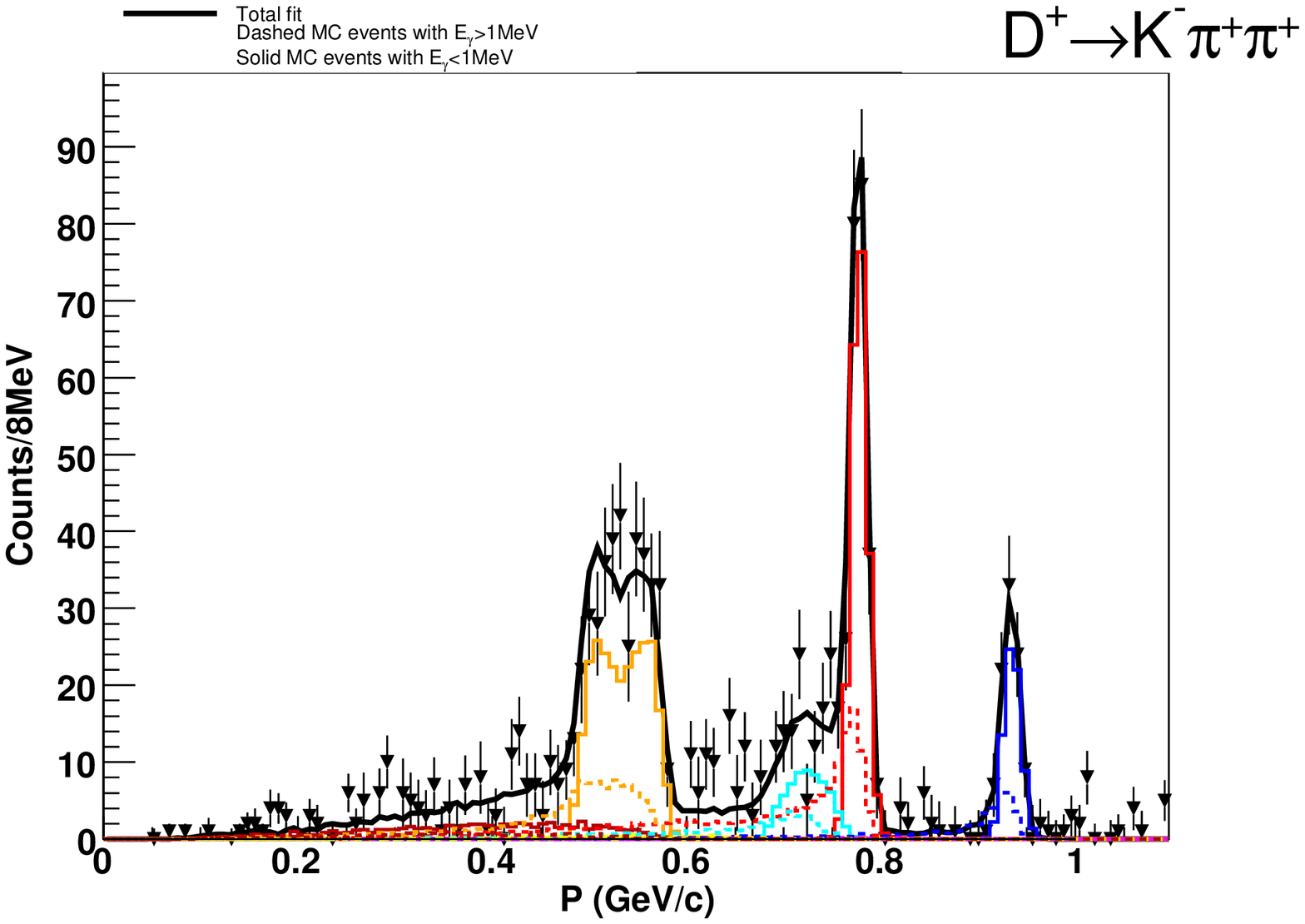}
\caption{Sideband-subtracted momentum spectrum for
$D^+\rightarrow{K^-}\pi^+\pi^+$ at 4180 MeV.  The data is the points
with error bars and the histograms are MC.}
\vspace{0.2cm}
\label{fig:Mom_Dp_4180}
\end{center}
\end{figure}

\begin{figure}[!p]
\begin{center}
\hspace{2.5pt}
\includegraphics[width=14.5cm]{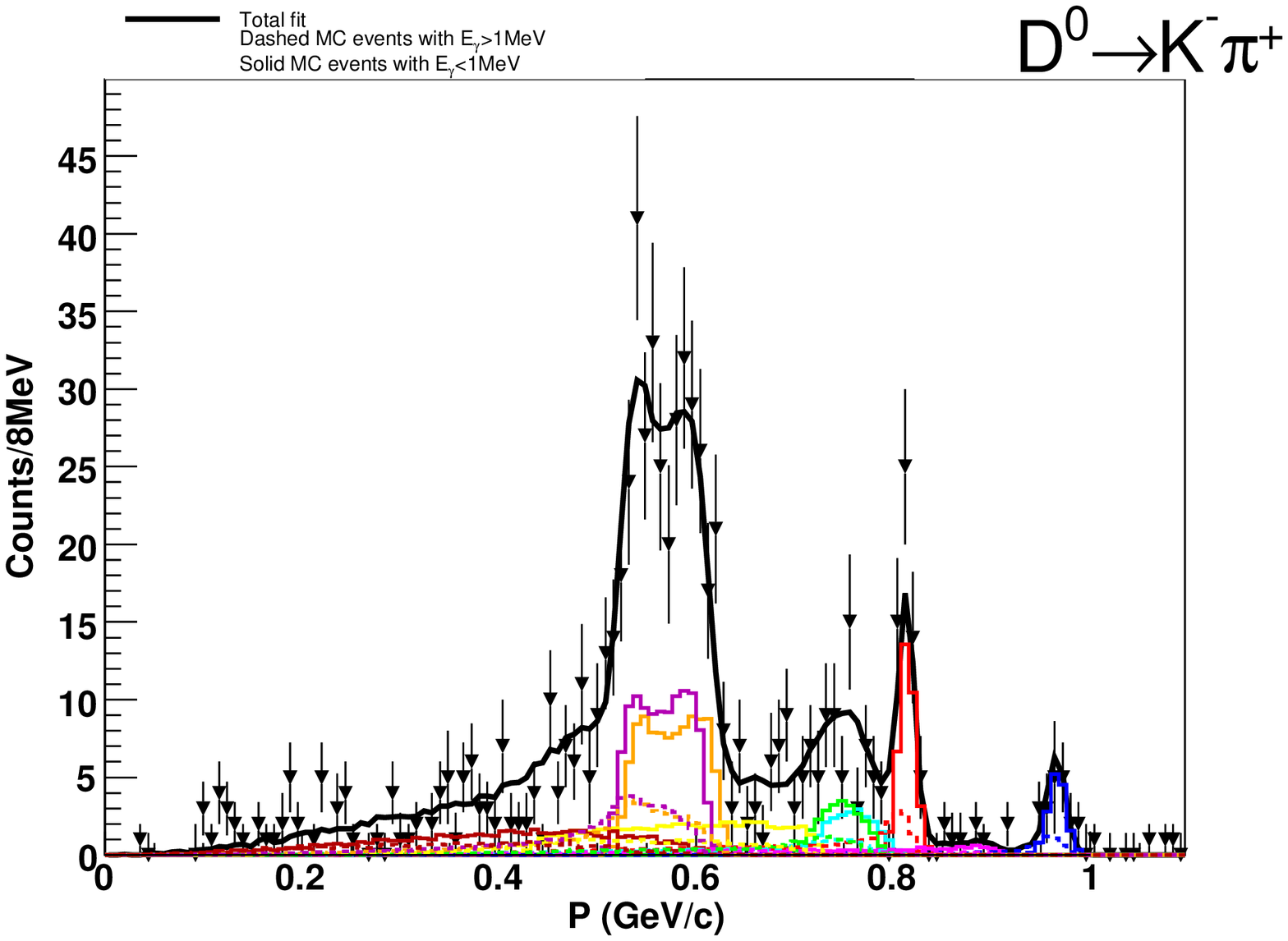}
\caption{Sideband-subtracted momentum spectrum for
$D^0\rightarrow{K^-}\pi^+$ at 4200 MeV.  The data is the points
with error bars and the histograms are MC.}
\vspace{0.2cm}
\label{fig:Mom_D0_4200}
\end{center}
\end{figure}

\begin{figure}[!p]
\begin{center}
\hspace{2.5pt}
\includegraphics[width=14.5cm]{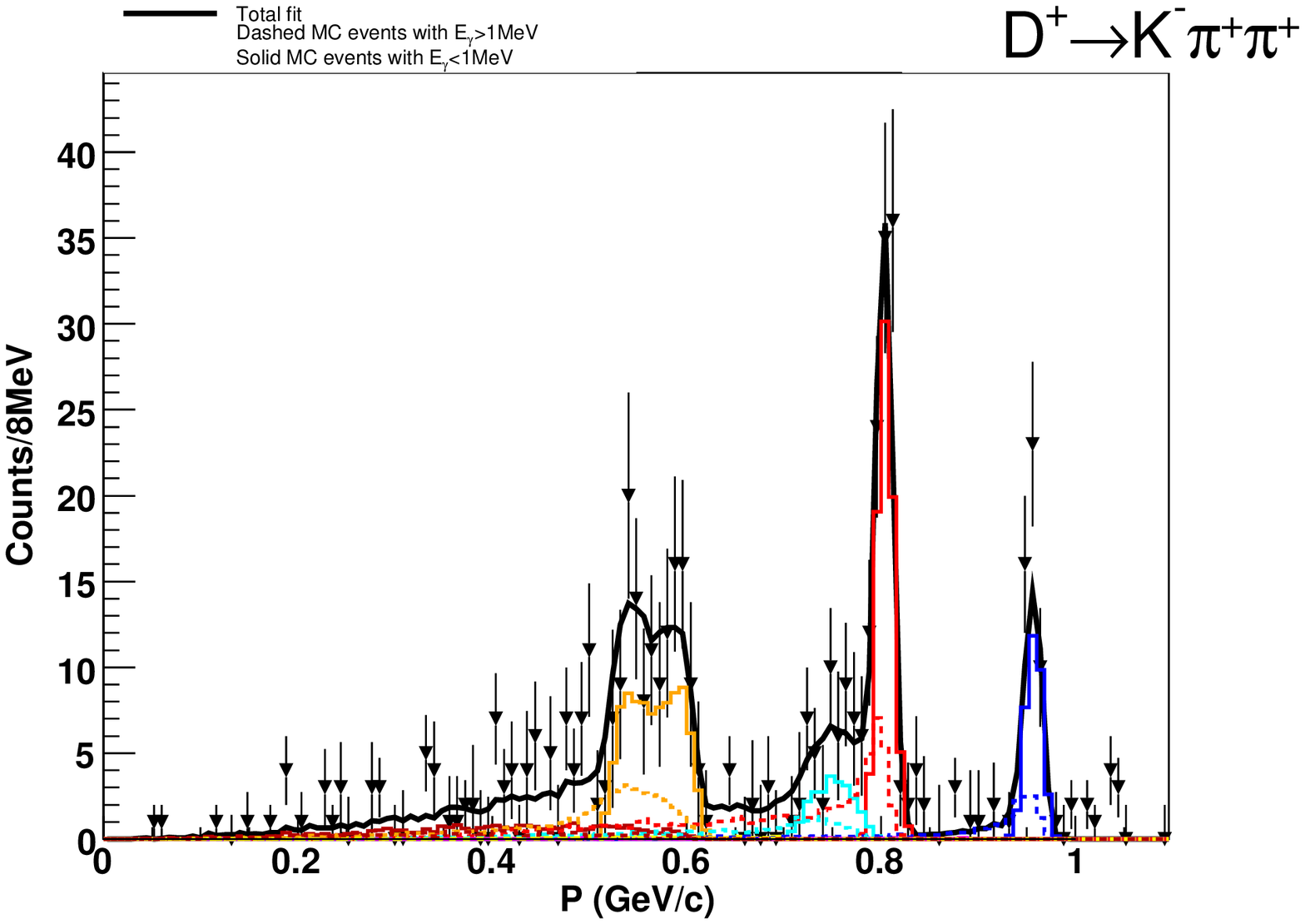}
\caption{Sideband-subtracted momentum spectrum for
$D^+\rightarrow{K^-}\pi^+\pi^+$ at 4200 MeV.  The data is the points
with error bars and the histograms are MC.}
\vspace{0.2cm}
\label{fig:Mom_Dp_4200}
\end{center}
\end{figure}

\begin{figure}[!p]
\begin{center}
\hspace{2.5pt}
\includegraphics[width=14.5cm]{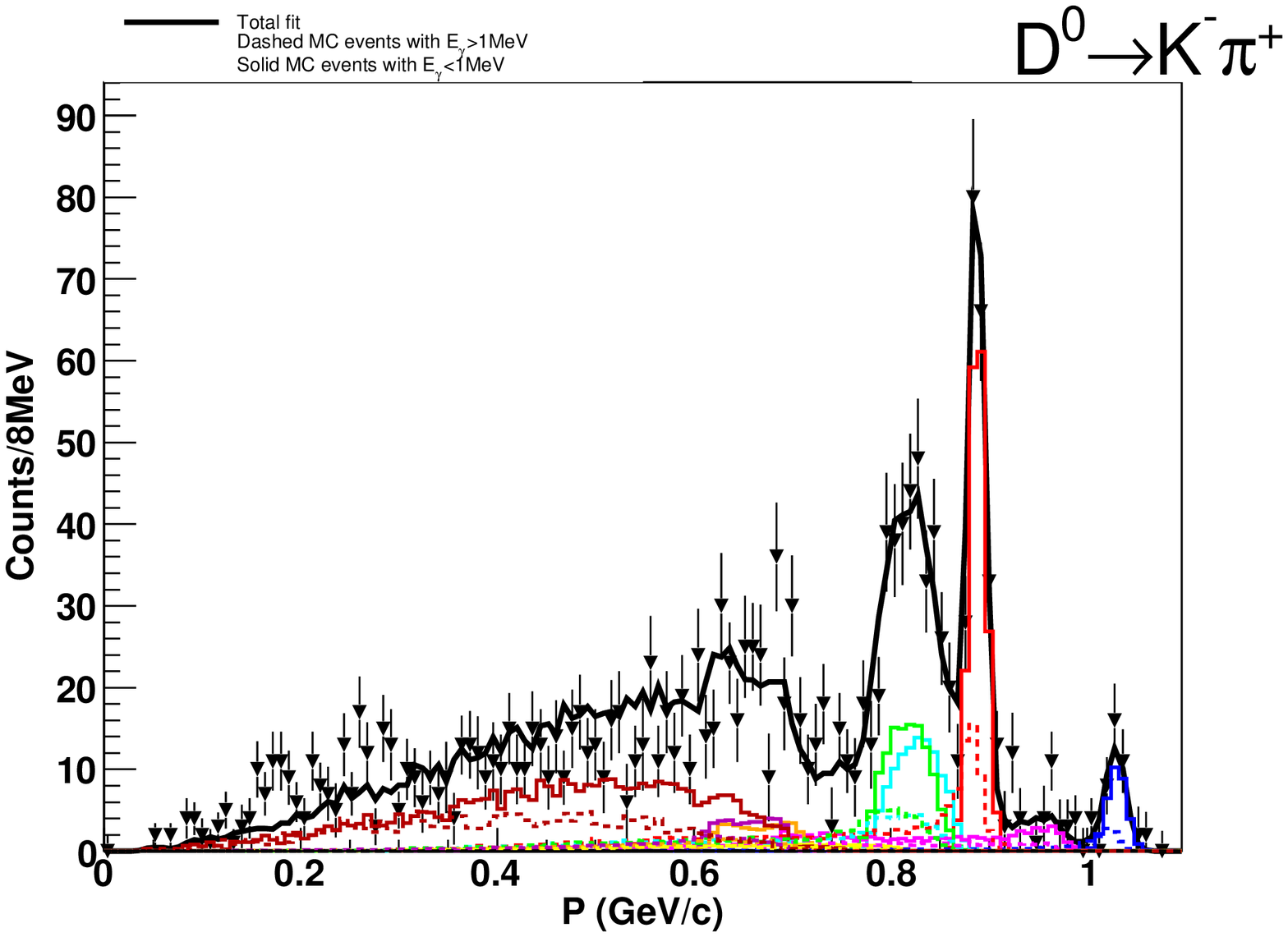}
\caption{Sideband-subtracted momentum spectrum for
$D^0\rightarrow{K^-}\pi^+$ at 4260 MeV.  The data is the points
with error bars and the histograms are MC.}
\vspace{0.2cm}
\label{fig:Mom_D0_4260}
\end{center}
\end{figure}

\begin{figure}[!p]
\begin{center}
\hspace{2.5pt}
\includegraphics[width=14.5cm]{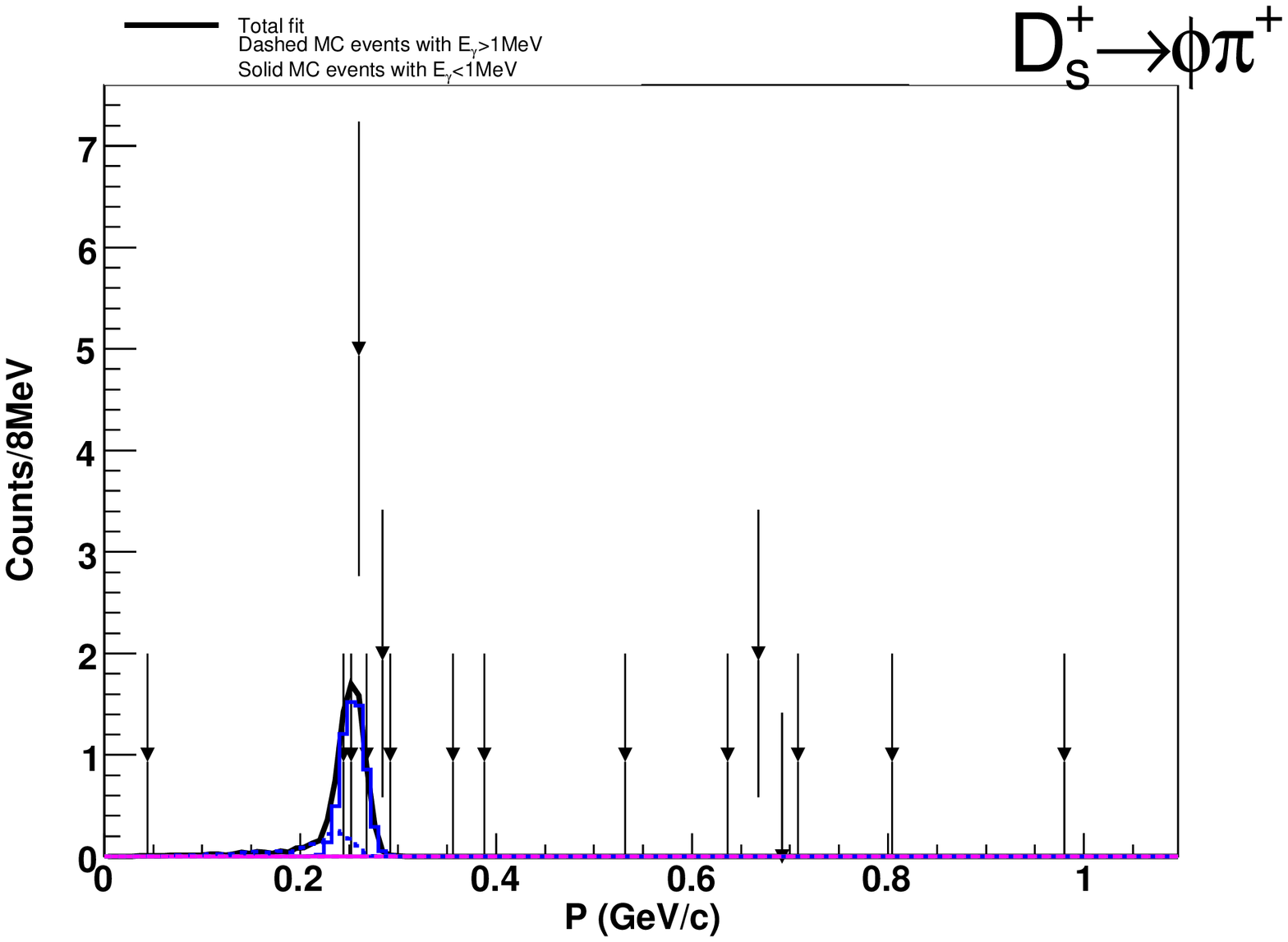}
\caption{Sideband-subtracted momentum spectrum for
$D_{s}^+\rightarrow\phi\pi^+$ at 3970 MeV.  The data is the points
with error bars and the histograms are MC.}
\vspace{0.2cm}
\label{fig:Mom_Ds_3970}
\end{center}
\end{figure}

\begin{figure}[!p]
\begin{center}
\hspace{2.5pt}
\includegraphics[width=14.5cm]{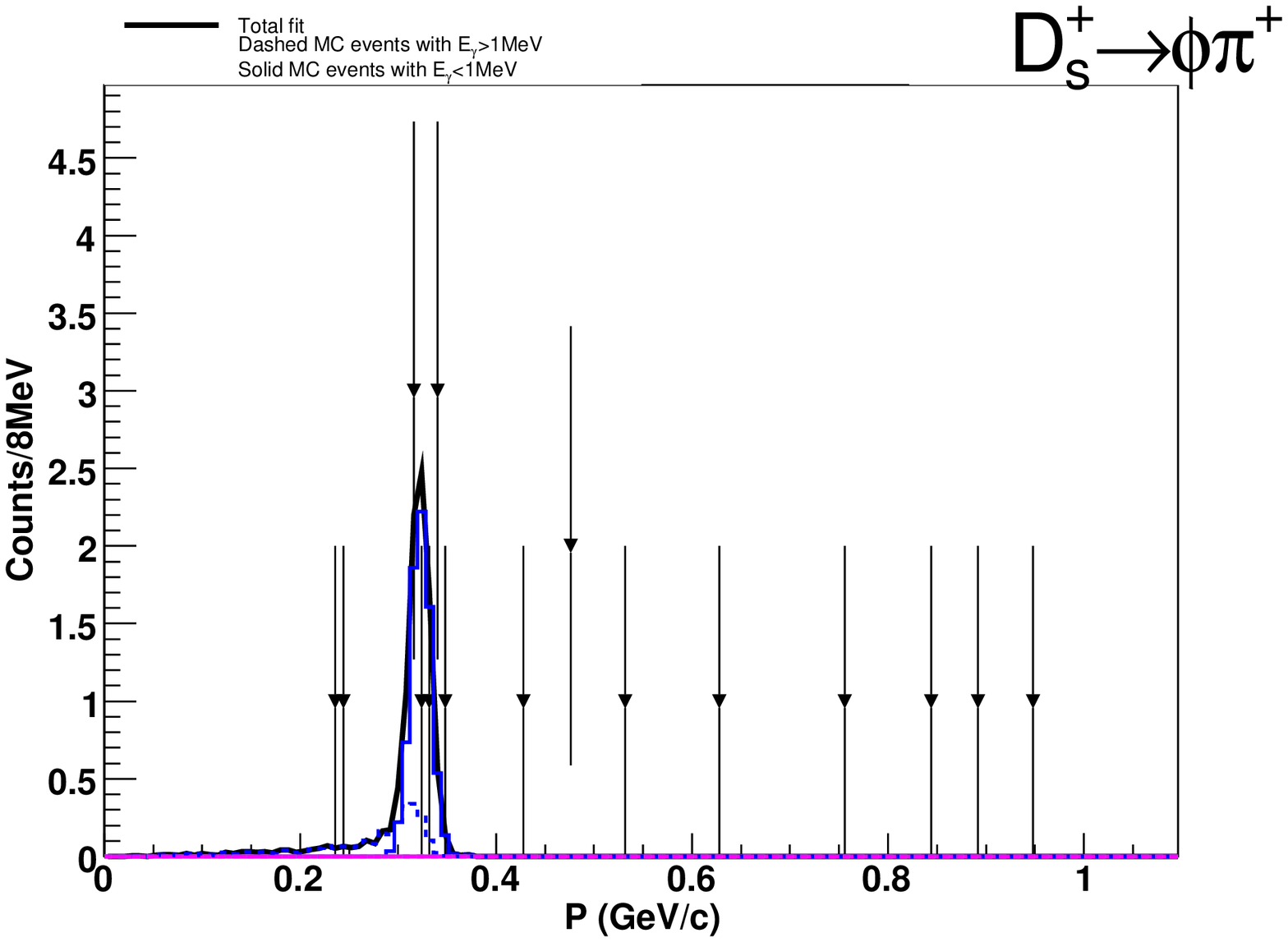}
\caption{Sideband-subtracted momentum spectrum for
$D_{s}^+\rightarrow\phi\pi^+$ at 3990 MeV.  The data is the points
with error bars and the histograms are MC.}
\vspace{0.2cm}
\label{fig:Mom_Ds_3990}
\end{center}
\end{figure}

\begin{figure}[!p]
\begin{center}
\hspace{2.5pt}
\includegraphics[width=14.5cm]{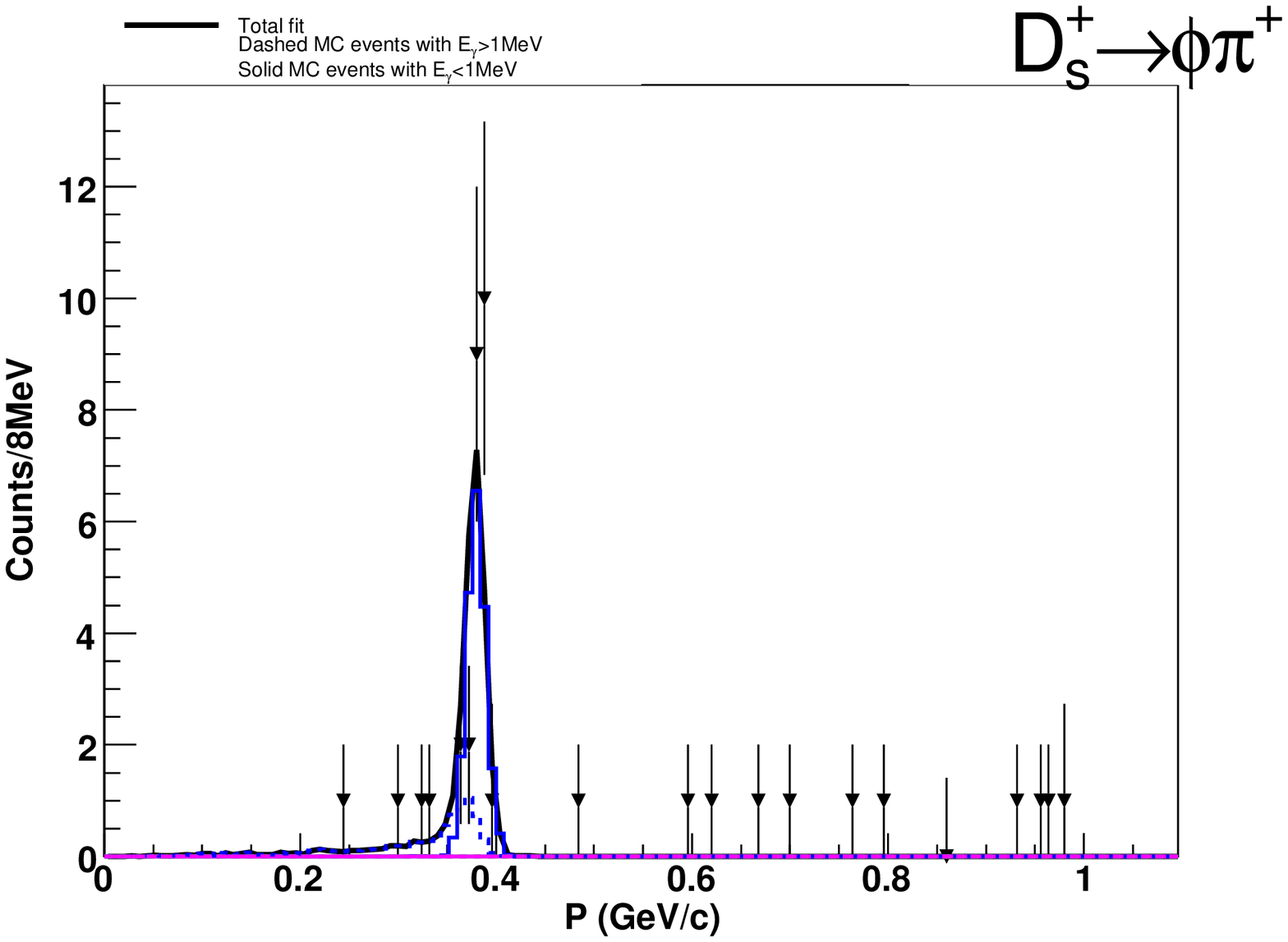}
\caption{Sideband-subtracted momentum spectrum for
$D_{s}^+\rightarrow\phi\pi^+$ at 4010 MeV.  The data is the points
with error bars and the histograms are MC.}
\vspace{0.2cm}
\label{fig:Mom_Ds_4010}
\end{center}
\end{figure}

\begin{figure}[!p]
\begin{center}
\hspace{2.5pt}
\includegraphics[width=14.5cm]{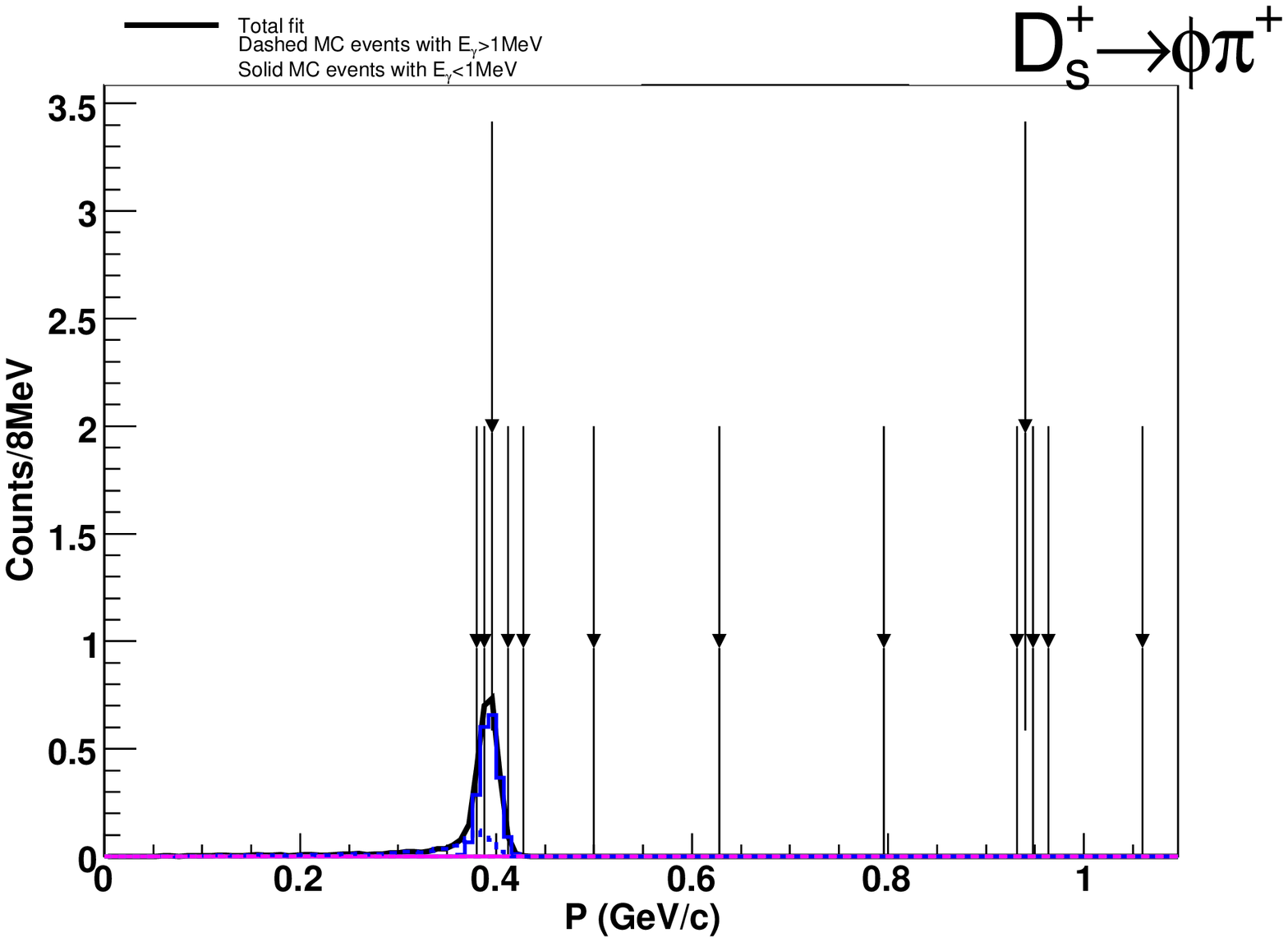}
\caption{Sideband-subtracted momentum spectrum for
$D_{s}^+\rightarrow\phi\pi^+$ at 4015 MeV.  The data is the points
with error bars and the histograms are MC.}
\vspace{0.2cm}
\label{fig:Mom_Ds_4015}
\end{center}
\end{figure}

\begin{figure}[!p]
\begin{center}
\hspace{2.5pt}
\includegraphics[width=14.5cm]{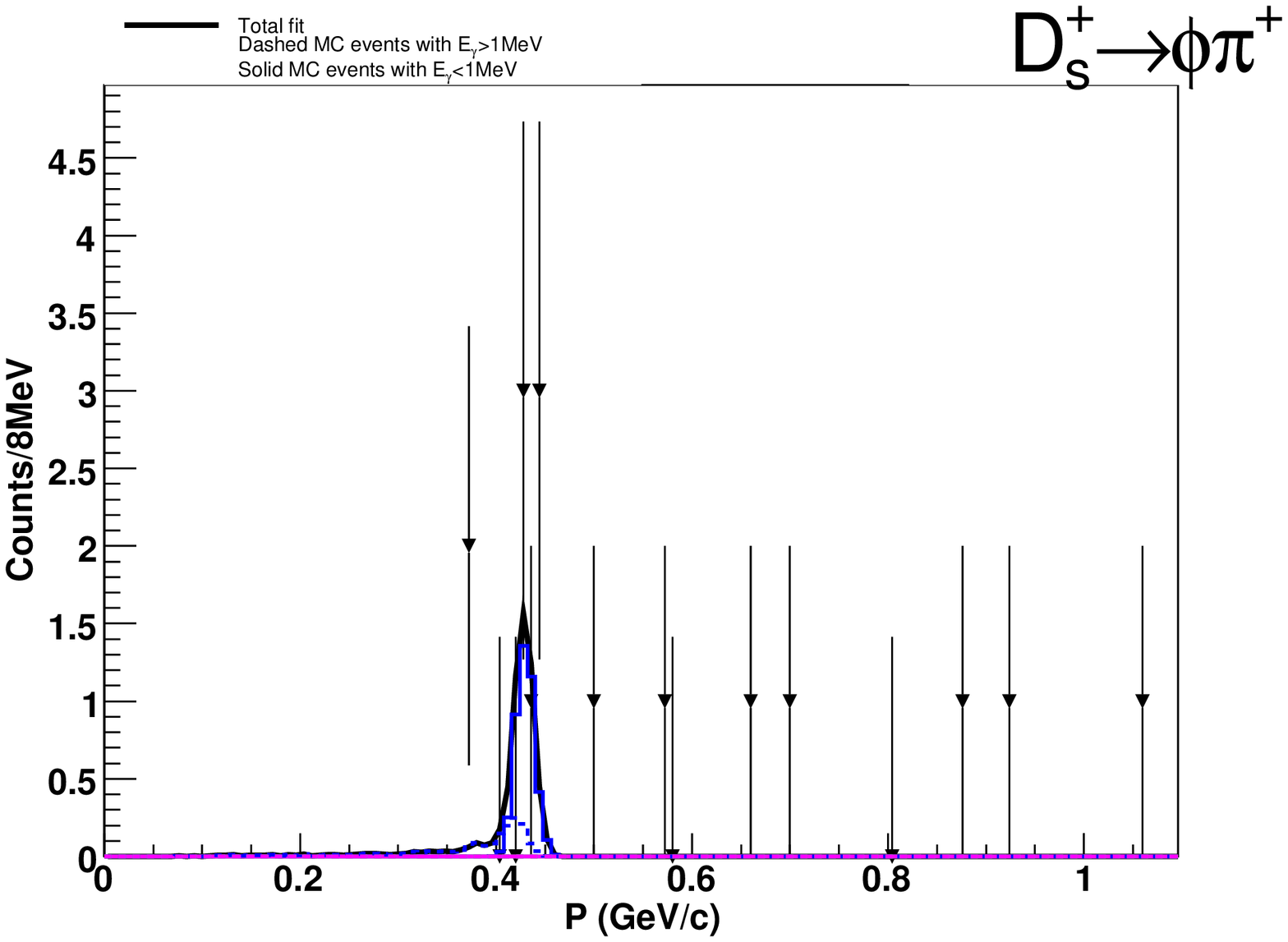}
\caption{Sideband-subtracted momentum spectrum for
$D_{s}^+\rightarrow\phi\pi^+$ at 4030 MeV.  The data is the points
with error bars and the histograms are MC.}
\vspace{0.2cm}
\label{fig:Mom_Ds_4030}
\end{center}
\end{figure}

\begin{figure}[!p]
\begin{center}
\hspace{2.5pt}
\includegraphics[width=14.5cm]{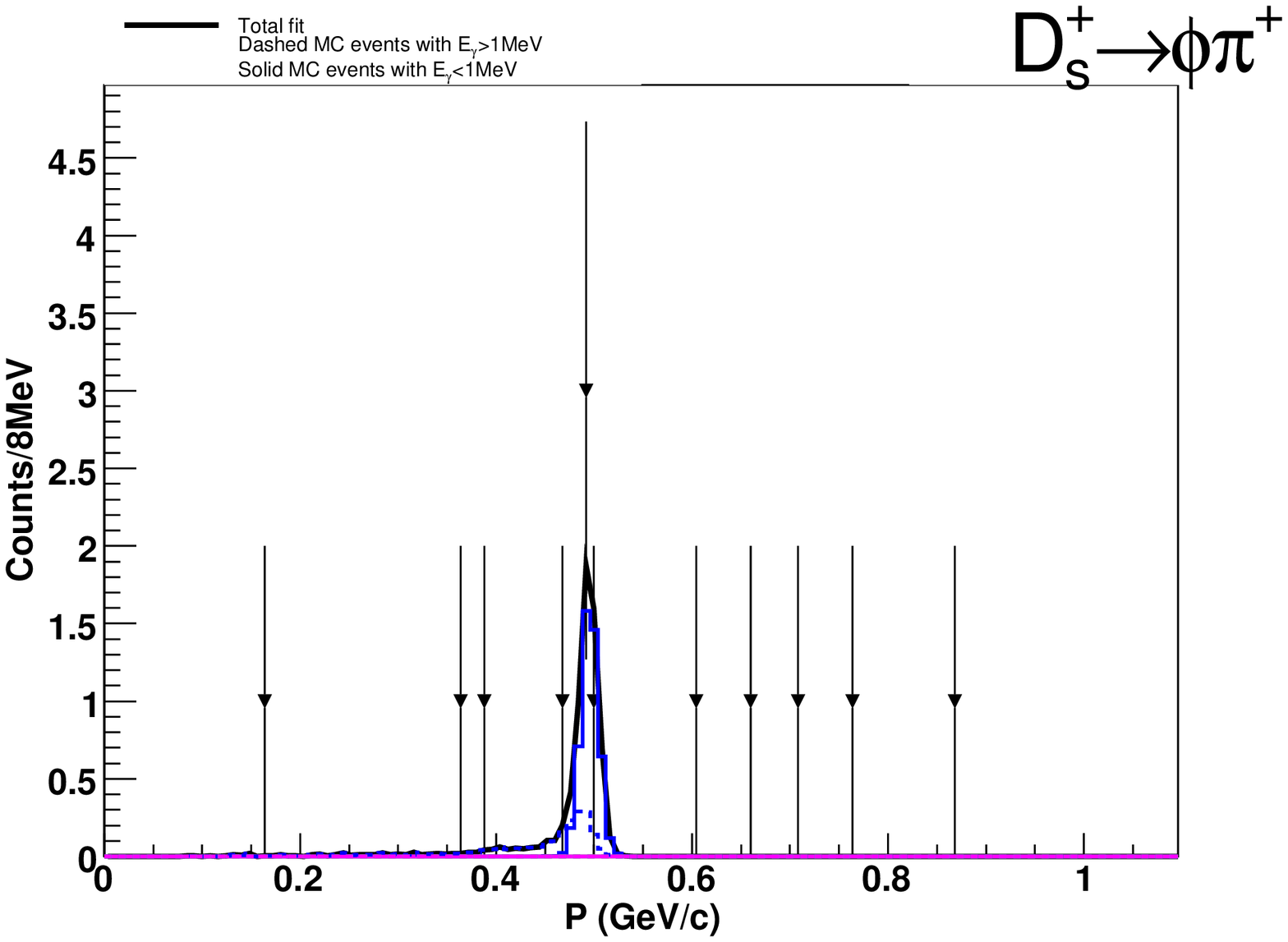}
\caption{Sideband-subtracted momentum spectrum for
$D_{s}^+\rightarrow\phi\pi^+$ at 4060 MeV.  The data is the points
with error bars and the histograms are MC.}
\vspace{0.2cm}
\label{fig:Mom_Ds_4060}
\end{center}
\end{figure}

\begin{figure}[!p]
\begin{center}
\hspace{2.5pt}
\includegraphics[width=14.5cm]{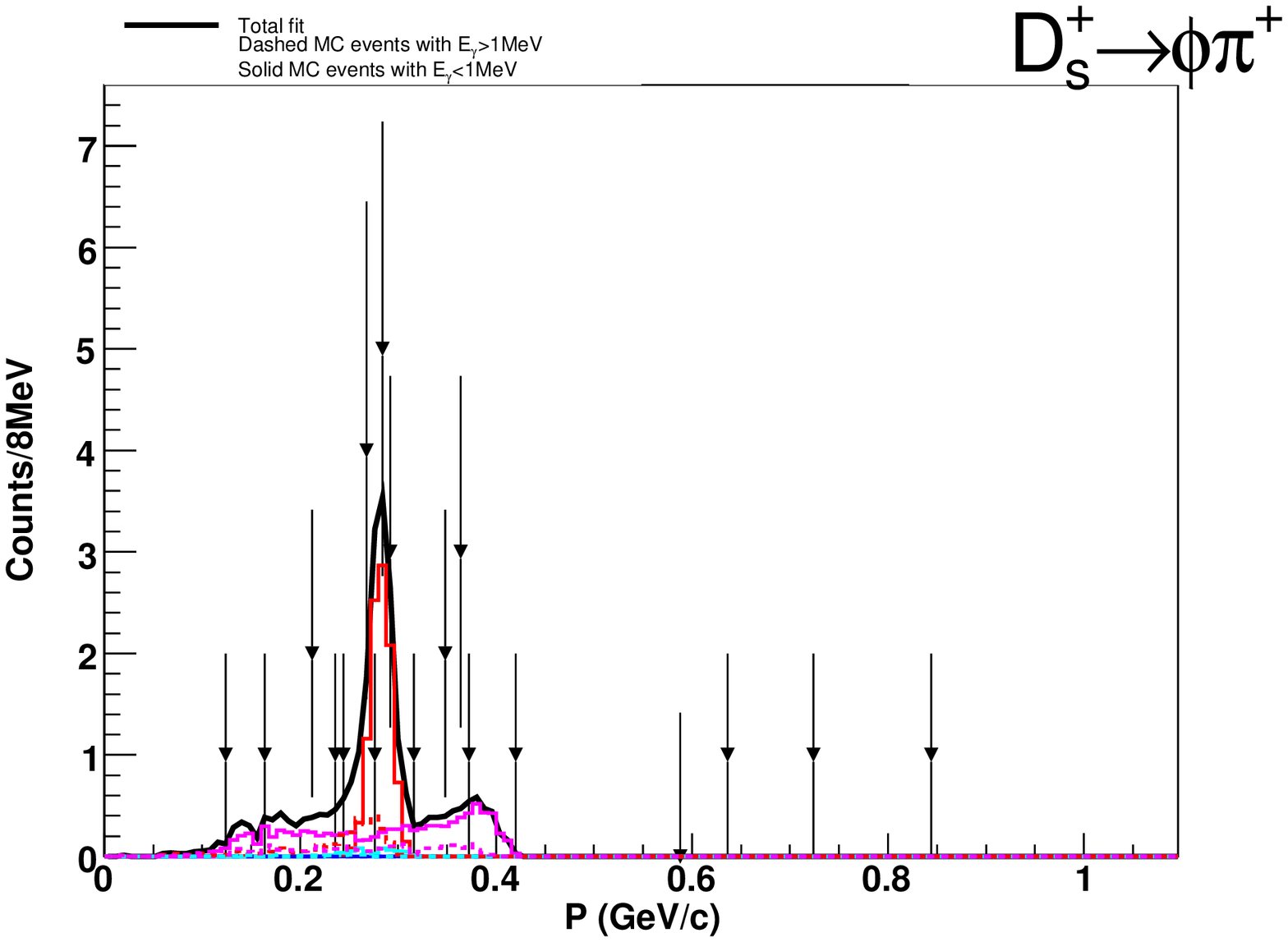}
\caption{Sideband-subtracted momentum spectrum for
$D_{s}^+\rightarrow\phi\pi^+$ at 4120 MeV.  The data is the points
with error bars and the histograms are MC.}
\vspace{0.2cm}
\label{fig:Mom_Ds_4120}
\end{center}
\end{figure}

\begin{figure}[!p]
\begin{center}
\hspace{2.5pt}
\includegraphics[width=14.5cm]{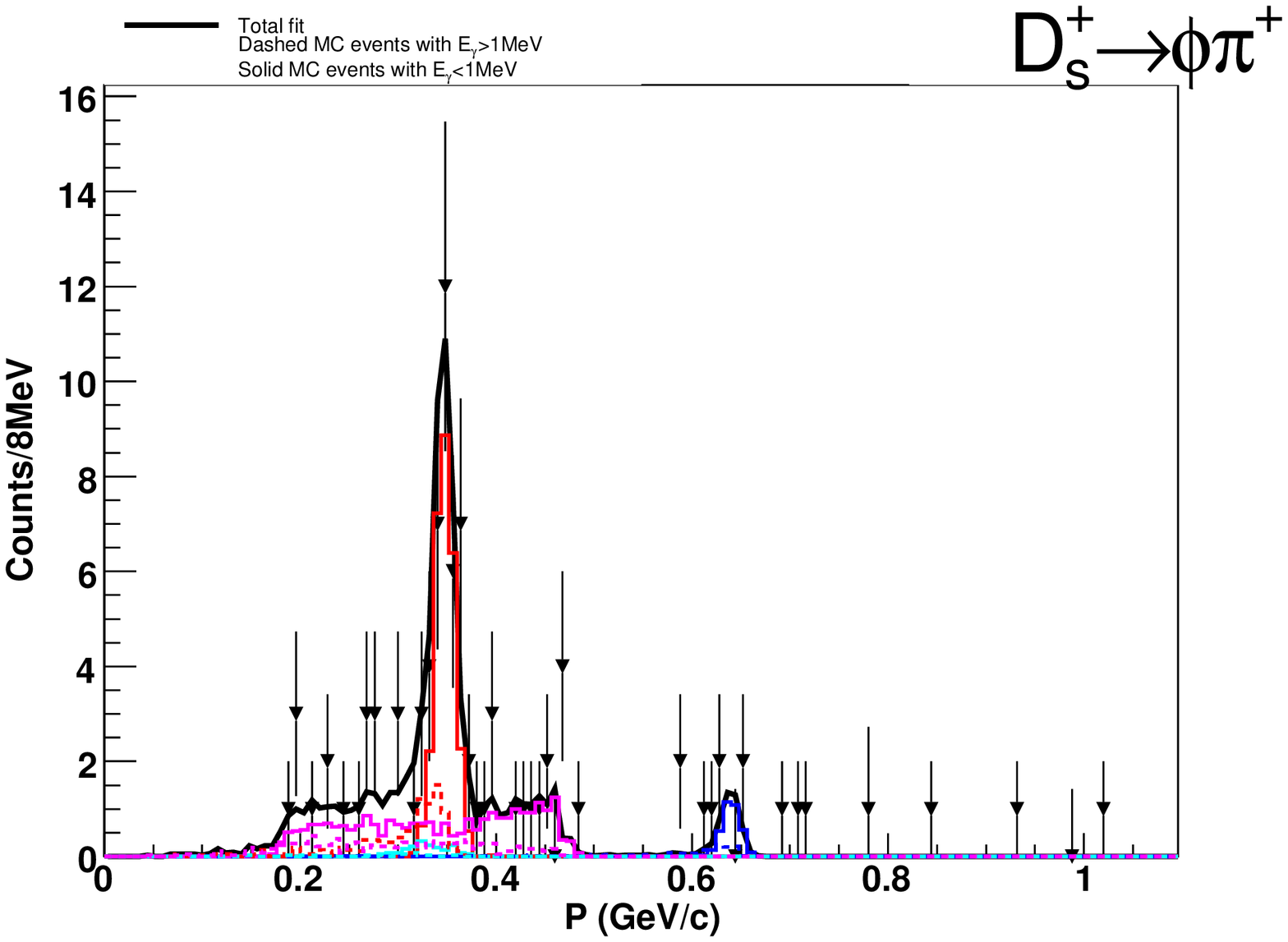}
\caption{Sideband-subtracted momentum spectrum for
$D_{s}^+\rightarrow\phi\pi^+$ at 4140 MeV.  The data is the points
with error bars and the histograms are MC.}
\vspace{0.2cm}
\label{fig:Mom_Ds_4140}
\end{center}
\end{figure}

\begin{figure}[!p]
\begin{center}
\hspace{2.5pt}
\includegraphics[width=14.5cm]{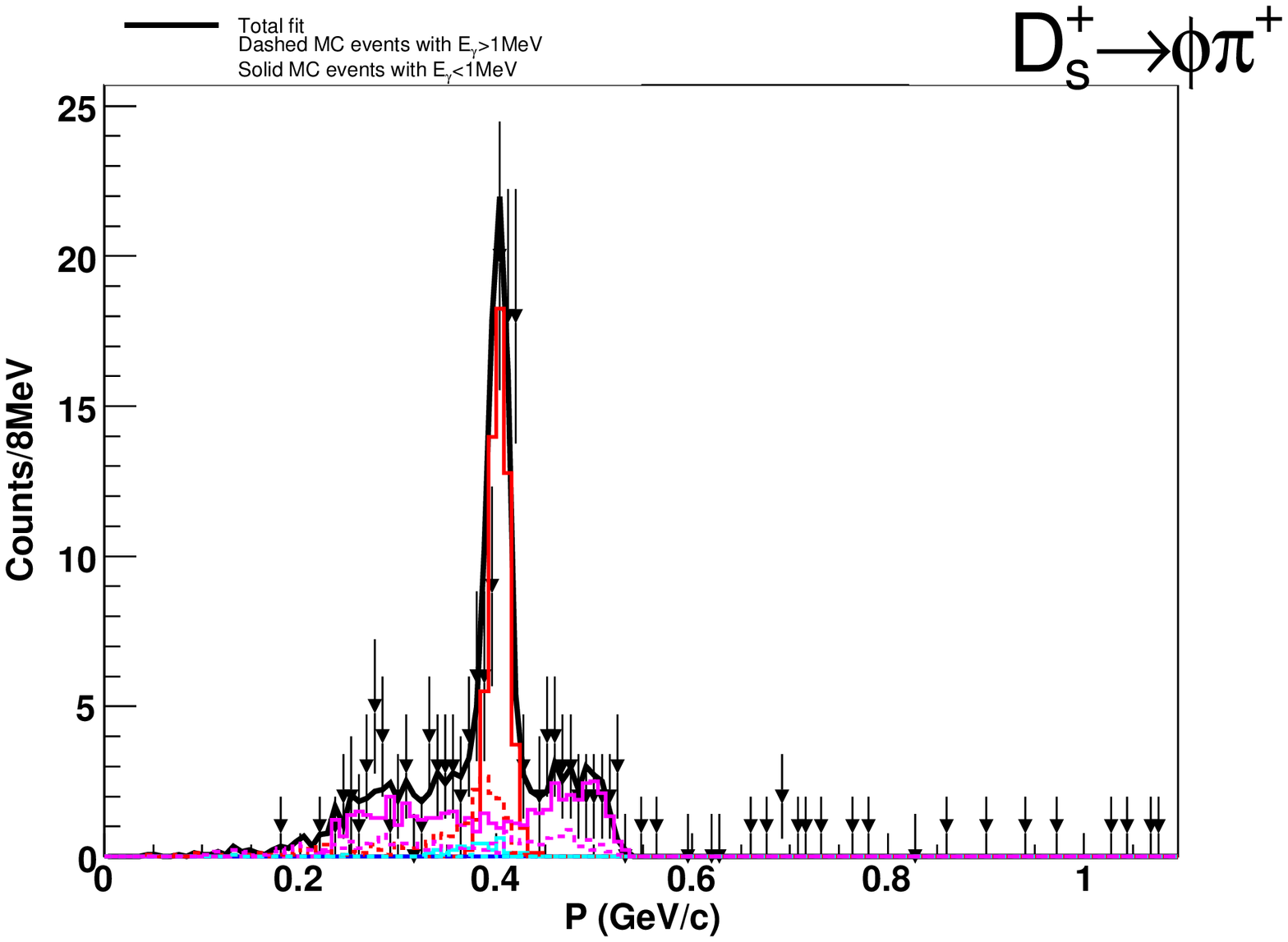}
\caption{Sideband-subtracted momentum spectrum for
$D_{s}^+\rightarrow\phi\pi^+$ at 4160 MeV.  The data is the points
with error bars and the histograms are MC.}
\vspace{0.2cm}
\label{fig:Mom_Ds_4160}
\end{center}
\end{figure}

\begin{figure}[!p]
\begin{center}
\hspace{2.5pt}
\includegraphics[width=14.5cm]{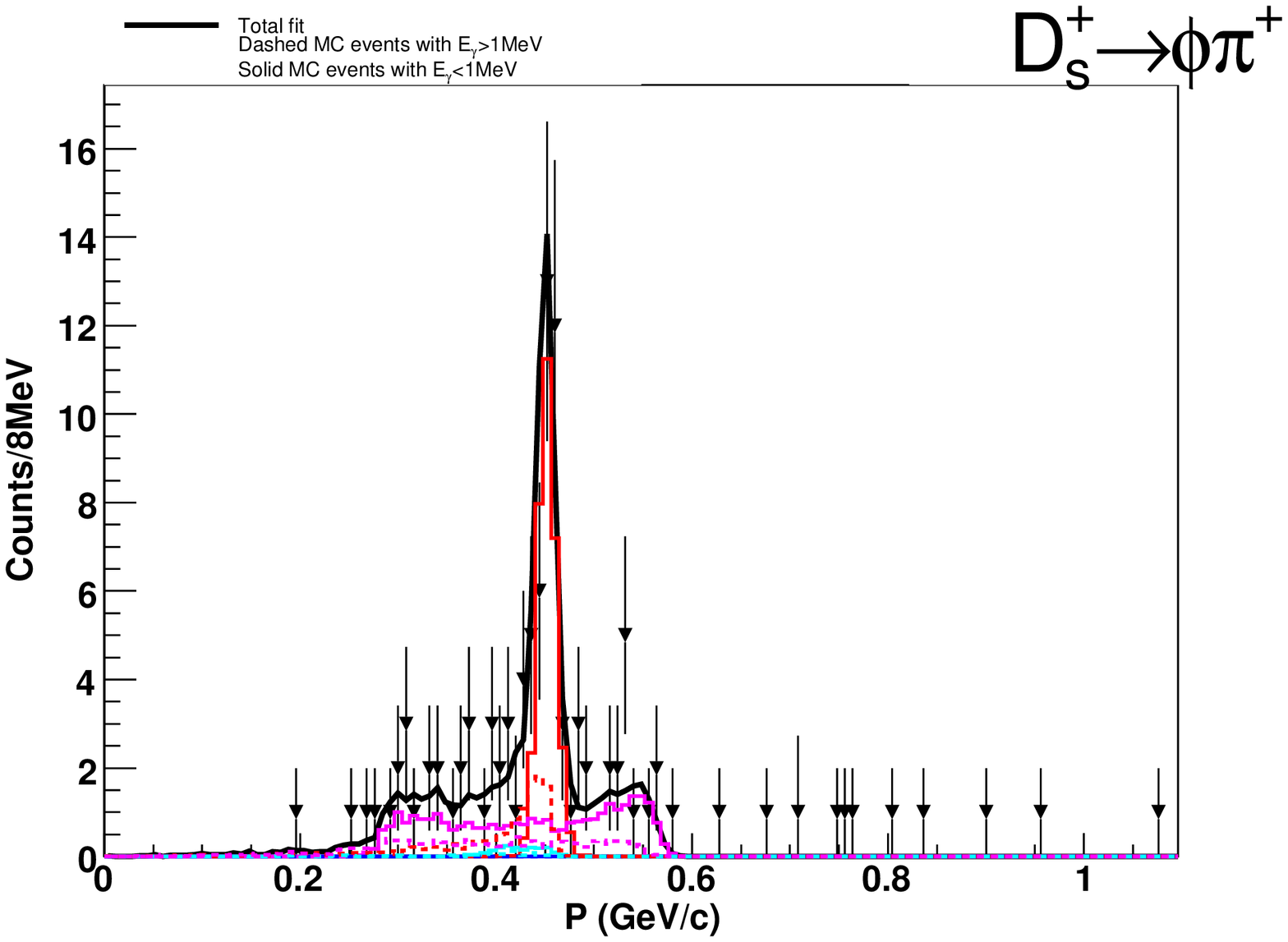}
\caption{Sideband-subtracted momentum spectrum for
$D_{s}^+\rightarrow\phi\pi^+$ at 4180 MeV.  The data is the points
with error bars and the histograms are MC.}
\vspace{0.2cm}
\label{fig:Mom_Ds_4180}
\end{center}
\end{figure}

\begin{figure}[!p]
\begin{center}
\hspace{2.5pt}
\includegraphics[width=14.5cm]{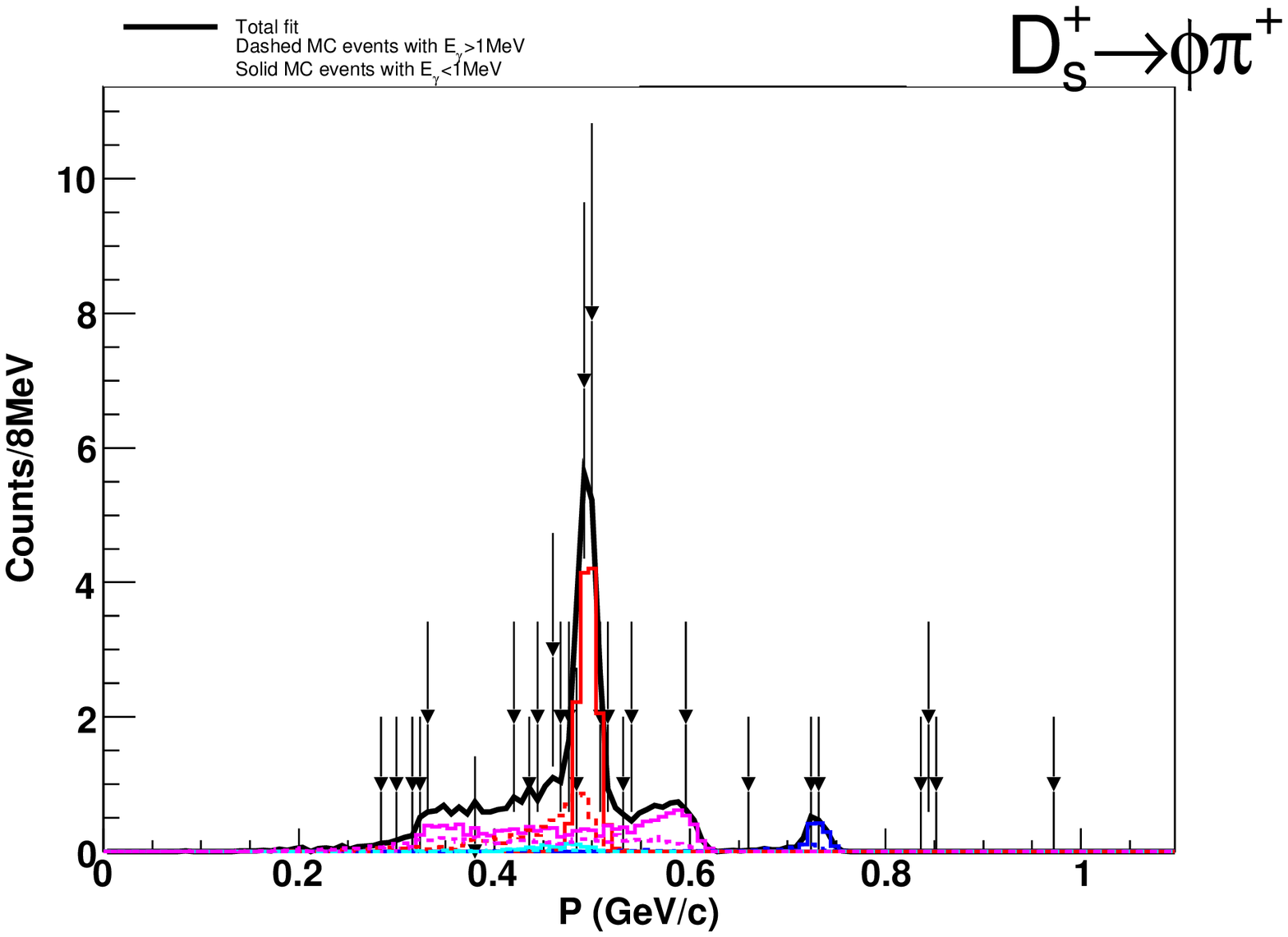}
\caption{Sideband-subtracted momentum spectrum for
$D_{s}^+\rightarrow\phi\pi^+$ at 4200 MeV.  The data is the points
with error bars and the histograms are MC.}
\vspace{0.2cm}
\label{fig:Mom_Ds_4200}
\end{center}
\end{figure}

\begin{figure}[!p]
\begin{center}
\hspace{2.5pt}
\includegraphics[width=14.5cm]{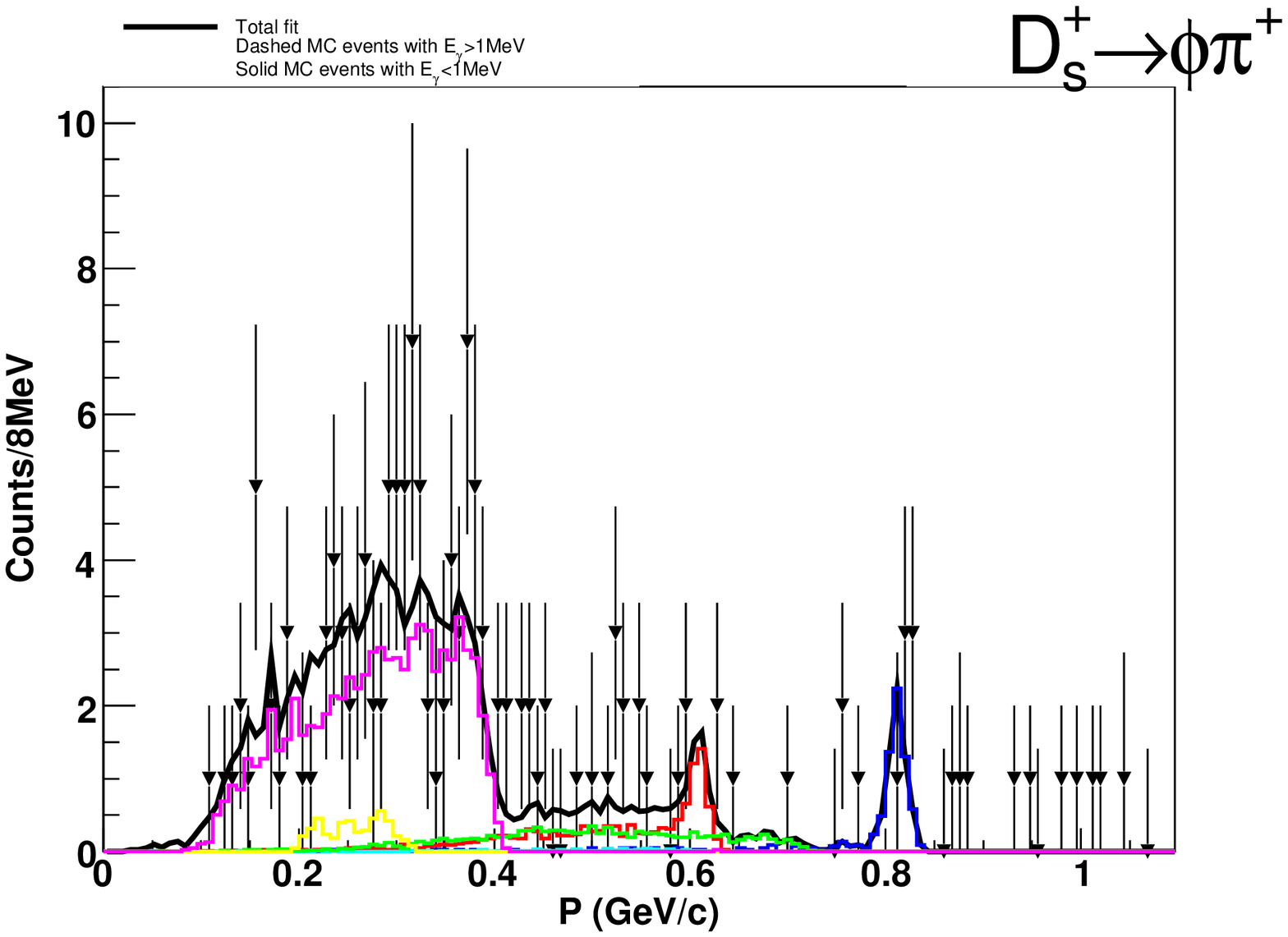}
\caption{Sideband-subtracted momentum spectrum for
$D_{s}^+\rightarrow\phi\pi^+$ at 4260 MeV.  The data is the points
with error bars and the histograms are MC.}
\vspace{0.2cm}
\label{fig:Mom_Ds_4260}
\end{center}
\end{figure}

Momentum-spectrum fits for data collected at other energy points are shown in \FIGS
\ref{fig:Mom_D0_3970}-\ref{fig:Mom_D0_4260} for $D^0$ and $D^+$ and \FIGS
\ref{fig:Mom_Ds_3970}-\ref{fig:Mom_Ds_4260} for $D_s$.  It is
interesting to see the emergence of the various final states as the center-of-mass energy is increased.

\begin{figure}[!htbp]
\begin{center}
\hspace{2.5pt}
\includegraphics[width=14.5cm]{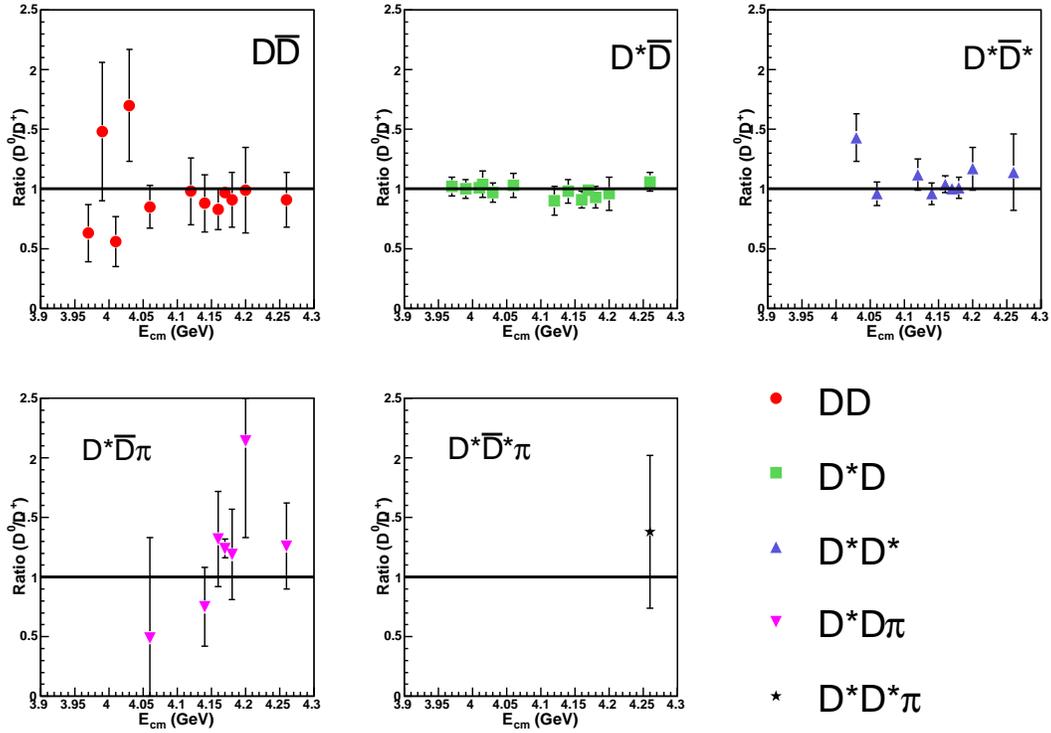}
\caption{The ratio of the cross section results from the
$D^{0}\rightarrow{K^-}\pi^+$ fit to the results from the
$D^{+}\rightarrow{K^-}\pi^+\pi^+$ fit.  Since the fits are independent,
the agreement in the ratios demonstrates the reliability of the
fitting procedure in extracting the mode-by-mode cross sections.  The
ratio, in all cases, should be one, indicated by the solid black line.
All event types are in good agreement with what is expected.} \vspace{0.2cm}
\label{fig:ratio_fit_check}
\end{center}
\end{figure}

\begin{figure}[!htbp]
\begin{center}
\hspace{2.5pt}
\includegraphics[width=14.5cm]{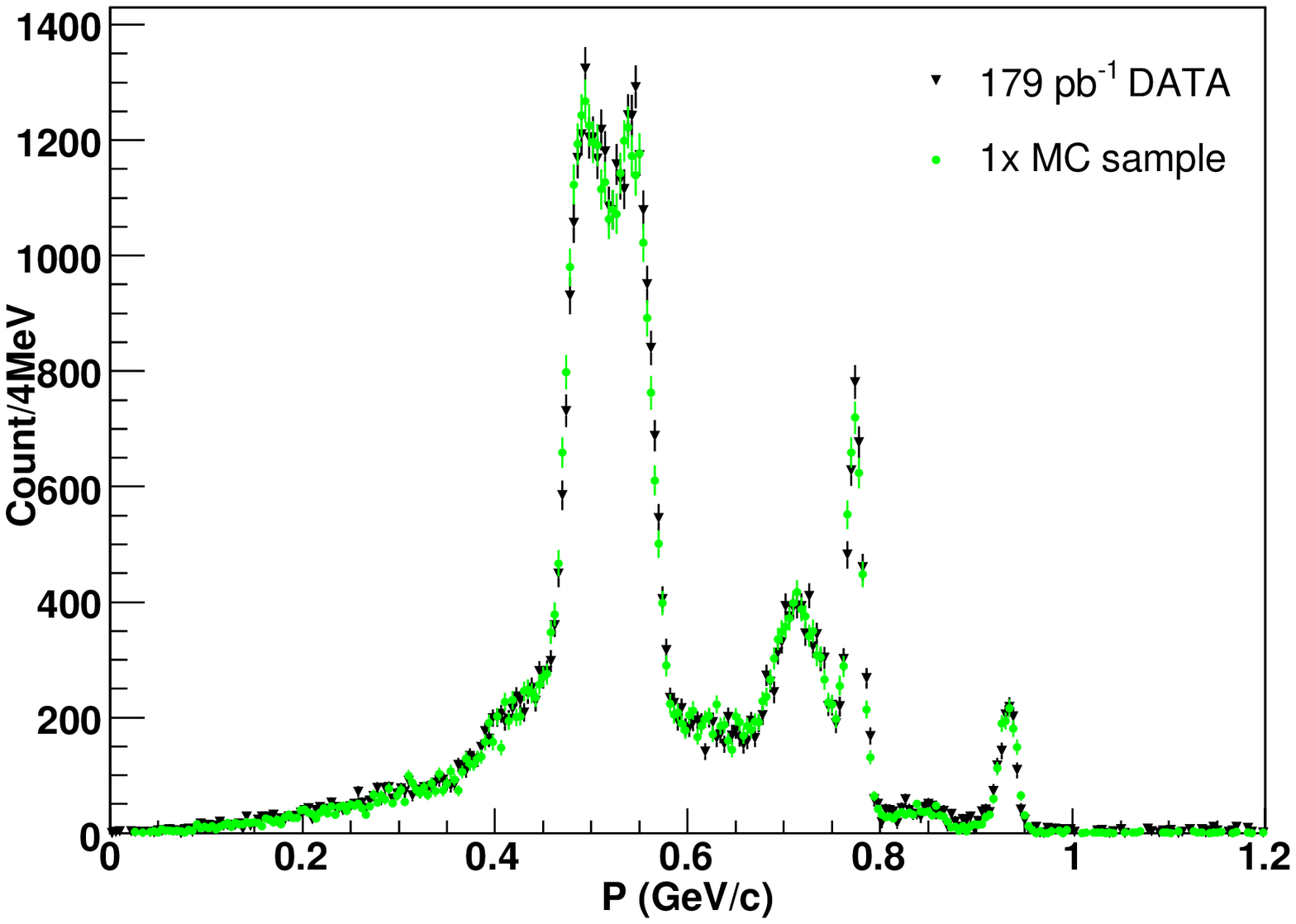}
\caption{Comparison between MC and data at 4170 MeV for
$D^{0}\rightarrow{K^{-}}\pi^+$.
Based on the momentum spectra fit results, an indepedent MC
sample with the same statistics as the 178 pb$^{-1}$ collected at 4170 MeV was generated using
all the knowledge gained about ISR, angular distributions (see the
App.) and cross sections.}
\vspace{0.2cm}
\label{fig:overlayD0}
\end{center}
\end{figure}

\begin{figure}[!htbp]
\begin{center}
\hspace{2.5pt}
\includegraphics[width=14.5cm]{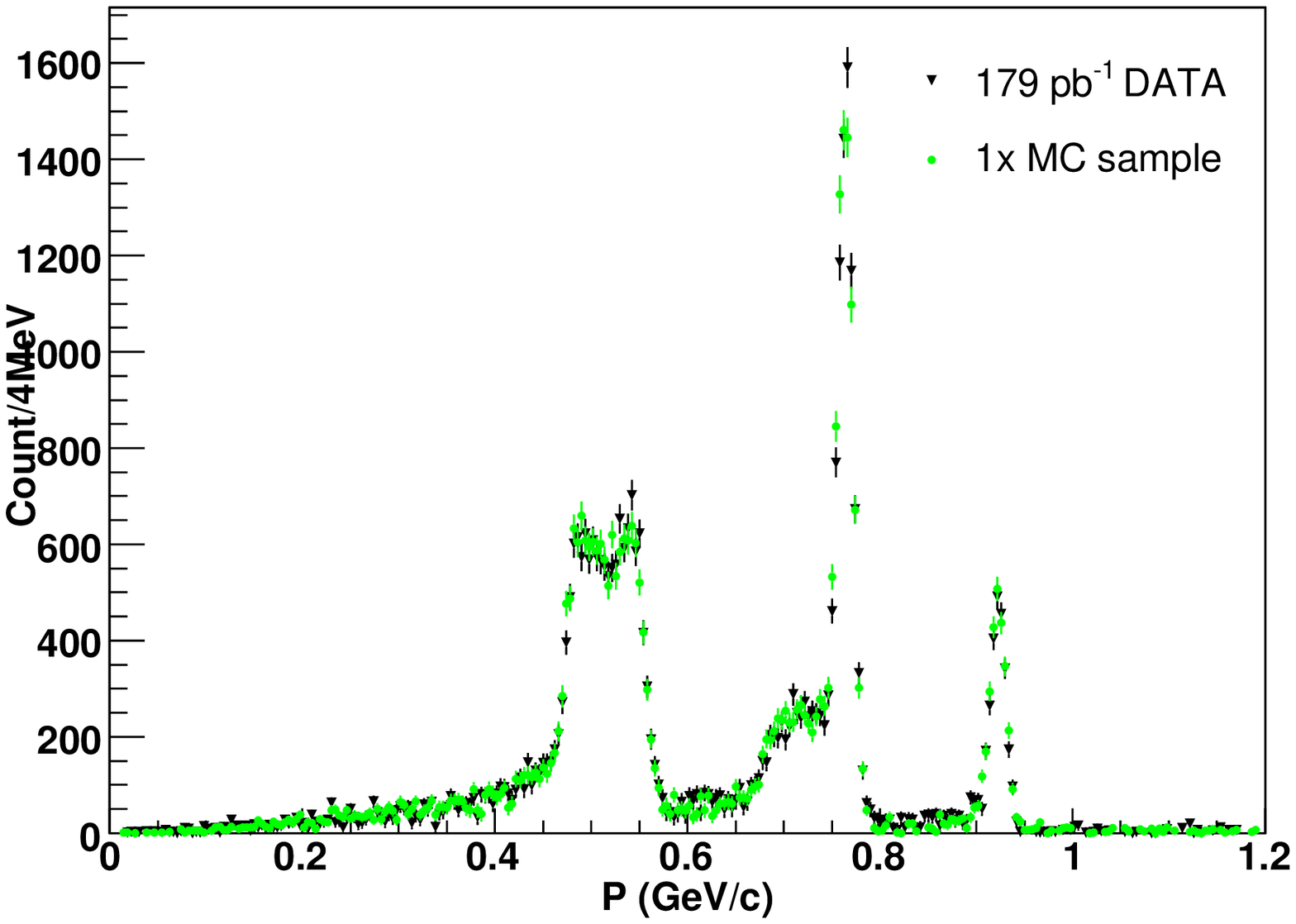}
\caption{Comparison between MC and data at 4170 MeV for
$D^{+}\rightarrow{K^{-}}\pi^+\pi^+$.
Based on the momentum spectrum fit results an independent MC
sample with the same statistics as the 178 pb$^-1$ collected at 4170 MeV was generated using
all the knowledge gained about ISR, angular distributions (see the App.) and cross sections.}
\vspace{0.2cm}
\label{fig:overlayDp}
\end{center}
\end{figure}

\begin{figure}[!htbp]
\begin{center}
\hspace{2.5pt}
\includegraphics[width=14.5cm]{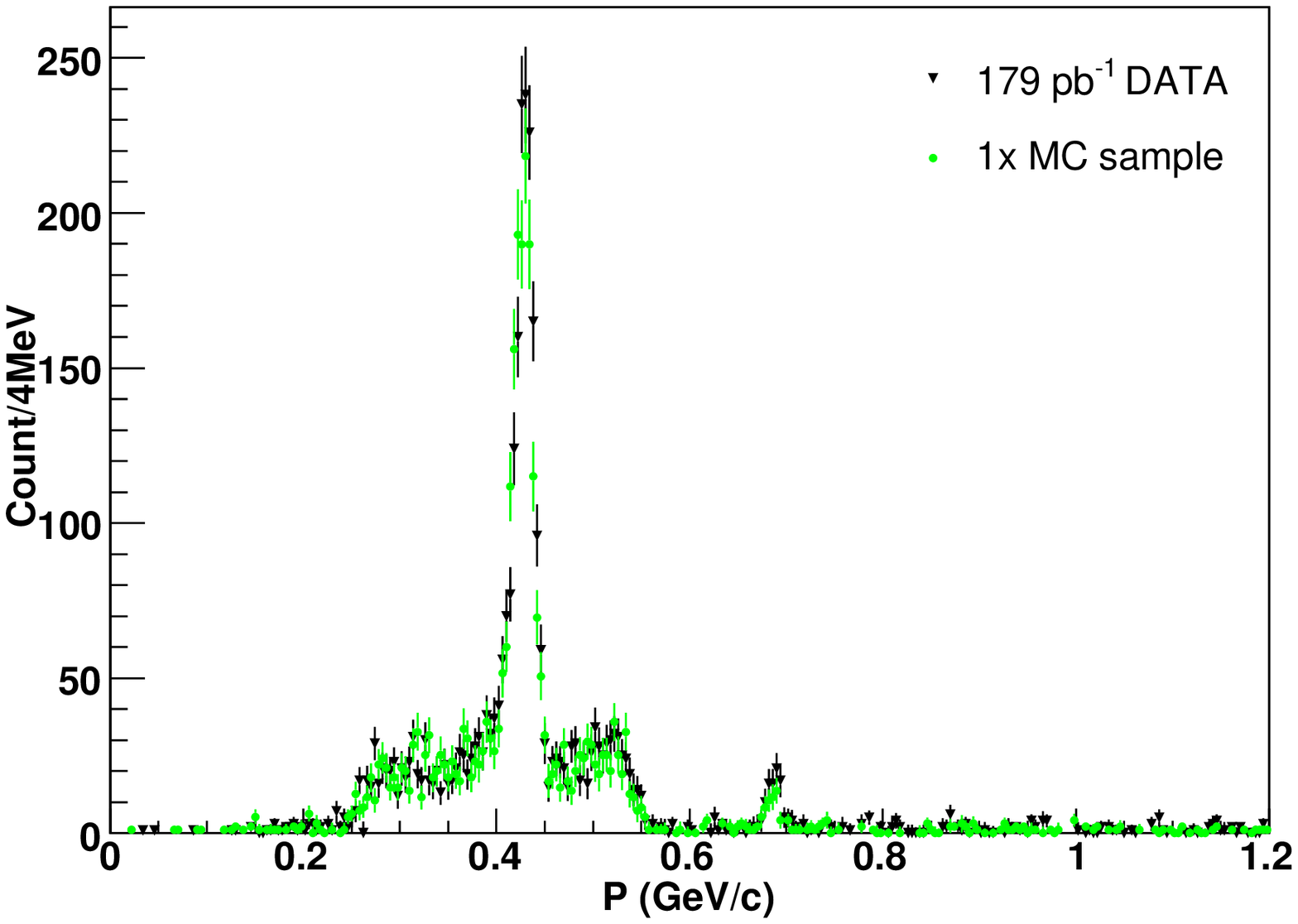}
\caption{Comparison between MC and data, at 4170 MeV, for
$D_{s}^{+}\rightarrow{\phi}\pi^+$.
Based on the momentum spectrum fit results an independent MC
sample with the same statistics as the 178 pb$^-1$ collected at 4170 MeV was generated using
all the knowledge gained about ISR, angular distributions (see the App.) and cross sections.}
\vspace{0.2cm}
\label{fig:overlayDs}
\end{center}
\end{figure}

Since the fit for $D^0$ is independent of $D^+$, a ratio of these fit
results, calculated for each event type, can be a good check of the
method.  The ratios for all event types and all energies are shown in \FIG
\ref{fig:ratio_fit_check}. At all energies the ratios are in good
agreement with what is expected (the solid
horizontal lines). As a further check, a MC sample with the same statistics as the $178$~pb$^{-1}$ of data was generated.  \FIGS \ref{fig:overlayD0}, \ref{fig:overlayDp}, and
\ref{fig:overlayDs} show the scaled momentum spectrum for $D^{0}\rightarrow{K^-}\pi^+$,
$D^{+}\rightarrow{K^-}\pi^+\pi^+$, and $D^{+}\rightarrow\phi\pi^+$,
respectively, for this sample overlaid with the data.  The agreement
is quite exceptional.

\begin{figure}[!htbp]
\begin{center}
\hspace{2.5pt}
\includegraphics[width=14.5cm]{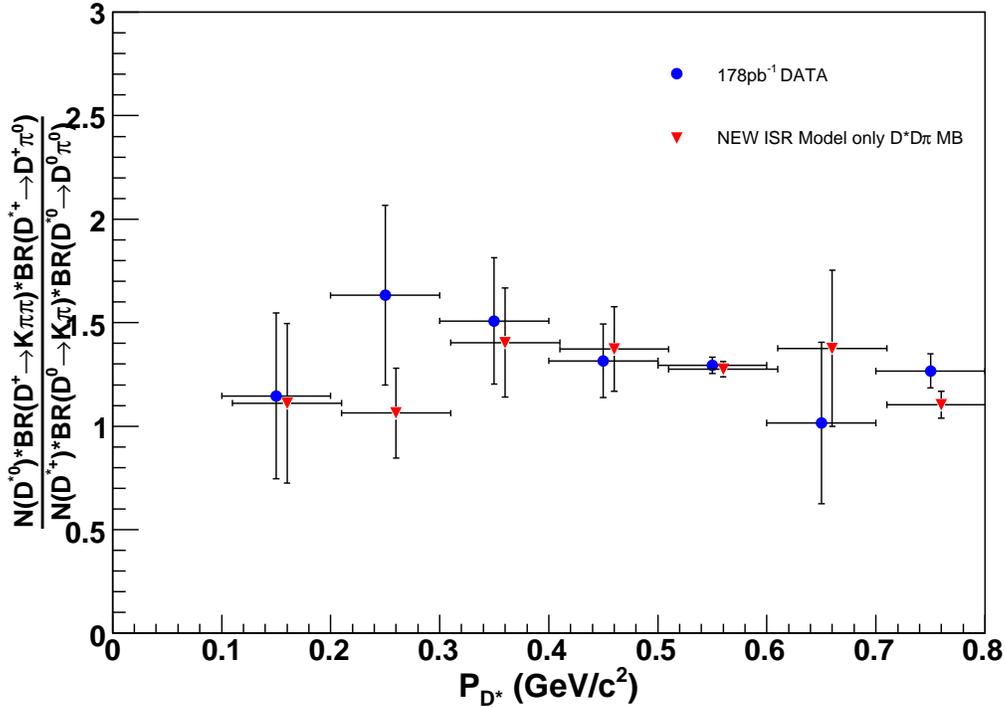}
\caption{Ratio of $D^{*0}$ to $D^{*+}$ in 4170 MeV data and MC as a
function of the reconstructed $D^{*}$ momentum. Both $D^*$ states are
detected with the $\pi^0$
channel, with the subsequent decays
$D^0\rightarrow{K^-}\pi^+$ and $D^+\rightarrow{K^-}\pi^+\pi^+$,
respectively.  The ratio is corrected for branching fractions, while 
efficiencies were assumed to be constant.  The MC includes
$D^*\bar{D}\pi$ multi-body and uses an updated ISR model that
incorporates individual two-body cross sections as described in the
text.  The updated MC sample is in great agreement
with the data at 4170 MeV.}
\vspace{0.2cm}
\label{fig:DStar_ratio_solved}
\end{center}
\end{figure}

\begin{figure}[!htbp]
\begin{center}
\hspace{2.5pt}
\includegraphics[width=14.5cm]{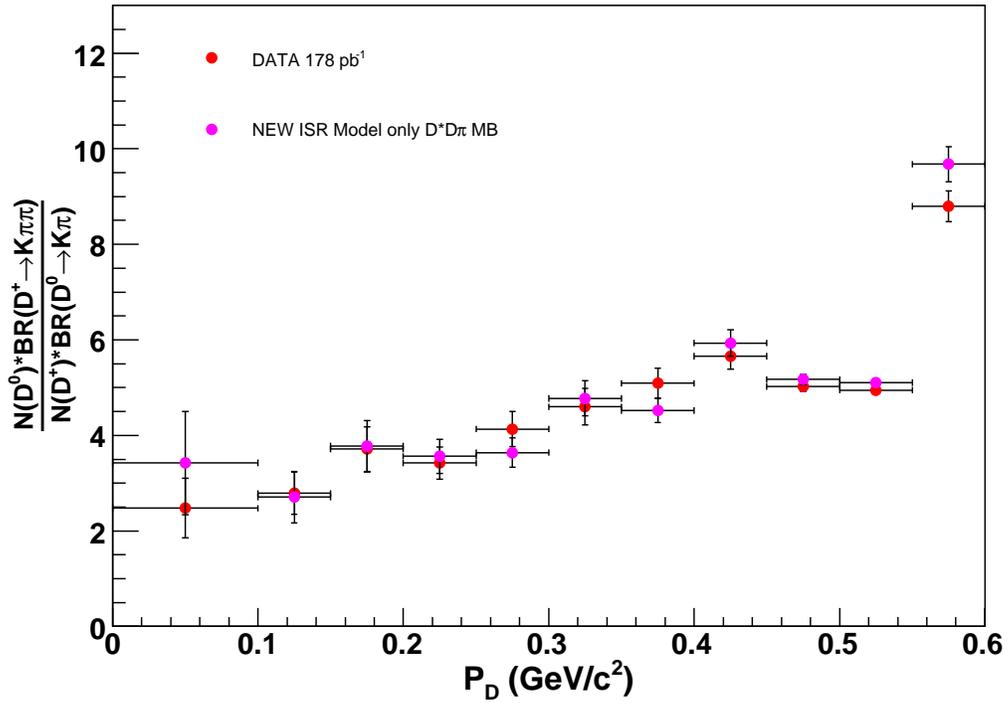}
\caption{Ratio of the number of $D^{0}$ to $D^{+}$ as a function of
reconstructed $D$ momentum at $E_{{\rm{cm}}}=4170$ MeV.  The
ratio is corrected for branching fractions, while efficiencies were
assumed to be constant. The MC includes $D^*\bar{D}\pi$ multi-body and
uses an updated ISR model that incorporates individual two-body cross sections as described in the
text.  The updated MC sample is in great agreement
with the data at 4170 MeV.}
\vspace{0.2cm}
\label{fig:D_ratio_solved}
\end{center}
\end{figure}

For another confidence test of the procedure we used this one-times
sample to verify the determination of the production ratios.
Comparisons of $D^{*0}$ to $D^{*+}$ and $D^{0}$ to $D^{+}$ for
the updated MC and data can be seen in \FIG
\ref{fig:DStar_ratio_solved} and \FIG \ref{fig:D_ratio_solved}.
Overall, the previously observed 
discrepancies between the data and MC have been resolved with the use
of a more accurate description of ISR in the MC.

\begin{figure}[!p]
\begin{center}
\hspace{2.5pt}
\includegraphics[width=14.5cm]{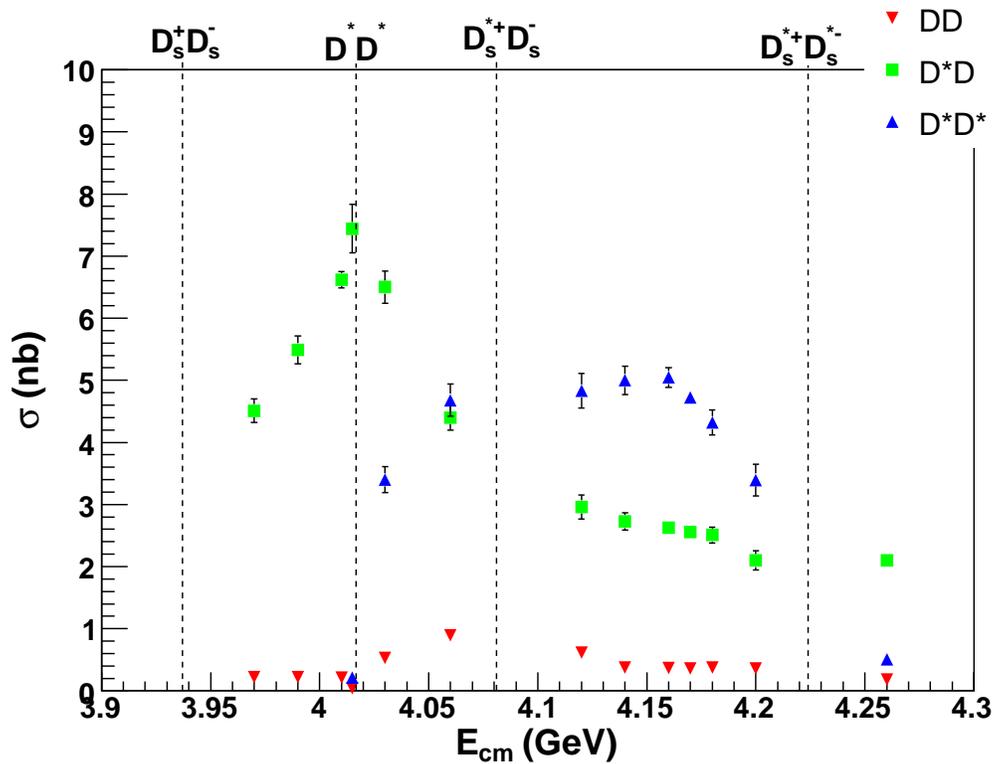}
\caption{Observed cross sections for $e^+e^-\rightarrow{D\bar{D}}$,
$D^{*}\bar{D}$ and $D^{*}\bar{D}^{*}$ as a function of center-of-mass
energy. They are determined by fitting the momentum spectrum for
$D^{0}\rightarrow{K^-}\pi^+$ and $D^{+}\rightarrow{K^-}\pi^+\pi^+$
with the updated ISR MC. Systematic errors are not included.}
\vspace{0.2cm}
\label{fig:DD_momentumfit_solved}
\end{center}
\end{figure}

\begin{figure}[!p]
\begin{center}
\hspace{2.5pt}
\includegraphics[width=14.5cm]{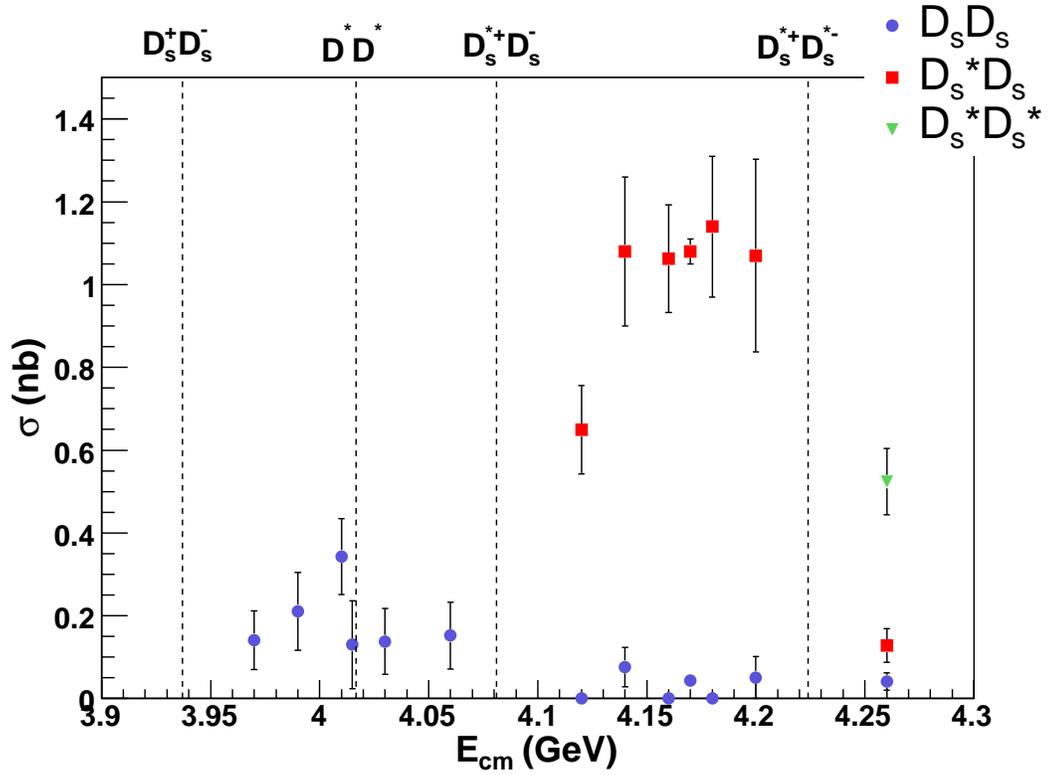}
\caption{Observed cross sections for
$e^+e^-\rightarrow{D_{s}\bar{D}_{s}}$, $D_{s}^{*}\bar{D}_{s}$ and
$D_{s}^{*}\bar{D}_{s}^{*}$ as a function of center-of-mass
energy. They are determined by fitting the momentum spectrum for
$D_{s}^{+}\rightarrow\phi\pi^+$ with the updated ISR model MC. A
branching fraction of 3.6\% was also used in the determination of the
cross section. Systematic errors are not included.} 
\vspace{0.2cm}
\label{fig:DsDs_momentumfit_solved}
\end{center}
\end{figure}

\begin{figure}[!p]
\begin{center}
\hspace{2.5pt}
\includegraphics[width=14.5cm]{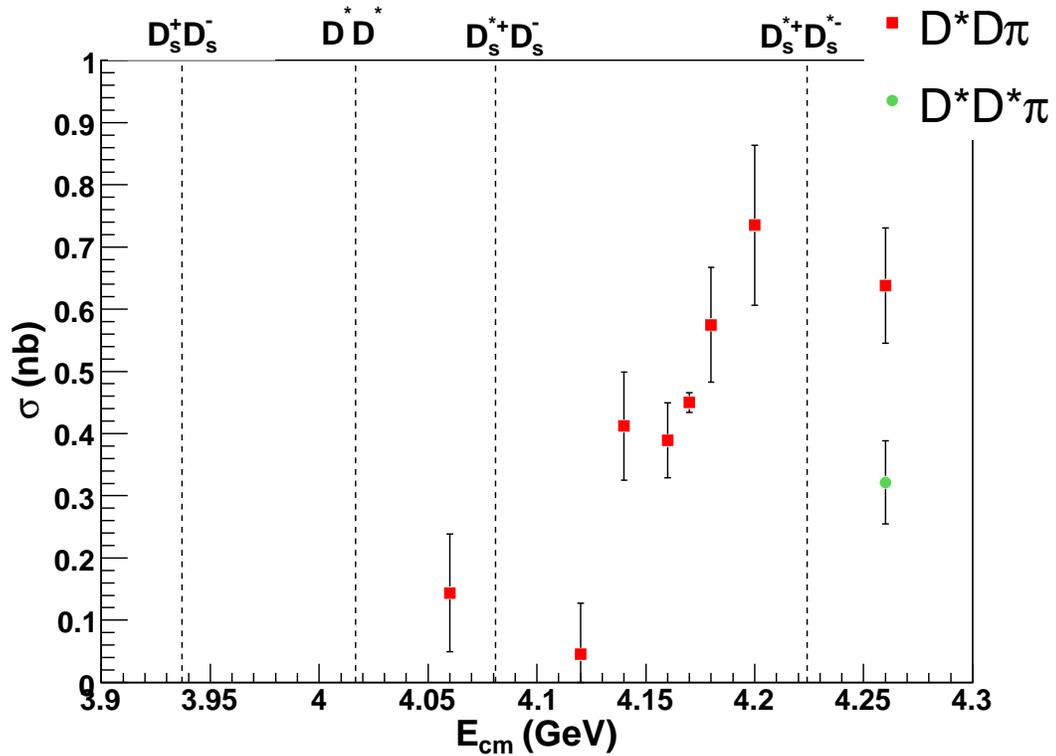}
\caption{Observed cross sections for
$e^+e^-\rightarrow{D^{*}\bar{D}\pi}$ and $D^{*}\bar{D}^{*}\pi$ as a
function of center-of-mass energy. They are determined
by fitting the momentum spectrum for $D^{0}\rightarrow{K^-}\pi^+$ and
$D^{+}\rightarrow{K^-}\pi^+\pi^+$ with the updated ISR model MC. The
multi-body contribution was modeled in MC using spin-averaged phase space. Systematic errors are not included.}  
\vspace{0.2cm}
\label{fig:MB_momentumfit_solved}
\end{center}
\end{figure}

\begin{figure}[!hp]
\begin{center}
\hspace{2.5pt}
\includegraphics[width=14.5cm]{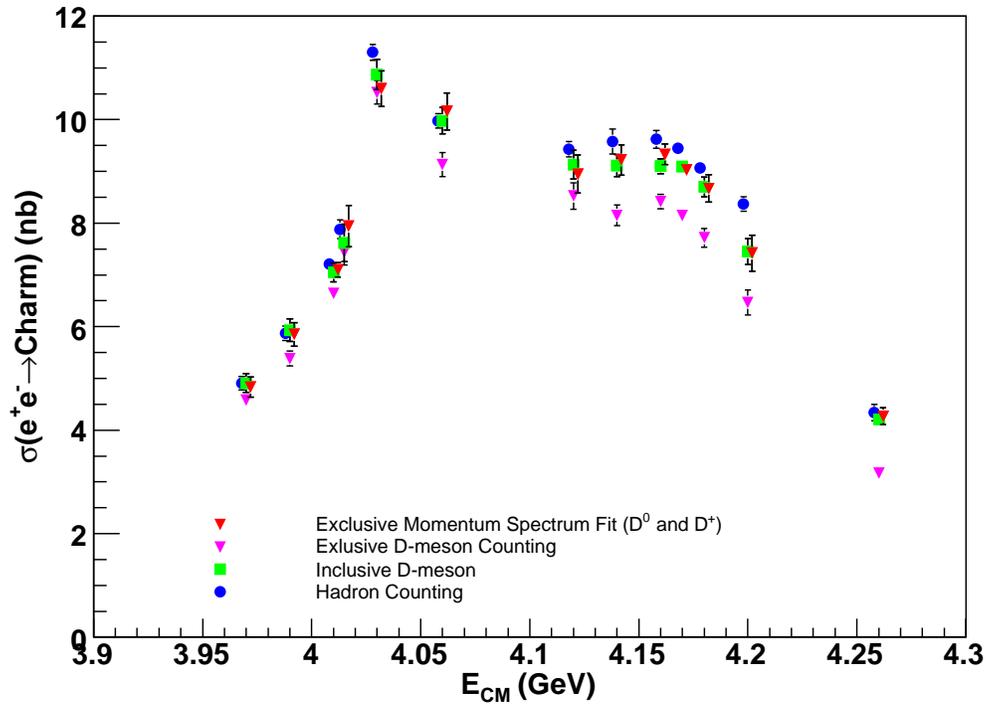}
\caption{Plot of the total charm cross section as calculated by each of the
three methods described in the text, in addition to the results from
the momentum-spectrum fit.  The reason for the discrepancy
between the inclusive and exclusive method is due to the presence of
multi-body background and ISR.  These two effects are taken into
account when a fit to the momentum spectrum is performed.  There is
good agreement between all three methods, the inclusive, the hadron
counting and the momentum fits. The previous exclusive result (ignoring multi-body) is shown, in
addition, for comparison.} 
\vspace{0.2cm}
\label{fig:All_method_solved}
\end{center}
\end{figure}

The final measurements of multi-body production at energies above 4030 MeV
are summarized in \TAB \ref{MB_table}.  The exclusive
cross sections for all modes studied from the momentum-spectrum fits
are shown in \TABS \ref{XS_results_fits_1} and
\ref{XS_results_fits_2}, as well as in \FIGS
\ref{fig:DD_momentumfit_solved}, \ref{fig:DsDs_momentumfit_solved}, and
\ref{fig:MB_momentumfit_solved}.  The total charm cross section for
all three methods, as discussed in the text, are shown in \FIG
\ref{fig:All_method_solved}.  It should be noted, unless
otherwise stated, that the weighted sum
technique from Sect. \ref{sec:sigmas} is used for the individual $D_{s}$
cross sections whose result is used in the determination of the
total exclusive charm cross section, e.g. \FIG \ref{fig:All_method_solved}.

\begin{table}[!htbp]
\begin{center}
\caption{The amount of multi-body present at energies above 4060 MeV
obtained by fitting the sideband-subtracted momentum
spectra for $D^0\rightarrow{K^-}\pi^+$ and
$D^+\rightarrow{K^-}\pi^+\pi^+$ with the two-body MC representation of
the various exclusive channels and a spin-averaged phase-space model
MC representation of multi-body. There is no evidence for the $DD\pi$ final state at any energy. Only statistical errors are shown.}
\label{MB_table}
\vspace{0.2cm}
\begin{tabular}{|c|c|c|}\hline
{E$_{cm}$ MeV}& {$\sigma(e^+e^-\rightarrow{D^*D\pi})$~nb}& {$\sigma(e^+e^-\rightarrow{D^*D^*\pi})$~nb}
\\ \hline
$4060$ & $0.14 \pm 0.09$& $- \pm -$\\ \hline
$4120$ & $0.05 \pm 0.08$& $- \pm -$\\ \hline
$4140$ & $0.41 \pm 0.09$& $- \pm -$\\ \hline
$4160$ & $0.39 \pm 0.06$& $- \pm -$\\ \hline
$4170$ & $0.44 \pm 0.01$& $- \pm -$\\ \hline
$4180$ & $0.58 \pm 0.09$& $- \pm -$\\ \hline
$4200$ & $0.74 \pm 0.13$& $- \pm -$\\ \hline
$4260$ & $0.64 \pm 0.09$& $0.32 \pm 0.07$\\ \hline
\end{tabular}
\end{center}
\end{table}

\clearpage
\pagebreak

%%%%%%%%%%%%%%%%%%%%%%%%%%%%%%%%%%%%%%%%%%%%%%%%%%%%%%%%%%%%%%%%%%
\section{Momentum Spectrum and Multi-Body Cross-Check}
%%%%%%%%%%%%%%%%%%%%%%%%%%%%%%%%%%%%%%%%%%%%%%%%%%%%%%%%%%%%%%%%%%

\begin{figure}[!htbp]
\begin{center}
\hspace{2.5pt}
\includegraphics[width=14.5cm]{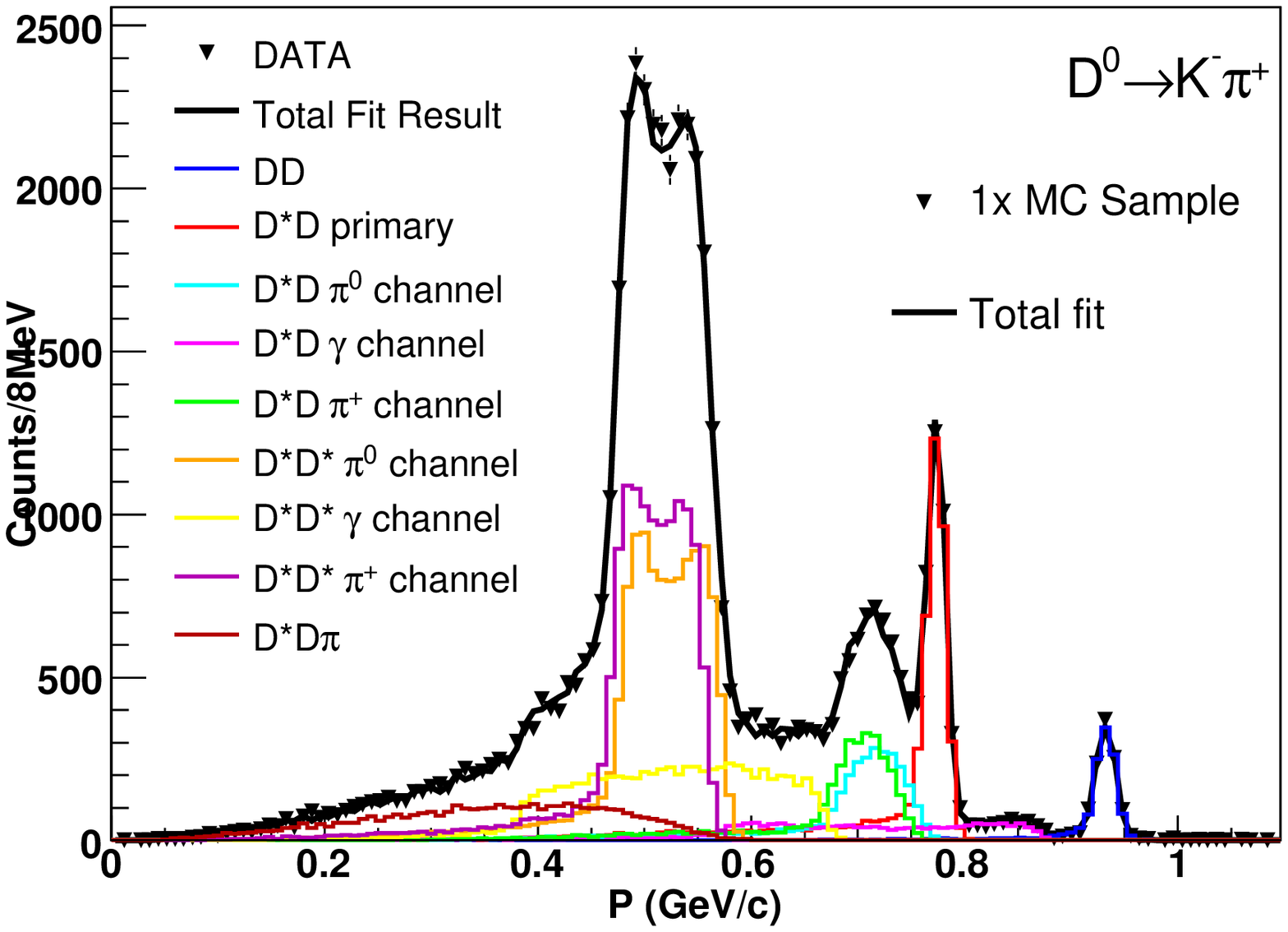}
\caption{Momentum-spectrum fit for the improved 178 pb$^{-1}$ MC sample for
$D^0\rightarrow{K^-}\pi^+$. The MC sample was treated just as the data
and used as a cross-check of the momentum-spectrum fit procedure.}
\vspace{0.2cm}
\label{fig:4170_D0_MC}
\end{center}
\end{figure}

\begin{figure}[!htbp]
\begin{center}
\hspace{2.5pt}
\includegraphics[width=14.5cm]{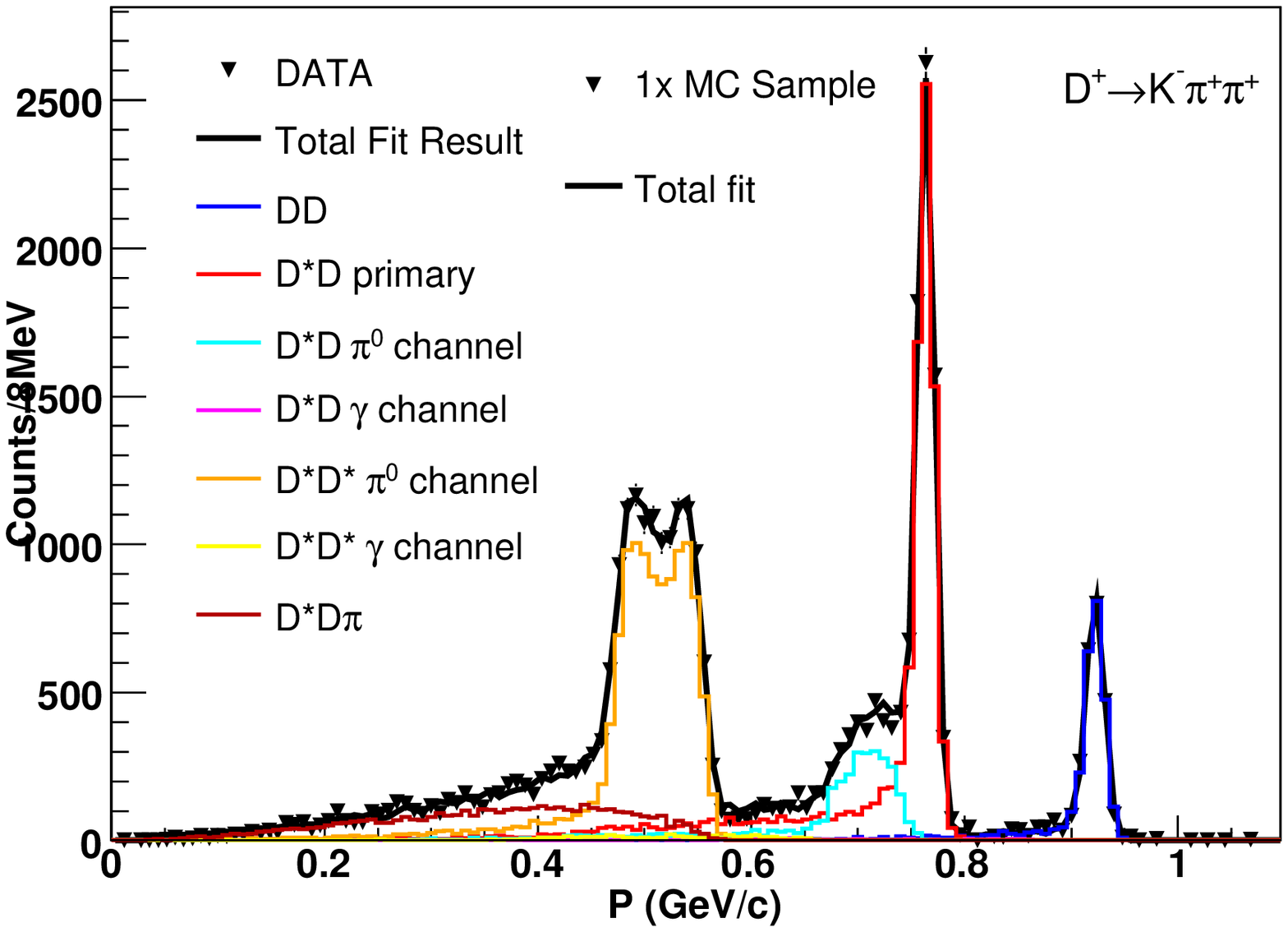}
\caption{Momentum-spectrum fit for the improved 178 pb$^{-1}$ MC sample for
$D^+\rightarrow{K^-}\pi^+\pi^+$. The MC sample was treated just as the data
and used as a cross-check of the momentum-spectrum fit procedure.} 
\vspace{0.2cm}
\label{fig:4170_Dp_MC}
\end{center}
\end{figure}

\begin{figure}[!htbp]
\begin{center}
\hspace{2.5pt}
\includegraphics[width=14.5cm]{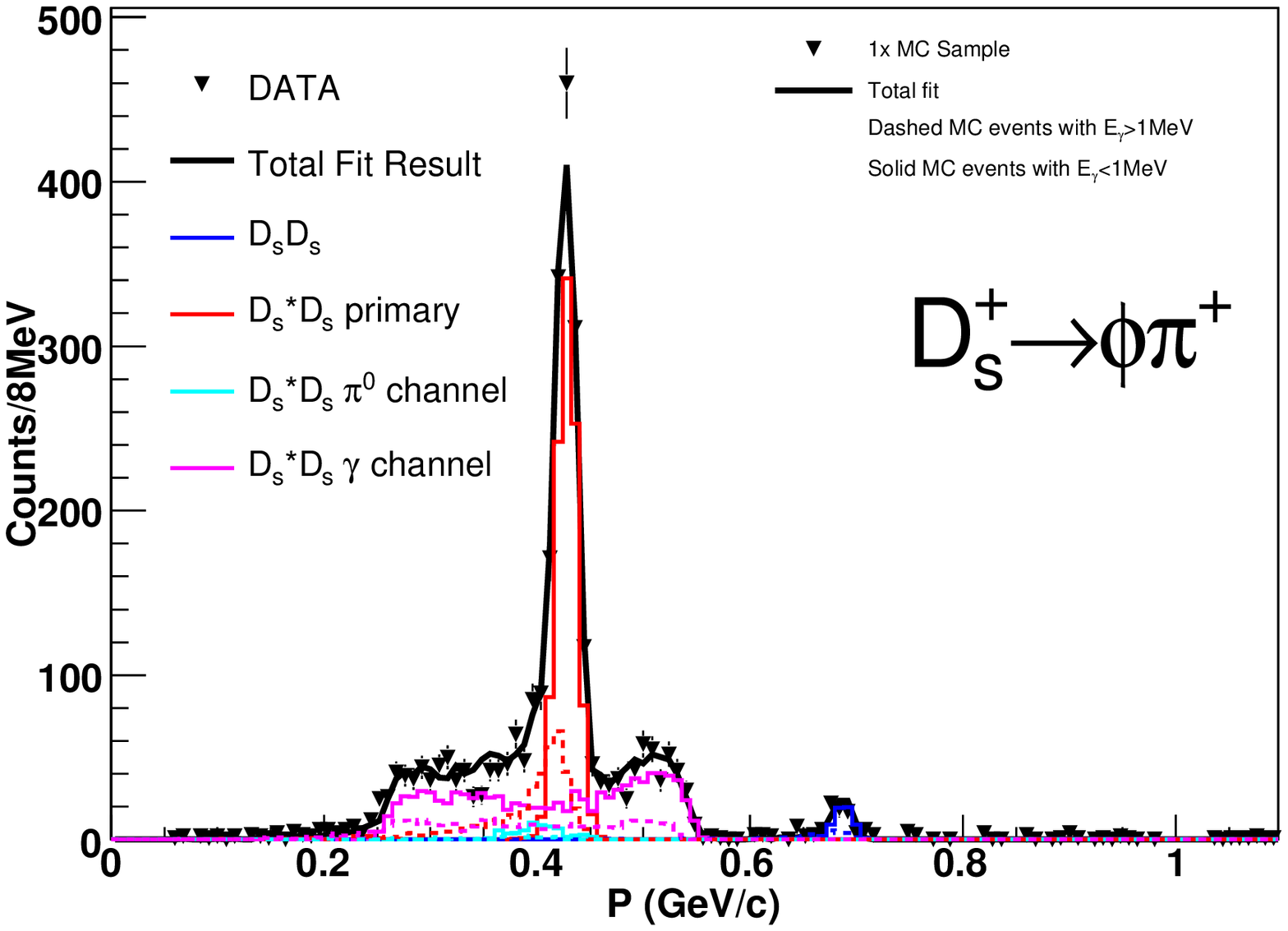}
\caption{Momentum-spectrum fit for the improved 178 pb$^{-1}$ MC sample for $D_{s}^+\rightarrow{\phi}\pi^+$. The MC sample was treated just as the data and used as a cross-check of the momentum-spectrum fit procedure.} 
\vspace{0.2cm}
\label{fig:4170_Ds_MC}
\end{center}
\end{figure}

Using the improved 178 pb$^{-1}$ MC sample that was generated to check the neutral-to-charged $D$-meson ratio, one can check the momentum-spectrum fit
procedure.  The momentum-spectrum fit results for
$D^0\rightarrow{K^-}\pi^+$, $D^0\rightarrow{K^-}\pi^+\pi^+$, and
$D_{s}^+\rightarrow{\phi}\pi^+$ are
shown in \FIGS \ref{fig:4170_D0_MC}, \ref{fig:4170_Dp_MC}, and \ref{fig:4170_Ds_MC}
respectively.  The comparison between what was generated and what is
obtained from the fit is shown in \TAB \ref{MC_results}.  The returned
fit results are in very good agreement with what was used to generate
the sample.

%ddstar_4170_farm_ISR_NEWModel_updatedXS.dec
\begin{table}[!htbp]
\begin{center}
\caption{Results of the momentum-spectrum fit of the improved 178 pb$^{-1}$ MC sample.
The agreement between the fitted and the generated values is very reasonable.}
\label{MC_results}
\vspace{0.2cm}
\begin{tabular}{|c|c|c|c|}\hline
{Parameter}& {Fitted Value (nb)}& {Generated Value (nb)} &{Difference}
\\ \hline
$D\bar{D}$ & $0.319 \pm 0.008$&$0.331$& $-1.5\sigma$\\ \hline
$D^{*}\bar{D}$ & $2.392 \pm 0.08$&$2.392$& $-0.1\sigma$\\ \hline
$D^{*}\bar{D}^{*}$ & $4.553 \pm 0.038$&$4.508$& $1.2\sigma$\\ \hline
$D^{*}\bar{D}\pi$ & $0.766\pm0.018 $&$0.780$& $-0.8\sigma$\\ \hline
$D_{s}\bar{D}_{s}$ & $ 0.036 \pm 0.005$&$0.040$& $-0.8\sigma$\\ \hline
$D^{*}_{s}\bar{D}_{s}$ & $1.039\pm0.028 $&$0.994$& $1.6\sigma$\\ \hline
\end{tabular}
\end{center}
\end{table}

\begin{figure}[!t]
\begin{center}
\hspace{2.5pt}
\includegraphics[width=14.5cm]{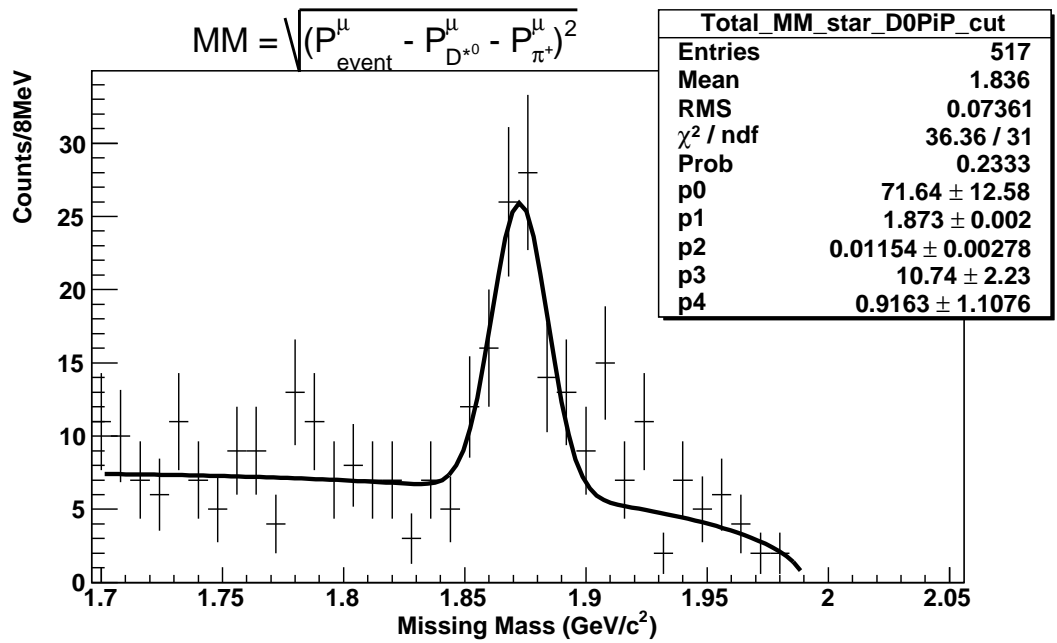}
\caption{Missing-mass spectrum using $D^0\rightarrow{K^{-}}\pi^+$ at
4170 MeV.  The fit function consists of a Gaussian and an Argus function.} 
\vspace{0.2cm}
\label{fig:4170_DS0PiP}
\end{center}
\end{figure}

\begin{figure}[!htbp]
\begin{center}
\hspace{2.5pt}
\includegraphics[width=14.5cm]{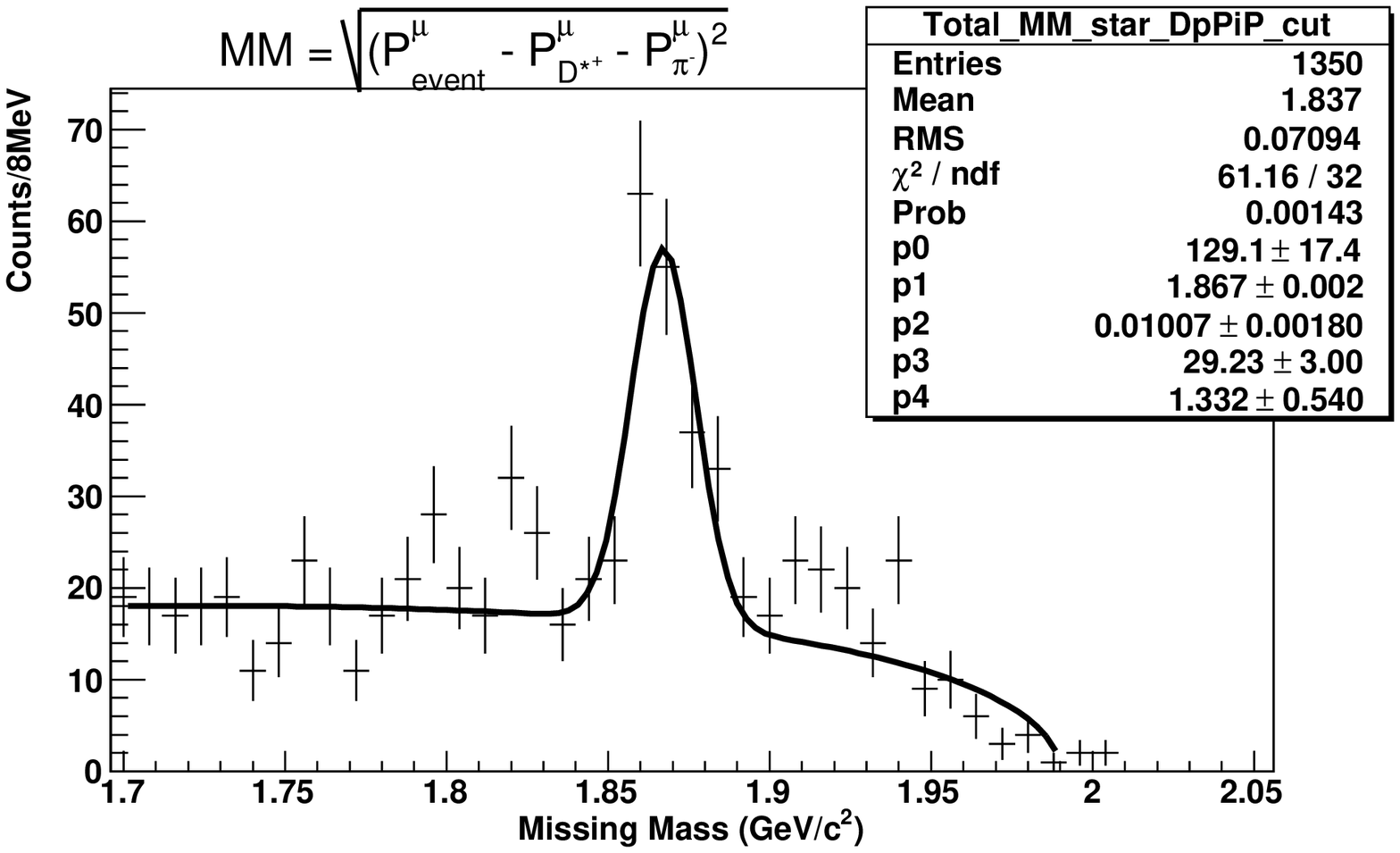}
\caption{Missing-mass spectrum using $D^+\rightarrow{K^{-}}\pi^+\pi^+$ at
4170 MeV.  The fit function consists of a Gaussian and an Argus function.} 
\vspace{0.2cm}
\label{fig:4170_DSPPiP}
\end{center}
\end{figure}

\begin{figure}[!htbp]
\begin{center}
\hspace{2.5pt}
\includegraphics[width=14.5cm]{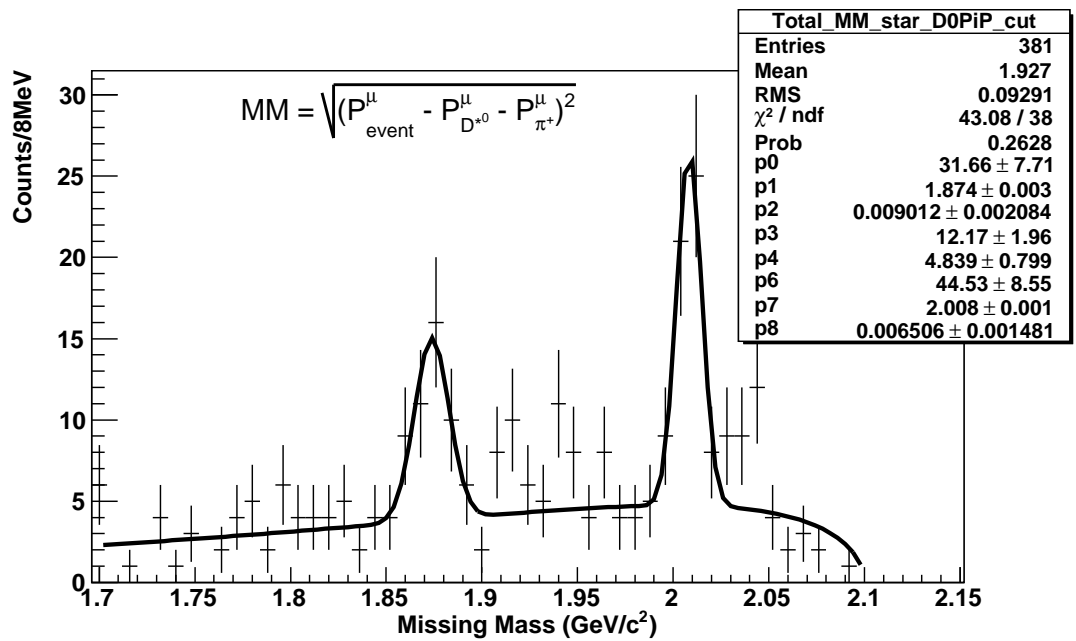}
\caption{Missing-mass spectrum using
$D^0\rightarrow{K^{-}}\pi^+$, $D^0\rightarrow{K^{-}}\pi^+\pi^0$, and
$D^0\rightarrow{K^{-}}\pi^+\pi^+\pi^-$ at 4260 MeV.  The fit function
consists of two Gaussians, one for each peak, and an Argus function.} 
\vspace{0.2cm}
\label{fig:4260_DS0PiP}
\end{center}
\end{figure}

As an additional check on the multi-body contributions at 4170 MeV and
4260 MeV, the two energies where either a large amount of data existed
or a large amount of multi-body is present, one can fit the missing-mass spectrum to obtain the cross section for multi-body.  The
fit results for $D^{*0}\rightarrow{D^{0}\pi^{0}}$ with
$D^{0}\rightarrow{K^-}\pi^+$, and $D^{*+}\rightarrow{D^{+}\pi^{0}}$ with
$D^{+}\rightarrow{K^-}\pi^+\pi^{+}$ plus an additional
charged pion at 4170 MeV are shown in \FIGS \ref{fig:4170_DS0PiP} and
\ref{fig:4170_DSPPiP}, respectively. The fit results for
$D^{*0}\rightarrow{D^{0}\pi^{0}}$ with $D^{0}\rightarrow{K^-}\pi^+$,
$D^{0}\rightarrow{K^-}\pi^+\pi^{0}$, or
$D^{0}\rightarrow{K^-}\pi^+\pi^{+}\pi^{-}$ plus an
additional charged pion at 4260 MeV is shown in \FIG \ref{fig:4260_DS0PiP}.
Using a spin-averaged phase-space model MC sample to determine
the corresponding efficiencies the cross sections can be determined.
The results from this method are shown, along with the momentum-spectrum fit results, in \TAB \ref{MB_results}.  The agreement between
these two different methods supplies more confidence in the results
that are being presented.

\begin{table}[!htbp]
\begin{center}
\caption{Comparison between the momentum-spectrum results and fits to
the missing-mass spectrum for multi-body at 4170 MeV and 4260 MeV.
The agreement between the two different methods instills confidence on
our handling of multi-body events.}
\label{MB_results}
\vspace{0.2cm}
\begin{tabular}{|c|c|c|c|}\hline
{Parameter}& {Energy (MeV)} & {Missing-Mass}& {Momentum-Spectrum} \\
&&Value (nb)& Value (nb)
\\ \hline
$D^{*}\bar{D}\pi (D^{*0})$ & $4170 $&$0.34\pm0.06$& $0.44 \pm 0.01$\\ \hline
$D^{*}\bar{D}\pi (D^{*+})$ & $4170 $&$0.42\pm0.06$& $0.44 \pm 0.01$\\ \hline
$D^{*}\bar{D}\pi (D^{*0})$ & $4260 $&$0.62\pm0.16$& $0.64 \pm 0.09$\\ \hline
$D^{*}\bar{D}^{*}\pi (D^{*0})$ & $4260 $&$0.23\pm0.05$& $0.32 \pm 0.07$\\ \hline
\end{tabular}
\end{center}
\end{table}

\chapter{SYSTEMATICS}
%\section{Systematics}
\label{sec:sys}
%%%%%%%%%%%%%%%%%%%%%%%%%%%%%%%%%%%%%%%%%%%%%%%%%%%%%%%%%%%%%%%%%%%%%%%%%%

%%%%%%%%%%%%%%%%%%%%%%%%%%%%%%%%%%%%%%%%%%%%%%%%%%%%%%%%%%%%%%%%%%
\section{Exclusive Cross Sections via Momentum \\Spectrum Fits}
%%%%%%%%%%%%%%%%%%%%%%%%%%%%%%%%%%%%%%%%%%%%%%%%%%%%%%%%%%%%%%%%%%
Using the published CLEO-c 
56~pb$^{-1}$ analysis \cite{CLEO_DHad_PRL,BFcbx}, at all energies we
apply a 0.7\%, 0.3\%, and 1.3\% systematic for tracking, pion PID,
and kaon PID respectively. In addition, we apply a 1\% systematic to
the luminosity \cite{LUM}.  Also, the error on the respective
branching ratios are 3.1\% for $D^0\rightarrow{K^-}\pi^+$, 3.9\%
for $D^+\rightarrow{K^-}\pi^+\pi^+$ \cite{CLEO_DHad_PRL,BFcbx}, 0.7\%
for $D^{*+}\rightarrow{D^{0}\pi^+}$ \cite{pdg} and the respective
errors associated with $D_{s}$ decays \cite{pdg,CLEO_DsBF}.

Besides the aforementioned systematic errors, the additional area where systematic
errors will be present is in the cross section shape assumed in
{\tt{EVTGEN}} \cite{EVTGEN}.  The systematic error resulting from the
shape of the cross section was determined by adjusting the two-body
exclusive cross sections from their nominal shapes (\FIG
\ref{fig:EvtGen_DD} and \FIG \ref{fig:EvtGen_DsDs}).  The
adjustments were determined by the existing data points as well as
taking some extremes to understand the effect of the shape
of the cross section assumed in {\tt{EVTGEN}}. The systematic errors
were investigated on the large data-set at 4170 MeV since statistical
errors are minimized.  Lastly, since an adjustment to the shape of single cross section
can effect the other cross sections in the fit, the total systematic error
for an individual cross section is determined by adding in
quadrature all cross section effects.

\subsection{$D\bar{D}$}
Two adjustments were investigated in regards to the shape of the
cross section.  First, a step was added starting at 4015 MeV and
extending to 4120 MeV where the cross section was doubled.  Since
the cross section in this range is larger, as compared to the nominal
assumption of a flat cross section across all energies, the
probability distribution will reflect this change.
Events which radiate a photon for this energy range are now more
probable and the fit results for this assumed cross section
will be larger as compared to the nominal assumption. The next adjustment
investigated is a decreasing cross section as a function of increasing
energy, the slope is 2 nb/GeV.  Following similar logic as before,
if the cross section at lower energies is larger, as compared to
the cross section at 4170 MeV and the nominal flat cross section then the
fit result will report a larger result for the $D\bar{D}$ and
$D_s\bar{D}_s$ cross sections as is shown in \TAB \ref{TAB:DD_sys}.  Similar
results are obtained for a cross section that is smaller at lower
energies, as compared to the cross section at 4170 MeV.  Since
increasing and decreasing the cross sections, at 2 nb/GeV, are extreme adjustments,
whereas the box is a conservative adjustment, the two are averaged and
used in the determination of the systematic errors.  The former is considered
extreme because the data does not suggest this type of behavior.

\begin{table}[!htb]
\begin{center}

\caption{Relative systematic errors for adjustments to the $D\bar{D}$
cross section.}\vspace{0.2cm}
\label{TAB:DD_sys}
\begin{tabular}{|c|c|c|c|}\hline 
{Event Type}& Box 4015-4120 MeV & Slope ($m=2$ nb/GeV) & Average
\\ \hline
\(D\bar{D}\)&\(+2.1\%\)&\(+4.2\%\) &\(+3.2\%\) 
\\ \hline
\(D^*\bar{D}\)&\(-0.5\%\)&\(-0.5\%\)&\(-0.5\%\)
\\ \hline
\(D^{*}\bar{D}^{*}\)&\(+0.1\%\)&\(+0.1\%\)&\(+0.1\%\)
\\ \hline
\(D^*\bar{D}\pi\)&\(+0.3\%\)&\(+0.2\%\)&\(+0.3\%\)
\\ \hline 
%\(D_{s}\bar{D}_{s}\)&\(+1.9\%\)&\(+4.1\%\)&\(+3.0\%\)
%\\ \hline
%\(D_{s}^*\bar{D}_{s}\)&\(-0.3\%\)&\(-0.4\%\)&\(-0.4\%\)
%\\ \hline \hline
\end{tabular}\end{center}\end{table}

\subsection{$D^{*}\bar{D}$}
\begin{figure}[!htbp]
\begin{center}
\hspace{2.5pt}
\includegraphics[width=14.5cm]{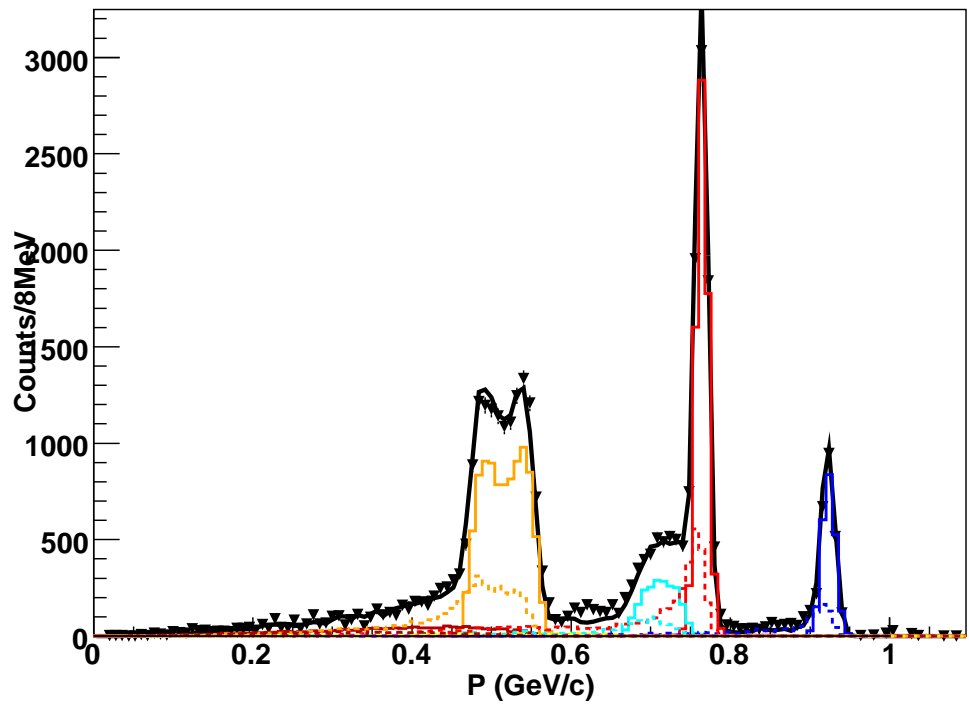}
\caption{The fit result after adjusting the $D^*\bar{D}$ cross section
from its nominal value to one that is flat between 4060-4170 MeV.  The
change results in a poor fit most notable at 600 MeV/$c$.} 
\vspace{0.2cm}
\label{FIG:DSD_sys}
\end{center}
\end{figure}

Two adjustments were investigated in regards to the shape of the
cross section.  First, the cross section was adjusted to
take out the kink at 4060 MeV allowing the cross section to decrease
constantly and continuously from 4015-4170 MeV.  Since the cross
section is increasing one would expect the fit result to reflect this
change and increase. However, since there is an overlap of both
$D\bar{D}$ and $D^{*}\bar{D}^*$, the simple explanation is no longer
straight forward.  The next adjustment was to assume a flat cross
section until 4060 MeV at which energy the cross section then increased to
4015 MeV. This assumption results in a poor fit to the momentum
spectrum, most notable in the $D^+\rightarrow{K^-}\pi^+\pi^+$
distribution as seen in \FIG \ref{FIG:DSD_sys}.  Therefore, because
the latter assumption produces a poor fit, systematic errors
of the former are used.  Since there is overlap between the various
two-body states present at this energy changes to the $D^{*}\bar{D}$
effects to the other cross sections result which are evident in \TAB \ref{TAB:DSD_sys}.
\begin{table}[!htb]
\begin{center}

\caption{Relative systematic errors for $D^{*}\bar{D}$.}
\label{TAB:DSD_sys}\vspace{0.2cm}
\begin{tabular}{|c|c|c|}\hline 
{Event Type}& Continuous Decrease From 4015 MeV & Flat to 4060 MeV
\\ \hline
\(D\bar{D}\)&\(+0.6\%\)&\(+0.8\%\)
\\ \hline
\(D^*\bar{D}\)&\(-1.1\%\)&\(-5.2\%\)
\\ \hline
\(D^{*}\bar{D}^{*}\)&\(+0.4\%\)&\(+3.0\%\)
\\ \hline
\(D^*\bar{D}\pi\)&\(+3.0\%\)&\(+2.0\%\)
\\ \hline
\end{tabular}\end{center}\end{table}

\subsection{$D^{*}\bar{D}^*$}
For these events we investigated changes to the shape of the cross
section in addition to changes in the respective helicity amplitudes.
First, the cross section for $D^{*}\bar{D}^*$ was adjusted such as to add
a dip in the plateau region of the nominal cross section.  In a similar
situation as $D^{*}\bar{D}$, the addition of other possible two-body
states along with the addition of multi-body a simple expectation is
difficult to ascertain.  Another adjustment was to remove the kink at
4030 MeV and to allow the cross section to decrease continuously to
zero at 4060 MeV. The effects of both adjustments on the
$D^{*}\bar{D}^*$ along with the other cross sections is shown in \TAB \ref{TAB:DSDS_sys}.

\begin{table}[!htb]
\begin{center}

\caption{Relative systematic errors for adjustments in the shape of
the cross section for $D^{*}\bar{D}^{*}$.}
\label{TAB:DSDS_sys}\vspace{0.2cm}
\begin{tabular}{|c|c|c|}\hline
{Event Type}& Dip & Decrease to Zero at 4070 MeV
\\ \hline
\(D\bar{D}\)&\(-\)&\(-\)
\\ \hline
\(D^*\bar{D}\)&\(-\)&\(+0.2\%\)
\\ \hline
\(D^{*}\bar{D}^{*}\)&\(+2.4\%\)&\(-2.4\%\)
\\ \hline
\(D^*\bar{D}\pi\)&\(-1.5\%\)&\(+9.1\%\)
\\ \hline
\end{tabular}\end{center}\end{table}

\begin{figure}[!htbp]
\begin{center}
\hspace{2.5pt}
\includegraphics[width=14.5cm]{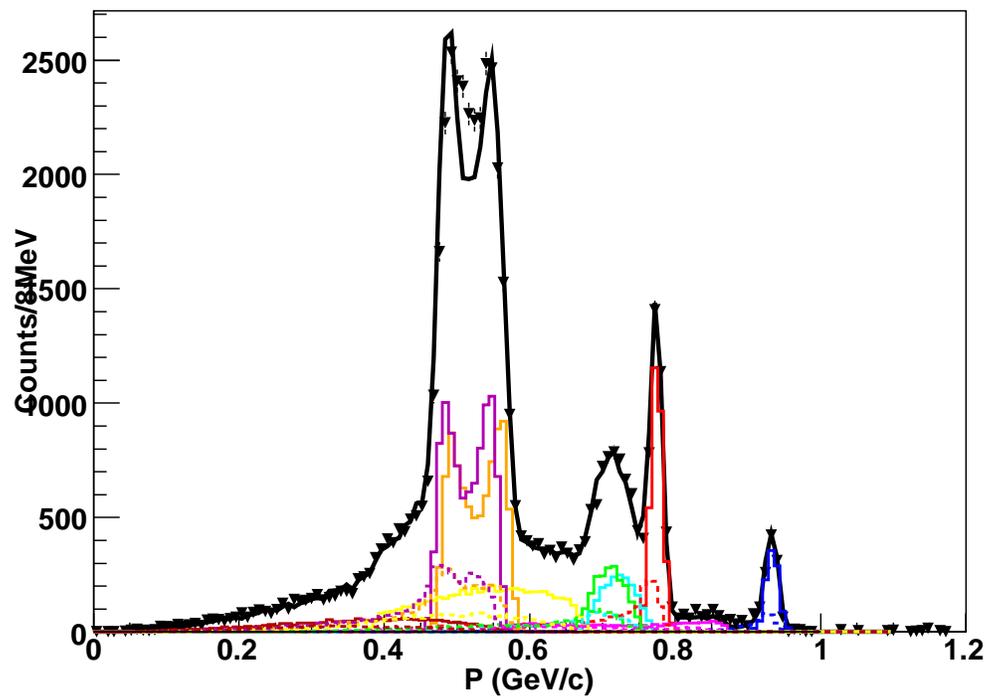}
\caption{The fit result after adjusting the angle between
the $D$ in the rest frame of the $D^{*}$ with respect to the momentum
of $D^{*}$ in the lab from its nominal value of 0.8 to 2.0. The fit is
poor most notably between 500-600 MeV/$c$.} 
\vspace{0.2cm}
\label{FIG:DSDS_sys2}
\end{center}
\end{figure}

The helicity amplitude affect the angular distributions of the
$D^{*}\bar{D}^*$ final state and therefore has an effect of the
resulting daughter $D$ (See Appendix for a discussion of angular
distributions).  This shows up, most notably, in the momentum spectrum
resulting from the $\pi$ decay channel, between 500-600 MeV/$c$.
Taking an extreme case we adjusted the coefficient in front of the
$\cos$ for the angle between the $D$ in the
rest frame of the $D^{*}$ with respect to the momentum of $D^{*}$ in
the lab from its nominal value of 0.8 to 2.0. The result 
is shown in \FIG \ref{FIG:DSDS_sys2} which clearly shows that such a
change results in a bad fit.  The effect on the cross sections is
shown in \TAB \ref{TAB:DSDS_sys2}.    
\begin{table}[!htb]
\begin{center}

\caption{Relative systematic errors for adjustments in the helicity
amplitudes for $D^{*}\bar{D}^{*}$.}
\label{TAB:DSDS_sys2}\vspace{0.2cm}
\begin{tabular}{|c|c|}\hline
{Event Type}& Angle ($\alpha^{'} = 2$)
\\ \hline
\(D\bar{D}\)&\(-\)
\\ \hline
\(D^*\bar{D}\)&\(+0.3\%\)
\\ \hline
\(D^{*}\bar{D}^{*}\)&\(-1.0\%\)
\\ \hline
\(D^*\bar{D}\pi\)&\(+6.0\%\)
\\ \hline
\end{tabular}\end{center}\end{table}

\subsection{$D^{*}\bar{D}\pi$}
Rather than assuming a flat cross section as a function of energy, an
adjustment that allows the cross section to increase as the energy
increases.  Since the cross section is now smaller at lower energies,
as compared to the cross section at 4170 MeV, the fit result will
return a larger result.  Because the momentum range of the multi-body
event is large, between 0-500 MeV/$c$, such a change in the cross
section will result in modest changes to the fit results which are
shown in \TAB  \ref{TAB:DSDpi_sys}.  
\begin{table}[!htb]
\begin{center}

\caption{Relative systematic errors for adjustments made to the shape
of the cross section for $D^{*}\bar{D}\pi$.}
\label{TAB:DSDpi_sys}\vspace{0.2cm}
\begin{tabular}{|c|c|}\hline
{Event Type}& Increasing Slope ($m = 4$ nb/GeV)
\\ \hline
\(D\bar{D}\)&\(-\)
\\ \hline
\(D^*\bar{D}\)&\(-\)
\\ \hline
\(D^{*}\bar{D}^{*}\)&\(-\)
\\ \hline
\(D^*\bar{D}\pi\)&\(+2.7\%\)
\\ \hline
\end{tabular}\end{center}\end{table}

\subsection{$D^{*}\bar{D}^{*}\pi$}
Since there is only one data point for
$D^{*}\bar{D}^{*}\pi$ a systematic of 25\% is assigned.

\subsection{Total Systematic Errors: Exclusive Momentum
Fit Method}

The total systematic errors for the exclusive cross sections by
fitting the momentum spectrum are shown in \TAB \ref{TAB:tot_sys}.

\begin{table}[!htb]
\begin{center}

\caption{Total relative systematic errors for the exclusive momentum
fit method.}
\label{TAB:tot_sys}\vspace{0.2cm}
\begin{tabular}{|c|c|}\hline
{Event Type}& Total
\\ \hline
\(D\bar{D}\)&\(4.5\%\)
\\ \hline
\(D^*\bar{D}\)&\(3.4\%\)
\\ \hline
\(D^{*}\bar{D}^{*}\)&\(4.7\%\)
\\ \hline
\(D^*\bar{D}\pi\)&\(12\%\)
\\ \hline
\(D^*\bar{D}^{*}\pi\)&\(25\%\)
%\\ \hline
%\(D_{s}\bar{D}_{s}\)&\(27\%\)
%\\ \hline
%\(D_{s}^{*}\bar{D}_{s}\)&\(27\%\)
%\\ \hline
%\(D_{s}^{*}\bar{D}_{s}^{*}\)&\(37\%\)
\\ \hline
\end{tabular}\end{center}\end{table}

\subsection{Systematic Errors: $D_{s}\bar{D}_{s}$,
$D_{s}^{*}\bar{D}_{s}$ and $D_{s}^{*}\bar{D}_{s}^{*}$}

For $D_{s}\bar{D}_{s}$, $D_{s}^{*}\bar{D}_{s}$ and 
$D_{s}^{*}\bar{D}_{s}^{*}$, a weighted sum technique is used to 
combine the eight $D_s$ decay modes and obtain the cross sections.
This technique was used to minimize the error, both statistical
and systematic.  Since multi-body type events involving a
$D_{s}$-meson are not expected by isospin which was confirmed by not observing
evidence of multi-body events in the momentum spectra of $D_{s}\rightarrow\phi\pi^+$ at any
of energies, the technique described in Sect. \ref{sec:sigmas} is used.
Besides the errors in regards to the branching
fractions, shown in \TAB \ref{Ds_BR}, are errors resulting from the
technique. It should be mentioned that because of the recently updated
$D_{s}$ branching fractions \cite{CLEO_DsBF} the weights used in the
calculation are not optimal because they were determined using the PDG 2005 \cite{pdg} values. However, the changes in the branching fractions have a minor, if no, effect on
the final result by virtue of the technique.  To demonstrate this we
look at the results for $D_{s}$ at 4170~MeV.  By comparing both the
individual mode results and the weighted sum result, \TAB \ref{Ds_method_compare},
one sees that the agreement between them is quite good and the updated
branching fractions reduce the associated systematic error without
effecting the quoted value for the $D_{s}$ cross sections stated in
Sect. \ref{sec:sigmas}.

\begin{table}[!htb]
\begin{center}
\caption{A comparison between the $D_{s}^{*}D_{s}$ cross sections at
4170~MeV determined by individual modes using the recently updated
branching fractions \cite{CLEO_DsBF} to that of the weighted sum
technique using branching fractions based on \cite{pdg}.  The
agreement between the individual modes and the weighted sum stresses
the fact that the result from the weighted sum is not effected by the
changes in the branching fractions.  The only effect of the updated
branching fractions is to decrease the systematic errors.}
\vspace{0.2cm}
\label{Ds_method_compare}
\begin{tabular}{|c|c|}\hline
{Mode}&Cross Section (nb)
\\ \hline
\(K_{s}K^{+}\)&\(0.85\pm0.05\)
\\ \hline
\(\pi^{+}\eta\)&\(0.99\pm0.05\)
\\ \hline
\(\pi^{+}\eta^{'}\)&\(0.90\pm0.04\)
\\ \hline
\(\pi^{+}\phi\)&\(0.96\pm0.03\)
\\ \hline
Weighted Sum&\(0.92\pm0.01\)
\\ \hline
\end{tabular}\end{center}\end{table}

\begin{table}[!tb]
\begin{center}
\caption{The $D_{s}$ branching fractions including the updated
branching fractions, in $10^{-2}$.}
\vspace{0.2cm}
\label{Ds_BR}
\begin{tabular}{|c|c|}
\hline
{Modes}& {Branching Fraction}
\\ \hline
\(\phi\pi^{+}\), 10~MeV cut on the Invariant
\(\phi\rightarrow{K}^{+}K^{-}\) Mass \cite{CLEO_DsBF}&\(1.98\pm0.15\)
\\ \hline
\(K^{*0}K^{+},K^{*0}\rightarrow{K^{-}\pi^{-}}\)~\cite{pdg}&\(2.2\pm0.6\)
\\ \hline
\(\eta\pi^{+},\eta\rightarrow{\gamma\gamma}\)~\cite{pdg,CLEO_DsBF}&\(0.58\pm0.07\)
\\ \hline
\(\eta\rho^{+},\eta\rightarrow{\gamma\gamma},\rho^{+}\rightarrow{\pi^{+}\pi^{0}}\)~\cite{pdg}&\(4.3\pm1.2\)
\\ \hline
\(\eta^{'}\pi^{+},\eta^{'}\rightarrow{\pi^{+}\pi^{-}\eta},\eta\rightarrow{\gamma\gamma}\)~\cite{pdg,CLEO_DsBF}&\(0.7\pm0.1\)
\\ \hline
\(\eta^{'}\rho^{+},\eta^{'}\rightarrow{\pi^{+}\pi^{-}\eta},\eta\rightarrow{\gamma\gamma},\rho^{+}\rightarrow{\pi^{+}\pi^{0}}\)~\cite{pdg}&\(1.8\pm0.5\)
\\ \hline
\(\phi\rho^{+},\phi\rightarrow{K^{+}K^{-}},\rho^{+}\rightarrow{\pi^{+}\pi^{0}}\)~\cite{pdg}&\(3.4\pm1.2\)
\\ \hline
\(K_{s}K^{+},K_{s}\rightarrow{\pi^{+}\pi^{-}}\)~\cite{pdg,CLEO_DsBF}&\(1.0\pm0.07\)
\\ \hline
\end{tabular}
\end{center}
\end{table}

The selection criteria, ${\rm{M_{bc}}}$ and $\Delta{\rm{E}}$, for the various exclusive cross
sections result in added systematic errors.
Having adjusted the $\Delta{\rm{E}}$ cut window by 10~MeV
the cross sections for $D_s\bar{D}_s$ changed by a maximum of 3\%
and therefore 3\% is applied to these cross sections at all
energies.  Next,  by adjusting the ${\rm{M_{bc}}}$ cut window by 10~MeV for $D_{s}^{*+}{D}_{s}^{-}$ the cross
sections changed by 2.5\% and therefore 2.5\% is applied to these
cross sections at all energies.
The charge in ${\rm{M_{bc}}}$ was asymmetric because the photon decay mode is included and allows for $D_{s}$ to be
spread out over a wide range of momenta. Lastly, by adjusting
${\rm{M_{bc}}}$ by 30~MeV for $D_{s}^{*+}\bar{D}_{s}^{*-}$ the result
changed by 5\%  therefore 5\% is applied to these cross sections at all
energies. The table of systematics for the exclusive method is shown in \TAB \ref{DsXS_SYS}.

\begin{table}[!tb]
\begin{center}
\caption{Total relative systematic errors for the $D_{s}$ cross sections.}
\label{DsXS_SYS}\vspace{0.2cm}
\begin{tabular}{|c|c|}\hline
{Event Type}& Total
\\ \hline
\(D_{s}\bar{D}_{s}\)&\(5.6\%\)
\\ \hline
\(D_{s}^{*}\bar{D}_{s}\)&\(5.3\%\)
\\ \hline
\(D_{s}^{*}\bar{D}_{s}^{*}\)&\(6.8\%\)
\\ \hline
\end{tabular}\end{center}\end{table}

%%%%%%%%%%%%%%%%%%%%%%%%%%%%%%%%%%%%%%%%%%%%%%%%%%%%%%%%%%%%%%%
\section{Inclusive $D$ Method}
%%%%%%%%%%%%%%%%%%%%%%%%%%%%%%%%%%%%%%%%%%%%%%%%%%%%%%%%%%%%%%%

Using the published CLEO-c 
56~pb$^{-1}$ analysis \cite{CLEO_DHad_PRL,BFcbx}, at all energies we
apply a 0.7\%, 0.3\%, and 1.3\% systematic for tracking, pion PID,
and kaon PID respectively. In addition, we apply a 1\% systematic to
the luminosity \cite{LUM}.  The largest systematic in regards to this
method is assigned to the fit function and the yields obtained by
a gaussian plus a polynomial of degree two.
Changing the background function to a linear function causes a 2\%
shift, which is the largest, is seen in the yields of $D^0$ and $D^+$. 
At energies above the $D_s\bar{D}_s^*$ threshold a 7\% shift, which is largest,
is seen in the $D_s$ yields and 2\% at all energies below threshold.
In all cases, except 4170 MeV, the change in yields is $<1\sigma$.  Lastly, the error on the respective
branching ratios are 3.1\% for $D^0\rightarrow{K^-}\pi^+$, 3.9\%
for $D^+\rightarrow{K^-}\pi^+\pi^+$ \cite{CLEO_DHad_PRL,BFcbx} and
6.4\% for $D_s^+\rightarrow{K^-}K^+\pi^+$ \cite{CLEO_DsBF}. Therefore, we quote a
systematic of 4.3\% for $D^0$ and 5.1\% for $D^+$ at all energies.
For $D_s$ an extra systematic error of 4\%, associated with the selection
criteria, is applied.  The determination of the latter systematic
error was taken from the investigation of the exclusive cross
sections for $D_{s}^{*+}{D}_{s}^{-}$ and $D_{s}^{+}{D}_{s}^{-}$.
Therefore, $D_s$ is quoted with a systematic error of 8.6\% for
energies below 4100 MeV and 10.8\% for those above.  The systematic
error associated with the branching fractions, as well as the PID and
tracking, should be improved in the near future with the updated $D$ and $D_{s}$
branching fractions.

%%%%%%%%%%%%%%%%%%%%%%%%%%%%%%%%%%%%%%%%%%%%%%%%%%%%%%%%%%%%%%%%%%%
\section{Hadron-Counting Method}
%%%%%%%%%%%%%%%%%%%%%%%%%%%%%%%%%%%%%%%%%%%%%%%%%%%%%%%%%%%%%%%%%%%

Since the method that was used here is identical to method used to
determine the cross section of
\(e^+e^-\rightarrow\psi(3770)\rightarrow{hadrons}\) at
\(E_{\rm{cm}}=3770\)~MeV \cite{Hajime,Hajime_update,Hajime_PRL}, most of the
systematics will be identical.  One difference that was investigated
was the systematic that is associated with the hadronic event selection
criteria, since this could vary with energy.  By applying tighter cuts,
THAD \cite{Hajime,Hajime_PRL}, the maximum change in the cross at any energy was
4.5\% and so is taken as the hadronic event
criteria systematic for all energies. In a similar fashion, as in the
reference \cite{Hajime,Hajime_update,Hajime_PRL}, a conservative error, to
account for the interference with the continuum and because they are
based on theoretical calculation, of 25\%
will be associated with $\psi(2S)$, $J/\psi$, and $\psi(3770)$
subtraction at each energy. Since the cross section varies for these
backgrounds, so does the associated error.  Errors which are common
between the energies are shown in \TAB \ref{Common_HAD_sys_table},
whereas the energy dependent errors are shown in \TAB
\ref{Dependent_HAD_sys_table}, including the total.

\begin{table}[t]
\begin{center}
\caption{Summary of various systematic errors that are common amongst
the scan energies for the hadron counting method.}
\label{Common_HAD_sys_table}
\vspace{0.2cm}
\begin{tabular}{|c|c|}\hline
{Source of error} & relative errors in $10^{-2}$ \\ \hline
QED/Two-Photons subtractions/suppressions & 0.3 \\ \hline
BeamGas/Wall/Cosmic subtraction & 0.5 \\ \hline
Track quality cuts & 0.5 \\ \hline
Luminosity & 1.0 \\ \hline
Continuum scaling & 2.1 \\ \hline
Hadronic event selection criteria & 4.5\\ \hline
Total common systematic error & 5.1 \\ \hline
\end{tabular}
\end{center}
\end{table}

\begin{table}[t]
\begin{center}
\caption{Summary of the energy dependent systematic errors, in $10^{-2}$, for the
hadron counting method.}
\label{Dependent_HAD_sys_table}
\vspace{0.2cm}
\begin{tabular}{|c|c|c|c|c|}\hline
{Energy (MeV)} & $J/\psi$ & $\psi(2S)$ & $\psi(3770)$ & Total relative
Systematic Errors\\ \hline
3970 & 3.1& 1.30& 0.44& 6.1 \\ \hline
3990 & 2.3& 1.13& 0.34& 5.7 \\ \hline
4010 & 1.8& 0.88& 0.25& 5.5 \\ \hline
4015 & 1.6& 0.80& 0.23& 5.4 \\ \hline
4030 & 1.1& 0.54& 0.14& 5.2 \\ \hline
4060 & 1.1& 0.61& 0.15& 5.3 \\ \hline
4120 & 0.99& 0.53& 0.12& 5.2 \\ \hline
4140 & 0.91& 0.51& 0.12& 5.2 \\ \hline
4160 & 0.89& 0.48& 0.10& 5.2 \\ \hline
4170 & 0.88& 0.49& 0.10& 5.2 \\ \hline
4180 & 0.88& 0.49& 0.11& 5.2 \\ \hline
4200 & 0.91& 0.50& 0.12& 5.2 \\ \hline
4260 & 1.53& 0.80& 0.19& 5.4 \\ \hline
\end{tabular}
\end{center}
\end{table}

\chapter{Radiative Corrections}
\label{sec:radcor}
%%%%%%%%%%%%%%%%%%%%%%%%%%%%%%%%%%%%%%%%%%%%%%%%%%%%%%%%%%%%%%%%%%%%%%%%%%
%
The Born or tree-level cross section is obtained by correcting the
observed cross sections for the effects of
initial-state radiation (ISR).
In high-energy electron-positron annihilation experiments, the incoming
particles can radiate.   The radiated photons can be quite energetic,
thereby changing the effective center-of-mass collision energy appreciably.
Therefore, the annihilation energy is not always
twice the beam energy and the observed cross sections obtained in the 
experiment correspond not to a single energy point but instead to a range of
energies.

The needed correction factors were calculated using two alternate procedures. First,
by following the method laid out by E.A. Kuraev and V.S. Fadin
\cite{KF} which states that the observed cross section $\sigma_{obs}(s)$ at 
any energy $\sqrt{s}$ can be written as
\begin{equation}
	\sigma_{obs}(s) = \int\limits_{0}^{1} dk \cdot f(k,s)\sigma_{B}(s_{eff}),
\end{equation}
\CONT where $\sigma_{B}$ is the Born cross section as a function
of the effective center-of-mass energy squared and $s_{eff} = s(1-x)$,
with $E_{\gamma}=xE_{\rm{beam}}$. The function, $f(x,s)$ is defined
as follows:
\begin{eqnarray}
\label{eq:KFEQ}
	f(x,s) =
	tx^{t-1}[1+\frac{\alpha}{\pi}(\frac{\pi^2}{3}-\frac{1}{2}) +
	\frac{3}{4}t - \frac{t^2}{24}(\frac{1}{3}\ln\frac{s}{m_e^2} +
	2\pi^2 - \frac{37}{4})] - t(1-\frac{x}{2}) \nonumber \\
	+ \frac{t^2}{8}[4(2-x)\ln\frac{1}{x}-\frac{1+3(1-x)^2}{x}\ln(1-x)-6+x],
\end{eqnarray}
\CONT where $t= \frac{2\alpha}{\pi}[\ln(\frac{s}{m_{e}^2}) -1]$. \EQ \ref{eq:KFEQ} is only used to first order in $t$.

The other method, by G. Bonneau and F. Martin \cite{BandM}, states
that the observed cross section can be written in terms of the Born cross
section as follows:
\begin{eqnarray}
\label{eq:BMEQ}
\sigma_{\rm{obs}}=\sigma_{B}[1+\frac{2\alpha}{\pi}\{(2\ln\frac{2E}{m_e}-1)(\ln{x_{min}}
+\frac{13}{12} +\nonumber \\
\int^{1}_{x_{min}}\frac{dx}{x}(1-x+\frac{x^2}{2})\frac{\sigma(s(1-x))}{\sigma{(s)}})-\frac{17}{36}+\frac{\pi^2}{6}\}].
\end{eqnarray}

\CONT Since the integral is infrared divergent, the $\ln{x_{min}}$ in
\EQ \ref{eq:BMEQ} corresponds to the low-energy cutoff. 

The only difference between the two methods is
that G. Bonneau and F. Martin include the vacuum polarization for
the electron, i.e. the electron bubble:
\begin{equation}
\label{eq:VP}
\delta_{vp} = \frac{2\alpha}{\pi}[-\frac{5}{9}+\frac{1}{3}\ln\frac{s}{m_{e}^{2}}].
\end{equation}
We confirmed this by subtracting \EQ \ref{eq:VP} from \EQ \ref{eq:BMEQ} and applying
to the simple test of a constant cross section as a function of center-of-mass energy.

Note that \EQS \ref{eq:KFEQ} and \ref{eq:BMEQ} are dependent on the tree-level
cross section, not only at $s$, but at all energies below $s$.
Therefore, the ISR correction relies on a theoretical model or
on already-radiatively-corrected experimental data to
describe the shape of the cross section at all relevant energies.

It was decided to approach determining the shape of the cross section at lower energies with two different
methods and to test on two different sets
of already-radiatively-corrected data.  The two data sets are from 
$R$ measurements in the region above $c\bar{c}$ threshold made by BES
\cite{BES_R} and Crystal Ball (CB) \cite{CB_R}. The two shape methods 
were a simple linear-interpolation procedure applied to
each data set, and a fit consisting of a sum of Breit-Wigners. Each of
these methods was applied to the data sets after the $uds$ background
had been subtracted using 
$\sigma_{uds} \sim \frac{196~{\rm{GeV^2}}~{\rm{nb}}}{s}$.

\begin{figure}[!htbp]
\begin{center}
\hspace{2.5pt}
\includegraphics[width=14.5cm]{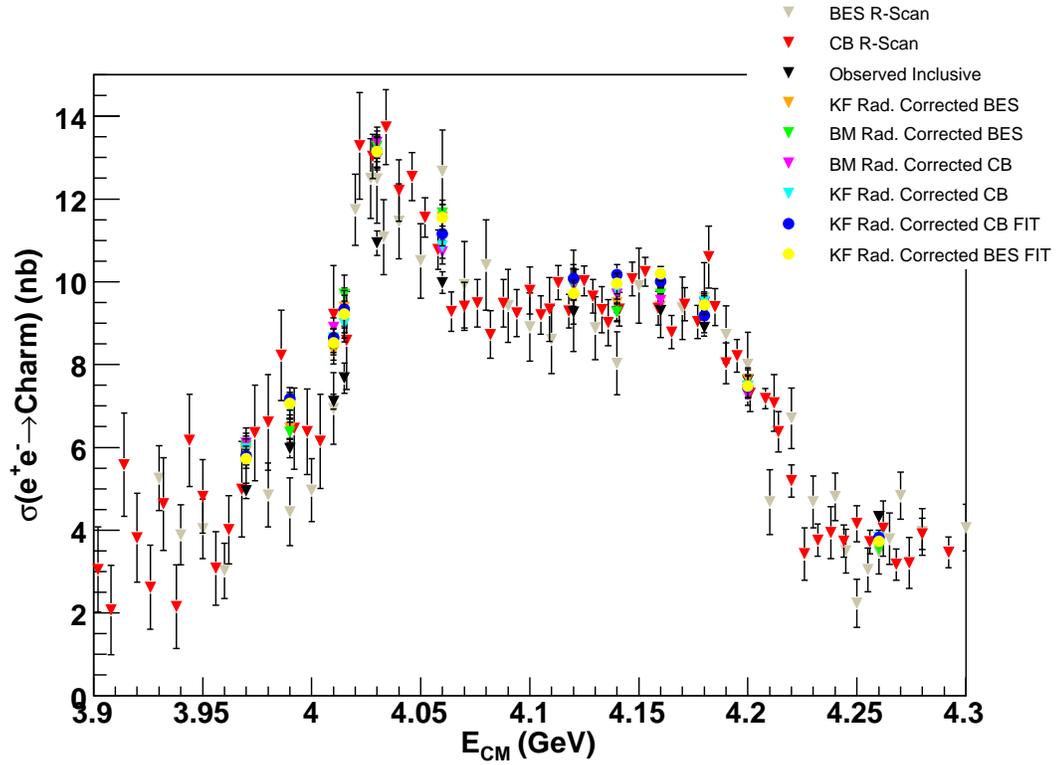}
\caption{The radiatively corrected inclusive cross sections using all
methods are compared to the BES and CB $uds$-subtracted $R$
data. Note that the corrected cross sections (including BES and CB) include the effect of 
vacuum polarization as defined in \EQ \ref{eq:VP} as well as the
bubbles due to $\mu$ and $\tau$ leptons.}
\vspace{0.2cm}
\label{fig:INXS_radcor}
\end{center}
\end{figure}

\begin{figure}[!htbp]
\begin{center}
\hspace{2.5pt}
\includegraphics[width=14.5cm]{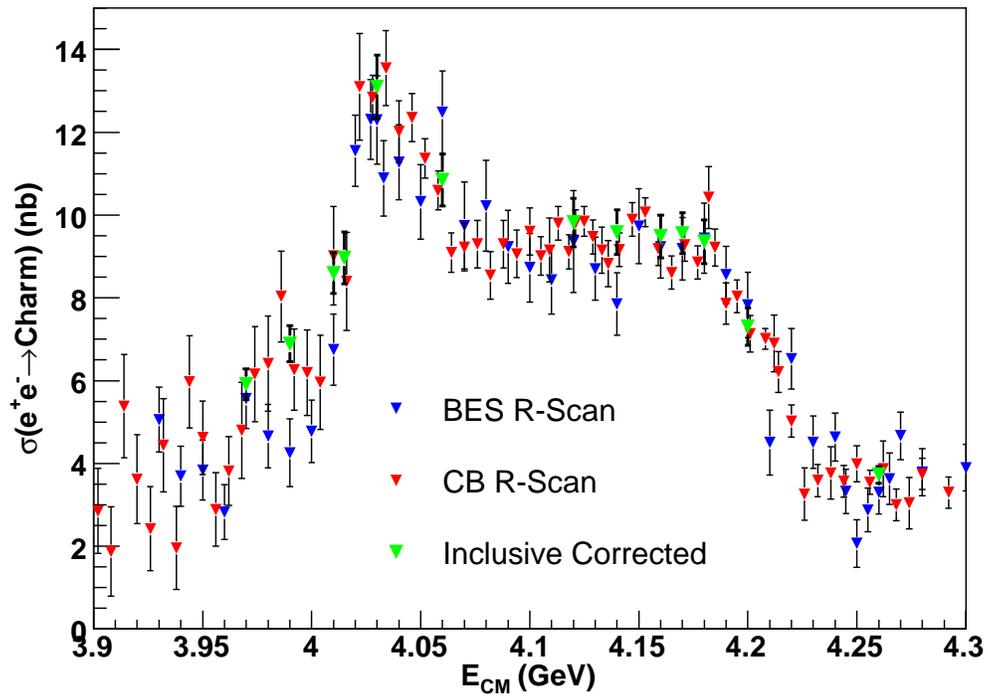}
\caption{The radiatively corrected inclusive cross sections using the KF
interpolation method on the CB data. The error bars include the
systematics as discussed in the text.}
\vspace{0.2cm}
\label{RadCor_INXS_fig}
\end{center}
\end{figure}

The results after radiatively correcting the inclusive cross section are shown in
\FIG \ref{fig:INXS_radcor}.  The difference in the corrected cross section
between the two shape-determining methods is mainly due to the
interpolation method being influenced by the ``jitteriness'' of the
data.  Since this method is a simple linear interpolation, the
fluctuations in the data play an important role, especially in the
region nearest to the point that is to be corrected. Even so, the
agreement is quite good, as seen in \TAB \ref{RadCor_table}. The
agreement between the two radiative correction methods, BM
and KF, is excellent, giving us confidence that both have been implemented
correctly.  The difference between the methods is taken to be the
systematic error for the ISR correction, which gives a 4\%
systematic error on the correction factor for all
energies.   Since all the methods are in good agreement, the KF
interpolation method on the CB data has been chosen for the nominal results.  The corrected
inclusive cross section for the energies investigated are shown in
\TAB \ref{RadCor_INXS} and graphically in \FIG \ref{RadCor_INXS_fig}.

\begin{figure}[!htbp]
\begin{center}
\hspace{2.5pt}
\includegraphics[width=14.5cm]{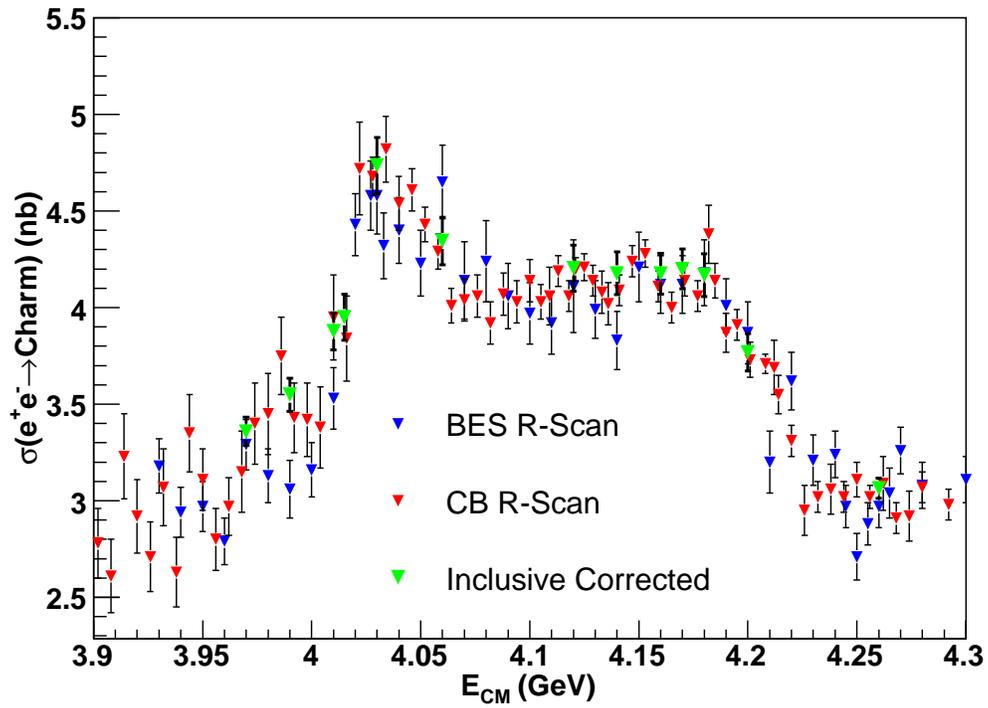}
\caption{R measurement obtained from these inclusive charm cross
sections, radiatively corrected, using the KF
interpolation method on the CB data. The error bars include the
systematics as discussed in the text.}
\vspace{0.2cm}
\label{RadCor_INR_fig}
\end{center}
\end{figure}

A measurement of $R_{{\rm{charm}}}$, sometimes referred to as
$R_{D}$, can be made by dividing $\sigma_{{\rm{charm}}}$ by the QED
tree-level cross section for muon pair production, \EQ \ref{R_eq}.
In addition to $R_{{\rm{charm}}}$, one needs to determine $R_{uds}$,
where $u$,$d$, and $s$ refer to the contribution of the light quarks to the
measurement of R.  By fitting $e^+e^-\rightarrow{{hadrons}}$, \FIG \ref{fig:PDG},
between 3.2 and 3.72 GeV, with a $\frac{1}{s}$ function we arrive at
$R_{uds}=2.285\pm0.03$.  The value of $R$ is then computed as the sum
of these contributions:
\begin{equation}
R = R_{uds} + R_{{\rm{charm}}}
\end{equation}
\CONT the results for $R$ are shown in \TAB \ref{RadCor_INR} and graphically in
\FIG \ref{RadCor_INR_fig}.

\begin{table}[!htbp]
\begin{center}
\caption{The radiative correction factors used in obtaining the
tree-level total charm cross section results.
The two different shape methods, calculation procedures, and the
two data sets used are shown. Note that the corrected cross sections
include the correction due to the vacuum polarization, as defined in
\EQ \ref{eq:VP} as well as the bubbles due to $\mu$ and $\tau$ leptons.}
\label{RadCor_table}
\vspace{0.2cm}
\begin{tabular}{|c|c|c|c|c|c|c|}\hline
{E$_{{\rm{cm}}}$ (MeV)} & BES$_{{\rm{KF}}}$&  BES$_{{\rm{BM}}}$&
BES$_{{\rm{KF}}}$& CB$_{{\rm{KF}}}$& CB$_{{\rm{BM}}}$&
CB$_{{\rm{KF}}}$ \\
& Inter. & Inter. & FIt & Inter & Inter & Fit
\\ \hline
3970 & 0.84& 0.82& 0.86& 0.83 &0.81 &0.85 \\ \hline
3990 & 0.92& 0.94& 0.85& 0.86 &0.86 &0.83 \\ \hline
4010 & 0.85& 0.84& 0.83& 0.82 &0.80 &0.82 \\ \hline
4015 & 0.82& 0.79& 0.83& 0.85 &0.84 &0.82 \\ \hline
4030 & 0.84& 0.82& 0.83& 0.83 &0.82 &0.83 \\ \hline
4060 & 0.86& 0.86& 0.86& 0.92 &0.93 &0.89 \\ \hline
4120 & 0.93& 0.94& 0.95& 0.93 &0.93 &0.92 \\ \hline
4140 & 0.98& 1.00& 0.93& 0.95 &0.96 &0.91 \\ \hline
4160 & 0.95& 0.96& 0.91& 0.96 &0.97 &0.93 \\ \hline
4170 & 0.94& 0.96& 0.92& 0.95 &0.96 &0.92 \\ \hline
4180 & 0.95& 0.96& 0.94& 0.93 &0.94 &0.97 \\ \hline
4200 & 1.00& 1.02& 1.02& 1.02 &1.05 &1.02 \\ \hline
4260 & 1.17& 1.22& 1.16& 1.13 &1.17 &1.13 \\ \hline
\end{tabular}
\end{center}
\end{table}

\begin{table}[!htbp]
\begin{center}
\caption{The total charm cross section as obtained by the inclusive
method after being radiatively corrected. The errors are statistical and systematic respectively.}
\label{RadCor_INXS}
\vspace{0.2cm}
\begin{tabular}{|c|c|}\hline
{E$_{cm}$ MeV}& {$\sigma(e^+e^-\rightarrow{D\bar{D}X})$~nb}
\\ \hline
3970 &$5.92\pm 0.22\pm 0.28$\\ \hline
3990 &$6.89\pm 0.25\pm 0.33$\\ \hline
4010 &$8.60\pm 0.22\pm 0.41$\\ \hline
4015 &$8.97\pm 0.42\pm 0.43$\\ \hline
4030 &$13.10\pm 0.35\pm 0.63$\\ \hline
4060 &$10.85\pm 0.28\pm 0.55$\\ \hline
4120 &$9.82\pm 0.30\pm 0.49$\\ \hline
4140 &$9.58\pm 0.22\pm 0.48$\\ \hline
4160 &$9.48\pm 0.15\pm 0.48$\\ \hline
4170 &$9.56\pm 0.07\pm 0.48$\\ \hline
4180 &$9.36\pm 0.21\pm 0.47$\\ \hline
4200 &$7.30\pm 0.25\pm 0.38$\\ \hline
4260 &$3.73\pm 0.08\pm 0.20$\\ \hline
\end{tabular}
\end{center}
\end{table}

\begin{table}[!htbp]
\begin{center}
\caption{$R$, after being radiatively corrected, as determined by the inclusive
method. The errors are statistical and systematic respectively. In
this method we use $R_{uds}=2.285\pm0.03$ as determined by a
$\frac{1}{s}$ fit to previous $R$ measurements between 3.2 and 3.72 GeV.}
\label{RadCor_INR}
\vspace{0.2cm}
\begin{tabular}{|c|c|}\hline
{E$_{cm}$ MeV}& {$\frac{\sigma(e^+e^-\rightarrow{hadrons})}{\sigma(e^+e^-\rightarrow{\mu^+\mu^-})}$}
\\ \hline
3970 &$3.36\pm 0.04\pm 0.05$\\ \hline
3990 &$3.55\pm 0.05\pm 0.06$\\ \hline
4010 &$3.88\pm 0.04\pm 0.08$\\ \hline
4015 &$3.95\pm 0.08\pm 0.08$\\ \hline
4030 &$4.74\pm 0.07\pm 0.12$\\ \hline
4060 &$4.34\pm 0.05\pm 0.10$\\ \hline
4120 &$4.21\pm 0.06\pm 0.10$\\ \hline
4140 &$4.18\pm 0.04\pm 0.10$\\ \hline
4160 &$4.18\pm 0.03\pm 0.10$\\ \hline
4170 &$4.20\pm 0.01\pm 0.10$\\ \hline
4180 &$4.17\pm 0.04\pm 0.10$\\ \hline
4200 &$3.77\pm 0.05\pm 0.08$\\ \hline
4260 &$3.06\pm 0.02\pm 0.04$\\ \hline
\end{tabular}
\end{center}
\end{table}

\pagebreak

\chapter{Interpretations and Conclusions}
\section{Comparisons}
\label{sec:comps}
%%%%%%%%%%%%%%%%%%%%%%%%%%%%%%%%%%%%%%%%%%%%%%%%%%%%%%%%%%%%%%%%%%%%%%%%%%
\subsection{Comparisons to Previous Measurements}
\begin{figure}[!htbp]
\begin{center}
\hspace{2.5pt}
\includegraphics[width=14.5cm]{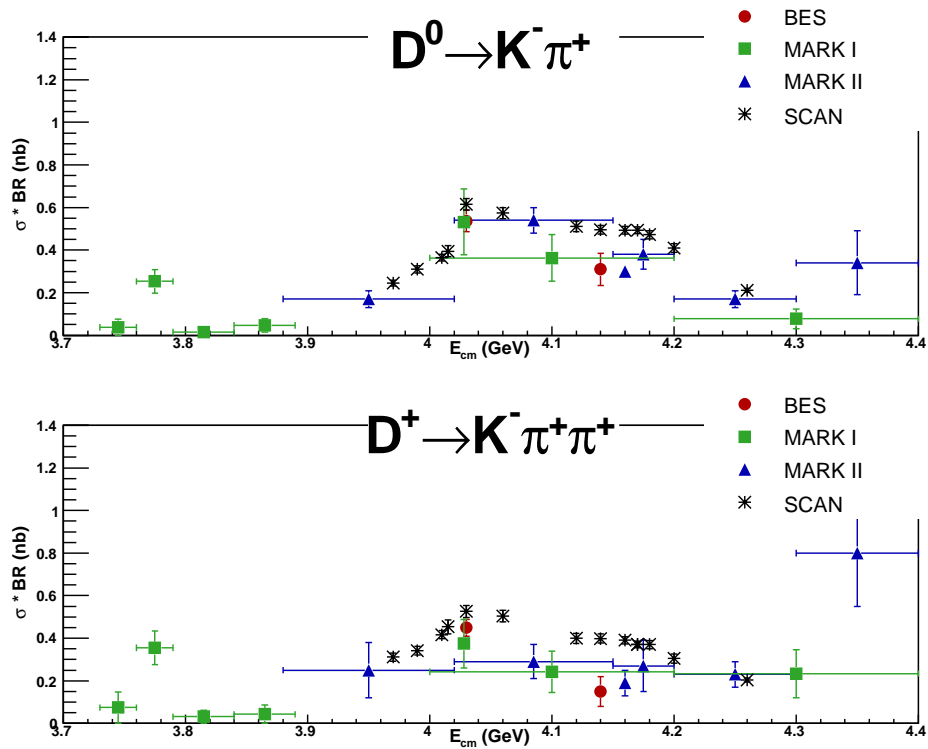}
\caption{The production cross section times branching ratio for
$D^{0}\rightarrow{K^-\pi^+}$ and $D^{0}\rightarrow{K^-\pi^+\pi^+}$ as
a function of energy for this analysis, as compared to previous
measurements. The results from the CLEO-c scan agree quite nicely 
with previous experiments.}
\vspace{0.2cm}
\label{fig:MARK_BES_results}
\end{center}
\end{figure}

A comparison of the measurements reported in this thesis of
$\sigma\cdot{\BR}(D^{0}\rightarrow{K\pi})$ and
$\sigma\cdot{\BR}(D^{+}\rightarrow{K\pi\pi})$
with previous measurements is shown in \FIG
\ref{fig:MARK_BES_results}. The results from the CLEO-c scan agree
quite nicely with previous experiments, and are much more precise.

In addition to the observed branching ratio times production cross
section for $D^{0}$ and $D^{+}$, both Mark III and BES have made similar
measurements involving the $D_{s}$.  These results, BES \cite{BES95} at center-of-mass
energy of 4030~MeV and Mark III \cite{MARK89} at center-of-mass of
4140~MeV, are shown along with the results of this analysis in \TAB
\ref{Ds_XStBR}.

\begin{table}[t]
\begin{center}
\caption{Measurements of observed branching ratio times production
cross sections. The first uncertainty in each case is statistical and the second (where given) is systematic.}
\label{Ds_XStBR}
\vspace{0.2cm}
\begin{tabular}{c|c|c}\hline
 Measurement & {Energy (MeV)} & Experiment
\\ \hline
$\sigma_{D_{s}^+D_{s}^-}\cdot{\BR}(D_{s}^+\rightarrow{\phi\pi^{+}})$ & &
\\ \hline
$11.2 \pm 2.0 \pm 2.5$~pb & 4030 & BES \\
$6.81 \pm 2.48$~pb & 4030 & This Analysis
\\ \hline
$\sigma_{D_{s}^{*+}D_{s}^-}\cdot{\BR}(D_{s}^+\rightarrow{\phi\pi^{+}})$ & &
\\ \hline
$26 \pm 6 \pm 5$~pb & 4140 & MARK III \\
$36.1 \pm 4.42$~pb & 4140 & This Analysis
\\ \hline
$\sigma_{D_{s}^{*+}D_{s}^-}\cdot{\BR}(D_{s}^+\rightarrow{\bar{K}^{0}K^{+}})$ & &
\\ \hline
$24 \pm 6 \pm 5$~pb & 4140 & MARK III \\
$23.1 \pm 4.9$~pb & 4140 & This Analysis
\\ \hline
$\sigma_{D_{s}^{*+}D_{s}^-}\cdot{\BR}(D_{s}^+\rightarrow{\bar{K}^{*0}K^{+}})$ & &
\\ \hline
$22 \pm 6 \pm 6$~pb & 4140 & MARK III \\
$21.7 \pm 4.0$~pb & 4140 & This Analysis 
\\ \hline
\end{tabular}
\end{center}
\end{table}

\begin{figure}[!htbp]
\begin{center}
\hspace{2.5pt}
\includegraphics[width=14.5cm]{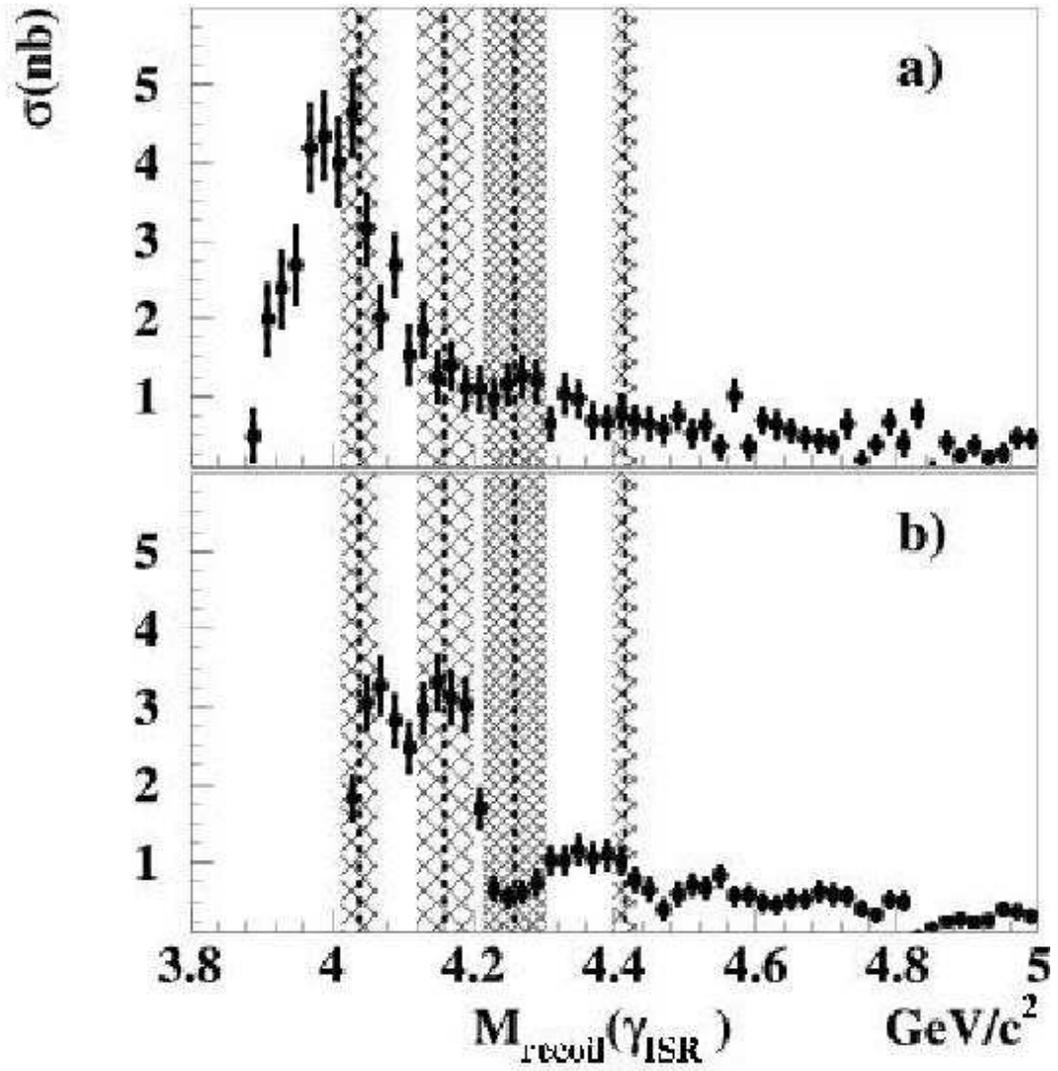}
\caption{The exclusive cross sections for $e^+e^-\rightarrow{D^{+}}D^{*-}$ (top) and $e^+e^-\rightarrow{D^{*+}}D^{*-}$ from Belle \cite{XS_Belle}.  The dashed lines correspond to the $\psi(4040)$, $\psi(4160)$, ${\rm{Y}}(4260)$, and $\psi(4415)$. }
\label{Belle_XSDSDSDSD}
\vspace{0.2cm}
\end{center}
\end{figure}

\begin{figure}[!htbp]
\begin{center}
\hspace{2.5pt}
\includegraphics[width=12.5cm]{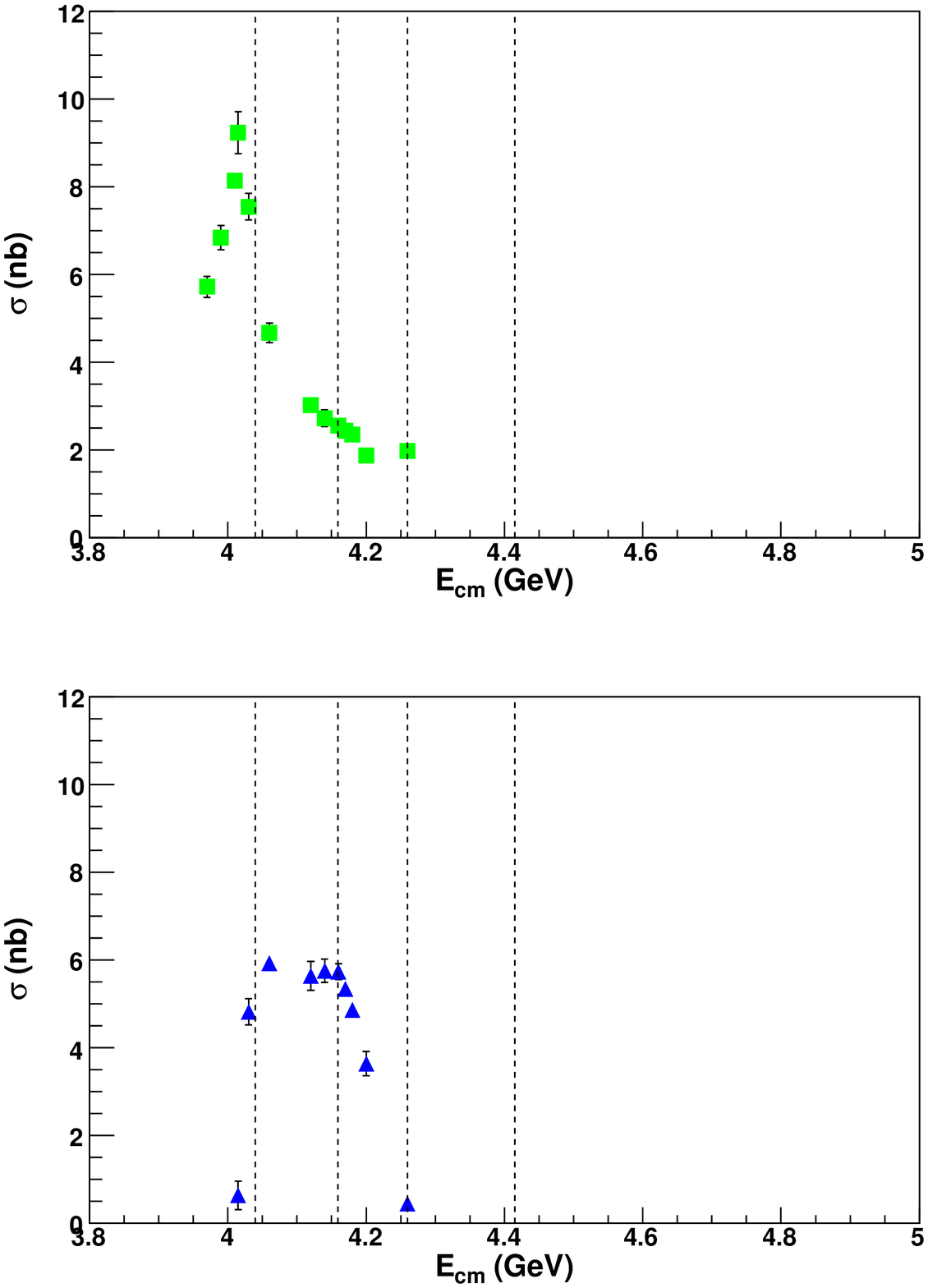}
\caption{The exclusive cross sections for $e^+e^-\rightarrow{D}\bar{D}^{*}$ (top) and $e^+e^-\rightarrow{D^{*}}\bar{D}^{*}$, after applying radiative corrections, for the work presented in this dissertation.  The dashed lines correspond to the $\psi(4040)$, $\psi(4160)$, ${\rm{Y}}(4260)$, and $\psi(4415)$.  There is a factor of two difference between these results and those of Belle in \FIG \ref{Belle_XSDSDSDSD}.}
\label{Our_RadCorDSDSDSD}
\vspace{0.2cm}
\end{center}
\end{figure}

During the final preparation of this dissertation the Belle Collaboration
\cite{XS_Belle} presented measurements of the cross sections for
$D^{+}\bar{D}^{*-}$ and $D^{*+}\bar{D}^{*-}$ in the center-of-mass energy range from 3.7 to 5.0 GeV using ISR events produced in $e^+e^-$ annihilation at the $\Upsilon(4S)$.
Their results are only for the charged $D$ pairs, so their results must be multiplied by a factor of 2 for comparison with the measurements presented in this thesis.  Also, they present, up to higher-order radiative effects, Born-level cross sections.  Based on a previous BaBar analysis \cite{Belle_early_Rad}, Belle estimated these higher-order corrections to be on the order of a percent.  These small corrections are well within the systematic errors, which are $\sim$10\%, and are therefore ignored.  To do a comparison, I radiatively corrected the exclusive $D^{*}\bar{D}$ and $D^{*}\bar{D}^{*}$ from \TABS \ref{XS_results_fits_1} and \ref{XS_results_fits_2} following the method described in Sect. \ref{sec:radcor}, where the shape of the cross section needed for the integral term was taken from \FIG \ref{fig:EvtGen_DD}.  As can be seen by comparing Belle's result (\FIG \ref{Belle_XSDSDSDSD}) to our result (\FIG \ref{Our_RadCorDSDSDSD}) the agreement is quite good, keeping in mind the factor of two difference.  It should be noted that their results cover a wider range of ${\rm{E_{CM}}}$, but are less precise than those present here.

In addition, the BaBar experiment \cite{XS_BaBar} has presented data on the
$D\bar{D}$ final state, also using the ISR technique in $e^+e^-$ annihilation at the $\Upsilon(4S)$.
BaBar's results are complicated by the fact that they combine several modes which have not been efficiency-corrected.  Therefore, a direct comparison to their results is of limited value, because, the BaBar efficiency is acknowledged not to be flat as a function of the invariant mass.  However, they do see a dip in the number of observed $D\bar{D}$ events in the neighborhood of 4015 MeV, which is in agreement with our observation. 

Taking the partial widths calculated at $4160$ MeV by T. Barnes
\cite{Barnes}, which are reproduced in \TAB \ref{tab:Barnes}, and
translating them into percentages, a comparison can be made to our
results.  This is shown in \TAB \ref{tab:Barnes_compare}. The disagreement between 
the measurements and Barnes's predictions are quite large, especially
for $D\bar{D}$, $D^*\bar{D}$ and $D_s\bar{D}_s$.
\begin{table}[!htbp]
\begin{center}
\caption{Percentages of event types present at center-of-mass energy
of $4160$ MeV for this analysis as compared to the T. Branes predictions \cite{Barnes}.}
\vspace{0.2cm}
\label{tab:Barnes_compare}
\begin{tabular}{|c|c|c|c|c|c|}
\hline
{Center-of-Mass}& {\(D\bar{D}\)} & {\(D^{*}\bar{D}\)}&
{\(D^{*}\bar{D}^{*}\)}& {\(D_{s}^{+}{D_{s}^{-}}\)}&
{\(D^{*+}_{s}{D^{-}_{s}}\)} \\
 Energy (MeV)&&&&&
\\ \hline
$4160$ (This Analysis)& $3.9\pm0.5$ & $28.2\pm1.8$ & $54.1\pm3.6$ &$-$ & $9.7\pm0.6$
\\ \hline
$4159$ (Barnes)& $21.6$ & $0.5$ & $47.3$ & $10.8$ & $18.9$
\\ \hline
\end{tabular}
\end{center}
\end{table}
  
\begin{figure}[!htbp]
\begin{center}
\hspace{2.5pt}
\includegraphics[width=14.5cm]{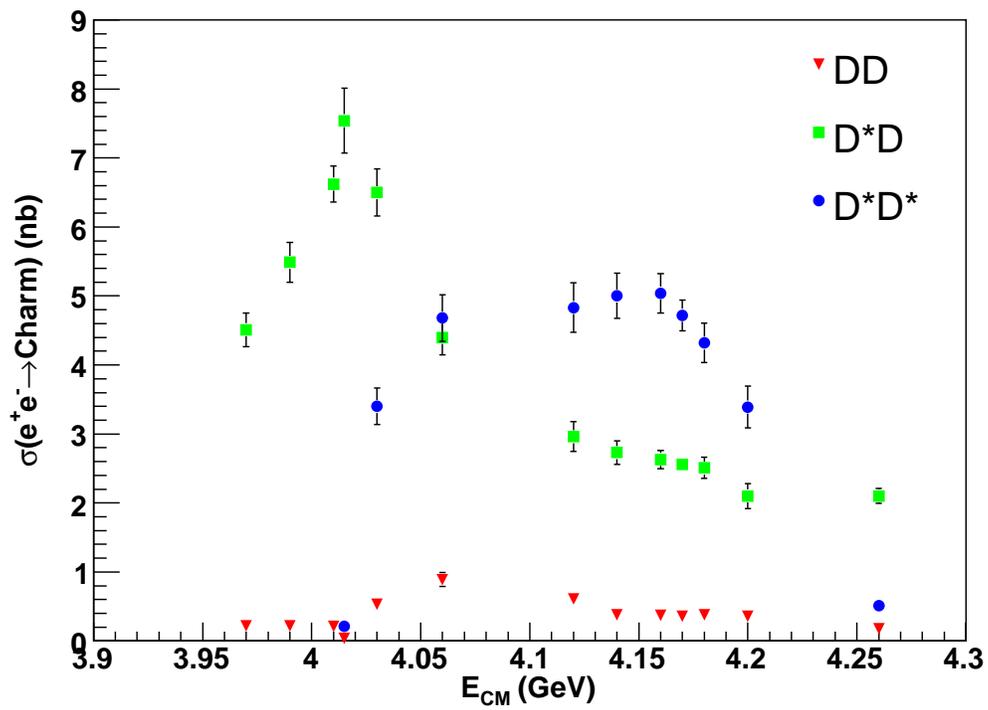}
\caption{Results for the non-strange $D$-meson cross sections.  The
error bars include both statistical and systematic errors.}
\label{FIG:Concl_sysDD}
\vspace{0.2cm}
\end{center}
\end{figure}

\begin{figure}[!htbp]
\begin{center}
\hspace{2.5pt}
\includegraphics[width=14.5cm]{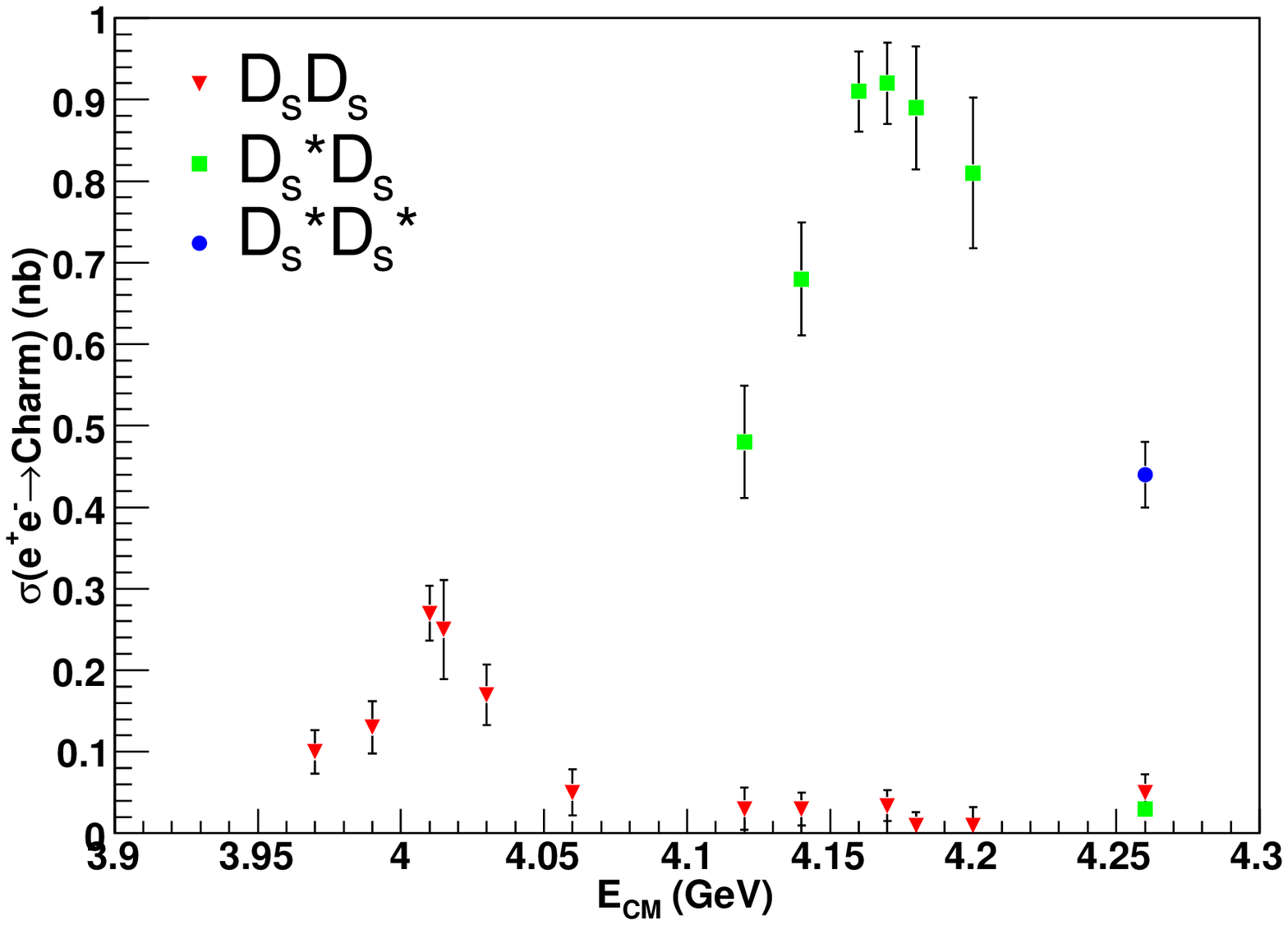}
\caption{Results for the $D_{s}$-meson cross sections.  The
error bars include both statistical and systematic errors.}
\label{FIG:Concl_sysDsDs}
\vspace{0.2cm}
\end{center}
\end{figure}

The analysis results for the cross sections two-body states (\FIGS \ref{FIG:Concl_sysDD}, \ref{FIG:Concl_sysDsDs}) are in
qualitative agreement with Eichten's \cite{Eichten} prediction, \FIGS
\ref{fig:Eichten_d} and \ref{fig:Eichten_ds}.  A more quantitative
comparison with Eichten's model awaits his updated calculations using the correct charmed-meson masses.  

\begin{figure}[p]
\begin{center}
\hspace{2.5pt}
\includegraphics[width=14.5cm]{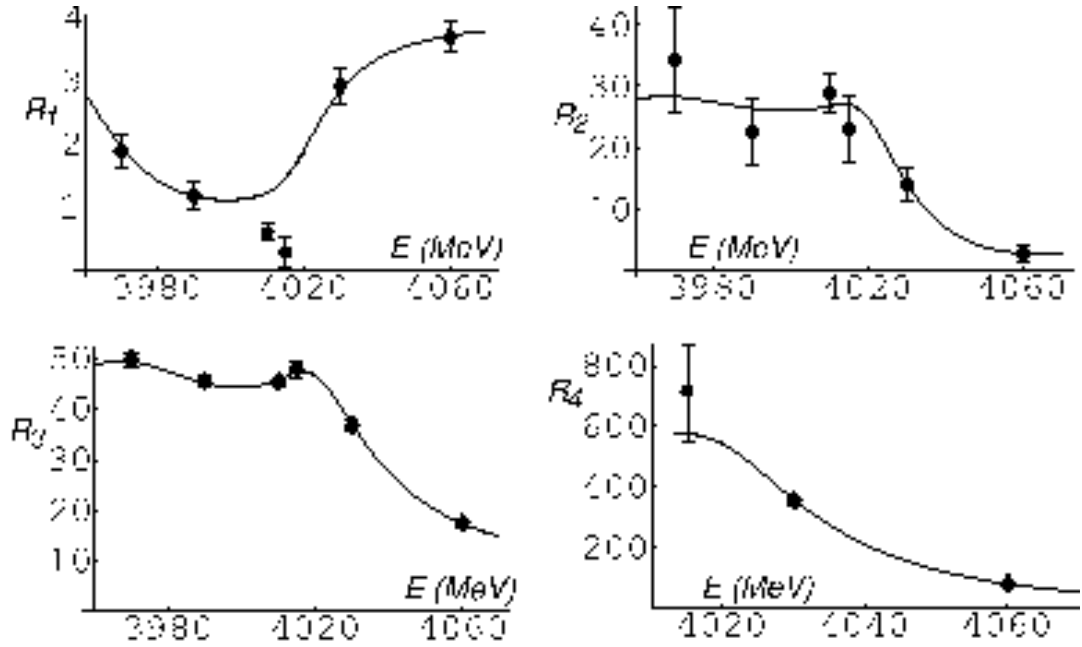}
\caption{The fit results from Ref. \cite{Vol_2} which used preliminary CLEO-c results. The data points that are excluded in the fit are shown with circles ($D\bar{D}$ at 4010 and 4015 MeV and $D^*\bar{D}^*$ at 4015 MeV). The fit are to dimensionless rate coefficients $R_{k}$ where $R_{k}\propto\sigma_{k}$ ($R_{1}$ (top left) corresponds to $D\bar{D}$, $R_{2}$ (top right) corresponds to $D_{s}\bar{D}_{s}$,  $R_{3}$ (bottom left) corresponds to $D\bar{D}^{*}$, and $R_{4}$ (bottom right) corresponds to $D^{*}\bar{D}^{*}$).}
\label{Voloshin_orig}
\vspace{0.2cm}
\end{center}
\end{figure}

\begin{figure}[p]
\begin{center}
\hspace{2.5pt}
\includegraphics[width=14.5cm]{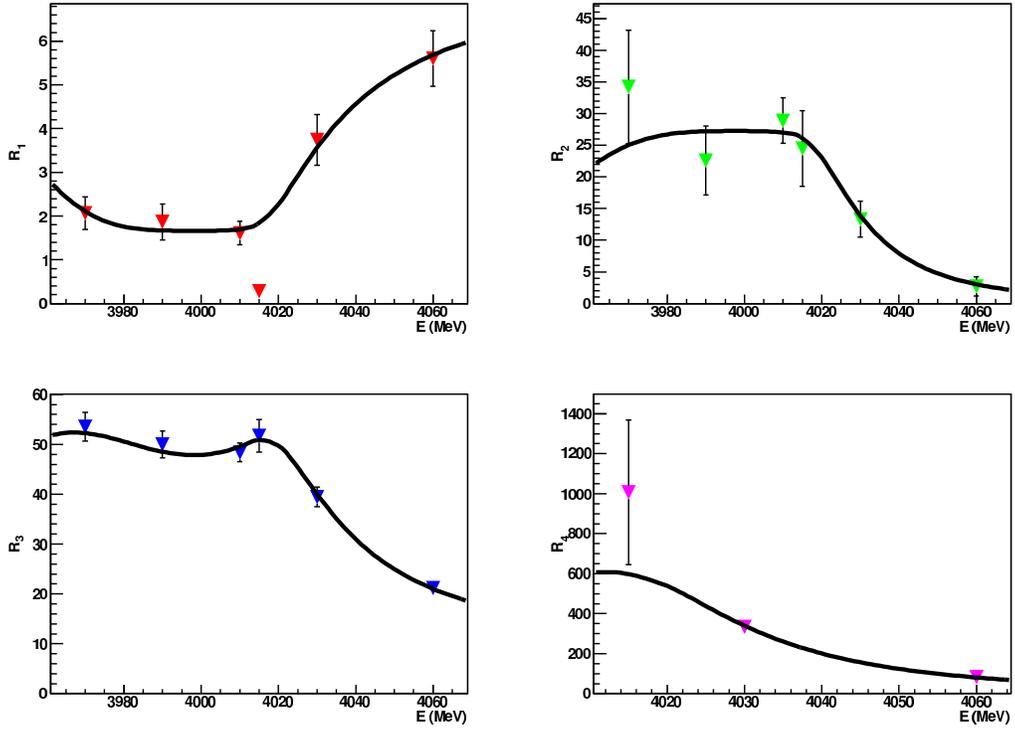}
\caption{Fit results, assuming a single resonance \cite{Vol_2}, using the cross sections presented in this dissertation. The fit are to dimensionless rate coefficients $R_{k}$ where $R_{k}\propto\sigma_{k}$ ($R_{1}$ (top left) corresponds to $D\bar{D}$, $R_{2}$ (top right) corresponds to $D_{s}\bar{D}_{s}$,  $R_{3}$ (bottom left) corresponds to $D\bar{D}^{*}$, and $R_{4}$ (bottom right) corresponds to $D^{*}\bar{D}^{*}$). }
\label{Voloshin_updated}
\vspace{0.2cm}
\end{center}
\end{figure}

Recently, there has been work on the theoretical side in regards to the cross sections in the neighborhood of the $D^*\bar{D}^*$ threshold.   Using very preliminary CLEO-c cross section results presented at the Flavor Physics and CP Violation Conference (FPCP) \cite{Ron_FPCP}, Voloshin \etal \cite{Vol_1, Vol_2}, using very general properties of amplitudes near a threshold, obtained resonance parameters assuming the cross sections between 3970 and 4060 MeV result from a single resonance.  In their investigation they determined that the preliminary cross sections can not be described by a single resonance, but in fact hint at the presence of two, with one of the resonances being narrow.  This can be found in their analysis when they struggled to describe the data by a single resonance which resulted in a $\chi^2/NDF=17.6/8$.  They determined that if they ignored the data points for the cross section at 4010 and 4015 MeV for $D\bar{D}$ as well as $D^*\bar{D}^*$ at 4015 MeV a statistically significant result is obtained, $\chi^2/NDF=3.0/5.0$ \FIG \ref{Voloshin_orig}.  The reason for the poor fit when including all data points is almost solely due to the data point at 4015 MeV in $D\bar{D}$, where the value for the cross section is at a minimum.  They suggest the existence of a new narrow resonance to describe this ``dip'' in the cross section.  Following the same fit procedure as they described \cite{Vol_2}, results are updated to reflect the cross sections presented in this dissertation.  These updated fits, shown in \FIG \ref{Voloshin_updated}, still agree with their initial assessment, that is because of the large dip in the $D\bar{D}$ cross section a single resonance is insufficient in describing the cross sections near the $D^*\bar{D}^*$ threshold.  The resonance parameters, with $D\bar{D}$ and $D^*\bar{D}^*$ at 4015 MeV left out the fit, are shown in \TAB \ref{Voloshin_updated_table}.  

\begin{table}[!htb]
\begin{center}
\caption{The resulting fit parameters from a single resonance fit to the cross sections, presented in this dissertation, around the $D^{*}\bar{D}^*$ threshold.  The $D\bar{D}$ and $D^*\bar{D}^*$ data points were excluded in the fit resulting in $\chi^2/{\rm{NDF}}=3.8/6$.  If all points were used the fit is quite poor, reflected in the resulting a $\chi^2/{\rm{NDF}}=36.3/8$, this is entirely because of $D\bar{D}$ at 4015 MeV.}
\label{Voloshin_updated_table}
\begin{tabular}{|c|c|}\hline
Fit parameters& Result from Fit
\\ \hline
$W_{0}$ & $4013 \pm 4$ MeV \\ \hline
$\Gamma_{0}$ & $66 \pm 8$ MeV \\ \hline
$w$ & $10.4 \pm 1.7$ MeV \\ \hline
$\Gamma_{D^*\bar{D}^*}$ & $15.4 \pm 1.3$ MeV \\ \hline
$\Gamma_{ee}$ & $1.9 \pm 0.7$ keV \\ \hline
\end{tabular}\end{center}\end{table}

\pagebreak
%%%%%%%%%%%%%%%%%%%%%%%%%%%%%%%%%%%%%%%%%%%%%%%%%%%%%%%%%%%%%%%%%%%%%%%%%%%%%%%
\section{Conclusion}
%%%%%%%%%%%%%%%%%%%%%%%%%%%%%%%%%%%%%%%%%%%%%%%%%%%%%%%%%%%%%%%%%%%%%%%%%%%%%%%
The CLEO-c scan was extremely successful in accomplishing the goals it put forth and as a result has provided a wealth of interesting results.

The location in center-of-mass energy that maximizes the yield of $D_{s}$ was determined by this analysis to be at 4170 MeV. The maximum occurs not with $D_{s}\bar{D}_{s}$ events, but rather with $D_{s}^*\bar{D}_{s}$ events.  It is with this information that CLEO-c has already been able to make significant progress, on the decay constant and hadronic branching fractions of the $D_s$ meson \cite{CLEO_DsBF,Ds_Decay}, with the data collected during the scan and in subsequent running at 4170 MeV.

\begin{figure}[!htbp]
\begin{center}
\hspace{2.5pt}
\includegraphics[width=14.5cm]{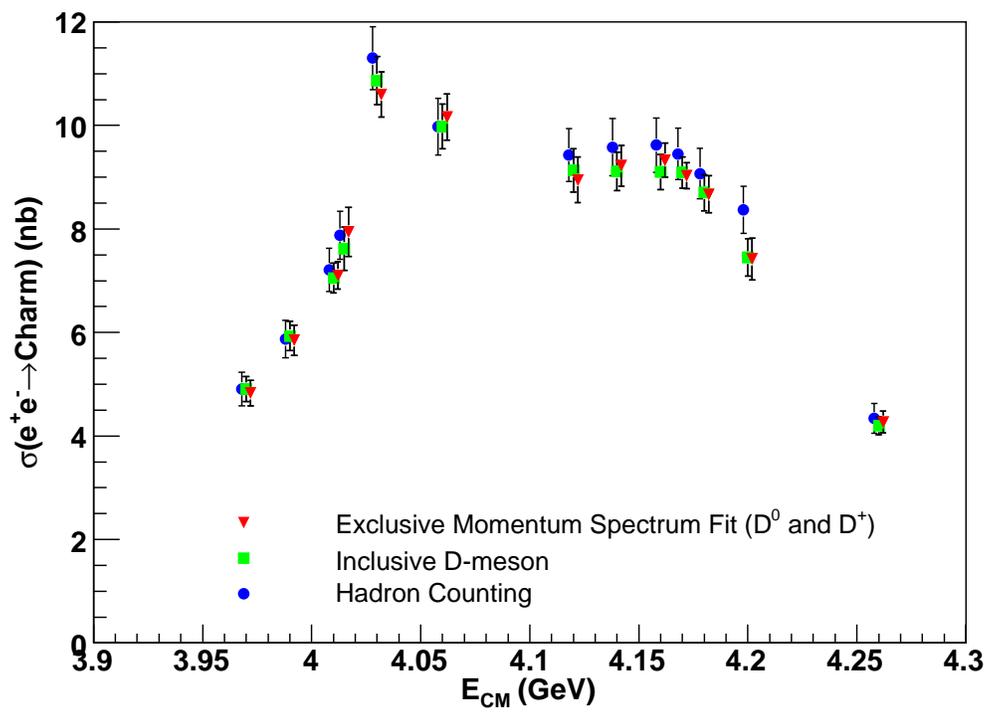}
\caption{Comparison of the total charm cross section using the 3
different methods. This is a graphical representation of Table
\ref{TAB:COM_SYS}.}
\label{FIG:Concl_sysAll}
\vspace{0.2cm}
\end{center}
\end{figure}

The total charm cross section between $3.97$ GeV and $4.26$ GeV has been measured and presented.
These results have been determined using three different techniques,
the exclusive $D$-meson, inclusive $D$-meson, and inclusive hadron
method, and are shown in \TAB \ref{TAB:COM_SYS} and graphically in \FIG \ref{FIG:Concl_sysAll}.  The agreement between
these three different methods is very good and provides a powerful
cross check on the results.   Radiative corrections were determined and
applied to the inclusive $D$ meson total charm result. The corrected cross sections are consistent with previous experimental results \cite{BES_R,CB_R} and are more precise (see \FIG \ref{RadCor_INXS_fig}).

\begin{figure}[!htbp]
\begin{center}
\hspace{2.5pt}
\includegraphics[width=14.5cm]{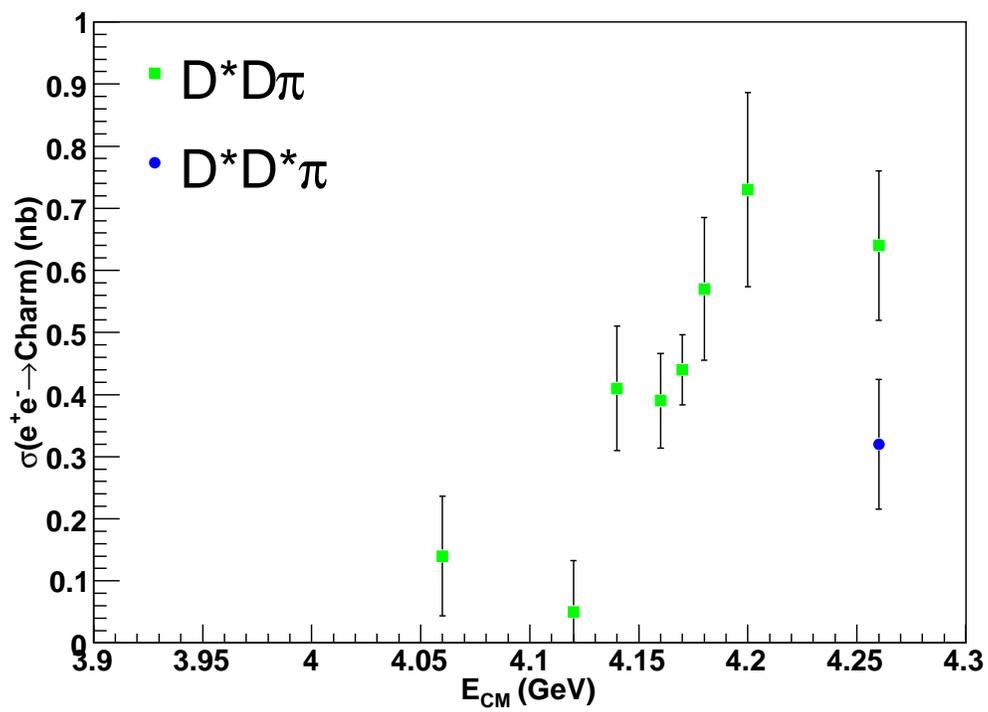}
\caption{Results for the multi-body cross sections.  The
error bars include both statistical and systematic errors.}
\label{FIG:Concl_sysDDpi}
\vspace{0.2cm}
\end{center}
\end{figure}

The composition of the total charm cross section, in terms of $D\bar{D}$,
$D^{*}\bar{D}$, $D^{*}\bar{D}^{*}$, $D_{s}\bar{D}_{s}$,
$D_{s}^{*}\bar{D}_{s}$, $D_{s}^{*}\bar{D}_{s}^{*}$, $D^*\bar{D}\pi$
and $D^*\bar{D}^*\pi$, at all energies that
were investigated, has been determined and presented. Results
are shown in \TABS \ref{XS_results_fits_1} and
\ref{XS_results_fits_2}, in addition to graphically in \FIGS
\ref{FIG:Concl_sysDD}, \ref{FIG:Concl_sysDsDs}, and
\ref{FIG:Concl_sysDDpi}.  There is qualitative agreement with the potential model put forth by Eichten \etal \cite{Eichten} for all modes excluding the multi-body production, which was not predicted. The existence of a peak in the $D^{*}\bar{D}$ and $D_{s}\bar{D}_{s}$ channels at the $D^{*}\bar{D}^*$ threshold, along with the observation that there is a minimum in the $D\bar{D}$, can be interpreted as a possible new narrow resonance \cite{Vol_2}.  Unfortunately, the available data sample is insufficient to investigate this latter idea further. 

The $D^{*}\bar{D}^{*}$ cross section shows a plateau region above its threshold, whereas the $D^*\bar{D}$ peaks right at the threshold, in agreement with the recently presented results by the Belle Collaboration \cite{XS_Belle}.  

Also, events containing $D$ mesons plus an additional pion, referred to as multi-body, have been shown to exists in the
energy region investigated. \FIGS \ref{fig:MultiBody_DSDPiPlus_4170}, \ref{fig:MultiBody_DSDPiPlus_4170_2}, and \ref{fig:MultiBody_DSDPiPlus_4260}, show the cross sections that have been determined.

Using the data sample collected at 4260 MeV we searched for insight into the nature of the $Y(4260)$
state.  By applying the techniques developed with the larger data sample collected at 4170 MeV we find that, besides the presence of the $D^*D^*\pi$ final state and larger overall production of multi-body, there is essentially nothing to distinguish the charm production at this energy from that at our lower-energy points.  Although studies are still ongoing, we have preliminarily concluded that our data sample is insufficient to determine the contribution of states including excited charmed meson states (e.g. $D_{1}\bar{D}$) that are expected under some $Y(4260)$ interpretations \cite{CLEO_Y}.

As for the future, it is unlikely that time will be available to allow further investigation of the Y(4260) and region around the $D^*\bar{D}^*$ to be done at CLEO-c.  However, the next stage in the BES experiment, BES-III \cite{BESIII}, is planning to investigate in-depth the region around $D^*\bar{D}^*$ and should potentially be able to add insight to this region of interest.  In addition to BES-III, both Belle and BaBar have shown that they too can investigate this region and we are impatiently awaiting their updated results.

\begin{table}[!htb]
\begin{center}
\caption{Comparison of the total charm cross section using the 3
different methods. First error is statistical and the second systematic.}
\label{TAB:COM_SYS}\vspace{0.2cm}
\begin{tabular}{|c|c|c|c|}\hline
{Energy}& Exclusive & Inclusive &
Hadron\\
(MeV)& $D$-meson (nb)& $D$-meson (nb)& Counting (nb)
\\ \hline
3970 & $ 4.83\pm 0.19 \pm 0.15$ & $ 4.91\pm 0.18 \pm 0.16$ & $ 4.91\pm 0.13 \pm 0.30$ \\ \hline
3990 & $ 5.85\pm 0.23 \pm 0.19$ & $ 5.93\pm 0.21 \pm 0.19$ & $ 5.87\pm 0.14 \pm 0.34$ \\ \hline
4010 & $ 7.10\pm 0.14 \pm 0.23$ & $ 7.05\pm 0.17 \pm 0.23$ & $ 7.21\pm 0.12 \pm 0.40$ \\ \hline
4015 & $ 7.94\pm 0.41 \pm 0.26$ & $ 7.62\pm 0.34 \pm 0.25$ & $ 7.88\pm 0.18 \pm 0.43$ \\ \hline
4030 & $ 10.60\pm 0.34 \pm 0.27$ & $ 10.87\pm 0.28 \pm 0.37$ & $ 11.30\pm 0.15 \pm 0.59$ \\ \hline
4060 & $ 10.16\pm 0.36 \pm 0.27$ & $ 9.98\pm 0.26 \pm 0.34$ & $ 9.98\pm 0.14 \pm 0.53$ \\ \hline
4120 & $ 8.95\pm 0.37 \pm 0.25$ & $ 9.13\pm 0.28 \pm 0.31$ & $ 9.43\pm 0.15 \pm 0.49$ \\ \hline
4140 & $ 9.22\pm 0.29 \pm 0.26$ & $ 9.11\pm 0.22 \pm 0.30$ & $ 9.58\pm 0.24 \pm 0.50$ \\ \hline
4160 & $ 9.33\pm 0.20 \pm 0.26$ & $ 9.10\pm 0.15 \pm 0.30$ & $ 9.62\pm 0.17 \pm 0.50$ \\ \hline
4170 & $ 9.03\pm 0.04 \pm 0.25$ & $ 9.09\pm 0.07 \pm 0.30$ & $ 9.45\pm 0.09 \pm 0.49$ \\ \hline
4180 & $ 8.67\pm 0.27 \pm 0.24$ & $ 8.70\pm 0.20 \pm 0.29$ & $ 9.07\pm 0.12 \pm 0.47$ \\ \hline
4200 & $ 7.42\pm 0.35 \pm 0.20$ & $ 7.45\pm 0.26 \pm 0.25$ & $ 8.37\pm 0.14 \pm 0.43$ \\ \hline
4260 & $ 4.27\pm 0.16 \pm 0.14$ & $ 4.20\pm 0.10 \pm 0.14$ & $ 4.34\pm 0.16 \pm 0.23$ \\ \hline
\end{tabular}\end{center}\end{table}

%%%%%%%%%%%%%%%%%%%%%%%%%%%%%%%%%%%%%%%%%%%%%%%%%%%%%%%%%%%%%%%%%%%%%%%%%%
%
%	REFERENCES
%
%%%%%%%%%%%%%%%%%%%%%%%%%%%%%%%%%%%%%%%%%%%%%%%%%%%%%%%%%%%%%%%%%%%%%%%%%%

\begin{table}[htbp]
\begin{center}
\caption{Efficiencies (units of $10^{-2}$) at each scan-energy point for 
selection of $D_s^+$ through various decay modes in the three exclusive
event types.}
\label{ExclusiveDs_eff_table}
\vspace{0.2cm}
\begin{tabular}{|c|c|c|c|c|}\hline
{Mode}&{E$_{cm}$ MeV}& {$D_{s}\bar{D}_{s}$}& {$D_{s}^{*}\bar{D}_{s}$}&
{$D_{s}^{*}\bar{D}_{s}^{*}$} \\ \hline
$K_{s}K^{+}$& $3970$ & $56.2 \pm1.7$ & $-\pm-$ & $-\pm-$\\
& $3990$ & $55.2 \pm1.7$ & $-\pm-$ & $-\pm-$\\
& $4010$ & $57.0 \pm1.7$ & $-\pm-$ & $-\pm-$\\
& $4015$ & $57.8 \pm1.4$ & $-\pm-$ & $-\pm-$\\
& $4030$ & $58.3 \pm1.5$ & $-\pm-$ & $-\pm-$\\
& $4060$ & $56.7 \pm1.5$ & $-\pm-$ & $-\pm-$\\
& $4120$ & $57.6 \pm1.9$ & $55.8 \pm 1.3$ & $-\pm-$\\
& $4140$ & $50.6 \pm1.4$ & $58.2 \pm 0.9$ & $-\pm-$\\
& $4160$ & $54.3 \pm1.9$ & $55.9 \pm 1.3$ & $-\pm-$\\
& $4180$ & $52.1 \pm1.9$ & $55.3 \pm 1.3$ & $-\pm-$\\
& $4200$ & $52.9 \pm1.9$ & $53.9 \pm 1.4$ & $-\pm-$\\
& $4260$ & $62.0 \pm4.4$ & $58.1 \pm 3.1$ & $56.4 \pm1.5$ \\ \hline

$\eta\pi^{+}$& $3970$ & $39.2 \pm2.3$ & $-\pm-$ & $-\pm-$\\
& $3990$ & $39.3 \pm2.3$ & $-\pm-$ & $-\pm-$\\
& $4010$ & $37.9 \pm2.2$ & $-\pm-$ & $-\pm-$\\
& $4015$ & $42.0 \pm1.9$ & $-\pm-$ & $-\pm-$\\
& $4030$ & $53.7 \pm2.1$ & $-\pm-$ & $-\pm-$\\
& $4060$ & $39.7 \pm2.0$ & $-\pm-$ & $-\pm-$\\
& $4120$ & $33.1 \pm2.4$ & $38.3 \pm 1.8$ & $-\pm-$\\
& $4140$ & $37.1 \pm1.8$ & $40.7 \pm 1.3$ & $-\pm-$\\
& $4160$ & $26.6 \pm2.3$ & $38.3 \pm 1.8$ & $-\pm-$\\
& $4180$ & $37.9 \pm2.5$ & $38.8 \pm 1.8$ & $-\pm-$\\
& $4200$ & $41.1 \pm2.6$ & $34.9 \pm 1.8$ & $-\pm-$\\
& $4260$ & $42.6 \pm6.0$ & $39.2 \pm 4.2$ & $40.0 \pm2.0$ \\ \hline
$\phi\pi^{+}$& $3970$ & $42.7 \pm1.4$ & $-\pm-$ & $-\pm-$\\
& $3990$ & $42.1 \pm1.4$ & $-\pm-$ & $-\pm-$\\
& $4010$ & $42.9 \pm1.4$ & $-\pm-$ & $-\pm-$\\
& $4015$ & $41.0 \pm1.1$ & $-\pm-$ & $-\pm-$\\
& $4030$ & $43.8 \pm1.3$ & $-\pm-$ & $-\pm-$\\
& $4060$ & $43.1 \pm1.3$ & $-\pm-$ & $-\pm-$\\
& $4120$ & $39.6 \pm1.6$ & $40.5 \pm 1.1$ & $-\pm-$\\
& $4140$ & $39.8 \pm1.1$ & $41.8 \pm 0.8$ & $-\pm-$\\
& $4160$ & $41.6 \pm1.6$ & $41.3 \pm 1.1$ & $-\pm-$\\
& $4180$ & $36.7 \pm1.5$ & $42.8 \pm 1.1$ & $-\pm-$\\
& $4200$ & $42.7 \pm1.6$ & $42.2 \pm 1.1$ & $-\pm-$\\
& $4260$ & $43.8 \pm3.7$ & $43.4 \pm 2.6$ & $41.8 \pm1.2$ \\ \hline
\end{tabular}
\end{center}
\end{table}
\begin{table}[htbp]
\begin{center}
%\caption{The efficiencies for $D_{s}^{+}$,
%which will be used to determine the exclusive cross sections.}
%\label{ExclusiveDs_eff_table}
%\vspace{0.2cm}
\begin{tabular}{|c|c|c|c|c|}\hline
{Mode}&{E$_{cm}$ MeV}& {$D_{s}\bar{D}_{s}$}& {$D_{s}^{*}\bar{D}_{s}$}&
{$D_{s}^{*}\bar{D}_{s}^{*}$} \\ \hline
$K^{*}K^{+}$& $3970$ & $39.1 \pm1.2$ & $-\pm-$ & $-\pm-$\\
& $3990$ & $40.0 \pm1.2$ & $-\pm-$ & $-\pm-$\\
& $4010$ & $42.5 \pm1.2$ & $-\pm-$ & $-\pm-$\\
& $4015$ & $41.3 \pm1.0$ & $-\pm-$ & $-\pm-$\\
& $4030$ & $42.0 \pm1.1$ & $-\pm-$ & $-\pm-$\\
& $4060$ & $41.5 \pm1.1$ & $-\pm-$ & $-\pm-$\\
& $4120$ & $40.1 \pm1.4$ & $40.8 \pm 1.0$ & $-\pm-$\\
& $4140$ & $39.2 \pm1.0$ & $42.0 \pm 0.7$ & $-\pm-$\\
& $4160$ & $42.7 \pm1.4$ & $42.0 \pm 1.0$ & $-\pm-$\\
& $4180$ & $43.3 \pm1.4$ & $42.6 \pm 1.0$ & $-\pm-$\\
& $4200$ & $37.7 \pm1.4$ & $41.6 \pm 1.0$ & $-\pm-$\\
& $4260$ & $41.2 \pm3.2$ & $35.0 \pm 2.2$ & $41.4 \pm1.1$ \\ \hline
$\eta\rho^{+}$& $3970$ & $19.2 \pm0.7$ & $-\pm-$ & $-\pm-$\\
& $3990$ & $17.1 \pm0.7$ & $-\pm-$ & $-\pm-$\\
& $4010$ & $16.8 \pm0.7$ & $-\pm-$ & $-\pm-$\\
& $4015$ & $16.9 \pm0.6$ & $-\pm-$ & $-\pm-$\\
& $4030$ & $24.0 \pm0.7$ & $-\pm-$ & $-\pm-$\\
& $4060$ & $16.5 \pm0.6$ & $-\pm-$ & $-\pm-$\\
& $4120$ & $16.7 \pm0.8$ & $16.9 \pm 0.5$ & $-\pm-$\\
& $4140$ & $17.1 \pm0.6$ & $16.0 \pm 0.4$ & $-\pm-$\\
& $4160$ & $14.9 \pm0.7$ & $16.1 \pm 0.5$ & $-\pm-$\\
& $4180$ & $16.3 \pm0.8$ & $16.2 \pm 0.5$ & $-\pm-$\\
& $4200$ & $16.9 \pm0.8$ & $17.1 \pm 0.6$ & $-\pm-$\\
& $4260$ & $15.4 \pm1.8$ & $18.7 \pm 1.3$ & $17.4 \pm0.6$ \\ \hline
$\eta^{'}\pi^{+}$& $3970$ & $31.9 \pm2.1$ & $-\pm-$ & $-\pm-$\\
& $3990$ & $33.2 \pm2.2$ & $-\pm-$ & $-\pm-$\\
& $4010$ & $30.9 \pm2.1$ & $-\pm-$ & $-\pm-$\\
& $4015$ & $34.8 \pm1.8$ & $-\pm-$ & $-\pm-$\\
& $4030$ & $32.4 \pm1.9$ & $-\pm-$ & $-\pm-$\\
& $4060$ & $26.7 \pm1.8$ & $-\pm-$ & $-\pm-$\\
& $4120$ & $21.2 \pm2.1$ & $28.8 \pm 1.7$ & $-\pm-$\\
& $4140$ & $27.6 \pm1.6$ & $25.8 \pm 1.1$ & $-\pm-$\\
& $4160$ & $29.2 \pm2.4$ & $29.9 \pm 1.7$ & $-\pm-$\\
& $4180$ & $29.9 \pm2.4$ & $32.8 \pm 1.7$ & $-\pm-$\\
& $4200$ & $29.5 \pm2.4$ & $29.8 \pm 1.7$ & $-\pm-$\\
& $4260$ & $16.1 \pm4.5$ & $35.0 \pm 4.1$ & $28.2 \pm1.8$ \\ \hline
\end{tabular}
\end{center}
\end{table}
\begin{table}[!htbp]
\begin{center}
%\caption{The efficiencies for $D_{s}^{+}$,
%which will be used to determine the exclusive cross sections.}
%\label{ExclusiveDs_eff_table}
%\vspace{0.2cm}
\begin{tabular}{|c|c|c|c|c|}\hline
{Mode}&{E$_{cm}$ MeV}& {$D_{s}\bar{D}_{s}$}& {$D_{s}^{*}\bar{D}_{s}$}&
{$D_{s}^{*}\bar{D}_{s}^{*}$} \\ \hline
$\eta^{'}\rho^{+}$&$3970$ & $16.0 \pm1.1$ & $-\pm-$ & $-\pm-$\\
& $3990$ & $12.9 \pm1.0$ & $-\pm-$ & $-\pm-$\\
& $4010$ & $13.4 \pm1.0$ & $-\pm-$ & $-\pm-$\\
& $4015$ & $13.7 \pm0.8$ & $-\pm-$ & $-\pm-$\\
& $4030$ & $13.0 \pm0.9$ & $-\pm-$ & $-\pm-$\\
& $4060$ & $11.6 \pm0.8$ & $-\pm-$ & $-\pm-$\\
& $4120$ & $11.7 \pm1.0$ & $14.2 \pm 0.8$ & $-\pm-$\\
& $4140$ & $12.0 \pm0.7$ & $12.5 \pm 0.5$ & $-\pm-$\\
& $4160$ & $11.1 \pm1.0$ & $11.8 \pm 0.7$ & $-\pm-$\\
& $4180$ & $15.0 \pm1.2$ & $12.1 \pm 0.7$ & $-\pm-$\\
& $4200$ & $14.0 \pm1.1$ & $13.3 \pm 0.8$ & $-\pm-$\\
& $4260$ & $14.2 \pm2.0$ & $14.9 \pm 1.9$ & $15.0 \pm0.9$ \\ \hline
$\phi\rho^{+}$& $3970$ & $13.6 \pm0.7$ & $-\pm-$ & $-\pm-$\\
& $3990$ & $13.4 \pm0.7$ & $-\pm-$ & $-\pm-$\\
& $4010$ & $13.4 \pm0.7$ & $-\pm-$ & $-\pm-$\\
& $4015$ & $12.9 \pm0.6$ & $-\pm-$ & $-\pm-$\\
& $4030$ & $15.5 \pm0.7$ & $-\pm-$ & $-\pm-$\\
& $4060$ & $12.6 \pm0.6$ & $-\pm-$ & $-\pm-$\\
& $4120$ & $13.7 \pm0.8$ & $12.5 \pm 0.6$ & $-\pm-$\\
& $4140$ & $14.2 \pm0.6$ & $11.8 \pm 0.4$ & $-\pm-$\\
& $4160$ & $14.9 \pm0.8$ & $12.3 \pm 0.5$ & $-\pm-$\\
& $4180$ & $15.6 \pm0.9$ & $13.4 \pm 0.6$ & $-\pm-$\\
& $4200$ & $15.8 \pm0.9$ & $13.6 \pm 0.6$ & $-\pm-$\\ 
& $4260$ & $15.2 \pm2.0$ & $15.5 \pm 1.4$ & $11.0 \pm0.6$ \\ \hline
\end{tabular}
\end{center}
\end{table}

\begin{table}[!htbp]
\begin{center}
\caption{Numbers of events at E$_{cm}=3671$ MeV for various event
types which are used for continuum subtraction in the Hadron Counting Method to determine the
total charm cross section.}
\label{HadCount_breakdown_off}
\vspace{0.2cm}
\begin{tabular}{c|c}\hline
Off-Res Data           &$   244663 \pm 495$ \\
B.W. Tail of $\psi(2S)$      &$           7201 \pm 493$ \\
Radiative Return to $J/\psi$   &$               10240 \pm 157$ \\
$\tau^+\tau^-$          &$           9227 \pm 55$ \\
$e^+e^-$               &$     3058 \pm 649$ \\
$\mu^+\mu^-$        &$            184 \pm 21$ \\ \hline
Efficiency Corrected Num. of $q\bar{q}$ &$ 378420 \pm 1934$ \\ \hline \hline
\end{tabular}
\end{center}
\end{table}

\begin{table}[!htbp]
\begin{center}
\caption{Numbers of events at the scan energies for various event
types which are used in the Hadron Counting Method to determine the
total charm cross section.}
\label{HadCount_breakdown}
\vspace{0.2cm}
\begin{tabular}{c|c}\hline
E$_{cm}=3970$ MeV \\ \hline
Number of Raw Hadronic Events                  &$   60811 \pm 247$\\
Scaled $q\bar{q}$              &$  38041 \pm 233$\\
Radiative Return to $\psi(2S)$                &$    2352 \pm 11$\\
Radiative Return to $J/\psi$               &$  1059 \pm 8$\\
Radiative Return to $\psi(3770)$               &$   332 \pm 1$\\
$\tau^+\tau^-$                    &$  2915 \pm 98$\\
$e^+e^-$                     &$  340 \pm 133$\\
$\mu^+\mu^-$                     &$  32 \pm 5$\\
Number of Beam Gas Events                     &$  439 \pm 102$\\ \hline
Efficiency Corrected Num.   &$   18926 \pm 485$\\ \hline \hline
& \\ 
E$_{cm}=3990$ MeV \\ \hline
Number of Raw Hadronic Events                  &$   54818 \pm 234$\\
Scaled $q\bar{q}$              &$  32649 \pm 201$\\
Radiative Return to $\psi(2S)$                &$    1869 \pm 9$\\
Radiative Return to $J/\psi$               &$  889 \pm 6$\\
Radiative Return to $\psi(3770)$               &$   264 \pm 1$\\
$\tau^+\tau^-$                    &$  2365 \pm 84$\\
$e^+e^-$                     &$  471 \pm 140$\\
$\mu^+\mu^-$                     &$  28 \pm 5$\\
Number of Beam Gas Events                     &$  341 \pm 122$\\ \hline
Efficiency Corrected Num.   &$   19696 \pm 458$\\ \hline \hline
& \\
E$_{cm}=4010$ MeV \\ \hline
Number of Raw Hadronic Events                  &$   97629 \pm 312$\\
Scaled $q\bar{q}$              &$  54635 \pm 327$\\
Radiative Return to $\psi(2S)$                &$    2929 \pm 13$\\
Radiative Return to $J/\psi$               &$  1429 \pm 10$\\
Radiative Return to $\psi(3770)$               &$   408 \pm 2$\\
$\tau^+\tau^-$                    &$  4128 \pm 144$\\
$e^+e^-$                     &$  635 \pm 212$\\
$\mu^+\mu^-$                     &$  46 \pm 8$\\
Number of Beam Gas Events                     &$  565 \pm 173$\\ \hline
Efficiency Corrected Num.   &$   40539 \pm 679$\\ \hline \hline
\end{tabular}
\end{center}
\end{table}

\begin{table}[!htbp]
\begin{center}
\begin{tabular}{c|c}\hline
E$_{cm}=4015$ MeV \\ \hline
Number of Raw Hadronic Events                  &$   26436 \pm 163$\\
Scaled $q\bar{q}$              &$  14240 \pm 96$\\
Radiative Return to $\psi(2S)$                &$    753 \pm 4$\\
Radiative Return to $J/\psi$               &$  371 \pm 3$\\
Radiative Return to $\psi(3770)$               &$   106 \pm 1$\\
$\tau^+\tau^-$                    &$  1137 \pm 39$\\
$e^+e^-$                     &$  165 \pm 55$\\
$\mu^+\mu^-$                     &$  12 \pm 2$\\
Number of Beam Gas Events                     &$  152 \pm 87$\\ \hline
Efficiency Corrected Num.   &$   11581 \pm 267$\\ \hline \hline
& \\
E$_{cm}=4030$ MeV \\ \hline
Number of Raw Hadronic Events                  &$   62354 \pm 250$\\
Scaled $q\bar{q}$              &$  29017 \pm 180$\\
Radiative Return to $\psi(2S)$                &$    1460 \pm 7$\\
Radiative Return to $J/\psi$               &$  733 \pm 5$\\
Radiative Return to $\psi(3770)$               &$   197 \pm 1$\\
$\tau^+\tau^-$                    &$  2252 \pm 78$\\
$e^+e^-$                     &$  593 \pm 145$\\
$\mu^+\mu^-$                     &$  25 \pm 4$\\
Number of Beam Gas Events                     &$  285 \pm 127$\\ \hline
Efficiency Corrected Num.   &$   33955 \pm 457$\\ \hline \hline
& \\
E$_{cm}=4060$ MeV \\ \hline
Number of Raw Hadronic Events                  &$   64117 \pm 253$\\
Scaled $q\bar{q}$              &$  31305 \pm 193$\\
Radiative Return to $\psi(2S)$                &$    1467 \pm 7$\\
Radiative Return to $J/\psi$               &$  797 \pm 6$\\
Radiative Return to $\psi(3770)$               &$   194 \pm 1$\\
$\tau^+\tau^-$                    &$  2582 \pm 87$\\
$e^+e^-$                     &$  277 \pm 108$\\
$\mu^+\mu^-$                     &$  26 \pm 4$\\
Number of Beam Gas Events                     &$  345 \pm 139$\\ \hline
Efficiency Corrected Num.   &$   32803 \pm 455$\\ \hline \hline
\end{tabular}
\end{center}
\end{table}

\begin{table}[!htbp]
\begin{center}
\begin{tabular}{c|c}\hline
E$_{cm}=4120$ MeV \\ \hline
Number of Raw Hadronic Events                  &$   51863 \pm 228$\\
Scaled $q\bar{q}$              &$  25694 \pm 161$\\
Radiative Return to $\psi(2S)$                &$    1027 \pm 5$\\
Radiative Return to $J/\psi$               &$  552 \pm 4$\\
Radiative Return to $\psi(3770)$               &$   126 \pm 1$\\
$\tau^+\tau^-$                    &$  2210 \pm 74$\\
$e^+e^-$                     &$  462 \pm 118$\\
$\mu^+\mu^-$                     &$  22 \pm 4$\\
Number of Beam Gas Events                     &$  280 \pm 131$\\ \hline
Efficiency Corrected Num.   &$   26016 \pm 411$\\ \hline \hline
& \\
E$_{cm}=4140$ MeV \\ \hline
Number of Raw Hadronic Events                  &$   91780 \pm 303$\\
Scaled $q\bar{q}$              &$  45202 \pm 220$\\
Radiative Return to $\psi(2S)$                &$    1705 \pm 7$\\
Radiative Return to $J/\psi$               &$  954 \pm 6$\\
Radiative Return to $\psi(3770)$               &$   221 \pm 1$\\
$\tau^+\tau^-$                    &$  3585 \pm 100$\\
$e^+e^-$                     &$  739 \pm 162$\\
$\mu^+\mu^-$                     &$  38 \pm 5$\\
Number of Beam Gas Events                     &$  518 \pm 163$\\ \hline
Efficiency Corrected Num.   &$   46910 \pm 548$\\ \hline \hline
& \\
E$_{cm}=4160$ MeV \\ \hline
Number of Raw Hadronic Events                  &$   190764 \pm 437$\\
Scaled $q\bar{q}$              &$  93833 \pm 398$\\
Radiative Return to $\psi(2S)$                &$    3467 \pm 11$\\
Radiative Return to $J/\psi$               &$  1876 \pm 11$\\
Radiative Return to $\psi(3770)$               &$   400 \pm 2$\\
$\tau^+\tau^-$                    &$  7761 \pm 192$\\
$e^+e^-$                     &$  1417 \pm 301$\\
$\mu^+\mu^-$                     &$ 78 \pm 9$\\
Number of Beam Gas Events                     &$  1019 \pm 212$\\ \hline
Efficiency Corrected Num.   &$   97750 \pm 879$\\ \hline \hline
\end{tabular}
\end{center}
\end{table}

\begin{table}[!htbp]
\begin{center}
\label{HadCount_breakdown_last}
\begin{tabular}{c|c}\hline
E$_{cm}=4170$ MeV \\ \hline
Number of Raw Hadronic Events                  &$   3301243 \pm 1817$\\
Scaled $q\bar{q}$              &$  1641982 \pm 9325$\\
Radiative Return to $\psi(2S)$                &$    59506 \pm 239$\\
Radiative Return to $J/\psi$               &$  33037 \pm 252$\\
Radiative Return to $\psi(3770)$               &$   7001 \pm 28$\\
$\tau^+\tau^-$                    &$  138524 \pm 4746$\\
$e^+e^-$                     &$  24731 \pm 7419$\\
$\mu^+\mu^-$                     &$  1361 \pm 227$\\
Number of Beam Gas Events                     &$  5256.47 \pm 205.89$\\ \hline
Efficiency Corrected Num.   &$   1683110 \pm 15884$\\ \hline \hline
& \\
E$_{cm}=4180$ MeV \\ \hline
Number of Raw Hadronic Events                  &$   103834 \pm 322$\\
Scaled $q\bar{q}$              &$  52037 \pm 310$\\
Radiative Return to $\psi(2S)$                &$    1806 \pm 8$\\
Radiative Return to $J/\psi$               &$  1009 \pm 8$\\
Radiative Return to $\psi(3770)$               &$   221 \pm 1$\\
$\tau^+\tau^-$                    &$  4744 \pm 156$\\
$e^+e^-$                     &$  840 \pm 234$\\
$\mu^+\mu^-$                     &$  43 \pm 7$\\
Number of Beam Gas Events                     &$  613 \pm 131$\\ \hline
Efficiency Corrected Num.   &$   51410 \pm 662$\\ \hline \hline
& \\
E$_{cm}=4200$ MeV \\ \hline
Number of Raw Hadronic Events                  &$   49296 \pm 222$\\
Scaled $q\bar{q}$              &$  25592 \pm 161$\\
Radiative Return to $\psi(2S)$                &$    852 \pm 4$\\
Radiative Return to $J/\psi$               &$  470 \pm 4$\\
Radiative Return to $\psi(3770)$               &$   109 \pm 1$\\
$\tau^+\tau^-$                    &$  2159 \pm 75$\\
$e^+e^-$                     &$  291 \pm 97$\\
$\mu^+\mu^-$                     &$  21 \pm 4$\\
Number of Beam Gas Events                     &$  329 \pm 131$\\ \hline
Efficiency Corrected Num.   &$   23509 \pm 397$\\ \hline \hline
\end{tabular}
\end{center}
\end{table}

\clearpage
\begin{table}[!htbp]
\begin{center}
\label{HadCount_breakdown_last}
\begin{tabular}{c|c}\hline
E$_{cm}=4260$ MeV \\ \hline
Number of Raw Hadronic Events                  &$   183192 \pm 427$\\
Scaled $q\bar{q}$              &$  117111 \pm 520$\\
Radiative Return to $\psi(2S)$                &$    3477 \pm 11$\\
Radiative Return to $J/\psi$               &$  1811 \pm 12$\\
Radiative Return to $\psi(3770)$               &$   424 \pm 2$\\
$\tau^+\tau^-$                    &$  10239 \pm 267$\\
$e^+e^-$                     &$  2178 \pm 426$\\
$\mu^+\mu^-$                     &$  97 \pm 12$\\
Number of Beam Gas Events                     &$  906 \pm 262$\\ \hline
Efficiency Corrected Num.   &$   56787 \pm 1066$\\ \hline \hline
\end{tabular}
\end{center}
\end{table}
\clearpage

\clearpage

\clearpage

\clearpage

\clearpage

\clearpage

\clearpage

\clearpage

\clearpage

\clearpage

\clearpage

\clearpage

\clearpage

\begin{figure}[!p]
\begin{center}
\hspace{2.5pt}
\includegraphics[width=14.5cm]{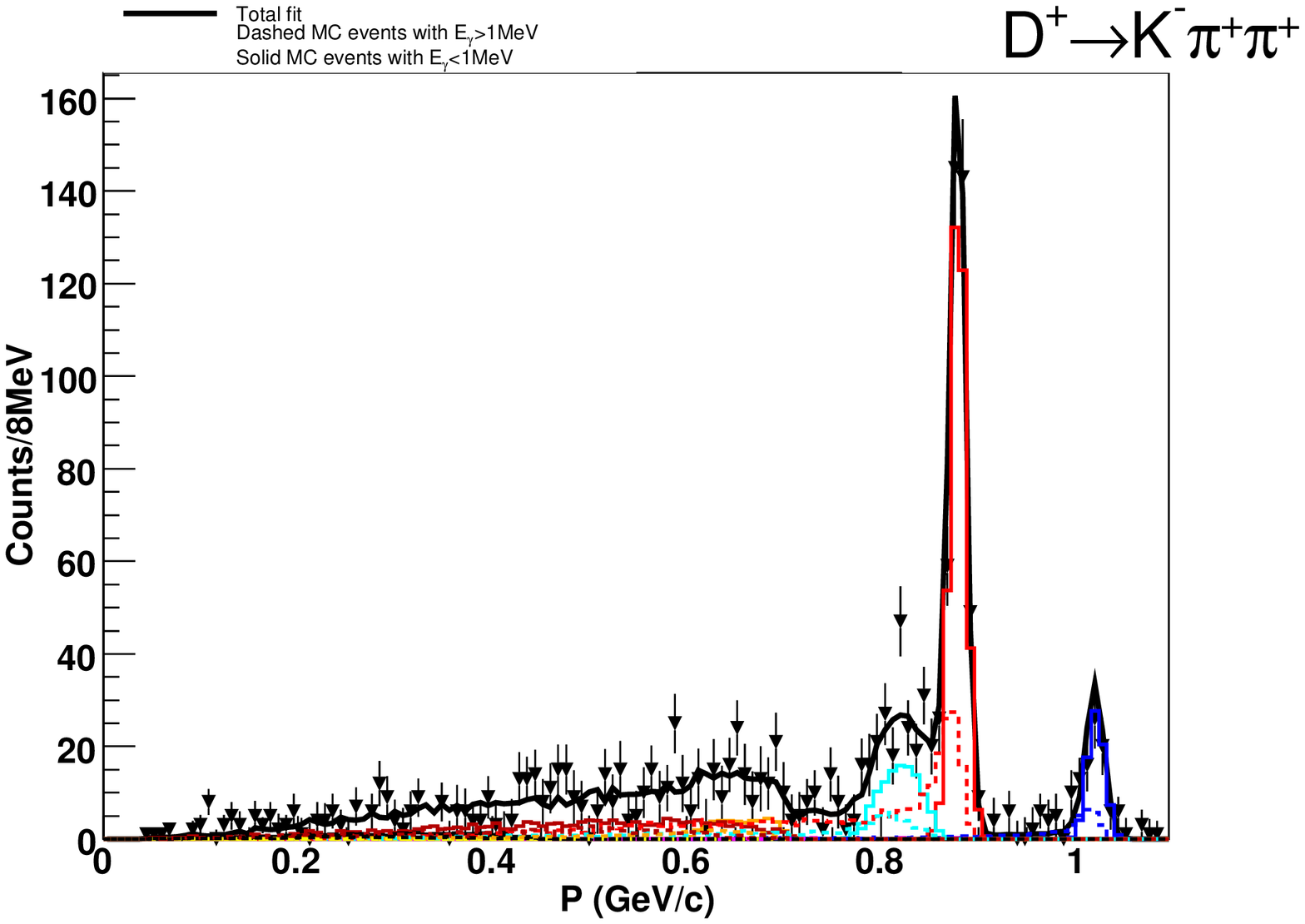}
\caption{Sideband-subtracted momentum spectrum for
$D^+\rightarrow{K^-}\pi^+\pi^+$ at 4260 MeV.  The data is the points
with error bars and the histograms are MC.}
\vspace{0.2cm}
\label{fig:Mom_Dp_4260}
\end{center}
\end{figure}

%%% ADD D_s mode fit
\clearpage

\clearpage

\clearpage

\clearpage

\clearpage

\clearpage

\begin{landscape}
\begin{center}
\begin{table}[!t]
\caption{Observed cross sections for various center of mass energies
as determined by the cut and count method described in the text.
All error are statistical and cross sections
shown are the weighted sums or averages for the event types.}
\label{XS_results_1}
\vspace{0.2cm}
\begin{center}
\begin{tabular}{|c|c|c|c|c|c|c|}\hline
{Cross Section}&\(3970\)~MeV&\(3990\)~MeV&\(4010\)~MeV&\(4015\)~MeV& \(4030\)~MeV &\(4060\)~MeV
\\ \hline
&${\cal{L}}$\(=3.85\)~pb\(^{-1}\)&${\cal{L}}$\(=3.36\)~pb\(^{-1}\)&${\cal{L}}$\(=5.63\)~pb\(^{-1}\)&${\cal{L}}$\(=1.47\)~pb\(^{-1}\)&${\cal{L}}$\(=3.01\)~pb\(^{-1}\)&${\cal{L}}$\(=3.29\)~pb\(^{-1}\)
\\ \hline
\(\sigma({D_{s}D_{s}})\) nb&\(0.102\pm0.026\)&\(0.133\pm0.031\)&\(0.269\pm0.030\)&\(0.250\pm0.059\)&\(0.174\pm0.036\)&\(0.051\pm0.028\)
\\ \hline
\(\sigma({D_{s}^{*}D_{s}})\) nb&\(-\pm-\)&\(-\pm-\)&\(-\pm-\)&\(-\pm-\)&\(-\pm-\)&\(-\pm-\)
\\ \hline
\(\sigma({D_{s}^{*}D_{s}^{*}})\) nb&\(-\pm-\)&\(-\pm-\)&\(-\pm-\)&\(-\pm-\)&\(-\pm-\)&\(-\pm-\)
\\ \hline
\(\sigma({D{D}})\) nb&\(0.194\pm0.027\)&\(0.146\pm0.027\)&\(0.085\pm0.020\)&\(0.037\pm0.033\)&\(0.410\pm0.037\)&\(0.615\pm0.041\)
\\ \hline
\(\sigma({D^{*}D})\) nb&\(4.289\pm0.117\)&\(5.105\pm0.136\)&\(6.285\pm0.116\)&\(6.999\pm0.236\)&\(6.200\pm0.163\)&\(3.687\pm0.129\)
\\ \hline
\(\sigma({D^{*}D^{*}})\) nb&\(-\pm-\)&\(-\pm-\)&\(-\pm-\)&\(0.153\pm0.035\)&\(3.745\pm0.151\)&\(4.779\pm0.189\)
\\ \hline
\(\sigma(Charm)\) nb &\(4.585\pm0.123\)&\(5.384\pm0.142\)&\(6.639\pm0.121\)&\(7.439\pm0.248\)&\(10.529\pm0.228\)&\(9.132\pm0.234\)
\\ \hline
\end{tabular}\end{center}\end{table}
\end{center}

\begin{center}
\begin{table}[!htbp]
\caption{Observed cross sections for various center of mass energies
as determined by the cut and count method described in the text.
All error are statistical and cross sections
shown are the weighted sums or averages for the event types.}
\label{XS_results_2}
\vspace{0.2cm}
\begin{center}
\begin{tabular}{|c|c|c|c|c|c|c|}\hline
{Cross Section}& \(4120\)~MeV &\(4140\)~MeV&\(4160\)~MeV&\(4180\)~MeV&\(4200\)~MeV&\(4260\)~MeV
\\ \hline
&${\cal{L}}$\(=2.76\)~pb\(^{-1}\)&${\cal{L}}$\(=4.87\)~pb\(^{-1}\)&${\cal{L}}$\(=10.16\)~pb\(^{-1}\)&${\cal{L}}$\(=5.67\)~pb\(^{-1}\)&${\cal{L}}$\(=2.81\)~pb\(^{-1}\)&${\cal{L}}$\(=13.11\)~pb\(^{-1}\)
\\ \hline
\(\sigma({D_{s}D_{s}})\) nb&\(0.026\pm0.026\)&\(0.025\pm0.020\)&\(-0.008\pm0.012\)&\(0.007\pm0.016\)&\(0.015\pm0.022\)&\(0.034\pm0.009\)
\\ \hline
\(\sigma({D_{s}^{*}D_{s}})\) nb&\(0.478\pm0.064\)&\(0.684\pm0.059\)&\(0.905\pm0.044\)&\(0.889\pm0.059\)&\(0.812\pm0.082\)&\(0.047\pm0.022\)
\\ \hline
\(\sigma({D_{s}^{*}D_{s}^{*}})\) nb&\(-\pm-\)&\(-\pm-\)&\(-\pm-\)&\(-\pm-\)&\(-\pm-\)&\(0.440\pm0.027\)
\\ \hline
\(\sigma({D{D}})\) nb&\(0.466\pm0.039\)&\(0.328\pm0.026\)&\(0.267\pm0.017\)&\(0.249\pm0.022\)&\(0.295\pm0.032\)&\(0.105\pm0.011\)
\\ \hline
\(\sigma({D^{*}D})\) nb&\(2.481\pm0.122\)&\(2.243\pm0.090\)&\(2.100\pm0.061\)&\(1.951\pm0.079\)&\(1.615\pm0.107\)&\(1.407\pm0.047\)
\\ \hline
\(\sigma({D^{*}D^{*}})\) nb&\(5.073\pm0.213\)&\(4.868\pm0.165\)&\(5.153\pm0.117\)&\(4.620\pm0.147\)&\(3.730\pm0.197\)&\(1.142\pm0.059\)
\\ \hline
\(\sigma(Charm)\) nb &\(8.524\pm0.258\)&\(8.148\pm0.200\)&\(8.417\pm0.141\)&\(7.716\pm0.179\)&\(6.467\pm0.242\)&\(3.175\pm0.084\)
\\ \hline
\end{tabular}\end{center}\end{table}
\end{center}
\end{landscape}

\begin{landscape}
\tiny{
\begin{table}[h]
\caption{Observed cross sections for various center of mass energies
as determined by fits to the $D^{0}$ and $D^{+}$ momentum spectra. The
$D_{s}$ cross sections are determined using a weighted sum techneque
discribed in Sect. \ref{sec:sigmas}.  The first error is statistical and
the second is systematic.}
\label{XS_results_fits_1}
\begin{center}
\begin{tabular}{c|c|c|c}
\hline
{Cross Section}&\(3970\)MeV&\(3990\)MeV&\(4010\)MeV
\\ \hline
&${\cal{L}}$\(=3.85\)pb\(^{-1}\)&${\cal{L}}$\(=3.36\)pb\(^{-1}\)&${\cal{L}}$\(=5.63\)pb\(^{-1}\)
\\ \hline
\(\sigma({D_{s}D_{s}})\) nb&\(0.102\pm0.026\pm0.006\)&\(0.133\pm0.031\pm0.007\)&\(0.269\pm0.030\pm0.015\)
\\ \hline
\(\sigma({D_{s}^{*}D_{s}})\) nb&\(-\pm-\pm-\)&\(-\pm-\pm-\)&\(-\pm-\pm-\)
\\ \hline
\(\sigma({D_{s}^{*}D_{s}^{*}})\) nb&\(-\pm-\pm-\)&\(-\pm-\pm-\)&\(-\pm-\pm-\)
\\ \hline
\(\sigma({D{D}})\) nb&\(0.223\pm0.039\pm0.010\)&\(0.223\pm0.047\pm0.010\)&\(0.211\pm0.033\pm0.009\)
\\ \hline
\(\sigma({D^{*}D})\) nb&\(4.510\pm0.187\pm0.153\)&\(5.490\pm0.221\pm0.187\)&\(6.620\pm0.133\pm0.225\)
\\ \hline
\(\sigma({D^{*}D^{*}})\) nb&\(-\pm-\pm-\)&\(-\pm-\pm-\)&\(-\pm-\pm-\)
\\ \hline
\(\sigma({D^{*}D\pi})\) nb&\(-\pm-\pm-\)&\(-\pm-\pm-\)&\(-\pm-\pm-\)
\\ \hline
\(\sigma({D^{*}D^{*}\pi})\) nb&\(-\pm-\pm-\)&\(-\pm-\pm-\)&\(-\pm-\pm-\)
\\ \hline
\(\sigma(Charm)\) nb &\(4.835\pm0.193\pm0.154\)&\(5.846\pm0.228\pm0.187\)&\(7.100\pm0.140\pm0.226\)
\\ \hline \hline
%\end{tabular}\end{center}\end{table}}

%\tiny{
%\begin{table}[h]
%\begin{center}
%\begin{tabular}{c|c|c|c}
&&& \\ 
{Cross Section}&\(4015\)MeV& \(4030\)MeV &\(4060\)MeV
\\ \hline
&${\cal{L}}$\(=1.47\)pb\(^{-1}\)&${\cal{L}}$\(=3.01\)pb\(^{-1}\)&${\cal{L}}$\(=3.29\)pb\(^{-1}\)
\\ \hline
\(\sigma({D_{s}D_{s}})\) nb&\(0.250\pm0.059\pm0.014\)&\(0.174\pm0.036\pm0.010\)&\(0.051\pm0.028\pm0.003\)
\\ \hline
\(\sigma({D_{s}^{*}D_{s}})\) nb&\(-\pm-\pm-\)&\(-\pm-\pm-\)&\(-\pm-\pm-\)
\\ \hline
\(\sigma({D_{s}^{*}D_{s}^{*}})\) nb&\(-\pm-\pm-\)&\(-\pm-\pm-\)&\(-\pm-\pm-\)
\\ \hline
\(\sigma({D{D}})\) nb&\(0.038\pm0.020\pm0.002\)&\(0.530\pm0.078\pm0.024\)&\(0.890\pm0.091\pm0.040\)
\\ \hline
\(\sigma({D^{*}D})\) nb&\(7.443\pm0.394\pm0.256\)&\(6.500\pm0.257\pm0.221\)&\(4.400\pm0.205\pm0.150\)
\\ \hline
\(\sigma({D^{*}D^{*}})\) nb&\(0.213\pm0.076\pm0.009\)&\(3.400\pm0.211\pm0.160\)&\(4.680\pm0.258\pm0.220\)
\\ \hline
\(\sigma({D^{*}D\pi})\) nbw&\(-\pm-\pm-\)&\(-\pm-\pm-\)&\(0.144\pm0.094\pm0.017\)
\\ \hline
\(\sigma({D^{*}D^{*}\pi})\) nb&\(-\pm-\pm-\)&\(-\pm-\pm-\)&\(-\pm-\pm-\)
\\ \hline
\(\sigma(Charm)\) nb &\(7.941\pm0.406\pm0.257\)&\(10.604\pm0.344\pm0.274\)&\(10.165\pm0.356\pm0.270\)
\\ \hline \hline
\end{tabular}\end{center}\end{table}}

\tiny{
\begin{table}[h]
\caption{Observed cross sections for various center of mass energies
as determined by fits to the $D^{0}$ and $D^{+}$ momentum spectra. The
$D_{s}$ cross sections are determined using a weighted sum techneque
discribed in Sect. \ref{sec:sigmas}.  The first error is statistical and
the second is systematic.}
\label{XS_results_fits_2}
\begin{center}
\begin{tabular}{c|c|c|c|c}
\hline
{Cross Section}& \(4120\)MeV &\(4140\)MeV&\(4160\)MeV&\(4170\)MeV
\\ \hline
&${\cal{L}}$\(=2.76\)pb\(^{-1}\)&${\cal{L}}$\(=4.87\)pb\(^{-1}\)&${\cal{L}}$\(=10.16\)pb\(^{-1}\)&${\cal{L}}$\(=178.9\)pb\(^{-1}\)
\\ \hline
\(\sigma({D_{s}D_{s}})\) nb&\(0.026\pm0.026\pm0.001\)&\(0.025\pm0.020\pm0.001\)&\(-0.008\pm0.012\pm0.000\)&\(0.034\pm0.003\pm0.002\)
\\ \hline
\(\sigma({D_{s}^{*}D_{s}})\) nb&\(0.478\pm0.064\pm0.025\)&\(0.684\pm0.059\pm0.036\)&\(0.905\pm0.011\pm0.048\)&\(0.916\pm0.011\pm0.049\)
\\ \hline
\(\sigma({D_{s}^{*}D_{s}^{*}})\) nb&\(-\pm-\pm-\)&\(-\pm-\pm-\)&\(-\pm-\pm-\)&\(-\pm-\pm-\)
\\ \hline
\(\sigma({D{D}})\) nb&\(0.613\pm0.086\pm0.028\)&\(0.377\pm0.049\pm0.017\)&\(0.367\pm0.035\pm0.017\)&\(0.360\pm0.010\pm0.016\)
\\ \hline
\(\sigma({D^{*}D})\) nb&\(2.960\pm0.192\pm0.101\)&\(2.726\pm0.140\pm0.093\)&\(2.628\pm0.098\pm0.089\)&\(2.560\pm0.030\pm0.087\)
\\ \hline
\(\sigma({D^{*}D^{*}})\) nb&\(4.830\pm0.278\pm0.227\)&\(4.999\pm0.228\pm0.235\)&\(5.045\pm0.159\pm0.237\)&\(4.720\pm0.030\pm0.222\)
\\ \hline
\(\sigma({D^{*}D\pi})\) nb&\(0.045\pm0.083\pm0.005\)&\(0.412\pm0.087\pm0.049\)&\(0.389\pm0.060\pm0.047\)&\(0.440\pm0.020\pm0.053\)
\\ \hline
\(\sigma({D^{*}D^{*}\pi})\) nb&\(-\pm-\pm-\)&\(-\pm-\pm-\)&\(-\pm-\pm-\)&\(-\pm-\pm-\)
\\ \hline
\(\sigma(Charm)\) nb &\(8.952\pm0.365\pm0.251\)&\(9.223\pm0.292\pm0.260\)&\(9.326\pm0.200\pm0.263\)&\(9.03\pm0.044\pm0.250\)
\\ \hline \hline
&&&& \\
{Cross Section}&\(4180\)MeV&\(4200\)MeV&\(4260\)MeV&
\\ \hline
&${\cal{L}}$\(=5.67\)pb\(^{-1}\)&${\cal{L}}$\(=2.81\)pb\(^{-1}\)&${\cal{L}}$\(=13.11\)pb\(^{-1}\)&
\\ \hline
\(\sigma({D_{s}D_{s}})\) nb&\(0.007\pm0.016\pm0.000\)&\(0.015\pm0.022\pm0.001\)&\(0.047\pm0.022\pm0.003\)&
\\ \hline
\(\sigma({D_{s}^{*}D_{s}})\) nb&\(0.889\pm0.059\pm0.047\)&\(0.812\pm0.082\pm0.043\)&\(0.034\pm0.009\pm0.002\)&
\\ \hline
\(\sigma({D_{s}^{*}D_{s}^{*}})\) nb&\(-\pm-\pm-\)&\(-\pm-\pm-\)&\(0.440\pm0.027\pm0.030\)&
\\ \hline
\(\sigma({D{D}})\) nb&\(0.376\pm0.047\pm0.017\)&\(0.361\pm0.066\pm0.016\)&\(0.180\pm0.022\pm0.008\)&
\\ \hline
\(\sigma({D^{*}D})\) nb&\(2.507\pm0.127\pm0.085\)&\(2.100\pm0.169\pm0.071\)&\(2.102\pm0.080\pm0.071\)&
\\ \hline
\(\sigma({D^{*}D^{*}})\) nb&\(4.318\pm0.201\pm0.203\)&\(3.394\pm0.256\pm0.160\)&\(0.506\pm0.069\pm0.024 \)&
\\ \hline
\(\sigma({D^{*}D\pi})\) nb&\(0.575\pm0.092\pm0.069\)&\(0.735\pm0.129\pm0.088\)&\(0.638\pm0.093\pm0.077 \)&
\\ \hline
\(\sigma({D^{*}D^{*}\pi})\) nb&\(-\pm-\pm-\)&\(-\pm-\pm-\)&\(0.322\pm0.067\pm0.080 \)&
\\ \hline
\(\sigma(Charm)\) nb &\(8.672\pm0.266\pm0.236\)&\(7.417\pm0.350\pm0.201\)&\(4.269\pm0.162\pm0.138\)&
\\ \hline \hline
\end{tabular}\end{center}\end{table}}
\end{landscape}

%%%%%%%%%%%%%%%%%%%%%%%%%%%%%%%%%%%%%%%%%%%%%%%%%%%%%%%%%%%%%%%%%%%%%%%%%%
%%%%%%%%%%%%%%%%%%%%%%%%%%%%%%%%%%%%%%%%%%%%%%%%%%%%%%%%%%%%%%%%%%%%%%%%%%
%
%	APPENDIX A
%
\appendix
%%%%%%%%%%%%%%%%%%%%%%%%%%%%%%%%%%%%%%%%%%%%%%%%%%%%%%%%%%%%%%%%%%%%%%%%%%
\chapter{Helicity Formalism and Angular Distributions}
%%%%%%%%%%%%%%%%%%%%%%%%%%%%%%%%%%%%%%%%%%%%%%%%%%%%%%%%%%%%%%%%%%%%%%%%%%%%%%%%%%%%
In the spirit of obtaining the best possible MC sample a
considerable amount of time
was spent on determining the final-state angular distributions for each of
the event types and the subsequent decays of the starred states. 
There are three different and distinct types of events that need to
be considered:
\begin{itemize}
\item {$1^{-}\rightarrow{0^-0^-}$, as in
      $e^+e^-\rightarrow\gamma^*\rightarrow{D\bar{D}}$}
\item {$1^{-}\rightarrow{0^-1^-}$, as in
      $e^+e^-\rightarrow\gamma^*\rightarrow{D\bar{D}^*}$}
\item {$1^{-}\rightarrow{1^-1^-}$, as in
      $e^+e^-\rightarrow\gamma^*\rightarrow{D^*\bar{D}^*}$}
\end{itemize}
In addition to the above, there are also the subsequent decays:
\begin{itemize}
\item {$1^{-}\rightarrow{1^-0^-}$, as in
      $D^*\rightarrow{\gamma{D}}$}
\item {$1^{-}\rightarrow{0^-0^-}$, as in
      $D^*\rightarrow{\pi{D}}$}
\end{itemize}

Each of the above cases, except $1^{-}\rightarrow{0^-0^-}$, since it
is already understood from $\psi(3770)$ running (for completeness its distribution is
$\sin^{2}\theta$), is described individually below using the
helicity formalism as discussed by J.D. Richman in Ref. \cite{RICHMAN}
%%%%%%%%%%%%%%%%%%%%%%%%%%%%%%%%%%%%%%%%%%%%%%%%%%%%%%%%%%%%%%%%%%%%%
\section{$e^+e^-\rightarrow\gamma^*\rightarrow{D\bar{D}^{*}}$}
%%%%%%%%%%%%%%%%%%%%%%%%%%%%%%%%%%%%%%%%%%%%%%%%%%%%%%%%%%%%%%%%%%%%%
The $J^{P}$ quantum numbers, as stated above, are
{$1^{-}\rightarrow{0^-1^-}$.  Since the reaction in question, and all
reactions that will be discussed henceforth, is an electromagnetic interaction,
parity must be conserved. As proved in Ref. \cite{RICHMAN}, parity
conservation can be stated as follows:
\begin{equation}
\label{eq:Parity}
	A_{\lambda_{D}\lambda_{D^{*}}}=(-1)^{J_{\gamma^{*}}-J_{D}-J_{D^{*}}}P_{\gamma^{*}}P_{D}P_{D^{*}}A_{-\lambda_{D}-\lambda_{D^{*}}}
\end{equation}
\CONT By \EQ \ref{eq:Parity} and the $J^{P}$ quantum numbers of the
reaction, it is seen that party conservation requires $A_{00}=-A_{00}$
and $A_{10}=-A_{-10}$.  Therefore, $A_{00}$ must be set to zero and
so the $D^{*}$ will always be produced with helicity of
$\lambda=\pm1$.  Next, the virtual photon will have a spin projection
along the beam axis.  This is because when $E_{cm}>>m_{e}$ electrons and positrons
will only couple to each other if there combined helicity is odd.
Using the above information we get the following final-state angular
distribution:
\begin{eqnarray}
	\frac{dN}{d\Omega} =
	\sum_{M_{\gamma}}|\sum_{\lambda_{D^{*}}}
	(D^{1}_{M_{\gamma}\lambda_{D^{*}}}A_{\lambda_{D^{*}}0})|^{2} =\nonumber\\ 
	\sum_{M_{\gamma}}|A_{10}|^{2}|(D^{1}_{M_{\gamma}-1} -
	D^{1}_{M_{\gamma}1})|^{2} =\nonumber\\
	\sum_{M_{\gamma}}|A_{10}|^{2}(|D^{1}_{M_{\gamma}-1}|^{2} +
	|D^{1}_{M_{\gamma}1}|^{2} -
	D^{1}_{M_{\gamma}1}D^{1*}_{M_{\gamma}-1} - D^{1}_{M_{\gamma}-1}D^{1*}_{M_{\gamma}1})
\end{eqnarray}
\CONT where $D^{1}_{M_{\gamma}1} =
e^{-iM_{\gamma}\phi}d^{1}_{M_{\gamma}1}(\theta)e^{i\phi}$ and  $D^{1}_{M_{\gamma}-1} =
e^{-iM_{\gamma}\phi}d^{1}_{M_{\gamma}-1}(\theta)e^{-i\phi}$. Substituting
and realizing that only $M_{\gamma}=1$ needs to be looked at, since
$d^{j}_{mm'}=d^{j}_{-m'-m}$, one arrives at by ignoring overall
constants and integrating over azimuth at,
\begin{equation}
\frac{dN}{d\cos\theta} \propto |A_{10}|^{2}(1+\cos^{2}\theta),
\end{equation}
\CONT where $\theta$ is defined as the angle between the \(D^{*}\) and
the beam axis.

\begin{figure}[!htbp]
\begin{center}
\hspace{2.5pt}
\includegraphics[width=14.5cm]{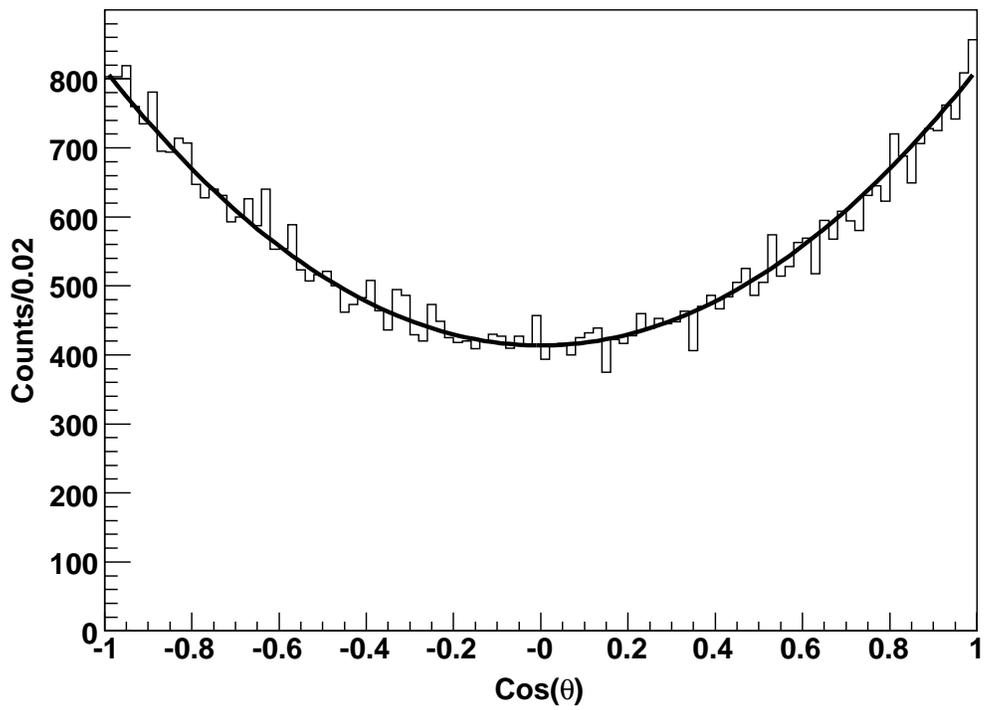}
\caption{Generator-level angular distribution for $D^{*}\bar{D}$ events.
$\theta$ is defined as the angle between the $D^{*}$ and the beam axis.}
\vspace{0.2cm}
\label{fig:DStarD}
\end{center}
\end{figure}

To generate the correct angular distributions for $D\bar{D}^{*}$
events, {\tt{EVTGEN}}'s Helicity Amplitude model
({\tt{HELAMP}}) \cite{EVTGEN} was
used and the following was added to the
decay.dec file for $D_{s}^{*+}{D}_{s}^{-}$:
\begin{center}
\texttt{D\_s*+ D\_s- HELAMP 1. 0. 0. 0. -1. 0. ;}\end{center}
%%%%%%%%%%%%%%%%%%%%%%%%%%%%%%%%%%%%%%%%%%%%%%%%%%%%%%%%%%%%%%%%%%%%
\section{$e^+e^-\rightarrow\gamma^*\rightarrow{D\bar{D}^{*}}$,
$D^{*}\rightarrow{D}{\gamma}$}
%%%%%%%%%%%%%%%%%%%%%%%%%%%%%%%%%%%%%%%%%%%%%%%%%%%%%%%%%%%%%%%%%%%%5
This decay is identical to $\gamma^*\rightarrow{D\bar{D}^{*}}$ except with
the $\gamma$ and the $D^{*}$ switched.  Thus the $D^{*}$ will have
a spin projection along its momentum vector and the $\gamma$ will be
produced with helicity of $\lambda=\pm1$.  Therefore, with $B_{10}$ being the helicity amplitude for
$D^{*}\rightarrow{D}{\gamma}$
\begin{equation}
	\frac{dN}{d\cos\theta^{'}} \propto
	|B_{10}|^2|A_{10}|^{2}(|d^{1}_{1-1}(\theta{'})|^{2} +
	|d^{1}_{11}(\theta^{'})|^{2}) \propto |B_{10}|^2|A_{10}|^{2}(1+\cos^{2}\theta^{'})
\end{equation}
\CONT where $\theta^{'}$ is defined as the angle between the $D$ in the
rest frame of the $D^{*}$ and the momentum vector of the
$D^{*}$ in the lab. The full derivation for this decay is given in
Appendix~\ref{app:full}.
\begin{figure}[!htbp]
\begin{center}
\hspace{2.5pt}
\includegraphics[width=14.5cm]{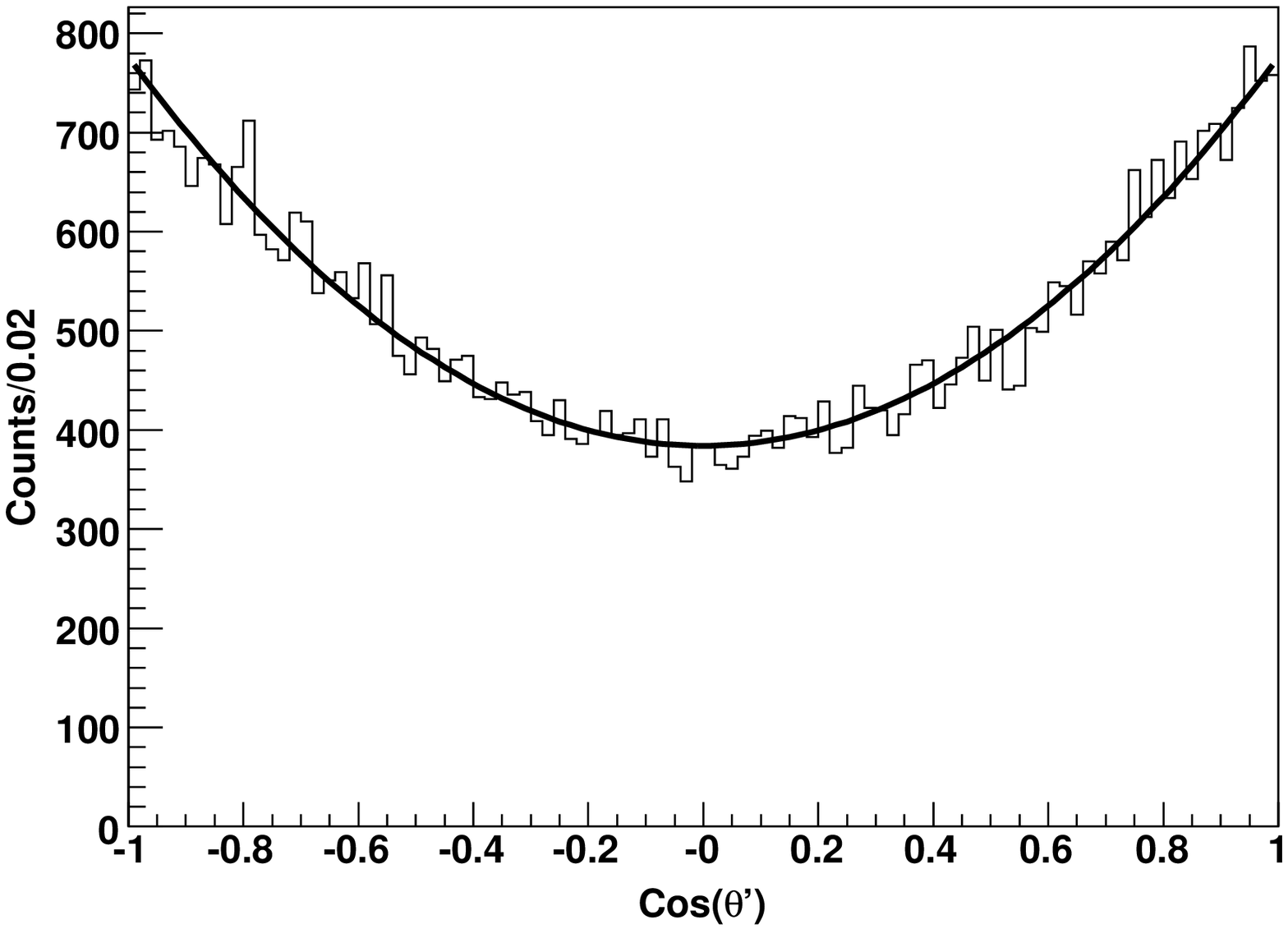}
\caption{Generator-level angular distribution for $D$ in
$D^{*}\rightarrow{D}{\gamma}$ for $D^{*}\bar{D}$ events. $\theta^{'}$ is defined as the angle between the $D$ in the
rest frame of the $D^{*}$ and the momentum vector of the
$D^{*}$ in the lab. }
\vspace{0.2cm}
\label{fig:DStarDgamma}
\end{center}
\end{figure}
%%%%%%%%%%%%%%%%%%%%%%%%%%%%%%%%%%%%%%%%%%%%%%%%%%%%%%%%%%%%%%%%%%55
\section{$e^+e^-\rightarrow\gamma^*\rightarrow{D\bar{D}^{*}}$,
$D^{*}\rightarrow{D}{\pi}$}
%%%%%%%%%%%%%%%%%%%%%%%%%%%%%%%%%%%%%%%%%%%%%%%%%%%%%%%%%%%%%%%%%%%
Since $\lambda_{\pi} = \lambda_{D} = 0$ the angular distribution is as
follows:
\begin{equation}
	\frac{dN}{d\cos\theta^{'}} \propto
	|A_{10}|^{2}|d^{1}_{10}(\theta{'})|^{2} \propto
	|B_{00}||A_{10}|^{2}(\sin^{2}\theta^{'})
\end{equation}
\CONT where $\theta^{'}$ is defined above and $B_{00}$ being the helicity amplitude for
$D^{*}\rightarrow{D}{\gamma}$.
\begin{figure}[!htbp]
\begin{center}
\hspace{2.5pt}
\includegraphics[width=14.5cm]{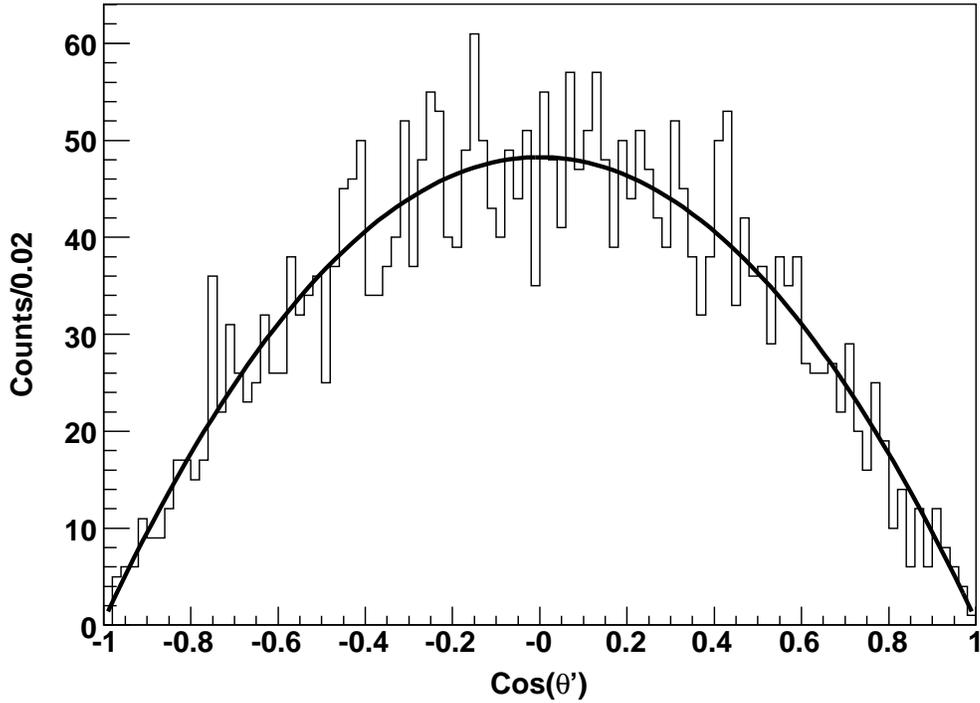}
\caption{Generator-level angular distribution for $D$ in
$D^{*}\rightarrow{D}{\pi}$ for $D^{*}\bar{D}$ events. $\theta^{'}$ is
defined as the angle between the $D$ in the rest frame of the $D^{*}$
to that of the momentum vector of the $D^{*}$ in the lab. }
\vspace{0.2cm}
\label{fig:DStarDpi0}
\end{center}
\end{figure}
%%%%%%%%%%%%%%%%%%%%%%%%%%%%%%%%%%%%%%%%%%%%%%%%%%%%%%%%%%%%%%%%%%%%%
\section{$e^+e^-\rightarrow\gamma^*\rightarrow{D^{*}\bar{D}^{*}}$}
%%%%%%%%%%%%%%%%%%%%%%%%%%%%%%%%%%%%%%%%%%%%%%%%%%%%%%%%%%%%%%%%%%%%%
There are seven helicity amplitudes that
determine the angular distribution of the final
state: $A_{00}$, $A_{0-1}$, $A_{-10}$, $A_{10}$, $A_{01}$, $A_{11}$, and $A_{-1-1}$. By the
conservation of parity and the symmetry of the final state,
$A_{\lambda_{D^{*}_{1}}\lambda_{D^{*}_{2}}}=A_{\lambda_{D^{*}_{2}}\lambda_{D^{*}_{1}}}=A_{-\lambda_{D^{*}_{1}}-\lambda_{D^{*}_{2}}}$,
so the only independent amplitudes are $A_{00}$, $A_{01}$ and
$A_{11}$. It therefore follows that
\begin{equation}
\label{eq:DSDS}
	\frac{dN}{d\cos{\theta}} = |A_{00}|^2(d^{1}_{10}(\theta))^{2} +
	2|A_{01}|^{2}((d^{1}_{11}(\theta))^2+(d^{1}_{1-1}(\theta))^2) + 2|A_{11}|^{2}(d^{1}_{10}(\theta))^{2}.
\end{equation}
\CONT The three independent helicity amplitudes have not been
measured.  For simplicity, in generating MC we assumed that all
helicity amplitudes are equal, $A_{00}=A_{01}=A_{11}$:
\begin{equation}
\label{DStarDStar_theta}
	\frac{dN}{d\cos{\theta}} \propto \frac{3}{2}(1-\cos^{2}\theta)
	+ (1+\cos^{2}\theta) \propto (1-\frac{1}{5}\cos^{2}\theta).
\end{equation}
\CONT The above helicity amplitude assumption is consistent with the
Mark I measurement of $e^+e^-\rightarrow\gamma^*\rightarrow{D^{*}\bar{D}^{*}}$ angular
distribution from 1977~\cite{spin}. However, it should be noted
that the error bar of that measurement is $\sim100\%$. Here, the error bar
refers to the fit parameter $\alpha$ in their fit function, $1+\alpha\cos^{2}\theta$.
\begin{figure}[!htbp]
\begin{center}
\hspace{2.5pt}
\includegraphics[width=14.5cm]{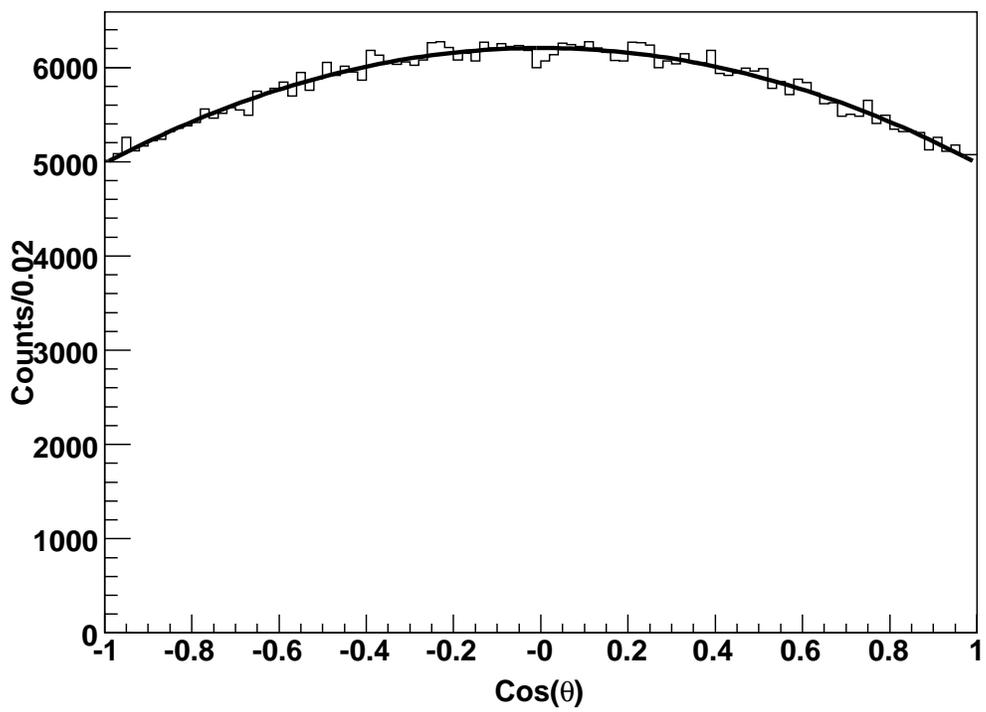}
\caption{Generator-level angular distribution for $D^{*}\bar{D}^{*}$
events. $\theta$ is defined at the angle between the $D^{*}$ and the
beam axis.}
\vspace{0.2cm}
\label{fig:DStarDStar}
\end{center}
\end{figure}
In order to generate this angular distribution for $D^{*}\bar{D}^{*}$
events the {\tt{HELAMP}} model was
used and the following was added to the
decay.dec file for $D^{*0}\bar{D}^{*0}$:
\begin{center}
\texttt{D*0 anti-D*0 HELAMP 1.0 0.0 1.0 0.0 1.0 0.0 1.0 0.0
                                    1.0 0.0 1.0 0.0 1.0 0.0;}
\end{center}

\CONT To check our assumption the helicity amplitudes have been determined
and are presented in Appendix B using the scan data above $D^{*}\bar{D}^{*}$
threshold.  However, it should be noted that the detection efficiency
does not depend on the coefficient of $\cos^2\theta$ in \EQ \ref{DStarDStar_theta}.

%%%%%%%%%%%%%%%%%%%%%%%%%%%%%%%%%%%%%%%%%%%%%%%%%%%%%%%%%%%%%%%%%%%%%%%%%5
\section{$e^+e^-\rightarrow\gamma^*\rightarrow{D^{*}\bar{D}^{*}}$,
$D^{*}\rightarrow{D}{\gamma}$}
%%%%%%%%%%%%%%%%%%%%%%%%%%%%%%%%%%%%%%%%%%%%%%%%%%%%%%%%%%%%%%%%%%%%%%%%555
It has already been shown in a previous section that the angular
distribution, when $D^{*}$ has helicity of 1, is
\begin{equation}
	\frac{dN}{d\cos\theta'} \propto (1+\cos^{2}\theta^{'}).
\end{equation}
\CONT Next, it can been shown that when $D^{*}$ has helicity of 0 the angular
distribution is as follows:
\begin{equation}
		\frac{dN}{d\cos\theta^{'}} \propto		
	(|d^{1}_{0-1}(\theta{'})|^{2} +
	|d^{1}_{01}(\theta^{'})|^{2}) \propto (1-\cos^{2}\theta^{'})
	= \sin^{2}\theta^{'}.
\end{equation}
Now when both helicities are allowed the angular distribution will be
a combination of these two.  There are three cases to consider:
\begin{itemize}
\item{Both $D^{*}$ are produced with helicity of 1.}
\item{Both $D^{*}$ are produced with helicity of 0.}
\item{One of $D^{*}$ is produced with helicity of 1 while the other is
produced with helicity of 0.}
\end{itemize}
Combining all the above situations, and being mindful to keep track of the needed
helicity amplitudes, leads to the following angular distribution:
\begin{equation}
		\frac{dN}{d\cos\theta^{'}} =
		|B_{01}|^2\{(3|A_{10}|^2+|A_{11}|^2+|A_{00}|^2) + (|A_{11}|^2-|A_{00}|^2-|A_{10}|^2)\cos^{2}\theta^{'}\},
\end{equation}
\CONT where $\theta^{'}$ is defined above and $B_{10}$ being the
helicity amplitude for $D^{*}\rightarrow{D}{\gamma}$. Again making the
same assumption as above, that is all the helicity amplitudes are
equal one arrives at the following:
\begin{equation}
		\frac{dN}{d\cos\theta^{'}} \propto (1-\frac{1}{5}\cos^{2}\theta^{'}).
\end{equation}

\begin{figure}[!htbp]
\begin{center}
\hspace{2.5pt}
\includegraphics[width=14.5cm]{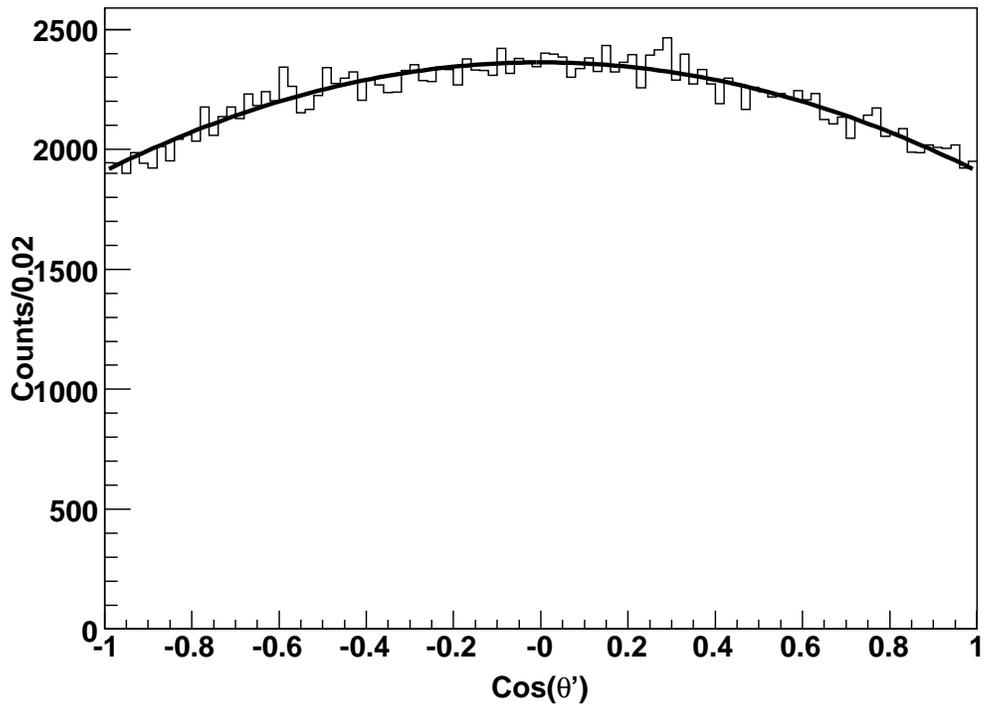}
\caption{Generator-level angular distribution for $D$ in
$D^{*}\rightarrow{D}{\gamma}$ for $D^{*}\bar{D}^{*}$
events. $\theta^{'}$ is defined as the angle between the $D$ in the
rest frame of the $D^{*}$ to that of the momentum vector of the $D^{*}$ in the lab. }
\vspace{0.2cm}
\label{fig:DStarDStargamma}
\end{center}
\end{figure}
%%%%%%%%%%%%%%%%%%%%%%%%%%%%%%%%%%%%%%%%%%%%%%%%%%%%%%%%%%%%%%%%%%%%%%%%%%%%%5
\section{$e^+e^-\rightarrow\gamma^*\rightarrow{D^{*}\bar{D}^{*}}$,
$D^{*}\rightarrow{D}{\pi}$}
%%%%%%%%%%%%%%%%%%%%%%%%%%%%%%%%%%%%%%%%%%%%%%%%%%%%%%%%%%%%%%%%%%%%%%%%%%%5
It has already been shown that the angular
distribution, for production of $D^{*}$ with helicity of 1 is
\begin{equation}
	\frac{dN}{d\cos\theta^{'}} \propto (1-\cos^{2}\theta^{'}).
\end{equation}
\CONT Next, it can been shown that when $D^{*}$ has helicity of 0 the angular
distribution is as follows:
\begin{equation}
		\frac{dN}{d\cos\theta^{'}} \propto		
	|d^{1}_{00}(\theta{'})|^{2} \propto \cos^{2}\theta^{'}.
\end{equation}
\CONT Again combining the three possible $D^{*}D^{*}$ helicity
cases, with their associated amplitudes, yields the following angular distribution:
\begin{equation}
		\frac{dN}{d\cos\theta^{'}} =
		|B_{00}|^2\{(|A_{11}|^2+|A_{01}|^2) + (|A_{00}|^2 +
		|A_{01}|^2 - |A_{11}|^2)\cos^2\theta^{'}\}.
\end{equation}
\CONT Applying the same assumption as above, that is
$A_{00}=A_{01}=A_{11}$, one arrives at the following distribution:
\begin{equation}
	\frac{dN}{d\cos\theta^{'}} \propto (1+\frac{1}{2}\cos^{2}\theta^{'}).
\end{equation}
\begin{figure}[!htbp]
\begin{center}
\hspace{2.5pt}
\includegraphics[width=14.5cm]{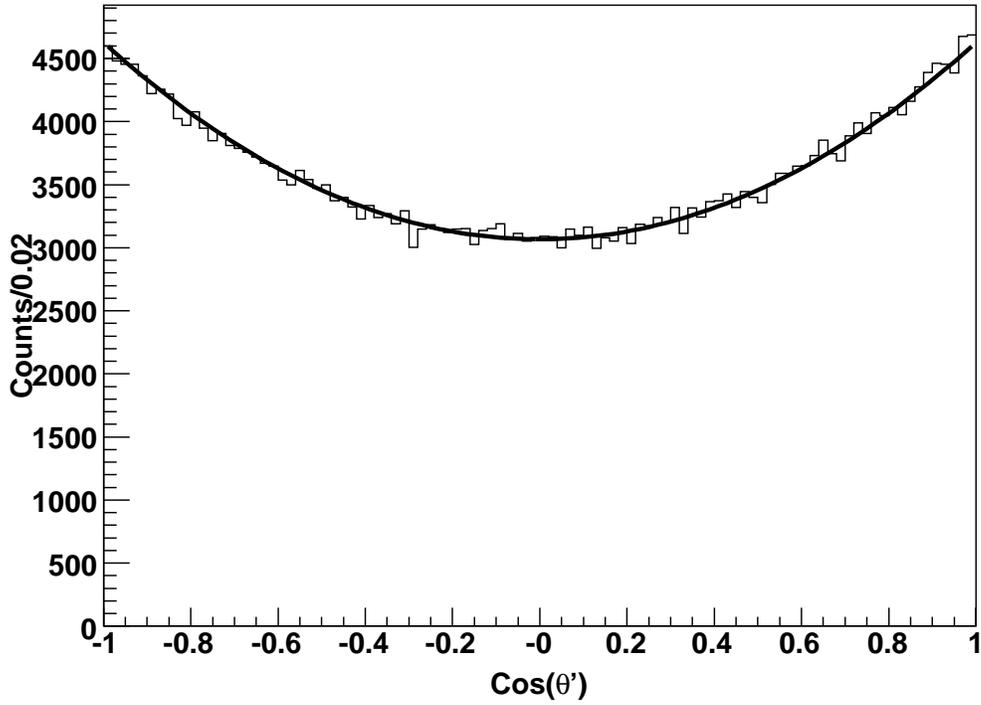}
\caption{Generator-level angular distribution for $D$ in
$D^{*}\rightarrow{D}{\pi^{0}}$ for $D^{*}\bar{D}^{*}$
events. $\theta^{'}$ is defined as the angle between the $D$ in the
rest frame of the $D^{*}$ to that of the momentum vector of the	$D^{*}$ in the lab. }
\vspace{0.2cm}
\label{fig:DStarDStarpi0}
\end{center}
\end{figure}

%%%%%%%%%%%%%%%%%%%%%%%%%%%%%%%%%%%%%%%%%%%%%%%%%%%%%%%%%%%%%
\subsection{Momentum Distributions}
%%%%%%%%%%%%%%%%%%%%%%%%%%%%%%%%%%%%%%%%%%%%%%%%%%%%%%%%%%%%%
\subsubsection{Momentum Distributions of $D$-mesons in
$e^+e^-\rightarrow\gamma^*\rightarrow{D^{*}\bar{D^{*}}}$ events}

Notice that in all cases the angular distribution of a $D$ meson can be
written as follows:
\begin{equation}
\frac{dN}{d\cos\theta} \propto 1+\alpha\cos^2(\theta)
\end{equation}
\CONT where $\alpha$ is a function of the various helicity amplitudes
present for the decay and event in question.

The momentum distribution for the decay $D$ mesons is as follows:
\begin{equation}
\frac{dN}{dP_{lab}} = \frac{dN}{d\cos\theta}\frac{d\cos\theta}{dP_{lab}}
\end{equation}
\CONT The energy of the decay $D$ in the lab can be written as follow:
\begin{equation}
 {E_{lab} = {\gamma}E^{o} + \beta{\gamma}P^{o}\cos(\theta)}
\end{equation}
\CONT thus one gets, using $P_{lab}^2 = E_{lab}^{2} - M^2$, the following expression:
\begin{equation}
 \cos(\theta) = \frac{\sqrt{P^{2}_{lab} + M^{2}} - {\gamma}E^{o}}{\beta\gamma{P^{0}}}
\end{equation}
\CONT and so:
\begin{equation}
 \frac{d\cos(\theta)}{dP_{lab}} = \frac{P_{lab}}{\beta\gamma{P^{o}}\sqrt{P^{2}_{lab} + M^{2}}}
\end{equation}
\CONT with $\beta\gamma = \frac{P_{D^*}}{M_{D^{*}}}$, $\gamma =
\frac{E_{D^*}}{M_{D^{*}}}$, and $E^{o} =
\frac{M_{D^{*}}^{2} + M_{X}^{2} - M^{2}}{2M_{D^{*}}}$, where $M$ and
$M_{X}$ are the mass of the daugther $D$ and other particle (either
a photon or pion) present in the decay, respectively. Therefore:
\begin{equation}
\label{eq:ang_momentum}
 \frac{dN}{dP_{lab}} = \frac{P_{lab}}{\beta\gamma{P^{o}}\sqrt{P^{2}_{lab} + M^{2}}}\{1+\alpha\{\frac{\sqrt{P_{lab}^{2}+M^{2}}-\gamma{E^{o}}}{\beta\gamma{P}^{o}}\}^2\}
\end{equation}

\CONT It is easily seen in the above expression the role that the angular
distribution, indicted by the parameter $\alpha$, has on the momentum
distribution.  Examples of two cases, at $E=4030~$MeV, are shown in
\FIGS \ref{fig:ang_momentum}-\ref{fig:ang_momentum_CON}.

\begin{figure}[!htbp]
\begin{center}
\hspace{2.5pt}
\includegraphics[width=14.5cm]{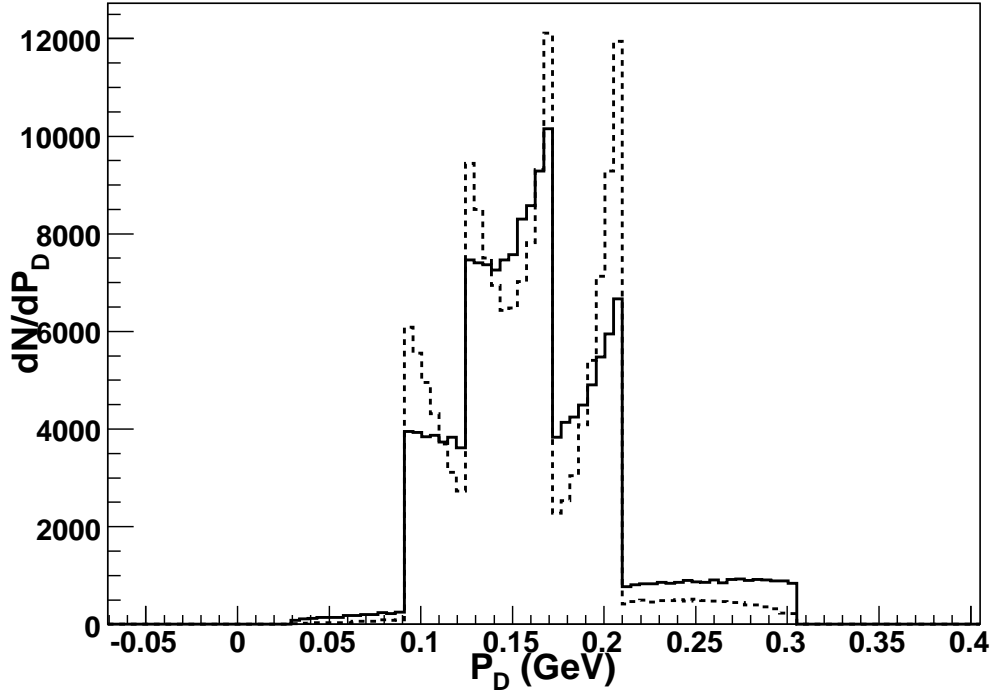}
\caption{Momentum distribution for $D^{0}$ in $D^{*}D^{*}$
events.  The different histograms correspond to different helicity
amplitudes, and therefore different values for $\alpha$ in \EQ
\ref{eq:ang_momentum}. The solid histogram has $\alpha = 0.65$ and
$\alpha = -0.24$ for the pion and gamma decay respectively.  The
dashed histogram has $\alpha = 5.0$ and $\alpha = -0.71$ for the pion
and gamma decay respectively.}
\vspace{0.2cm}
\label{fig:ang_momentum}
\end{center}
\end{figure}

\begin{figure}[!htbp]
\begin{center}
\hspace{2.5pt}
\includegraphics[width=14.5cm]{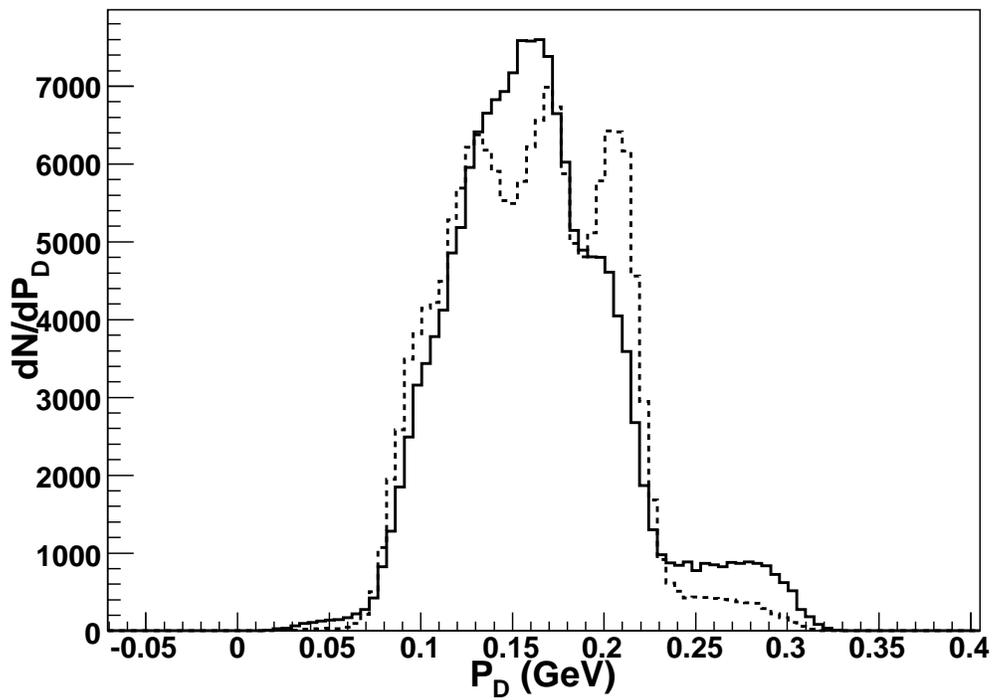}
\caption{This has the same parameters and values as \FIG
\ref{fig:ang_momentum} but has been convoluted with a gaussian of
10~MeV to account for detector resolution. This distribution should be compared
to \FIG \ref{fig:Mom_D0_4030}.}
\vspace{0.2cm}
\label{fig:ang_momentum_CON}
\end{center}
\end{figure}

%%%%%%%%%%%%%%%%%%%%%%%%%%%%%%%%%%%%%%%%%%%%%%%%%%%%%%%%%%%%%%%%%%%%%%%%%%%%%%%%
\section{Complete Derivation of $e^+e^-\rightarrow\gamma^*\rightarrow{D^{*}\bar{D}}$,
$D^{*}\rightarrow{D}{\gamma}$}
\label{app:full}
%%%%%%%%%%%%%%%%%%%%%%%%%%%%%%%%%%%%%%%%%%%%%%%%%%%%%%%%%%%%%%%%%%%%%%%%%
To help curb confusion let us define $e^+e^-\rightarrow\gamma^*\rightarrow{D^{*}\bar{D}}$,
$D^{*}\rightarrow{D}{\gamma}$ as $\gamma^{*}\rightarrow12$ and
$1\rightarrow34$ and define $A_{\lambda_{1}\lambda_{2}}$ to be the helicity
amplitudes for $\gamma^{*}\rightarrow12$ and
$B_{\lambda_{3}\lambda_{4}}$ to be the helicity
amplitudes for $1\rightarrow34$. Therefore
\begin{equation}
	A(M_{\gamma}, \lambda_{1}, \lambda_{2}, \lambda_{3},
	\lambda_{4}) = \sum_{\lambda_{1}} D^{1}_{\lambda_{1}\lambda'} D^{1}_{M_{\gamma}\lambda}B_{\lambda_3\lambda_4}A_{\lambda_1\lambda_2}
\end{equation}
\CONT where $\lambda'=\lambda_{3}-\lambda_{4}$ and
$\lambda=\lambda_{1}-\lambda_{2}$ and since $\lambda_{2}=\lambda_{4}=0$, $\lambda'=\lambda_{3}$ and
$\lambda=\lambda_{1}$ therefore:
\begin{eqnarray}
	A(M_{\gamma}, \lambda_{1},0, \lambda_{3},0) =
	D^{1}_{0\lambda_{4}}D^{1}_{M_{\gamma}0}B_{0\lambda_4}A_{00} +
	\nonumber\\
	D^{1}_{1\lambda_{4}}D^{1}_{M_{\gamma}1}B_{0\lambda_4}A_{10} +
	\nonumber\\
	D^{1}_{-1\lambda_{4}}D^{1}_{M_{\gamma}-1}B_{0\lambda_4}A_{-10}
\end{eqnarray}

\CONT By \EQ \ref{eq:Parity} and the $J^{P}$ quantum numbers of the
reaction, it is seen that party conservation requires $A_{00}=-A_{00}$
and $A_{10}=-A_{-10}$.  Therefore, $A_{00}$ must be set to zero and
so the $D^{*}$ will always be produced with helicity of
$\lambda=\pm1$. So:
\begin{equation}
	A(M_{\gamma}, \lambda_{1},0, \lambda_{3},0) =
	B_{0\lambda_4}A_{10}(D^{1}_{1\lambda_{4}}D^{1}_{M_{\gamma}1} -
	D^{1}_{-1\lambda_{4}}D^{1}_{M_{\gamma}-1})
\end{equation}
\CONT The angular distribution is therefore:
\begin{eqnarray}
	\frac{dN}{d\Omega}=\sum_{M_{\gamma}}\sum_{\lambda_4}|A(M_{\gamma}, \lambda_{1},0, \lambda_{3},0)|^{2} \nonumber \\
	= \sum_{M_{\gamma}}\sum_{\lambda_4}|A_{10}|^{2}|B_{0\lambda_{4}}|^{2}|D^{1}_{1\lambda_{4}}D^{1}_{M_{\gamma}1} -
	D^{1}_{-1\lambda_{4}}D^{1}_{M_{\gamma}-1}|^{2}
\end{eqnarray}
\CONT
where $D^{1}_{M_{\gamma}1} =
e^{-iM_{\gamma}\phi}d^{1}_{M_{\gamma}1}(\theta)e^{i\phi}$,
$D^{1}_{M_{\gamma}-1} =
e^{-iM_{\gamma}\phi}d^{1}_{M_{\gamma}-1}(\theta)e^{-i\phi}$, 

\CONT $D^{1}_{1\lambda_{4}} = e^{-i\phi_{4}}d^{1}_{1\lambda_{4}}(\theta_{4})e^{i\lambda_{4}\phi_{4}}$ and $D^{1}_{-1\lambda_{4}} = e^{i\phi_{4}}d^{1}_{-1\lambda_{4}}(\theta_{4})e^{i\lambda_{4}\phi_{4}}$
where $\theta$ is defined as the angle between the \(D^{*}\) and
the beam axis and $\theta_{4}$ is defined as the angle between the $D$ in the
rest frame of the $D^{*}$ to that of the momentum vector of the
$D^{*}$ in the lab. By substitution of the $D$ functions one arrives
at:
\begin{eqnarray}
	\frac{dN}{d\Omega}
	=
	\sum_{M_{\gamma}}\sum_{\lambda_4}|A_{10}|^{2}|B_{0\lambda_{4}}|^{2}(|d^{1}_{1\lambda_{4}}(\theta_{4})d^{1}_{M_{\gamma}1}(\theta)|^{2} + \nonumber \\
	|d^{1}_{-1\lambda_{4}}(\theta_{4})d^{1}_{M_{\gamma}-1}(\theta)|^{2} \nonumber \\
 - (d^{1}_{1\lambda_{4}}(\theta_{4})d^{1}_{M_{\gamma}1}(\theta)d^{1}_{-1\lambda_{4}}(\theta_{4})d^{1}_{M_{\gamma}-1}(\theta))2\cos(2\delta\phi))
\end{eqnarray}
\CONT where $\delta\phi = \phi_{4}-\phi$. Next, we sum over
$\lambda_{4}$ where 
\(\lambda_{4}=\pm1\) since it represents a photon for the present
case. Therefore $|B_{01}|^{2}=|B_{0-1}|^{2}$ meaning:
\begin{eqnarray}
	\frac{dN}{d\Omega}
	= |A_{10}|^{2}|B_{01}|^{2}(|d^{1}_{11}(\theta_{4})d^{1}_{M_{\gamma}1}(\theta)|^{2} + \nonumber \\
	|d^{1}_{-11}(\theta_{4})d^{1}_{M_{\gamma}-1}(\theta)|^{2} \nonumber \\
 -
   (d^{1}_{11}(\theta_{4})d^{1}_{M_{\gamma}1}(\theta)d^{1}_{-11}(\theta_{4})d^{1}_{M_{\gamma}-1}(\theta))2\cos(2\delta\phi) + \nonumber \\
	|d^{1}_{1-1}(\theta_{4})d^{1}_{M_{\gamma}1}(\theta)|^{2} + \nonumber \\
	|d^{1}_{-1-1}(\theta_{4})d^{1}_{M_{\gamma}-1}(\theta)|^{2} \nonumber \\
 -
   (d^{1}_{1-1}(\theta_{4})d^{1}_{M_{\gamma}1}(\theta)d^{1}_{-1-1}(\theta_{4})d^{1}_{M_{\gamma}-1}(\theta))2\cos(2\delta\phi))
\end{eqnarray}
\CONT again since we are not concerned by overall constants we set
$M_{\gamma}=1$ and using the fact that
$d^{1}_{11}=d^{1}_{-1-1}=\frac{1+\cos\theta}{2}$ and
$d^{1}_{1-1}=d^{1}_{-11}=\frac{1-\cos\theta}{2}$ one arrives at the
following:
\begin{equation}
\frac{dN}{d\Omega} \propto
|A_{10}|^{2}|B_{01}|^{2}((1+\cos^{2}\theta_{4})(1+\cos^{2}\theta) +
\sin^{2}\theta\sin^{2}\theta_{4}\cos(2\delta\phi))
\end{equation}
\CONT therefore by integrating over the azimuthal angle, the $D^{*}$
has a $(1+\cos^{2}\theta)$ distribution with respect to the beam
axis. In addition, one sees that the $D$, in
$D^{*}\rightarrow{D}{\gamma}, $will have a
$(1+\cos^{2}\theta_{4})$ distribution in the rest frame of the $D^{*}$.

In repeating the above for $D^{*}\rightarrow{D}{\pi}$, that is setting
$\lambda_{4}=0$, one arrives at:
\begin{equation}
\frac{dN}{d\Omega} \propto
|A_{10}|^{2}|B_{00}|^{2}((1+\cos^{2}\theta) +
\sin^{2}\theta\cos(2\delta\phi))\sin^{2}\theta_{4}
\end{equation}
\CONT therefore one sees that this time the $D$, in
$D^{*}\rightarrow{D}{\pi}$,  will have a
$\sin^{2}\theta_{4}$ distribution in the rest frame of the $D^{*}$.
Also, it is clear that there is also azimuthal angular dependence present.
\begin{figure}[!htbp]
\begin{center}
\hspace{2.5pt}
\includegraphics[width=14.5cm]{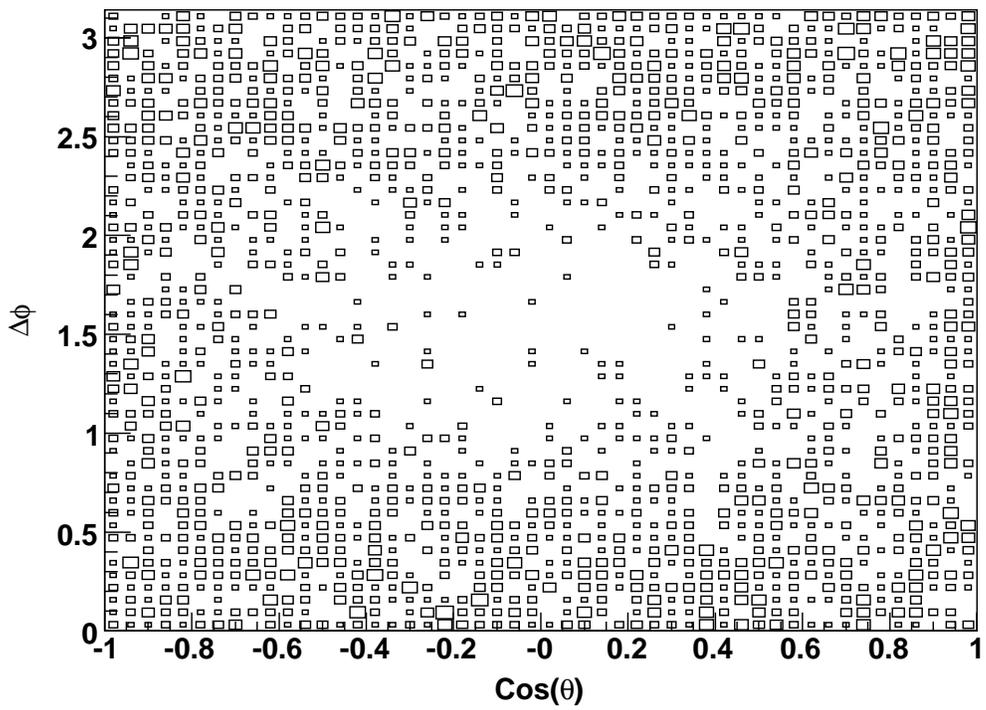}
\caption{Generator level scatter plot showing the expected $\phi$
dependence for  $e^{+}e^{-}\rightarrow\gamma^{*}\rightarrow{D^{*}\bar{D}}$
where $D^{*}\rightarrow{D}{\pi}$.}
\vspace{0.2cm}
%\label{fig:DStarDStargamma}
\end{center}
\end{figure}
%%%%%%%%%%%%%%%%%%%%%%%%%%%%%%%%%%%%%%%%%%%%%%%%%%%%%%%%%%%%%%%%%%%%%%%%%%
%
%	APPENDIX b
%
%%%%%%%%%%%%%%%%%%%%%%%%%%%%%%%%%%%%%%%%%%%%%%%%%%%%%%%%%%%%%%%%%%%%%%%%%%
\chapter{Results of Angular Distributions \\ of $D^*$ in $D^*\bar{D}^*$ 
Events}
\label{sec:ang}
\section{$e^+e^-\rightarrow\gamma^*\rightarrow{D^{*}\bar{D}^{*}}$}
%%%%%%%%%%%%%%%%%%%%%%%%%%%%%%%%%%%%%%%%%%%%%%%%%%%%%%%%%%%%%%%%%%%%%%%%%%
The method used to reconstruct $D^*$ in $D^*\bar{D}^*$ events is to first
find $D$ mesons that
populate the appropriate region in $M_{\rm{bc}}$, see \TAB
\ref{D0Dp_cut_table}.  After doing so, a
pion, either charged or neutral, is added to the $D$ and a cut applied
to $\Delta{\rm{M}}=|{\rm{M}}_{D^*}-{\rm{M}}_{D}|$ to purify the sample. For this study only
$D^{*0}\rightarrow{D^{0}\pi^0}$ with $D^0\rightarrow{K^-}{\pi^+}$ and
$D^{*+}\rightarrow{D^{+}\pi^0}$ with $D^+\rightarrow{K^-}{\pi^+}\pi^+$ are
used.  These $D$ modes represent the cleanest
modes available and the $\pi^0$ decay modes for $D^*$ have an
efficiency that is flat as a function of $\cos\theta$, which is
not the case of the charged pion because of the opening angle in the
$D^{*+}$ decays.

This method was first applied to Monte-Carlo and in addition to
looking at $D^*$ in $D^*\bar{D}^*$ we also looked at $D$ in
$D\bar{D}$ since its angular distribution is well-known and therefore used
as a control.  The MC results for $\alpha$, where $\alpha$ is defined
in the general formula $1+\alpha\cos^2\theta$, as a function of center-of-mass
energy is shown in \FIG \ref{Alpha_MC}, where the solid black lines
correspond to the MC input value.  The data result of $\alpha$ as a function of center-of-mass
energy is shown in \FIG \ref{Alpha_Data} where the red and blue
lines correspond to the average value for $D^0\rightarrow{K^-}{\pi^+}$ and
$D^+\rightarrow{K^-}{\pi^+}\pi^+$ respectively.  In both cases the top plot
presents the results of $D$ in $D\bar{D}$ and the bottom plot of $D^*$ in $D^*\bar{D}^*$.
Lastly, $\theta$ is defined as the angle between the $D$ or $D^*$
momentum vector and the beam axis, its polar angle.

\begin{figure}[!htbp]
\begin{center}
\hspace{2.5pt}
\includegraphics[width=14.5cm]{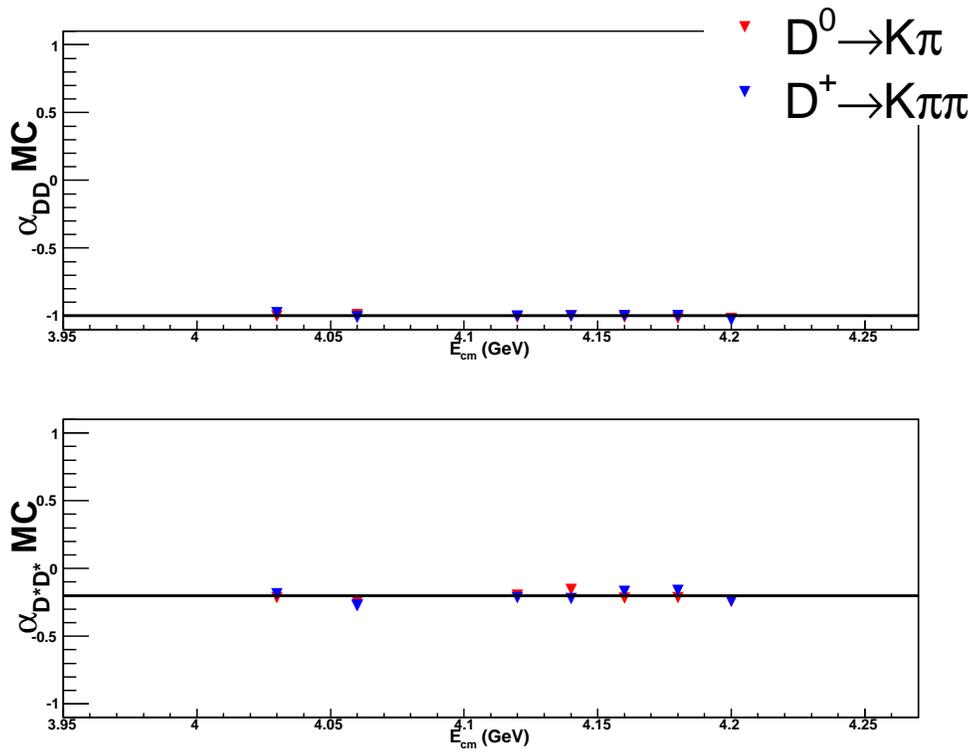}
\caption{$\alpha$, defined in $1+\alpha\cos^2\theta$, as a function of
center-of-mass energy from MC. The solid black lines
correspond to the MC input value. Top plot presents the results of $D$
in $D\bar{D}$ and the bottom plot of $D^*$ in $D^*\bar{D}^*$.}
\vspace{0.2cm}
\label{Alpha_MC}
\end{center}
\end{figure}

\begin{figure}[!htbp]
\begin{center}
\hspace{2.5pt}
\includegraphics[width=14.5cm]{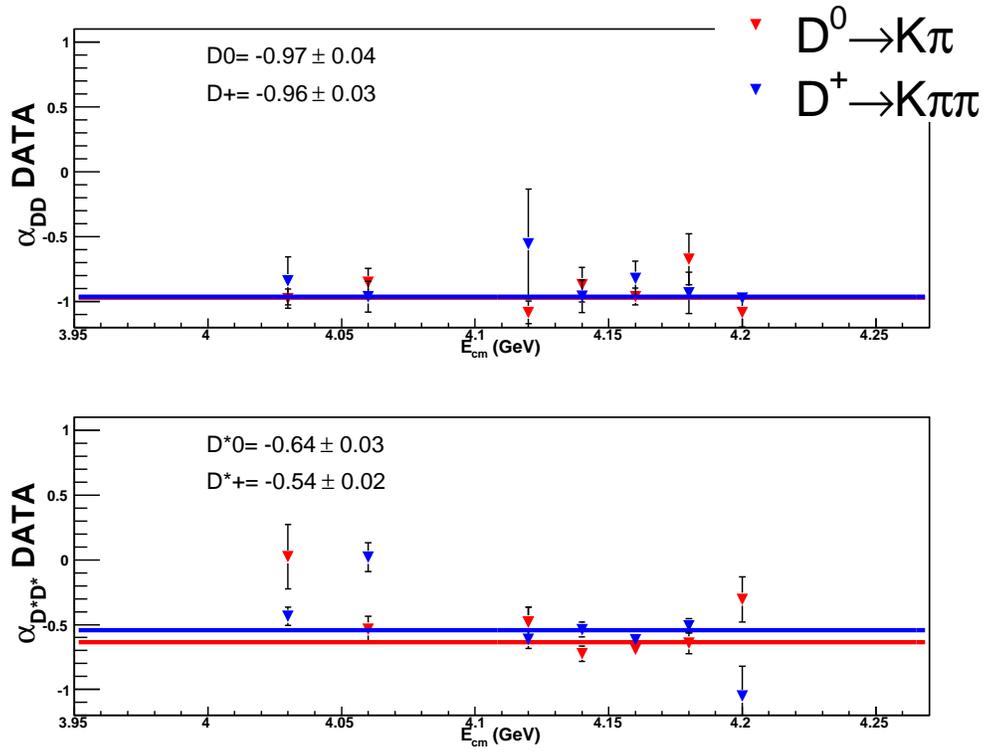}
\caption{$\alpha$, defined in $1+\alpha\cos^2\theta$, as a function of
center-of-mass energy from Data. The red and blue
lines correspond to the average value for $D^0\rightarrow{K^-}{\pi^+}$ and
$D^+\rightarrow{K^-}{\pi^+}\pi^+$ respectively. Top plot
presents the results of $D$ in $D\bar{D}$ and the bottom plot of $D^*$ in $D^*\bar{D}^*$.}
\vspace{0.2cm}
\label{Alpha_Data}
\end{center}
\end{figure}

To help in our understanding it is beneficial to translate the helicity amplitudes into
partial wave amplitudes using the Jacob-Wick Transformation:
\begin{eqnarray}
\label{eq:JacobWick}
	A_{\lambda_{D_{1}^*}\lambda_{D_{2}^*}} =
	\sum_{{\rm{LS}}}\sqrt{\frac{2{\rm{L}}+1}{2{\rm{J}}+1}}({\rm{L}}0,S(\lambda_{D_{1}^*}-\lambda_{D_{2}^*})|J(\lambda_{D_{1}^*}-\lambda_{D_{2}^*}))\nonumber \\(s_{D_1^*}\lambda_{D_1^*}, s_{D_2^*}-\lambda_{D_2^*}|S(\lambda_{D_{1}^*}-\lambda_{D_{2}^*})){\rm{M(L,S)}}
\end{eqnarray}

\CONT where $(S m_s, T m_t|J m_j)$ correspond to Clebsh-Gordon coefficients.
Either representation is valid, however, the Partial Wave representation
is somewhat more intuitive since it is written in terms of orbital
angular momentum and spin.  For the case at hand, $e^+e^-\rightarrow{D^*D^*}$, there are three
independent partial wave amplitudes, $M(1,0)$, $M(1,2)$,
and $M(3,2)$, analogous to the three helicity amplitudes. By using \EQ
\ref{eq:JacobWick} the following relations are obtained:

\begin{equation}
\label{A00}
	A_{00} = -\sqrt{\frac{1}{3}} M(1,0) - \sqrt{\frac{4}{15}}
	M(1,2) + \sqrt{\frac{2}{5}}M(3,2)
\end{equation}
\begin{equation}
	A_{10} = -\sqrt{\frac{3}{20}} M(1,2) - \sqrt{\frac{1}{10}}M(3,2)
\end{equation}
\begin{equation}
\label{A11}
	A_{11} = \sqrt{\frac{1}{3}} M(1,0) - \sqrt{\frac{1}{15}}
	M(1,2) + \sqrt{\frac{1}{10}}M(3,2)
\end{equation}

Since in \EQ \ref{eq:DSDS} the amplitudes are always in terms of the
modulus square \EQS \ref{A00}-\ref{A11} are more useful if written as
follows:
\begin{eqnarray}
\label{A002}
	|A_{00}|^2 = \frac{1}{3} |M(1,0)|^2 + \frac{4}{15}
	|M(1,2)|^2 + \frac{2}{5}|M(3,2)|^2 + \nonumber \\
	4\sqrt{\frac{1}{45}}|M(1,2)||M(1,0)|\cos\phi_{M_{12}M_{10}} -\nonumber \\
	2\sqrt{\frac{2}{15}}|M(1,0)||M(3,2)|\cos\phi_{M_{10}M_{32}} -\nonumber \\
	\frac{4}{5}\sqrt{\frac{2}{3}}|M(1,2)||M(3,2)|\cos\phi_{M_{12}M_{32}}
\end{eqnarray}
\begin{eqnarray}
	|A_{10}|^2 = \frac{3}{20} |M(1,2)|^2 + \frac{1}{10} |M(3,2)|^2
	+ \nonumber \\
	\frac{1}{5}\sqrt{\frac{3}{2}}|M(1,2)||M(3,2)|\cos\phi_{M_{12}M_{32}}
\end{eqnarray}
\begin{eqnarray}
\label{A112}
	|A_{11}|^2 = \frac{1}{3} |M(1,0)|^2 + \frac{1}{15} |M(1,2)|^2 +\frac{1}{10}|M(3,2)|^2\nonumber \\
	- 2\sqrt{\frac{1}{45}}|M(1,2)||M(1,0)|\cos\phi_{M_{12}M_{10}}
	  \nonumber \\
        + 2\sqrt{\frac{1}{30}}|M(1,0)||M(3,2)|\cos\phi_{M_{10}M_{32}}\nonumber \\
	- \frac{2}{5}\sqrt{\frac{1}{6}}|M(1,2)||M(3,2)|\cos\phi_{M_{12}M_{32}}	
\end{eqnarray}

\CONT where $\phi_{M_{12}M_{10}}$ is the $\arg(M(1,2)M^*(1,0))$,
$\phi_{M_{10}M_{32}}$ is the $\arg(M(1,0)M^*(3,2))$, and $\phi_{M_{12}M_{32}}$ is the $\arg(M(1,2)M^*(3,2))$.  By applying
the above equations to \EQ \ref{eq:DSDS} one arrives at:
\begin{eqnarray}
\label{eq:DSDSPW}
	\frac{dN}{d\cos{\theta}} =
	\frac{3}{2}|M(1,0)|^2(1-z) +
	\frac{21}{20}|M(1,2)|^2(1-\frac{1}{7}z) +\nonumber \\
	\frac{6}{5}|M(2,3)|^2(1-\frac{1}{2}z) -
        \frac{3}{5}\sqrt{\frac{3}{2}}|M(1,2)||M(3,2)|(3z-1)\cos\phi_{M_{12}M_{32}}
\end{eqnarray}
\CONT where $z=\cos(\theta)^2$. Now the $L=3$ F-wave amplitude
contains two extra powers of the center-of-mass momentum, relative to
a P-wave, and therefore should be small near the threshold \cite{voloshin}.  By
setting $|M(3,2)|=0$ in \EQ \ref{eq:DSDSPW} one arrives at the
following:
\begin{equation}
\label{eq:DSDSPW2}
	\frac{dN}{d\cos{\theta}} = \frac{3}{2}|M(1,0)|^2(1-z) +
	\frac{21}{20}|M(1,2)|^2(1-\frac{1}{7}z).
\end{equation}
\CONT
Notice that this distribution is independent of any phase. By setting
$|M(1,2)|^2=|M(1,0)|^2$ we get:
\begin{equation}
\label{eq:DSDSPW2}
	\frac{dN}{d\cos{\theta}} \propto
	1-\frac{11}{17}\cos^2\theta = 1- 0.65\cos^2\theta.
\end{equation}

It is seen that the translation to the partial wave basis is
only beneficial for the case near threshold and adds another unknown to
our equation at all other energies. Therefore, we will stick to the
helicity basis and will rewrite \EQ \ref{eq:DSDS} as follows:
\begin{equation}
	\frac{dN}{d\cos{\theta}} \propto 1+\alpha\cos^{2}\theta
\end{equation}
\CONT where $\alpha$ is defined as:
\begin{equation}
	\alpha =
	\frac{2|A_{01}|^2-2|A_{11}|^2-|A_{00}|^2}{2|A_{01}|^2+2|A_{11}|^2+|A_{00}|^2} = \frac{1-2b}{1+2b}
\end{equation}
\CONT with $b=\frac{|A_{00}|^2+2|A_{11}|^2}{4|A_{01}|^2}$.  The best
fit to the $D^*$ production distribution is shown in \FIG \ref{fig:data_fit_angular}.

\begin{figure}[!htbp]
\begin{center}
\hspace{2.5pt}
\includegraphics[width=14.5cm]{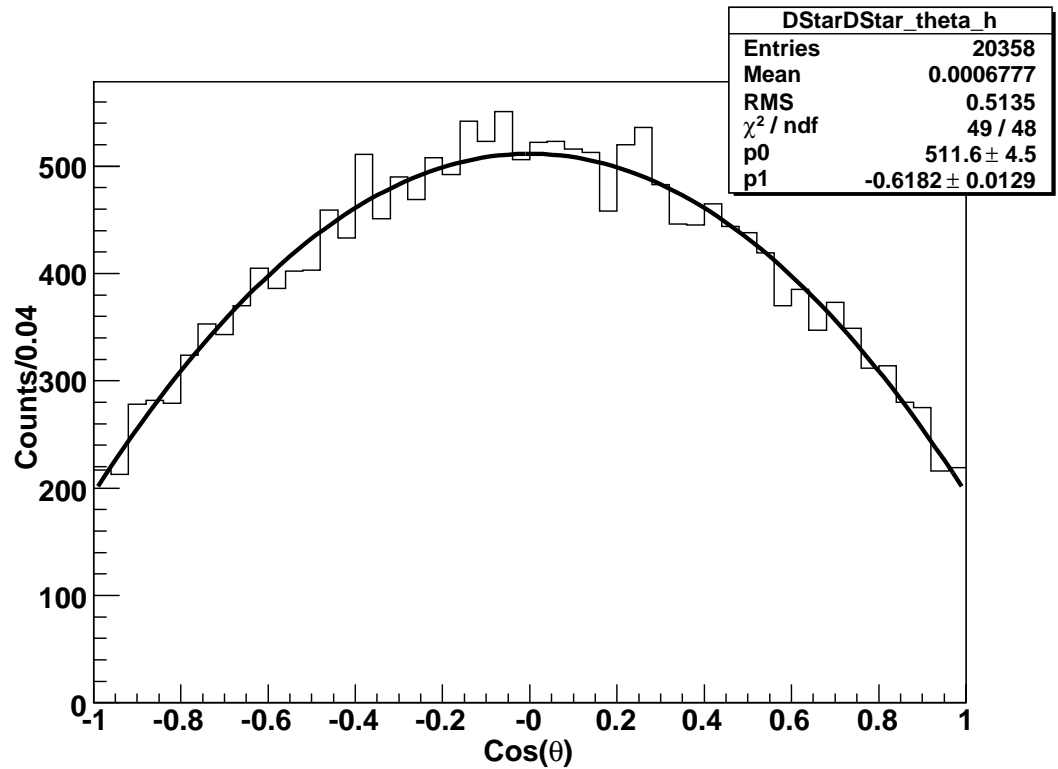}
\caption{Plot of the production angle for $D^{*}$ in $D^*D^*$ events for
data at 4170 MeV.  The solid line shows the best fit to the data using
the fit function $1+\alpha\cos^2(\theta)$, where $\theta$ is the
production angle. Based on the fit $\alpha = -0.62 \pm 0.01$.}
\vspace{0.2cm}
\label{fig:data_fit_angular}
\end{center}
\end{figure}

%%%%%%%%%%%%%%%%%%%%%%%%%%%%%%%%%%%%%%%%%%%%%%%%%%%%%%%%%%%%%%%%%%%%%%%%%%
\section{$e^+e^-\rightarrow\gamma^*\rightarrow{D^{*}\bar{D}^{*}}$,
$D^{*}\rightarrow{D}{\pi}$}
%%%%%%%%%%%%%%%%%%%%%%%%%%%%%%%%%%%%%%%%%%%%%%%%%%%%%%%%%%%%%%%%%%%%%%%%%%
It has been shown earlier that the angular distribution for $D$-mesons
in $D^*\rightarrow{D}{\pi}$ decays is as follows:
\begin{equation}
\label{eq:DSDSpi}
		\frac{dN}{d\cos\theta^{'}} =
		|B_{00}|^2\{(|A_{11}|^2+|A_{01}|^2) + (|A_{00}|^2 +
		|A_{01}|^2 - |A_{11}|^2)\cos^2\theta^{'}\}.
\end{equation}
where $\theta^{'}$ is defined as the angle between the $D$ in the
rest frame of the $D^{*}$ and the momentum vector of the
$D^{*}$ in the lab and $B_{00}$ is the helicity amplitude for
$D^{*}\rightarrow{D}{\pi}$. By substituting \EQS
\ref{A002}-\ref{A112} one arrives at:
\begin{eqnarray}
\label{eq:DSDSpiPW}
		\frac{dN}{d\cos\theta^{'}} =
		\frac{1}{3}|M(1,0)|^2 +
		\frac{21}{60}|M(1,2)|^2(z+\frac{13}{21}) +
		\frac{1}{5}|M(3,2)|^2(1+2z) \nonumber \\
		+ 2\sqrt{\frac{1}{5}}|M(1,2)||M(1,0)|\cos\phi_{M_{12}M_{10}}(z-\frac{1}{3})
		\nonumber \\ 
		+ 2\sqrt{\frac{1}{30}}|M(1,0)||M(3,2)|\cos\phi_{M_{10}M_{32}}(1-3z) \nonumber \\
		+ \frac{1}{5}\sqrt{\frac{1}{6}}|M(1,2)||M(3,2)|\cos\phi_{M_{12}M_{32}}(1-3z)
\end{eqnarray}
\CONT Now assuming that the $D$-mesons are produced close to threshold,
that is $L=3$ F-wave amplitude is small, we can set $|M(3,2)|=0$ in
\EQ \ref{eq:DSDSpiPW} and arrive at the following:
\begin{eqnarray}
		\frac{dN}{d\cos\theta^{'}} =
		\frac{13}{60}|M(1,2)|^2 + \frac{1}{3}|M(1,0)|^2 -
		2\sqrt{\frac{1}{45}}|M(1,2)||M(1,0)|\cos\phi_{02}
		\nonumber \\ +
		(\frac{21}{60}|M(1,2)|^2 + 6\sqrt{\frac{1}{45}}|M(1,2)||M(1,0)|\cos\phi_{02})\cos^2\theta^{'}
\end{eqnarray}
\CONT Notice that this distribution is NOT independent of
$\phi_{M_{10}M_{12}}$. By setting $|M(1,2)|^2=|M(1,0)|^2$ and
$\phi_{M_{10}M_{12}}=\frac{\pi}{2}$ we get:
	\begin{equation}
		\frac{dN}{d\cos\theta^{'}} \propto
	1+\frac{21}{33}\cos^2\theta^{'} = 1 + 0.64\cos^2\theta^{'}
\end{equation}
\CONT and with $\phi_{M_{10}M_{12}}=0$ we get:
	\begin{equation}
		\frac{dN}{d\cos\theta^{'}} \propto
	1+\frac{\frac{21}{60}+6\sqrt{\frac{1}{45}}}{\frac{33}{60}-2\sqrt{\frac{1}{45}}}\cos^2\theta^{'} = 1 + 5.0\cos^2\theta^{'}.
\end{equation}
\CONT showing that, even in the case where $|M(3,2)|=0$, a wide range
of angular distributions are allowed. 

Again, we see that transforming to the partial wave basis is only
useful near threshold, therefore we will stick to the
helicity basis and will rewrite \EQ \ref{eq:DSDSpi} as follows:
\begin{equation}
	\frac{dN}{d\cos{\theta^{'}}} \propto 1+\alpha^{'}\cos^{2}\theta^{'}
\end{equation}
\CONT where $\alpha^{'}$ is defined as:
\begin{equation}
	\alpha^{'} =
	\frac{|A_{00}|^2+|A_{01}|^2-|A_{11}|^2}{|A_{11}|^2+|A_{01}|^2} = \frac{1+\frac{1+a}{4b}-\frac{a}{2}}{\frac{a}{2}+\frac{1+a}{4b}}
\end{equation}
\CONT with $a=\frac{2|A_{11}|^2}{|A_{00}|^2}$. Using \EQ
\ref{eq:ang_momentum} convoluted with a Crystal Ball resolution function,
the width of the resolution function was chosen to be $\sim8.5$ MeV which is
based on MC $D\bar{D}$ events generated with a momentum of
500 MeV/$c$, one can obtain the
coefficient $\alpha^{'}$. A Crystal Ball function \EQ
\ref{eq:crystal_ball} is used in an attempt to account for
the ISR effects which are present. The Crystal Ball function is
defined as follows:

\begin{equation}
\label{eq:crystal_ball}
		f(x;\alpha,n,\bar{x},\sigma)=N\begin{cases}
		e^{-\frac{(x-\bar{x})^{2}}{2\sigma^{2}}}, &
		{\rm{for}}\frac{x-\bar{x}}{\sigma}>-~\alpha \\
                A\cdot(B-\frac{x-\bar{x}}{\sigma})^{-n}, &
		{\rm{for}}\frac{x-\bar{x}}{\sigma}\le{-~}\alpha \end{cases}
\end{equation}

\CONT where $A = (\frac{n}{\alpha})^{n}\cdot{e}^{-\frac{|\alpha|^2}{2}}$,
$B=\frac{n}{|\alpha|}-|\alpha|$ and N, $\alpha$, $n$, $\bar{x}$, $\sigma$ are
the fit parameters which are determined by the generated MC and then
fixed in the convolution.  The fit to the $D$ momentum spectrum is shown in \FIG
\ref{fig:data_mom_fit_angular}.

\begin{figure}[!htbp]
\begin{center}
\hspace{2.5pt}
\includegraphics[width=14.5cm]{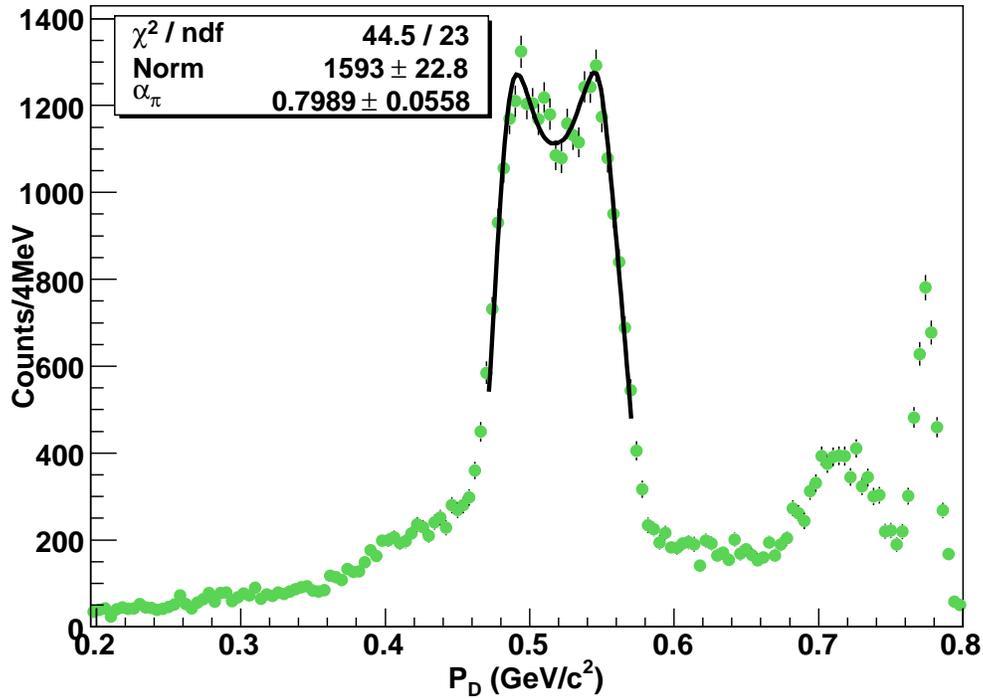}
\caption{The $D^{0}$ momentum spectrum.  The points with error bars is
the data collected at 4170 MeV and the solid line is the best fit
result.  The fit function consisted of \EQ \ref{eq:ang_momentum}
convoluted with a gaussian resolution function.  The result of the fit
yields $\alpha^{'}=0.80 \pm 0.06$.}
\vspace{0.2cm}
\label{fig:data_mom_fit_angular}
\end{center}
\end{figure}

Using the definitions of $\alpha$ and $\alpha^{'}$ along with the
normalization condition that $|A_{00}|^2+4|A_{10}|^2+2|A_{11}|=1$, one
is able to arrive at the following:
\begin{equation}
\label{A00_result}
	|A_{00}|^2 = 0.314 \pm 0.009 
\end{equation}
\begin{equation}
	|A_{10}|^2 = 0.080 \pm 0.003
\end{equation}
\begin{equation}
\label{A11_result}
	|A_{11}|^2 = 0.183 \pm 0.003
\end{equation}
\CONT Since in the equations \EQ \ref{eq:DSDS} and \ref{eq:DSDSpi} the
helicity amplitudes enter as the modulus squared, one needs not to
worry about the relative phase or the sign of the amplitudes.
Therefore, in order to generate the observed angular distributions for $D^*D^*$
events using {\tt{EVTGEN}}'s Helicity Amplitude Model ({\tt{HELAMP}}) one needs the following:
\begin{center}
\small
{\tt{D*0 anti-D*0 HELAMP 0.43 0.0 0.28 0.0 0.28 0.0 0.56 0.0 0.28 0.0 0.28 0.0 0.43 0.0;}}
\end{center}
\CONT
Comparison between the data and a generated MC sample, using the above
{\tt{HELAMP}} model, is shown in \FIGS \ref{fig:DStarDStar_data_MC}
and \ref{fig:data_anglerest}.

\begin{figure}[!htbp]
\begin{center}
\hspace{2.5pt}
\includegraphics[width=14.5cm]{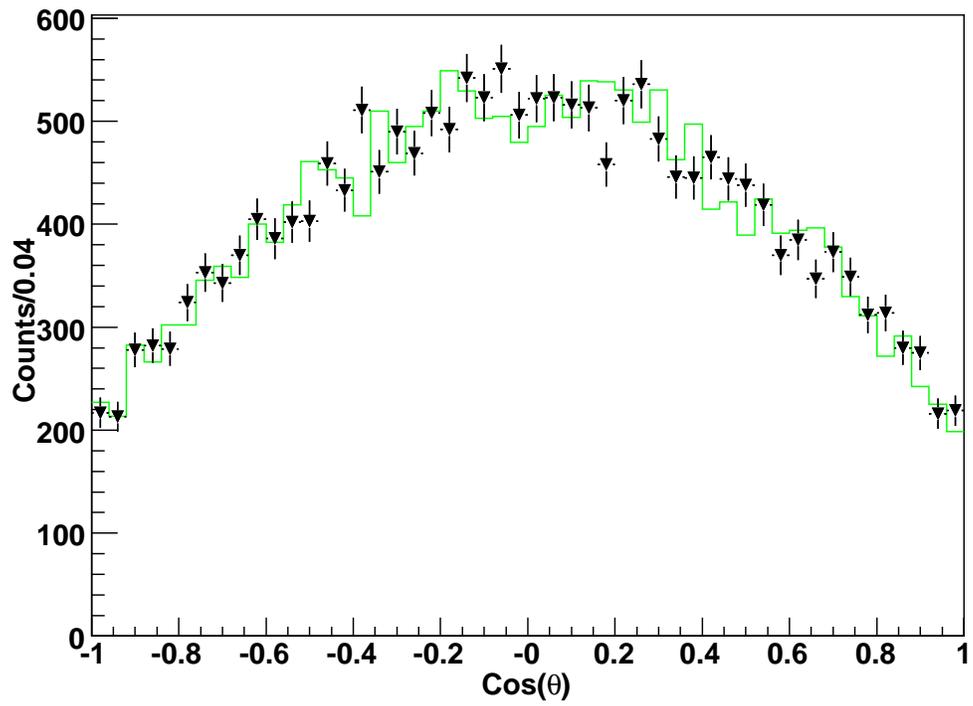}
\caption{Plot of the production angle for $D^*$ at 4170
MeV. Data is shown along with a generated MC sample using the helicity
amplitudes described in the text.}
\vspace{0.2cm}
\label{fig:DStarDStar_data_MC}
\end{center}
\end{figure}

\begin{figure}[!htbp]
\begin{center}
\hspace{2.5pt}
\includegraphics[width=14.5cm]{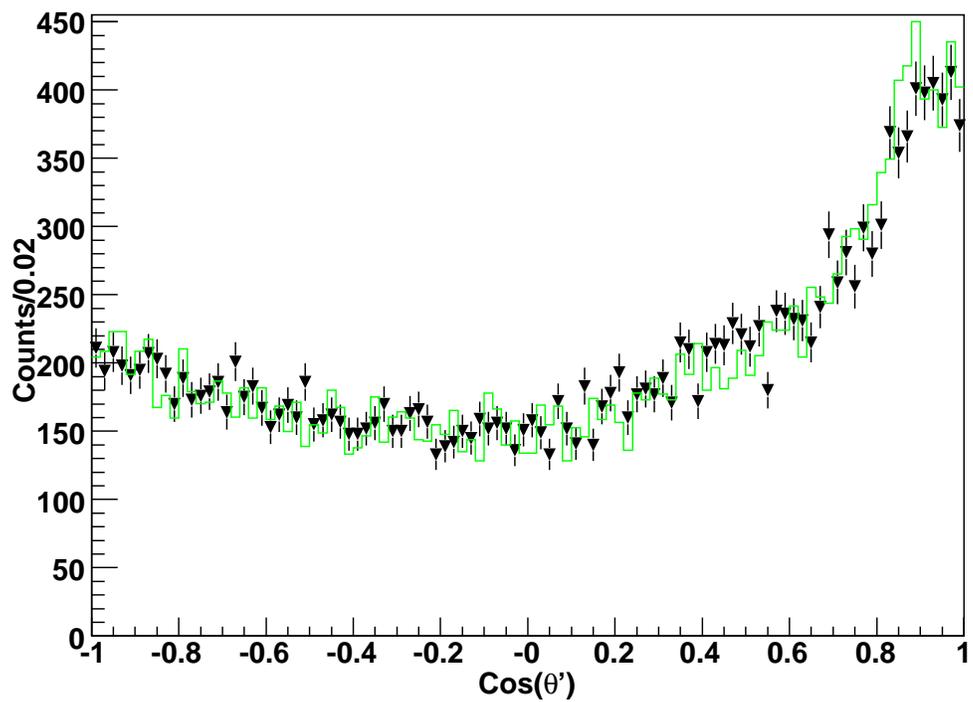}
\caption{Plot of the angle between the $D$ in the rest frame of the $D^*$ and the
momentum of the $D^*$ in the lab for $D^*D^*$ events at $4170$~MeV.
Data is shown along with a generated MC sample using the helicity
amplitudes described in the text.}
\vspace{0.2cm}
\label{fig:data_anglerest}
\end{center}
\end{figure}

	\end{document}